université
**PARIS-SACLAY**

# Magnetochemical coupling effects on thermodynamics, point-defect formation and diffusion in Fe-Ni alloys: a theoretical study

*Effets du couplage magnétochimique sur la thermodynamique, la formation de défauts ponctuels et la diffusion dans les alliages Fe-Ni : une étude théorique*

## Kangming LI

**Titre:** Effets du couplage magnétochimique sur la thermodynamique, la formation de défauts ponctuels et la diffusion dans les alliages Fe-Ni : une étude théorique

**Mots clés:** magnétisme, alliages Fe-Ni alloys, diffusion atomique, défauts ponctuels, théorie de la fonctionnelle de la densité, méthode de Monte-Carlo

**Résumé:** Cette thèse porte sur les propriétés thermo-dynamiques, de formation des défauts ponctuels et de diffusion dans les alliages Fe-Ni cfc sur toute la gamme de composition et en fonction de la température, avec un intérêt particulier pour les effets magnétochimique. Les résultats sont obtenus à partir de calculs de la théorie fonctionnelle de la densité (DFT) et de sim-ulations de Monte Carlo atomique incluant un mod-èle d'interaction effective (EIM) ajusté sur les calculs DFT. Ce modèle traite de façon explicite les interac-tions chimiques et magnétiques, et leur couplage.

La première partie de ce travail est centrée sur la thermodynamique. Nous calculons via la méthode DFT, les propriétés énergétiques, magnétiques et vi-brationnelles, ainsi que le diagramme de phase cc-cfc en-dessous de la température de Curie. Nous mettons en évidence les contributions relatives des effets magné-tiques et vibrationnels. A l'aide des simulations Monte Carlo, nous obtenons le diagramme de phase cfc en-dessous et au-dessus des températures de Curie. Nous mettons en évidence un fort couplage entre les ordres magnétiques et chimiques. Enfin, nous étudions les ef-fets d'un ajout de Mn ou de Cr sur le magnétisme et la stabilité des phases de l'alliage Fe-Ni.

La deuxième partie du travail est consacrée à l'étude des propriétés des défauts ponctuels. Nous développons des algorithmes Monte Carlo pour cal-culer l'énergie libre de formation de la lacune dans les alliages. Nous montrons que la formation de la-cune dans Fe et Ni cfc, présente des caractéristiques bien distinctes de celles dans Fe cc, soulignant ainsi l'importance des fluctuations de spin longitudinales. Nous montrons que dans les alliages, le désordre mag-nétique a tendance à augmenter l'énergie libre de for-mation de lacunes, tandis que le désordre chimique a un effet opposé. Enfin, nous étudions par la méth-ode DFT, les effets magnétiques sur les propriétés des atomes auto-interstitiels dans Fe et Ni cfc.

La dernière partie du travail est consacrée à la dif-fusion atomique. Nous calculons les propriétés de diffu-sion de traceur et des défauts ponctuels en fonction de la température, sur la totalité de l'intervalle en com-position de la phase austénitique. Nous mettons en évidence les effets des transitions magnétiques et chim-iques sur les propriétés de diffusion et nous fournissons des données inaccessibles aux expériences.

Ce travail prend pleinement en compte l'impact des fluctuations de spin transversales et longitudinales, ainsi que le couplage entre le magnétisme et l'ordre chimique. Il fournit une modélisation précise et co-hérente des propriétés thermodynamiques, de forma-tion des défauts et de diffusion du système Fe-Ni. Il contribue ainsi à une meilleure compréhension des ef-fets du magnétisme dans les aciers austénitiques.

**Title:** Magnetochemical coupling effects on thermodynamics, point-defect formation and diffusion in Fe-Ni alloys: a theoretical study

**Keywords:** magnetism, Fe-Ni alloys, atomic diffusion, point defects, density functional theory, Monte Carlo method

**Abstract:**

This thesis is a theoretical study on the temperature-dependent thermodynamic, point-defect formation and diffusion properties in fcc Fe-Ni alloys over the whole range of composition, with a focus on the magnetochemical effects. The results are derived from density functional theory (DFT) calculations and Monte Carlo simulations using a DFT-parametrized effective interaction model (EIM) with explicit atomic and spin variables.

The first part of this work is focused on thermodynamics. We compute via DFT energetic, magnetic and vibrational properties and the bcc-fcc phase diagram below the Curie points, revealing the relative importance between magnetic and vibrational entropies. Combining Monte Carlo simulations with the EIM, we obtain an fcc phase diagram across the Curie points and demonstrate a close interplay between magnetic and chemical orders. Finally, we also discuss the effects of Mn and Cr additions on phase stability.

The second part of the work is dedicated to point-defect properties. We develop Monte Carlo schemes to compute vacancy formation free energy in alloys. We show that vacancy formation in fcc Fe and Ni exhibits features that are well distinct from those in bcc Fe, pointing out the relevance of longitudinal spin fluctuations. The results in fcc Fe-Ni alloys reveal that magnetic disorder tends to increase vacancy formation free energy, while chemical disorder shows an opposite effect. Finally, we study via DFT the magnetic effects on the properties of self-interstitial atoms in fcc Fe and Ni.

The final part of the work is devoted to vacancy-mediated diffusion. We perform kinetic Monte Carlo simulations to obtain temperature-dependent diffusion properties over the whole concentration range. The results allow to probe into the effects of magnetic and chemical transitions on diffusion properties and provide details inaccessible from experiments.

This work fully takes into account the impacts of transversal and longitudinal spin fluctuations and the interplay between magnetism and chemical order. It provides an accurate and consistent prediction of thermodynamic, defect formation and diffusion properties in the Fe-Ni system, and contributes to a better understanding of effects of magnetism in austenitic steels. The applied approach is also transferable to the investigation of other magnetic alloys.

# Acknowledgements

First of all, I would like to thank the reviewers of this thesis, Fabienne Ribeiro and Ralf Drautz for spending their time in reading the manuscript and writing the reports. I would also like to thank the jury members Michel Perez and André Thiaville for participating in my thesis defense.

I would like to express my deepest gratitude to my thesis supervisor, Chu-Chun Fu. Thank you for your guidance and advice, for your patience and encouragement, for your availability for discussion, for many many things. Thank you for bringing me into this fascinating field, which I wish I could explore more within these three years. I would like to extend my sincere thanks to my thesis director Maylise for her guidance and advice. I really enjoy our discussion, during which you are always enthusiastic about helping me to discover and convincing me to work on different problems.

I would like to thank all my colleagues in the SRMP for their kindness. In particular, I thank our former group members Anton Schneider, Elric Barbé and Van-Truong Tran for their help and experience in simulations. I thank Frédéric Soisson and Yimi Wang for the discussion and collaboration on Fe-Ni alloys. I thank Océane Buggenhoudt for our daily chat and mutual support. I also would like to thank the head of SRMP, Jean-Luc Béchade, who is very kind and always full of smiles.

I desire to tender my heartfelt thanks to all my friends (once) in Paris for their accompany and help: Yang Li, Yanjun Wang, Kan Ma, Liangzhao Huang, Ziling Peng, Lu Liu, Guodong Gai, Yanshu Wang, Nan Jiang, Weiying Feng, Xian Huang, Yang Song, Jiali Liang, Chengming Shang, Hantao Lin.

I have not been home in China for more than two years due to the Covid travel restrictions. I really miss my family there and I would like to thank them for their support and understanding. Last but not least, I give my special thanks to my wife, Yiling Li. I feel truly lucky to have you by my side, loving and taking very good care of me during these three years in France.

This project has received funding from the European Union's Horizon 2020 research and innovation programme under grant agreement No 800945—NUMERICS—H2020-MSCA-COFUND-2017.

# Contents

# List of Abbreviations

| | |
|---|---|
| **AF** | AntiFerromagnetic |
| **AFD** | AntiFerromagnetic Double-layer |
| **AFS** | AntiFerromagnetic Single-layer |
| **ALRO** | Atomic Long-Range Order |
| **ASRO** | Atomic Song-Range Order |
| **BCC** | Body-Centered Cubic |
| **BCT** | Body-Centered Tetragonal |
| **CALPHAD** | CALculation of PHAse Diagrams |
| **CMC** | spin-atom Canonical Monte Carlo |
| **CVM** | Cluster Variation Method |
| **DFT** | Density Functional Theory |
| **DLM** | Disordered Local Moment |
| **DOS** | Density Of State |
| **EAM** | Embedded Atom Model |
| **EIM** | Effective Interaction Model |
| **FCC** | Face-Centered Cubic |
| **FCT** | Face-Centered Tetragonal |
| **FM** | FerroMagnetic |
| **GGA** | Generalized Gradient Approximation |
| **LDA** | Local Density Approximation |
| **LAPW** | Linearised Augmented-Plane-Wave |
| **MC** | Monte Carlo |
| **MCE** | Magnetic Cluster Expansion |
| **MLRO** | Magnetic Long-Range Order |
| **MSRO** | Magnetic Song-Range Order |
| **NEB** | Nudged Elastic Band |
| **NM** | NonMagnetic |
| **NN** | Nearest Neighbor |
| **PAW** | Projector Augmented Wave |
| **PDOS** | Projected Density Of State |
| **PM** | ParaMagnetic |
| **SGCMC** | Semi-Grand Canonical Monte Carlo |

| **SIA** | **S**elf-**I**nterstitial **A**tom |
| **SLD** | **S**pin-**L**attice **D**ynamics |
| **SMC** | **S**pin **M**onte **C**arlo |
| **SQS** | **S**pecial **Q**uasi-random **S**tructure |

# 1 Introduction

Steel plays an important role in the human history and has a broad industrial application in our modern society. It is an Fe-based alloy with typically less than 2% of carbon to improve its strength and fracture resistance. Steel as applied in the industry is also alloyed with other metal elements to introduce desired characteristics. For instance, Cr in the stainless steel can lead to the formation of a protective oxide layer on the steel surface, whereas Ni can improve the corrosion resistance in acid environments and workability [1]. High-entropy alloys, comprising multiple principal elements, have been gaining increasing interest thanks to their novel properties such as superior mechanical performance at high temperatures, exceptional ductility and fracture toughness at cryogenic temperatures, superparamagnetism, and superconductivity [2].

Magnetism is an indispensable ingredient for predicting the properties in Fe and Fe-based alloys. It plays a crucial role in the relative stability of different phases of Fe [3, 4] and in the bcc-fcc ($\alpha$-$\gamma$) phase transition in Fe [5–7]. In bcc Fe, the magnetization is shown to have significant impacts on other properties such as vacancy formation energy [8–18], self- and solute diffusion coefficients and activation energies [13–15, 19–22]. In Fe-based alloys magnetism can have a strong interplay with chemical orders, resulting in a change of the chemical order-disorder transition temperature or local clusering or unmixing tendency [23–29].

Modelling multicomponent alloy systems is a desirable but highly challenging task, since the complexity and the amount of possible configurations increase dramatically with the number of elements. Despite the ever-growing speed and capability of our computers, parameter-free first principles calculations are still limited to tens or hundreds of atoms, and also too computationally expensive to explore the configurational space. Meanwhile, accurate model Hamiltonians and empirical potentials for the upper-scale atomistic approaches such as Monte Carlo or molecular dynamics simulations are difficult to obtain for multicomponent alloys, especially when magnetism is considered explicitly. They are often built upon the descriptions of simpler binary systems, where there are still many open issues regarding some fundamental properties.

In this work, we focus on the thermodynamics, defect formation and atomic diffusion in the binary Fe-Ni alloys. The Fe-Ni system exhibits distinct magnetic and mechanical properties depending on the chemical composition. For instance, permalloy, with about 80% Ni, is used in magnetic shielding thanks to its high magnetic permeability [30]. Another well-known example is the Invar Fe-Ni alloys, which possess vanishingly low thermal expansion coefficients [31], and are widely in precision tools, laser sources and seismographic devices where high dimensional stability with temperature is required [32]. The fcc Fe-Ni alloys exhibit respectively an antiferromagnetic and ferromagnetic tendency in the Fe-rich and Ni-rich regime, with a nonlinear composition dependence of Curie temperatures. Meanwhile, the alloys with around 50% and 75% Ni are ferromagnetic and chemically ordered at low temperatures. With increasing temperature, the successive chemical and magnetic transitions occur within a small temperature window, therefore presenting a strong magneto-chemical coupling. Experimentally, most of the measurements on thermodynamic, vacancy formation and diffusion properties were performed at high temperatures, where the alloys are already paramagnetic and chemically disordered. The effects of magnetic and chemical orders on phase stability and defect properties remain largely unexplored experimentally and theoretically.

The major challenge of modelling magnetic effects at finite temperatures is to properly describe the magnetic excitations and paramagnetism, as well as their coupling with other degrees of freedom. One of the commonly used approaches is the disordered local moment (DLM) method, usually combined with coherent potential approximation calculations or a supercell approach, which consists in randomly distributing an equal portion of spin-up and spin-down atoms to mimic a fully paramagnetic state [16, 33–36]. Partially disordered local moment (PDLM) models, using different portions of randomly distributed spin-up and spin-down atoms, have also been proposed in studies simulating the partially disordered ferromagnetic state below the Curie point [23, 24, 37–39]. One of the shortcomings of the DLM/PDLM method is that the effects of magnetic short-range order are often neglected. Also, when employing the PDLM method to study a given property as a function of temperature, one has to know beforehand the magnetization value to be simulated for each temperature. In addition, the conventional DLM approach does not take into account longitudinal spin fluctuations, resulting in vanishing local moments on elements such as Ni and Cr [23, 24, 35]. Although recently the DLM method has been extended to consider longitudinal spin fluctuations, it is further complicated by the use of a sophisticated statistical treatment combined with constrained-moment calculations [40–43]. Finally, the DLM/PDLM method, along with other first principles approaches such as the spin wave method [36, 44, 45] and the dynamical mean-field theory [46–48], is too computationally demanding to be employed for a systematic investigation of magnetic effects as functions of temperature and composition and with the presence of defects. Therefore, it is highly desirable to combine accurate *ab initio* methods with upper-scale approaches to efficiently explore a broad range of spin-atom configurations of alloys.

As a generalization to molecular dynamics, spin-lattice dynamics (SLD) is a recently developed formalism to simulate atomic displacements as well as thermal spin fluctuations within a unified framework [15, 49–53]. Due to the inclusion of the spin variables, the number of degrees of freedom in SLD is twice that of molecular dynamics. Consequently, a significantly greater amount of data need to be generated and selected for the parametrization of a sufficiently accurate many-body spin-dependent potential. The development of SLD potentials for alloys with the presence of defects are not yet available and remains a grand challenge [53]. In addition, simulating vacancy-mediated diffusion, which is already computationally heavy to obtain sufficient statistics in molecular dynamics, is expected to be even more challenging in SLD simulations.

Another approach enabling large-scale modelling of magnetic systems is the use of effective interaction models, which include explicit spin and atomic variables, in combination with on-lattice Monte Carlo simulations [22, 24, 29, 54–59]. Most of these studies are focused on magnetic properties and phase stability of defect-free systems [24, 29, 54–58], whereas the magnetic effects on the vacancy formation and diffusion in bcc Fe have been investigated only very recently [22].

Several issues remain to be addressed in the modelling of the fcc Fe-Ni alloys. The first one is the absence of models that describe accurately the magnetic and chemical phase stability over the whole composition range. Indeed, this task is highly non-trivial. For instance, the existing classical interatomic potential for molecular dynamics applications overestimates the experimental chemical transition temperatures by more than 1000 K [60]. Meanwhile, there is no available SLD potential for fcc Fe-Ni alloys. Constructing such an SLD potential is expected to be even more challenging than the classical ones, considering the presence of both magnetic and chemical transitions. Regarding the effective interaction models, they were previously parametrized only for very limited composition regions (e.g. fcc Fe and Ni [54], the alloys with 70-80% Ni [24]). Recently such models have been developed for fcc Fe-Ni alloys over the whole composition [56, 57], but their phase stability predictions are found to contradict with the established experimental phase diagram. Despite the difficulties in the parametrization, accurate magnetic models are highly desirable because they enable to study, for example, the mutual influence between magnetic and chemical orders, and relative importance of vibrational and magnetic effects on the phase transition. In addition, the models of defect-free Fe-Ni alloys constitute the basis of more sophisticated ones that include defects and other elements.

The second issue is the lack of understanding for the effects of longitudinal spin fluctuations on defect properties. Most of the theoretical efforts have been devoted to elucidating the magnetic effects on defect properties in bcc Fe [13–17, 22, 61], where the dominant magnetic effects are attributed to transversal spin fluctuations. Very little is known for other more itinerant magnetic systems (e.g., bcc Cr, fcc Fe and Ni) that exhibit important longitudinal spin fluctuations at finite temperatures [54]. Such fluctuations below and above the magnetic transition temperature could have effects distinct from those of transversal spin fluctuations.

The third one is related to the fact that even in the absence of magnetism, there is no established formalism to compute the defect formation properties in alloys irrespective the underlying chemical orders. Indeed, the existing approaches are dedicated exclusively to either low-temperature ordered structures with a minimal amount of antisites [62–70], or high-temperature random alloys with vanishing chemical short-range order [71–73]. A unified framework addressing the defect formation properties as functions of temperature across the chemical and magnetic transition temperatures, is still missing in the literature.

Finally, there is a lack of physical understanding as to how the defect formation and diffusion properties are influenced by the underlying magnetic and chemical orders in alloys. Due to the lack of experimental data, it would be interesting to demonstrate the effects of chemical and magnetic transitions on defect properties, and to compare the relative importance of chemical and magnetic effects. It is also unclear, for instance, how the magnetic effects on defect properties may differ between the alloys with distinct chemical orders and compositions.

In this thesis, we investigate the magnetochemical effects on thermodynamic, point-defect formation and diffusion properties in fcc Fe-Ni alloys over the whole range of composition. We address the above-mentioned issues by combining density functional theory (DFT) calculations, effective interaction models and Monte Carlo simulations. The computational details and the model parametrization procedure are given in Chapter 2, in which we also present the formalism to compute vacancy formation free energy in magnetic alloys.

In Chapter 3, we investigate the thermodynamics of defect-free Fe-Ni alloys. We first present a systematic DFT study in bcc and fcc Fe-Ni alloys in magnetically ordered states. In particular, we discuss the vibrational effects on phase stability that are often neglected in this system. We then couple the DFT-parametrized model with spin-atom Monte Carlo simulations to reveal the magnetic effects on phase stability, the relative importance of vibrational and magnetic contributions, and the mutual influence between magnetic and chemical orders. Finally, we also discuss how the additions of Mn and Cr may impact on the phase stability of Fe-Ni based alloys, which is important for designing multicomponent steels.

In Chapter 4, we study the point-defect properties in fcc Fe-Ni alloys. In the comparative study in bcc Fe, fcc Fe and Ni, we elucidate the system-dependent magnetic effects on vacancy formation properties and the implications of longitudinal spin fluctuations. Based on our proposed formalism, we study the temperature and concentration dependences of vacancy formation properties in fcc Fe-Ni alloys, revealing the distinct effects of magnetic and chemical orders. In the end of this Chapter, we also discuss the magnetic effects on the formation and migration of self-interstitial atom (SIA) in fcc Fe and Ni. Such effects on SIA properties are far less known theoretically and experimentally compared to vacancy.

Finally, we dedicate Chapter 5 to the vacancy-mediated diffusion in fcc Fe-Ni alloys. We perform kinetic Monte Carlo simulations to calculate diffusion properties as functions of temperature. We investigate the effects of magnetic excitations and transitions on self- and solute diffusion in fcc Fe and Ni. We demonstrate how the diffusion properties evolve with changing magnetic and chemical orders in the alloys, providing details that are inaccessible from experiments.

# 2 Methods

## 2.1 Modelling strategy

The modelling in this work is based on density functional theory (DFT). As a quantum mechanical method, DFT requires no adjustable parameters or external inputs. It has been proven to be accurate and reliable for predicting various properties of materials. Therefore, DFT calculations are used to provide key properties related to phase stability and defect formation and diffusion.

Performing DFT calculations is also the first step towards the upper-scale modelling. Indeed, DFT calculations are computationally expensive, performed in systems with no more than a few hundreds of atoms. It is also difficult to extrapolate the obtained results at 0 K to intermediate and high temperatures. Therefore, we perform DFT calculations to generate input data. Based on these data, we parametrize a magnetic model Hamiltonian developed within this thesis that can reproduce the key properties found from DFT calculations.

Then, the model is implemented for Monte Carlo simulations in order to predict the thermodynamic, point-defect formation and diffusion properties as functions of temperature and alloy concentration. In Monte Carlo simulations, the temperature-dependent effects of magnetic and chemical configurations on these properties can be properly taken into account. We use the conventional algorithms for the Monte Carlo calculations of thermodynamic and point-defect diffusion properties. On the other hand, to the best of our knowledge, there is no existing algorithm in the literature to obtain temperature-dependent point-defect formation properties in magnetic alloys. In this work, we develop Monte Carlo algorithms for this purpose which are presented in Sec. 2.4.2 and 2.4.3.

## 2.2 Density functional theory

### 2.2.1 Principle

First principles calculations allow to investigate the electronic structure of of many-body systems on the basis of quantum mechanics, which mainly consist of solving the time-independent Schrödinger equation:

$$\widehat{H}\psi = E\psi \tag{2.1}$$

with

$$[\widehat{T_e} + \widehat{T_N} + \widehat{V}_{ee} + \widehat{V}_{NN} + \widehat{V}_{eN}]\psi = E\psi \tag{2.2}$$

where $\widehat{T_e}$ and $\widehat{T_N}$ respectively represent the kinetic energy of electrons and nuclei, and $\widehat{V}_{ee}$, $\widehat{V}_{NN}$ and $\widehat{V}_{eN}$ respectively denote the electron-electron, nuclei-nuclei, electron-nuclei interactions.

The Born-Oppenheimer approximation is the first simplification of the problem. It recognizes the fact that nuclei are much heavier, and move much slower than electrons. Nuclei can therefore be considered as immobile with respect to electrons. Within this approximation, the kinetic term of nuclei $\widehat{T_N}$ is zero, the nuclei-nuclei interaction term $\widehat{V}_{NN}$ becomes a constant, and the system consists of electrons moving under the influence of the external potential $\widehat{V}_{ext}$ of nuclei. Namely, the Schrödinger equation becomes

$$[\widehat{T_e} + \widehat{V}_{ee} + \widehat{V}_{ext}]\psi = E\psi \tag{2.3}$$

For a system of N electrons, Eq. 2.3 is an equation of 3N variables (neglecting spin degrees of freedom). Its solution remains extremely difficult, if not impossible. Density functional theory (DFT) is established on the idea of using electron density $n(\boldsymbol{r})$, which is a function of only three variables, as the central quantity in describing electron interactions. In the framework of the Hohenberg-Khon theorems [74], Kohn and Sham [75] reformulated the problem by replacing the interacting many-electron system under the real external potential as a system of non-interacting electrons moving within an effective potential. This leads to the Kohn-Sham equations, in the same form as the single-particle Schrödinger equation with an effective potential:

$$[-\frac{\hbar^2}{2m}\nabla^2 + V_{eff}(\boldsymbol{r})]\phi_i(\boldsymbol{r}) = \epsilon_i\phi_i(\boldsymbol{r}) \tag{2.4}$$

with

$$V_{eff}(\boldsymbol{r}) = V_{ext}(\boldsymbol{r}) + \int \frac{n(\boldsymbol{r}')}{|\boldsymbol{r}' - \boldsymbol{r}|}d\boldsymbol{r}' + V_{xc}(\boldsymbol{r}) \tag{2.5}$$

where the first term is the external potential, the second is the Hartree potential describing the electron-electron Coulomb repulsion, and the third is the non-classical exchange-correlation potential. Note that the effective potential itself contains the electron density $n(\boldsymbol{r})$, which is calculated from the single-particle wave functions $\phi_i(\boldsymbol{r})$. Therefore, the Kohn-Sham equations are solved iteratively.

The exchange-correlation potential $V_{xc}(\boldsymbol{r})$ is formally defined as the derivative (with respect to $n(\boldsymbol{r})$) of the exchange-correlation functional $E_{xc}[n(\boldsymbol{r})]$, whose general form is simply unknown. But in the case of uniform electron gas (UGS), namely $n(\boldsymbol{r})$ is constant in the space, $E_{xc}^{UGS}[n(\boldsymbol{r})]$ can be derived exactly. The simplest approximation is to assume the exchange-correlation functional at each point to be the known $E_{xc}^{UGS}$ with the local electron density observed at that position. Another commonly used class of functional after the LDA is the generalized gradient approximation (GGA), which uses not only local electronic density but also its gradient to approximate the true functional [76, 77].

It is well known that the LDA functional fails to correctly reproduce the ground state of Fe: the nonmagnetic hcp and fcc phases of Fe are predicted to be lower in energy than the ferromagnetic bcc Fe, which is the experimental ground state [3]. On the other hand, the GGA functional is considered more robust for the study of Fe based alloys, especially for magnetic properties. Furthermore, the equilibrium lattice parameters of Fe and Ni are usually underestimated in the LDA, while the GGA predictions are close to the experimental values [3, 78].

In DFT calculations using plane wave basis sets, a very large number of plane waves are needed to expand the wave functions of core electrons. These electrons generally have only minor effects on chemical bonding and thus on materials properties, compared to less tightly bound valence electrons. Several approaches have been proposed to reduce the computational burden due to core electrons. One of the commonly employed method is the use of pseudopotentials, which replace the core wave functions with a pseudo wave function able to represent the effective potential of the core electrons. Despite their formal simplicity and improved efficiency, pseudopotentials prevent the recovery of full core electron wave functions, and the information on the charge density and wave functions near the nucleus is lost [79], leading to less accurate predictions for some properties such as magnetism [80]. The projector augmented wave (PAW) method is a generalization of the pseudopotential approach and of the linearised augmented-plane-wave (LAPW) method [79], the latter being the most accurate and general method for electronic structure at the present time and commonly used as the benchmark for other methods [77]. The PAW method introduces projectors and auxiliary wave functions, and allows to recover the full core-electron wave function as in the LAPW method, but at a lower computational cost closer to that of pseudopotentials [81]. It is considered to be more robust than pseudopotentials in systems with strong magnetic moments [76]. Therefore, the PAW method is proposed as the optimal choice when studying properties of magnetic transition metals in terms of accuracy and computational cost [80].

### 2.2.2 General computational setup

The computational setup for our DFT calculations is as follows. The calculations are performed using the projector augmented wave (PAW) method [79, 81] as implemented in the Vienna Ab-initio Simulation Package (VASP) code [82–84]. $3d$ and $4s$ electrons are considered as valence electrons. The plane-wave basis cutoff is set to 400 eV. The Methfessel-Paxton broadening scheme with a smearing width of 0.1 eV was used [85]. The convergence cutoff for the electronic self-consistency loop was set to $10^{-6}$ eV. The $k$-point grids were adjusted according to the cell size, to achieve a sampling density equivalent to a cubic unit cell with a $16^3$ shifted grid following the Monkhorst-Pack scheme [86]. We mainly adopt the generalized gradient approximation (GGA) for the exchange-correlation functional in the Perdew-Burke-Ernzerhof (PBE) parametrization [87]. Some local density approximation (LDA) calculations are also performed to examine the effects of the exchange-correlation functional on the stability of Invar alloys, and the vacancy-related properties in Ni-rich alloys.

All DFT calculations are spin polarized and are performed within the collinear approximation, unless otherwise stated. In calculations without magnetic constraint, the atomic magnetic moments are fully optimized for a given magnetic ordering. In constrained local moment calculations within the collinear approximation, the orientations (spin up or spin down) and the magnitudes of some magnetic moments are constrained.

Random chemical solutions are represented by the special quasirandom structures (SQSs) [88] with minimized atomic short-range order parameters [89, 90]. Magnetic random structures are represented similarly by magnetic SQSs, by treating the spin-up and spin-down atoms as different species.

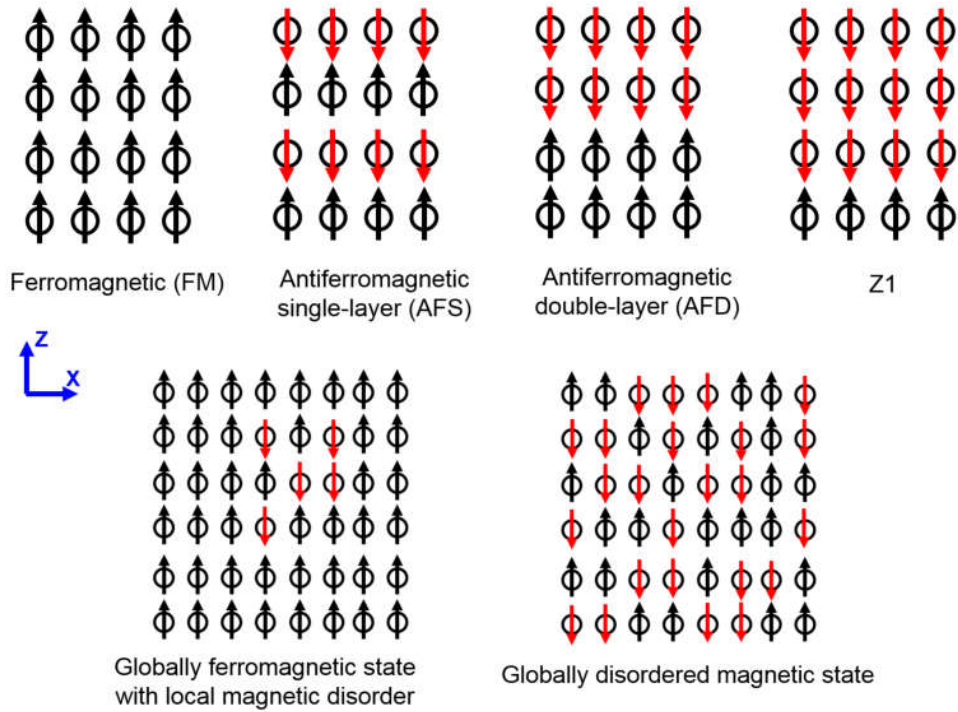

FIGURE 2.1: Illustration of the considered magnetic states.

The magnetically ordered states considered in this study are illustrated in Fig. 2.1. The following relaxation setting is adopted, unless specified otherwise. For the magnetically ordered states, the atomic positions, cell shape and volume of supercells are fully optimized in DFT calculations, to achieve a maximum residual force of 0.02 eV/Å and a maximum residual stress of 10 kbar. For the structures presenting local or global magnetic disorder as sketched in Fig. 2.1, the atomic positions are fixed to those obtained in the (collinear) magnetic ground states, while the shape and volume of supercells are optimized.

Vibrational properties of various structures are calculated based on the finite displacement method within the harmonic approximation [91]. In this method, each atom is displaced in each independent direction, and the Hessian matrix is constructed by calculating the second derivatives of the energies with respect to atomic displacements. The finite displacement method is natively implemented in VASP, but only the vibrational frequencies at the $\Gamma$ point are calculated. We use the Phonopy code [92] as a convenient tool to extract additional information and perform further analysis based on the force constants obtained from VASP. In practice, we first optimize the structure of a given system, to ensure a maximum residual force of 0.001 eV/Å and a maximum residual stress of 1 kbar. Then from the optimized structure, a number of displacements are generated by displacing one atom by 0.01 Å at a time. The number of displacements can be reduced for the structures with some degrees of symmetry.

### 2.2.3 Point-defect formation

In the pure system, the vacancy and self-interstitial atom (SIA) formation energies $E_{f,V}$ and $E_{f,SIA}$ can be calculated in a straightforward way:

$$E_{f,V} = E_{tot,V} - \frac{N-1}{N} E_{tot,0} \qquad (2.6)$$

$$E_{f,SIA} = E_{tot,SIA} - \frac{N+1}{N} E_{tot,0} \qquad (2.7)$$

where $E_{tot,0}$ is the energy of the perfect system with $N$ atoms, $E_{tot,V}$ and $E_{tot,SIA}$ are respectively the energies of the defective systems with a vacancy and a SIA. For simplicity, we will use $E_f$ to denote the point-defect formation energy when the type of the point defect being discussed is evident from the context.

In this work, the SIA formation properties are investigated only in the pure fcc Fe and Ni systems, while the vacancy formation properties are studied over the whole composition range of fcc Fe-Ni alloys. Therefore, the following discussion focuses on the vacancy formation energy in alloys.

#### Vacancy formation energies in disordered alloys

In a disordered alloy, the local vacancy formation energy is site dependent. It can be defined as [71, 93–95]:

$$E_f^i = E_{tot,V_i} - E_{tot,0} + \mu \qquad (2.8)$$

where $E_{tot,V_i}$ is the energy of the system with a vacancy at site $i$, and $\mu$ is the chemical potential of the removed atom.

At 0 K, the chemical potentials for a substitutional alloy can be solved from the following equations:

$$N(x_A \mu_A + x_B \mu_B) = E_{tot,0} \qquad (2.9)$$

$$\mu_A - \mu_B = \Delta\mu(B \rightarrow A) \qquad (2.10)$$

This requires the knowledge of the chemical potential difference $\Delta\mu$, which usually can not be obtained in one single DFT calculation.

A general method to calculate $\Delta\mu$ is the Widom-type substitution technique [96, 97]:

$$\mu_A - \mu_B = -k_B T \cdot \ln < \frac{N_B}{N_A + 1} \exp[-\frac{\Delta E(B \to A)}{k_B T}] > \qquad (2.11)$$

$$\mu_B - \mu_A = -k_B T \cdot \ln < \frac{N_A}{N_B + 1} \exp[-\frac{\Delta E(A \to B)}{k_B T}] > \qquad (2.12)$$

where $\Delta E(B \to A)$ is the energy change by substituting a $B$ atom by an $A$ atom, and the angle bracket denotes the ensemble average. These two equations are in principle the same and provide a way to check the results.

For disordered alloys, we may estimate $\Delta\mu$ in two extreme cases based on the SQS approach. The first one [93] considers the low-temperature limit, since the DFT calculations are performed at 0 K. In this case, Eq. 2.11 becomes finding the minimum energy change among all the substitutions. The second one [94] considers the high-temperature limit, since SQS is a representation of the ideal disordered alloy at high temperatures. Therefore, Eq. 2.11 is equivalent to calculate the arithmetic average of the energy changes of all the substitutions. Both require to perform DFT calculations in the configurations with atom substitutions at multiple sites [93, 94], which is computationally expensive. Furthermore, the vacancy formation energies of the disordered alloys at 0 K are not expected to be accurate, because the disordered Fe-Ni alloys are stable only above the chemical transition temperatures where magnetic disorder should be taken into account. Therefore, in this work the vacancy formation properties are not directly computed from DFT, but from Monte Carlo simulations using the procedure presented in Sec. 2.4.3.

**Vacancy formation energies in ordered alloys**

As the ordered structures L1$_0$-FeNi and L1$_2$-FeNi$_3$ are the chemical ground states of the disordered fcc Fe-Ni alloys, it is of interest to calculate the vacancy formation energies in these ordered structures. A general method to calculate the point-defect equilibrium concentrations, point-defect formation energies and chemical potentials in ordered phases has been proposed by Hagen and Finnis based on a statistical formalism [62]. This formalism has been applied to study the point defect properties in various ordered structures such as the Ni-Al, Al-Sc and Ti-Al systems [63–66]. In the following we briefly review the essential points of this formalism.

For an ordered phase $A_m B_n$, where $m$ and $n$ denote respectively the numbers of $A$ and $B$ atoms in the unit cell, we can calculate the following energies for supercells consisting of $N$ unit cells:

- $E_{tot,0}$: the energy of the perfect stoichiometric structure, consisting of $Nm$ $A$ atoms and $Nm$ $B$ atoms.

- $E_{tot,V_A}$: the energy of the structure with a vacancy in the $A$ sublattice, consisting of $(Nm-1)$ atoms $A$, $Nm$ atoms $B$ and 1 vacancy.
- $E_{tot,V_B}$: the energy of the structure with a vacancy in the $B$ sublattice, consisting of $(Nm)$ atoms $A$, $(Nm-1)$ atoms $B$ and 1 vacancy.
- $E_{tot,B_A}$: the energy of the structure with a $B$ antisite in the $A$ sublattice, consisting of $(Nm-1)$ atoms $A$ and $(Nm+1)$ atoms $B$.
- $E_{tot,A_B}$: the energy of the structure with a $A$ antisite in the $B$ sublattice, consisting of $(Nm+1)$ atoms $A$ and $(Nm-1)$ atoms $B$.

From them, we can obtain the following five quantities:

- $e_0 = \frac{E_{tot,0}}{N(m+n)}$: the energy per atom of the stoichiometric ordered structure.
- $e_{PD} = E_{tot,PD} - E_{tot,0}$: the energy difference between the systems with a point defect ($PD=V_A$, $V_B$, $B_A$ or $A_B$) and without defect.

These quantities are used to express the free energy of the ordered structure, based on two assumptions: The first one is the noninteracting-defect approximation, where the concentrations of point defects are assumed to be sufficiently small so that the interaction between them is negligible. Note that for a canonical system with a fixed number of $A$ and $B$ atoms, the numbers of lattice sites are not constant due to the creation of the vacancies. The second assumption states that the numbers of the two types of sublattice sites always follow the stoichiometric ratio (even if the nominal concentration is off-stoichiometric). Based on these assumptions, the free energy of the system can be expressed as:

$$G = N_0(m+n)e_0 + N_{V_A}e_{V_A} + N_{V_B}e_{V_B} + N_{A_B}e_{A_B} + N_{B_A}e_{B_A} - T*S_{conf} \qquad (2.13)$$

where $N_0$ is the number of unit cells (so $mN_0$ and $nN_0$ are respectively the numbers of $A$ and $B$ sublattice sites), $S_{conf}$ is the configurational entropy of the system:

$$S_{conf} = - k_B[N_0 m(c_{A_A}\ln c_{A_A} + c_{B_A}\ln c_{B_A} + c_{V_A}\ln c_{V_A})$$
$$+ N_0 n(c_{A_B}\ln c_{A_B} + c_{B_B}\ln c_{B_B} + c_{V_B}\ln c_{V_B})] \qquad (2.14)$$

Here $c_{i_A}$ and $c_{i_B}$ are the concentration of lattice sites occupied by $i$ in the $A$ and $B$ sublattices, respectively:

$$c_{i_A} = \frac{N_{i_A}}{N_0 m} \qquad (2.15)$$

$$c_{i_B} = \frac{N_{i_B}}{N_0 n} \qquad (2.16)$$

Therefore, the free energy $G$ is a function of $N_0$, and the six sublattice concentrations $c_{i_A}$ and $c_{i_B}$.

The equilibrium sublattice concentrations can be found by minimizing $G$ with respect to the seven variables. However, the sublattice concentrations are not mutually independent, and are constrained by the relations:

$$c_{A_A} + c_{B_A} + c_{V_A} = 1 \tag{2.17}$$

$$c_{A_B} + c_{B_B} + c_{V_B} = 1 \tag{2.18}$$

$$\frac{m \cdot c_{A_A} + n \cdot c_{A_B}}{m \cdot c_{B_A} + n \cdot c_{B_B}} = \frac{x}{1-x} \tag{2.19}$$

where $x$ is the nominal concentration of $A$, which is equal to $\frac{m}{m+n}$ if the system has a stoichiometric composition. These constraints are considered in the minimization of $G$ by introducing the corresponding Lagrange multipliers.

The precedure of minimizing $G$ leads to a set of nonlinear equations, from which the equilibrium sublattice concentrations can be solved numerically [62]. Mishin *et al.* [63] further showed that simple analytical expressions can be derived for the systems where the equilibrium antisite concentrations are much higher than the vacancy ones, which is the case in L1$_0$-FeNi and L1$_2$-FeNi$_3$. For the stoichiometric composition, the point-defect formation energies at $T \to 0K$ are expressed as follows [63]:

$$E_f(V_A) = e_{V_A} + e_0 + \frac{n}{2(m+n)}(e_{A_B} - e_{B_A}) \tag{2.20}$$

$$E_f(V_B) = e_{V_B} + e_0 + \frac{m}{2(m+n)}(e_{B_A} - e_{A_B}) \tag{2.21}$$

$$E_f(B_A) = E_f(A_B) = \frac{1}{2}(e_{A_B} + e_{B_A}) \tag{2.22}$$

$$\tag{2.23}$$

We note that the point-defect formation energies defined here are related to the sublattice concentrations instead of the overall concentrations of the point defects [62, 63]. For example, if neglecting entropic contribution, $\exp\left[-\frac{E_f(B_A)}{k_B T}\right]$ gives the concentration of $B$ antisites in the $A$ sublattice.

The above analytical expressions are derived based on the statistical formalism [62]. In the following, we derive these expressions following the same idea of the Widom technique as applied in the disordered alloys. This is important from the methodological point of view: the Monte Carlo algorithms we develop in Sec. 2.3.2 to calculate the vacancy formation properties are based on the Widom technique and are applied for any chemical structure over the whole composition and temperature ranges. Consequently it is important to verify that the Widom technique yields the same predictions as those from the statistical formalism [62], which are applicable only in the ordered structures with few antisites (namely valid only at low temperatures).

First, it is natural to express the point-defect formation energies as follows:

$$E_f(V_A) = e_{V_A} + \mu_A \tag{2.24}$$

$$E_f(V_B) = e_{V_B} + \mu_B \tag{2.25}$$

$$E_f(B_A) = e_{B_A} + \mu_A - \mu_B \tag{2.26}$$

$$E_f(A_B) = e_{A_B} + \mu_B - \mu_A \tag{2.27}$$

where $\mu_A$ and $\mu_B$ are the chemical potentials at the 0 K limit. As in the cases of disordered alloys, $\mu_A$ and $\mu_B$ in the ordered structure can also be solved from the following relations

$$m\mu_A + n\mu_B = (m+n)e_0 \tag{2.28}$$

$$\mu_A - \mu_B = \Delta\mu(B \to A) = -\Delta\mu(A \to B) \tag{2.29}$$

One may want to compute $\mu_A - \mu_B$ using the Widom technique by considering the two possible substitutions

$$\mu_A - \mu_B = \Delta E(B_B \to A_B) = e_{A_B} \tag{2.30}$$

$$\mu_B - \mu_A = \Delta E(A_A \to B_A) = e_{B_A} \tag{2.31}$$

However, $\mu_A - \mu_B$ calculated from these two equations are not equal, because $e_{A_B} + e_{B_A}$ is the formation energy of a pair of isolated antisites. This seeming contradiction arises from the fact that the substitutions on the antisites are not considered. Indeed, there are always antisites created at any finite temperature. Before discussing these relevant details when applying the Widom technique, we would like to first present and comment on the final conclusions of it. From the Widom technique, we can show that

$$\mu_A - \mu_B = \frac{1}{2}(e_{A_B} - e_{B_A}) \tag{2.32}$$

Combining this with Eq. 2.28, the chemical potentials can be solved as

$$\mu_A = e_0 + \frac{n}{2(m+n)}(e_{A_B} - e_{B_A}) \tag{2.33}$$

$$\mu_B = e_0 + \frac{m}{2(m+n)}(e_{B_A} - e_{A_B}) \tag{2.34}$$

Entering these expressions in Eq. 2.24-2.27, we recover the same expressions of the point-defect formation energies derived the statistical formalism.

The remaining part of this section is to show the derivation of Eq. 2.32 from the Widom technique. For the substitution type $B \to A$ at $T \to 0$, we have

$$\mu_A - \mu_B = \lim_{T \to 0} -k_B T \cdot \ln < \frac{N_A}{N_B + 1} \exp[-\frac{\Delta E(B \to A)}{k_B T}] > \tag{2.35}$$

$$= \lim_{T \to 0} -k_B T \cdot \ln < \exp[-\frac{\Delta E(B \to A)}{k_B T}] > \tag{2.36}$$

$$= \lim_{T \to 0} -k_B T \cdot \ln(c_{A_B} \exp[-\frac{E(B_A \to A)}{k_B T}] + c_{B_B} \exp[-\frac{E(B_B \to A)}{k_B T}]) \tag{2.37}$$

$$= \lim_{T \to 0} -k_B T \cdot \ln(c_{A_B} \exp[-\frac{-e_{B_A}}{k_B T}] + \exp[-\frac{e_{A_B}}{k_B T}]) \tag{2.38}$$

The first equality is the application of the Widom technique by definition. The second simply reflects the fact that $\lim_{T \to 0} T \ln \frac{N_A}{N_B + 1} = 0$. For the third equality, we consider the $B \to A$ substitutions, which can take place on the $A$ or $B$ sublattice site occupied by $B$, with a probability of $c_{A_B}$ and $c_{B_B}$ respectively. The fourth equality uses the definitions of $e_{B_A}$ and $e_{A_B}$, and the fact that $\lim_{T \to 0} c_{B_B} = 0$. We note that the fourth equality becomes Eq. 2.30, if the term $c_{A_B} \exp[-\frac{-e_{B_A}}{k_B T}]$ is neglected. However, this term cannot be neglected because $\lim_{T \to 0} \exp[-\frac{-e_{B_A}}{k_B T}] \to +\infty$ (since $e_{B_A}$ is positive).

Using the relation

$$c_{A_B} = \exp[-\frac{E_f(A_B)}{k_B T}] \tag{2.39}$$

Eq. 2.38 can be written as

$$\mu_A - \mu_B = \lim_{T \to 0} -k_B T \cdot \ln(\exp[-\frac{E_f(A_B) - e_{B_A}}{k_B T}] + \exp[-\frac{e_{A_B}}{k_B T}]) \tag{2.40}$$

We can use Eq. 2.27 to express $E_f(A_B)$, but the further derivation becomes complicated in this way. To simplify the technical details, we use the expression of $E_f(A_B)$ derived from the statistical formalism. Namely, using Eq. 2.22, we have

$$\mu_A - \mu_B = \lim_{T \to 0} -k_B T \cdot \ln(\exp[-\frac{\frac{1}{2}(e_{A_B} + e_{B_A}) - e_{B_A}}{k_B T}] + \exp[-\frac{e_{A_B}}{k_B T}]) \tag{2.41}$$

$$= \lim_{T \to 0} -k_B T \cdot \ln(\exp[-\frac{\frac{1}{2}(e_{A_B} - e_{B_A})}{k_B T}] + \exp[-\frac{e_{A_B}}{k_B T}]) \tag{2.42}$$

$$= \lim_{T \to 0} -k_B T \cdot \ln(\exp[-\frac{\frac{1}{2}(e_{A_B} - e_{B_A})}{k_B T}] \cdot (1 + \exp[-\frac{\frac{1}{2}(e_{A_B} + e_{B_A})}{k_B T}])) \tag{2.43}$$

$$\tag{2.44}$$

As $e_{A_B} + e_{B_A}$ is positive, we have

$$\lim_{T \to 0} -k_B T \cdot \ln(1 + \exp[-\frac{\frac{1}{2}(e_{A_B} + e_{B_A})}{k_B T}]) = 0 \tag{2.45}$$

Therefore, we have

$$\mu_A - \mu_B = \lim_{T \to 0} -k_B T \cdot \ln(\exp[-\frac{\frac{1}{2}(e_{A_B} - e_{B_A})}{k_B T}]) \tag{2.46}$$

$$= \frac{1}{2}(e_{A_B} - e_{B_A}) \tag{2.47}$$

Combining this equation with Eq. 2.28, the chemical potentials can be solved:

$$\mu_A = e_0 + \frac{n}{2(m+n)}(e_{A_B} - e_{B_A}) \tag{2.48}$$

$$\mu_B = e_0 + \frac{m}{2(m+n)}(e_{B_A} - e_{A_B}) \tag{2.49}$$

With the expressions of $\mu_A$ and $\mu_B$, Eq. 2.24-2.27 become Eq. 2.20-2.22 from the statistical formalism.

### 2.2.4   Point-defect migration

The nudged elastic band (NEB) method [98] as implemented in VASP is used to find the saddle point and the minimum energy path for the vacancy and SIA migrations. It consists of inserting along the reaction path several images, commonly constructed by a linear interpolation of the initial and final stable states. The initial path built from the linear interpolation is usually different from the minimum energy path. A simple relaxation scheme for these images is not appropriate as the images will slide into the closest energy minimum. In the NEB method, the images are connected with springs, which constrain each image to stay in the middle of its neighbouring images. The lowest-energy images are found by minimizing the atomic forces normal to the reaction path and the tangential components of the spring forces between the images. To ensure finding the precise saddle point, the climbing NEB method implemented in VASP is also used. It allows the highest energy image to be freed from spring forces and to maximize its energy along the band. Then, the migration energy is computed as the energy difference between the saddle point and the initial state:

$$E_m = E_{SP} - E_{ini} \tag{2.50}$$

For the vacancy migration, we perform phonon calculations in the saddle-point and the intial stable configurations to obtain the vibrational frequencies at the $\Gamma$ point. From them, the attempt frequency $\nu$ is determined according to Vineyard's harmonic transition-state theory [99]:

$$\nu = \frac{\prod_{i=1}^{3N-3} v_i^{stable}}{\prod_{i=1}^{3N-4} v_i^{saddle}} \tag{2.51}$$

where $N$ is the number of atoms, and there are $3N-3$ positive vibrational frequencies in the initial stable configuration and $3N-4$ positive vibrational frequencies in the saddle-point configuration.

For vacancy-mediated diffusion, the atoms exchange with the nearest neighbour vacancy. The self-diffusion coefficient in the fcc $A$ lattice is calculated as [100]

$$D_A^{A*} = a^2 f_0 C_V^A w_0 \tag{2.52}$$

where $a$ is the lattice parameter, $f_0 = 0.7815$ is the self-diffusion correlation factor for the fcc lattice, $C_V^A$ is the vacancy concentration in the $A$ lattice, and $w_0$ is the jump frequency of $A$ atoms calculated as

$$w_0 = \nu_0 \exp\left(-\frac{E_{m,0}}{k_B T}\right) \tag{2.53}$$

where $\nu_0$ is the attempt frequency of $A$ in the $A$ lattice, and $E_{m,0}$ is the migration barrier of an $A$ atom exchanging with the first-nearest-neighbour vacancy. In this study, $E_{m,0}$ should be replaced by the temperature-dependent magnetic free energy of migration $G_{m,0}$ with the magnetic entropic contribution, and this is also applied to the case of $w_2$ for the solute diffusion presented below.

In the presence of a solute $B$ as the first-nearest neighbour of the vacancy, the migration barriers of $A$ and $B$ are different from that of $A$ atoms away from the solute $B$. According to the analytical model proposed by LeClaire [100], five jump frequencies as illustrated in Fig. 2.2 are required to calculate the solute diffusion coefficient $D_A^{B*}$, expressed as

$$D_A^{B*} = a^2 f_2 C_V^A w_2 \exp\left(-\frac{\Delta G_b}{k_B T}\right) \tag{2.54}$$

where $f_2$ is the solute diffusion correlation factor, $w_2$ is the jump frequency of the solute, and $\Delta G_b$ is the nearest-neighbour solute-vacancy binding free energy, which may include the electronic, magnetic and vibrational entropies. The correlation factor $f_2$ is a function of $w_i$:

$$f_2 = \frac{2w_1 + 7w_3 F}{2w_1 + 2w_2 + 7w_3 F} \tag{2.55}$$

where $F$ is a function of $\alpha = \frac{w_4}{w_0}$ [101, 102]:

$$F = 1 - \frac{10\alpha^4 + 180.5\alpha^3 + 927\alpha^2 + 1341\alpha}{7(2\alpha^4 + 40.2\alpha^3 + 254\alpha^2 + 597\alpha + 436)} \tag{2.56}$$

We note that this analytical model for the solute diffusion is valid only if the solution concentration is extremely dilute and that the vacancy-solute interaction beyond the first-nearest-neighbour distance is negligible [100].

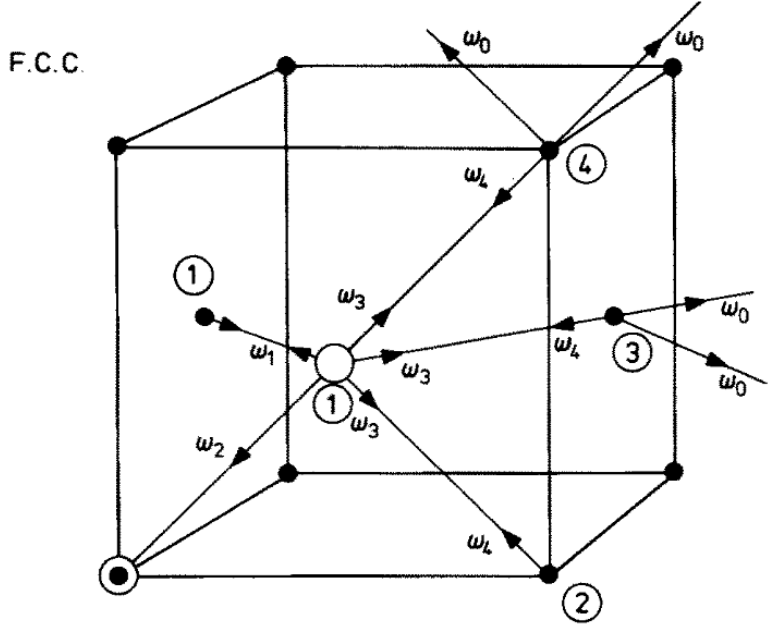

FIGURE 2.2: Vacancy jumps involved in the five-frequency model [100]. The unfilled circle is the vacancy, the partially filled circle is the solute *B* and filled circles are the *A* atoms. The figure is taken from Ref. [100].

## 2.3  Effective interaction model

In this section, we present the effective interaction models (EIMs) and the corresponding parametrization procedures. The resulting model parameters and the quality of the EIMs are detailed in Appendix A.

### 2.3.1  Parametrization for the defect-free system

We parametrize a new effective interaction model for the fcc Fe-Ni system. The EIM form is similar to the previous ones used to investigate magnetic properties, phase stability and vacancy formation and diffusion properties of the Fe alloys (e.g. Fe-Ni [56], Fe-Cr [103], Fe-Ni-Cr [57], Fe-Co [58], Fe-Mn [22, 59]). The total energy of the defect-free fcc Fe-Ni system is expressed in the present EIM as follows:

$$
\begin{aligned}
E_{tot} = \sum_i \sigma_i \cdot \underbrace{(A_i M_i^2 + B_i M_i^4 + \sum_j \sigma_j \cdot J_{ij} \boldsymbol{M}_i \boldsymbol{M}_j)}_{\text{magnetic}} + \\
\sum_i \sigma_i \cdot \underbrace{(\epsilon_i + \sum_j \sigma_j \cdot (V_{ij} + \alpha_{ij} T))}_{\text{chemical (nonmagnetic)}}
\end{aligned}
\tag{2.57}
$$

Here $i$ denotes the $i$-th lattice site. $\sigma_i$ is the occupation variable, and is equal to 1 (or 0) for an occupied (or empty) lattice site. Namely, in a defect-free system all $\sigma_i$ are equal to 1. The sum $\sum_j$ goes over all the neighbouring sites up to the fourth-neighbour shell. In the magnetic part, $M_i$ is the local magnetic moment of atom $i$, $A_i$ and $B_i$ are the on-site magnetic parameters controlling longitudinal spin variations, $J_{ij}$ are the exchange interaction parameters controlling transversal spin variations. In the nonmagnetic part, $\epsilon_i$ is the on-site nonmagnetic parameter, $V_{ij}$ and $\alpha_{ij}$ are the nonmagnetic interaction parameters, and $T$ is the absolute temperature. The parameters $B_i$, $\epsilon_i$ and $V_{ij}$ are the constants depending on the type of atom $i$ ($V_{ij}$ also depends on the type of atom $j$). The parameters $A_i$, $J_{ij}$ and $\alpha_{ij}$ have linear dependence on the local Ni concentration around the site $i$ ($J_{ij}$ also depends on the type of atom $j$). The details of these parameters are given in Appendix A.

In the following, we present briefly the procedure to fit the EIM parameters to the DFT results. The procedure is divided in several steps, by fitting first to the energy difference due to the variation in the magnetic configurations in the pure and binary systems with given compositions, and then to the energy difference due to the different chemical configurations.

First, the magnetic parameters $A_i$, $B_i$ and $J_{ij}$ for pure Fe and Ni are fitted to the energy differences between various magnetic structures. For this, we consider not only the structures with the different magnetic orderings (i.e. different alignments of spins), but also the ones having the same magnetic orderings but different magnetic-moment magnitudes. The latter is important to accurately describe the longitudinal spin fluctuations. In total, we consider more than 120 and 60 different magnetic structures for fcc Fe and Ni, respectively, and 12 parameters (6 for Fe and 6 for Ni) are fitted in this step.

In the second step, the additional magnetic parameters are fitted to the energy differences between different magnetic states having the same chemical configurations. The considered chemical configurations include the disordered structures with 25% to 90% Ni, and the ordered structures $L1_0$-FeNi, $L1_2$-FeNi$_3$ and $L1_2$-Fe$_3$Ni. In total, more than 130 DFT data are used to fit 9 parameters.

In the third step, the nonmagnetic interaction parameters $V_{ij}$ are fitted to the mixing enthalpies (with respect to fcc Fe and Ni) of various chemical configurations in the magnetic ground states. The considered chemical configurations include the disordered and ordered structures with 5% to 95% Ni. Then, the on-site nonmagnetic parameters $\epsilon_i$ are fitted to the energy per atom of Fe and Ni in the collinear magnetic ground states. In total, 70 different chemical structures are used to fit 14 parameters in this step.

In the fourth and final step, the nonmagnetic interaction parameters $\alpha_{ij}$ are fitted. Indeed, our DFT study shows that the vibrational entropy has significant effects on the chemical order-disorder transition temperatures in the Fe-Ni system [104]. In the present rigid-lattice EIM, we choose a very simple way to incorporate these effects: we introduce the nonmagnetic parameters $\alpha_{ij}$ to fit the vibrational entropies of mixing of the ferromagnetic structures. This simple treatment neglects the possible magnon-phonon coupling, and amounts to integrating the contribution from the fast vibrational degrees of freedom into the nonmagnetic pair interactions. The nonmagnetic interactions thus become the pair Gibbs free energies [105–107], instead of the simple pair energies of the usual models, due to the inclusion of the entropic contribution. We are aware that the characteristic time scales of the magnon and phonon excitations are not very different [108, 109], and a more sophisticated model treating explicitly both vibrational and magnetic degrees of freedom may be envisaged for a future study.

For the fitting of $\alpha_{ij}$, we set $\alpha_{ij}$ for the Fe-Fe and Ni-Ni pairs to zero, and fit $\alpha_{ij}$ only for the Fe-Ni pairs to the vibrational entropies of mixing of the ordered and disordered structures (with respect to pure Fe and Ni in the ground states). Consequently, the EIM does not predict the absolute vibrational entropy but only the vibrational entropy of mixing.

### 2.3.2 Parametrization for vacancy formation properties

The total energy of the system containing vacancies has the same form as that of the defect-free system:

$$
E_{tot} = \sum_i \sigma_i \cdot \underbrace{(A_{i,nV} M_i^2 + B_{i,nV} M_i^4 + \sum_j \sigma_j \cdot J_{ij,nV} \boldsymbol{M}_i \boldsymbol{M}_j)}_{\text{magnetic}} +
$$
$$
\sum_i \sigma_i \cdot \underbrace{(\epsilon_{i,nV} + \sum_j \sigma_j \cdot V_{ij})}_{\text{chemical (nonmagnetic)}}
$$

(2.58)

where we introduce the subscript $nV$ to denote the distance between the site $i$ and its nearest vacancy. For the atoms with no vacancy within the second-neighbour shell ($nV=0$), the model parameters are the same as those in the defect-free system. For the atoms with a vacancy in the first-neighbour shell ($nV = 1$) or in the second-neighbour shell ($nV = 2$), some of the model parameters are modified to incorporate the effects of vacancies.

The fitting procedure is similar to the one for the defect-free system. First, we modify the magnetic parameters to fit the variations of the total energies and of the vacancy formation energies in pure Fe and Ni, due to the change in the magnetic states. The magnetic parameters $A_{i,nV}$, $B_{i,nV}$ and $J_{ij,nV}$ ($nV = 1$ and 2) for pure Fe and Ni are fitted to the energies differences and the vacancy formation energies in various magnetic states. For the latter, we consider not only distinct magnetic orderings but also different magnetic-moment magnitudes for the same magnetic orderings, so that the EIM can well capture the dependence of vacancy formation energy on the longitudinal spin variations. In total, 16 new parameters (8 for Fe and 8 for Ni) are fitted to more than 140 and 90 DFT data in fcc Fe and Ni, respectively.

In the second and third steps, we fit the magnetic and nonmagnetic parameters to the DFT data obtained in the different magnetic states. We use for the fitting the solute-vacancy binding energies in the dilute limit, and the vacancy formation energies in the concentrated structures, including the disordered structures with 25% to 75% Ni, and the ordered structures L1$_0$-FeNi and L1$_2$-FeNi$_3$. The vacancy formation energies are calculated using the chemical potential of the removed atom in the pure phase, instead of the one in the alloy, which is difficult to obtain in DFT. In total, more than 500 DFT data are used to fit 6 new magnetic parameters ($A_{i,nV}$ and $J_{ij,nV}$) and 4 new nonmagnetic parameters ($\epsilon_{i,nV}$).

It should be noted that the present EIM does not allow to predict the vacancy vibrational formation entropies. This is not only due to the fact that the EIM is fitted to the vibrational entropy of mixing of the defect-free system instead of its absolute vibrational entropy. The latter is in principle possible to be fitted into the EIM. More importantly, it is also because calculating vacancy vibrational formation entropies in concentrated alloys from DFT are very difficult (such calculations are not available in the literature to the best of the author's knowledge). The vacancy vibrational formation entropies in alloys are approximated by a linear interpolation of the values in the pure systems obtained from DFT, and are added to the results from Monte Carlo simulations using the EIM.

### 2.3.3  Parametrization for vacancy migration barriers

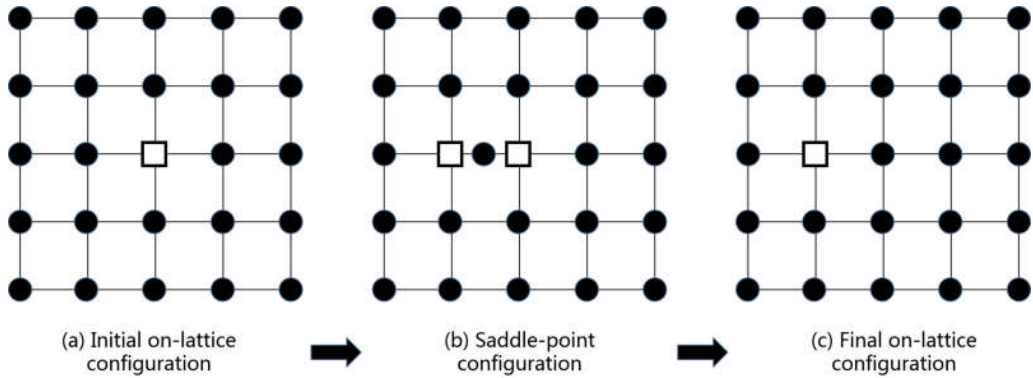

FIGURE 2.3: Illustration of the vacancy-mediated migration process on a two-dimensional lattice. The filled circles represent the atoms and the unfilled squares represent the vacancies. In the initial and final on-lattice configurations, all the atoms are on the lattice sites; in the saddle-point configuration, the migrating atom is on the off-lattice saddle point while the rest of atoms are still on the lattice sites.

The vacancy-mediated diffusion consists of the nearest-neighbour atoms of the vacancy jumping to that vacant lattice site, as illustrated in Fig. 2.3. The key quantity of this process is the migration barrier, calculated as the energy difference between the saddle-point (SP) configuration and the initial on-lattice (OL) configuration. The EIM parametrized in the previous sections allows to calculate the energy of the OL configurations with one vacancy (or isolated non-interacting vacancies). As illustrated in Fig. 2.4, we express the total energy of the SP configuration as the sum of two parts:

$$E_{tot} = E_{OL} + E_{SP} \tag{2.59}$$

where $E_{OL}$ is the energy due to the interactions between the on-lattice atoms, and $E_{SP}$ is the energy due to the interactions of the SP atom with its neighbouring atoms on the lattice. Here $E_{OL}$ is calculated with the EIM parametrized in the last section for the vacancy-containing system, and $E_{SP}$ is the part to be parametrized in this section. Note that $E_{OL}$ calculated from the EIM may not predict correctly the energy of the system with two adjacent vacancies, but we choose to include this effect in the $E_{SP}$ so that $E_{tot}$ of the SP configuration, or equivalently the migration barrier, is correctly predicted by the EIM. Note that this splitting of the energy is just one possibility, and the central quantity to reproduce is the migration barrier.

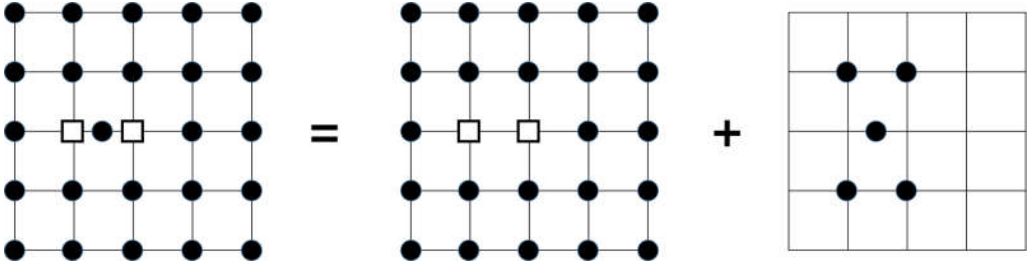

FIGURE 2.4: Decomposition of the energy of the saddle-point configuration.

The expression of $E_{SP}$ is as follows:

$$E_{SP} = A_{SP}M_{SP}^2 + B_{SP}M_{SP}^4 + \sum_j J_{SP,j} \boldsymbol{M}_{SP}\boldsymbol{M}_j + \epsilon_{SP} + \sum_j V_{SP,j} \tag{2.60}$$

where $M_{SP}$ is the magnetic moment of the saddle-point atom, and $A_{SP}$, $B_{SP}$, $J_{SP,j}$, $\epsilon_{SP}$ and $V_{SP,j}$ are the corresponding magnetic and nonmagnetic parameters. The sum $\sum_j$ goes over the atoms within the fourth-neighbour shell of the saddle-point atom. Note that the number of neighbours and the distance from neighbours for the saddle-point atom are different from those of the on-lattice atoms (see Table 2.1).

TABLE 2.1: Number of neighbours and distance (in $a_0$) from the neighbours for the on-lattice and the saddle-point atoms. The distances obtained in the fully relaxed structures may differ by 0.03 $a_0$ from the values given here.

| Neighbour shell | On-lattice atom | | Saddle-point atom | |
|---|---|---|---|---|
| | *num-nb* | distance | *num-nb* | distance |
| 1st | 12 | 0.71 | 4 | 0.61 |
| 2nd | 6 | 1.00 | 4 | 0.79 |
| 3rd | 24 | 1.22 | 8 | 0.94 |
| 4th | 12 | 1.41 | 6 | 1.06 |

The fitting procedure is similar to the previous ones. First, we fit the parameters for pure Fe and Ni to the migration barriers in the pure systems in different magnetic states. In this step, 14 parameters are fitted to around 350 DFT data. Then, we fit the parameters for the Fe-Ni structures to the migration barriers in the alloys in different magnetic states.

## 2.4   Monte Carlo simulations

### 2.4.1   Thermodynamic properties

Temperature-dependent properties are determined from the EIM combined with on-lattice Monte Carlo (MC) simulations with periodic boundary conditions. Some of these properties are defined as follows.

The mixing energy per atom of a given structure is calculated with respect to fcc Fe and Ni:

$$\Delta E_{mix}(T) = E_{alloy}(T) - (1-x)E_{Fe}(T) - xE_{Ni}(T) \tag{2.61}$$

where $x$ is the atomic concentration of Ni, and $E_{alloy}$, $E_{Fe}$ and $E_{Ni}$ are the energies per atom of the fcc alloy and pure systems at the same temperature.

Following the Warren-Cowley formulation [89], the atomic short-range order (ASRO) parameter for the $n^{th}$ coordination shell is defined as follows:

$$\text{ASRO}_n = \frac{1}{N_{\text{Fe}}} \sum_{i=1}^{N_{\text{Fe}}} \alpha_{\text{Fe}_i}^n \tag{2.62}$$

$$\alpha_{\text{Fe}_i}^n = 1 - \frac{x_{\text{Ni}}^n}{x_{\text{Ni}}} \tag{2.63}$$

where $x_{\text{Ni}}$ is the global Ni concentration, and $x_{\text{Ni}}^n$ is the local Ni concentration in the $n^{th}$ coordination shell of the Fe atom $i$. The atomic long-range order (ALRO) parameter is defined as

$$\text{ALRO} = \frac{N_{\text{Fe}}^{\text{Fe}}}{N_{\text{Fe}}} - \frac{N_{\text{Fe}}^{\text{Ni}}}{N_{\text{Ni}}} \tag{2.64}$$

where $N_{\text{Fe}}$ and $N_{\text{Ni}}$ are the total numbers of Fe and Ni, respectively, and $N_{\text{Fe}}^{\text{Fe}}$ and $N_{\text{Ni}}^{\text{Fe}}$ are the numbers of Fe in the Fe and Ni sublattices, respectively. In practice, we observe that the ALRO parameters range either between 0.4 and 1, or between 0.0 and 0.05. Therefore, the structures are considered to be ordered if ALRO$\in$[0.4, 1], and disordered if otherwise. The chemical transition temperature is defined as the lowest temperature at which the ALRO parameter is below 0.4.

The magnetic short-range order (MSRO) parameter for the $n^{th}$ coordination shell is defined as

$$\text{MSRO}_n = < \frac{\mathbf{M}_i \cdot \mathbf{M}_j}{|\mathbf{M}_i| \cdot |\mathbf{M}_j|} > \tag{2.65}$$

which is averaged over all the pairs of the atoms $i$ and $j$ separated by the $n^{th}$ nearest-neighbour distance. The Curie temperature $T_c$ of a ferromagnetic (FM) system is estimated as the inflection point of the following function [103] fitted to the obtained magnetization (namely magnetic moment per atom):

$$\frac{M(T)}{M(T=1K)} = (1 - aT)\frac{1 + \exp(-\frac{b}{c})}{1 + \exp(\frac{T-b}{c})} \tag{2.66}$$

We verify that the thermodynamic properties (magnetization, mixing enthalpies,

magnetic and atomic short-range order parameters, and magnetic and chemical transition temperatures in alloys with 25%, 50% and 75% Ni, etc.) calculated with the 4000-atom fcc lattice are the same (within the desired precision) as the ones with the 16384-atom fcc lattice. Some tests are also performed for larger fcc lattices containing up to 62500 atoms, and the results remain the same regardless of the cell size. Therefore, the effects of cell size are estimated to be rather small on the investigated properties in the fcc Fe-Ni system. The results reported in this study are obtained with the 16384-site fcc lattice.

For the defect-free bulk system, we use three types of MC schemes for different purposes: spin Monte Carlo (SMC), spin-atom canonical Monte Carlo (CMC) and semi-grand canonical Monte Carlo (SGCMC).

In SMC simulations, the atomic configuration is frozen, and only the spin configuration evolves with temperature by performing Metropolis SMC steps. In each SMC step, a random variation of the magnetic moment of a randomly chosen atom is attempted. From SMC simulations, the equilibrium spin configuration of a fixed chemical structure can be obtained, and the corresponding magnetic properties (i.e. magnetization, MSRO and magnetic transition temperature) can be calculated.

In CMC simulations, the chemical composition is fixed, while both the spin and the atomic configurations are equilibrated, leading to the equilibrium spin-atom configuration at a given Ni concentration. CMC simulations are used to obtain the magnetic properties, and the ALRO and ASRO parameters for the equilibrium configuration as functions of temperature. From the results of the ALRO parameters, the chemical transition temperatures at 50% and 75% Ni can then be determined.

In SGCMC simulations, the chemical potential difference $\Delta\mu$ between Fe and Ni is fixed, which determines the equilibrium chemical composition and spin-atom configuration. By varying $\Delta\mu$, the equilibrium phase boundary can be deduced from the semi-grand canonical isotherm [107]. SGCMC simulations are used as a convenient way to construct the fcc Fe-Ni phase diagram.

**Quantum statistics for the spin system**

It is customary to employ classical statistics in atomistic simulations. For its application in the spin system, however, there are some known limitations. For instance, the simulated magnetic specific heat is nonzero at the low-temperature limit [6, 110]. More importantly, the obtained magnetization decreases too rapidly with increasing temperature [6, 110]. This can be an issue for the predictions of the magnetization-dependent properties such as chemical transition temperatures [24], vacancy formation and migration properties [22].

The quantum statistics can be treated within quantum MC [111], which is much more complicated and computationally heavier than classical MC [112, 113]. To incorporate quantum Bose-Einstein statistics in the classical MC simulation framework, we use in this work the procedure proposed by Bergqvist and Bergman [110] as summarized below.

In SMC simulations, a random variation in the spin is attempted with the acceptance probability $\min[1, \exp(\Delta E/\eta)]$. Within classical statistics, the scaling factor $\eta^c$ is given by

$$\eta^c(T) = k_B T \tag{2.67}$$

where the superscript $c$ denotes the use of classical statistics. To incorporate the Bose-Einstein statistics for the spin system, $\eta^c$ should be replaced by $\eta^q$ [110, 114]:

$$\eta^q(T) = \int_0^{E_c(T)} \frac{E}{\exp(\frac{E}{k_B T}) - 1} g(E, T) dE \tag{2.68}$$

where $T$ is the absolute temperature, $E$ is the magnon energy, $g(E, T)$ is the magnetic density of states (mDOS), and $E_c(T)$ is the cut-off magnon energy. Therefore, the central quantity to be determined is the temperature-dependent mDOS $g(E, T)$. In this work, the ground-state mDOS $g(E, 0)$ is obtained by integrating the magnon dispersion $E(\boldsymbol{k})$ over the $k$-space. The latter is calculated by solving the dynamic equation for the spin system [115] using our model parameters $J_{ij}$ up to the fourth-nearest neighbour:

$$E(\boldsymbol{k}) = \sum_{i \neq 0} J(|\boldsymbol{R}_0 - \boldsymbol{R}_i|) \cdot (1 - e^{i\boldsymbol{k}(\boldsymbol{R}_0 - \boldsymbol{R}_i)}) \tag{2.69}$$

Then, based on the quasiharmonic approximations (QHA) as outlined in Ref. [110, 114], the temperature-dependent mDOS $g(E, T)$ below the magnetic transition temperature $T_c$ can be obtained by retaining the shape of $g(E, 0)$, while rescaling the cut-off magnon energy as

$$E_c(T) = E_c(0) \cdot (1 - \frac{T}{T_c})^\beta \tag{2.70}$$

where $\beta$ is the critical exponent associated with the magnetization in the three-dimensional Heisenberg model.

We note that in MC simulations applied with the above quantum statistical treatment of spins, the quantum scaling factor $\eta^q$ is only used to update the spin system, while the chemical evolution is still controlled by using the classical scaling factor $\eta^c$.

### 2.4.2  Vacancy formation in nonmagnetic alloys

In this section, we first review how vacancy formation free energy is related to equilibrium vacancy concentration. Then we propose a Widom-type Monte Carlo scheme to calculate vacancy formation free energy in nonmagnetic alloys. This scheme is in principle also applicable in magnetic alloys, but it is inconvenient in practice due to the difficulty in calculating the magnetic entropy. The problem in magnetic alloys is addressed in the next section.

**Relation between vacancy formation free energy and equilibrium concentration**

To begin with, consider a pure system containing $N_A$ $A$ atoms and $N_V$ isolated mono-vacancies. Note that we only consider isolated monovacancy in this work. The free energy of the system is written as

$$G(N_A, N_V) = G(N_A, 0) + \Delta G(N_V) \tag{2.71}$$

where $G(N_A, 0)$ is the free energy of the perfect system taken as a reference, and $\Delta G(N_V)$ is the free energy change resulting from the creation of $N_V$ vacancies

$$\Delta G(N_V) = N_V G_f - T S_{conf} \tag{2.72}$$

Here $G_f$ is the (mono)vacancy formation free energy that includes all the non-configurational entropies (e.g., electronic, vibrational and magnetic entropies). $S_{conf}$ is the configurational entropy change of creating $N_V$ vacancies

$$S_{conf} = k_B \ln W \tag{2.73}$$

where $W = \binom{N_A + N_V}{N_V} = \frac{(N_A + N_V)!}{N_A! N_V!}$ is the number of distinguishable ways of distributing $N_V$ vacancies among the $(N_A + N_V)$ lattice sites. In the thermodynamic limit ($N_A \rightarrow \infty$, $N_V \rightarrow \infty$), we can apply the Stirling formula $\ln x! \approx x \ln x - x$, so that

$$S_{conf} \approx k_B [(N_A + N_V) \ln(N_A + N_V) - N_A \ln N_A - N_V \ln N_V] \tag{2.74}$$

With Eq. 2.72 and 2.74, we have

$$\mu_V = \frac{\partial G(N_A, N_V)}{\partial N_V}$$

$$= G_f - k_B T \ln \frac{N_A + N_V}{N_V} \tag{2.75}$$

$$= G_f + k_B T \ln C_V \tag{2.76}$$

One may find that Eq. 2.76 resembles the expression of the species $i$ in an ideal solution ($\mu_i = \mu_i^\circ + k_B T \ln x_i$).

The number of vacancies in the thermal equilibrium is found by minimizing the free energy of the system, namely

$$\frac{\partial G(N_A, N_V)}{\partial N_V} = 0 \tag{2.77}$$

We arrive at the well-known expression [72, 73, 116, 117]

$$C_V^{eq} = \frac{N_V}{N_A + N_V} = \exp(-\frac{G_f}{k_B T}) \tag{2.78}$$

This shows that the equilibrium vacancy concentration is related to the vacancy formation free energy $G_f$ which does not contain the configurational entropy of vacancy formation.

A similar derivation can be done for the binary system. The free energy of the system can be written as

$$G(N_A, N_B, N_V) = G(N_A, N_B, 0) + N_V G_f - T S_{conf} \tag{2.79}$$

with

$$
\begin{aligned}
S_{conf} =& k_B \ln \frac{(N_A + N_B + N_V)!}{(N_A + N_B)! N_V!} \\
\approx& k_B [(N_A + N_B + N_V) \ln(N_A + N_B + N_V) \\
& - (N_A + N_B) \ln(N_A + N_B) - N_V \ln N_V]
\end{aligned} \tag{2.80}
$$

Therefore,

$$
\begin{aligned}
\mu_V &= \frac{\partial G(N_A, N_B, N_V)}{\partial N_V} \\
&= G_f - k_B T \ln \frac{N_A + N_B + N_V}{N_V}
\end{aligned} \tag{2.81}
$$

The equilibrium vacancy concentration is found by minimizing $G(N_A, N_B, N_V)$ with respect to $N_V$, leading to

$$C_V^{eq} = \frac{N_V}{N_A + N_B + N_V} = \exp\left(-\frac{G_f}{k_B T}\right) \tag{2.82}$$

which has the same form as the one for the unary system. Note that $G_f$ does not contain any configurational entropy of the alloy, which has been emphasized recently for ideally random alloys [72, 73, 95].

**Evaluation of vacancy formation free energy**

In the following we show how to calculate $G_f$ using the Widom-type technique. Here we recall Eq. 2.11 used to calculate the chemical potential difference in a binary *A-B* alloy

$$\mu_A - \mu_B = -k_B T \ln < \frac{N_B}{N_A + 1} \exp[-\frac{\Delta E(B \rightarrow A)}{k_B T}] > \tag{2.83}$$

In practice, this consists of calculating the energy change of virtually substituting a *B* atom by an *A* atom. The substitution does not actually take place and the number of each species remains the same before and after the Widom substitution step.

To calculate $G_f$ in the unary system *A*, the Widom substitution is performed in the defect-free system by replacing an *A* atom by a vacancy. This leads to the following

equation:

$$\mu_V - \mu_A = -k_B T \ln < N_A \exp[-\frac{\Delta E(A \to V)}{k_B T}] > \qquad (2.84)$$

Namely,

$$\mu_V = -k_B T \ln < \exp[-\frac{\Delta E(A \to V)}{k_B T}] > + \mu_A - k_B T \ln N_A \qquad (2.85)$$

Recognizing that this $\mu_V$ by definition is equal to $\frac{\partial G(N_A, N_V)}{\partial N_V}|_{N_V=0}$, Eq. 2.75 becomes

$$\mu_V = G_f - k_B T \ln N_A \qquad (2.86)$$

Finally, we have

$$\begin{aligned} G_f &= \mu_V + k_B T \ln N_A \\ &= -k_B T \ln < \exp[-\frac{\Delta E(A \to V)}{k_B T}] > + \mu_A \end{aligned} \qquad (2.87)$$

Eq. 2.87 is the equation used in the Monte Carlo simulation for the pure system.

To calculate $G_f$ in the binary $A - B$ system, the Widom substitution can be performed in the defect-free system by replacing *always* an $A$ atom by a vacancy. Following the same derivation as in the pure system, we have exactly the same expression of $\mu_V$ as Eq. 2.85. Then, recognizing this $\mu_V$ is equal to $\frac{\partial G(N_A, N_B, N_V)}{\partial N_V}|_{N_V=0}$, Eq. 2.81 becomes

$$\mu_V = G_f - k_B T \ln(N_A + N_B) \qquad (2.88)$$

Finally, Eq. 2.85 and 2.88 lead to

$$\begin{aligned} G_f &= \mu_V + k_B T \ln(N_A + N_B) \\ &= -k_B T \ln < \exp[-\frac{\Delta E(A \to V)}{k_B T}] > + \mu_A - k_B T \ln \frac{N_A}{N_A + N_B} \\ &= -k_B T \ln < \exp[-\frac{\Delta E(A \to V)}{k_B T}] > + \mu_A - k_B T \ln x_A \end{aligned} \qquad (2.89)$$

where $x_A = \frac{N_A}{N_A + N_B}$ is the concentration of $A$ in the binary system. In an ideal solid solution, $\mu_A$ contains a configurational entropy of mixing $k_B T \ln x_A$, which cancels with $-k_B T \ln x_A$ in Eq. 2.89. This is reasonable, otherwise the calculated $G_f$ always varies with temperature in the usual pair-interaction models.

It should be noted that Eq. 2.89 and Eq. 2.82 are expected to hold in any alloy, as their derivations are not based on the assumption that the alloy is an ideal solution. Indeed, they are based on the validity of Eq. 2.81, which expresses $\mu_V$ as the sum of $G_f$ and $k_B T \ln C_V$. The latter is not related to the mixing behaviour between $A$ and $B$, but to the ideal mixing between vacancies and atoms, which is justified by the extremely dilute vacancy concentration.

We note that the Widom substitution in the binary system can be performed either always on $A$ atoms, or always on $B$ during the simulation. The results $G_f$ of two types of substitutions should be the same.

In practice, the calculation of $G_f$ consists of two parts. The first part is the evaluation of $-k_B T \ln < \exp[-\frac{\Delta E(A \to V)}{k_B T}] >$, following the procedure below:

1. Perform the relaxations of the magnetic and chemical structures for the defect-free system.

2. Randomly choose an atom $A$ and calculate $\Delta E(A \to V)$ by virtually deleting it.

3. Repeat steps 1 and 2, calculate the average of $\exp[-\frac{\Delta E(A \to V)}{k_B T}]$ and evaluate $-k_B T \ln < \exp[-\frac{\Delta E(A \to V)}{k_B T}] >$.

The second part is to calculate $\mu_A$. This can be solved from the following equations

$$\mu_A - \mu_B = \Delta \mu(B \to A) \tag{2.90}$$

$$x_A \mu_A + x_B \mu_B = \mu_{avg} \tag{2.91}$$

The right hands of these two equations, $\Delta \mu(B \to A)$ and $\mu_{avg}$, also need to be calculated in Monte Carlo simulations. The chemical potential difference $\Delta \mu(B \to A)$ can be computed by the usual Widom substitution for atoms by applying Eq. 2.83. The evaluation of the average chemical potential $\mu_{avg}$ is more complicated. One may obtain it from the thermodynamic integration as follows:

- Perform Monte Carlo simulations at different temperatures and compute the energy per atom $e_{avg}$ at each temperature.

- Calculate the specific heat $c_P = \frac{\partial e_{avg}}{\partial T}$ by the numerical differentiation.

- Calculate the entropy per atom $s(T) = \int_0^T \frac{c_P(T')}{T'} dT'$ by the numerical integration. In our study, this entropy is the sum of the magnetic and configurational entropies.

- Calculate $\mu_{avg}(T) = e_{avg} - T \cdot s(T)$.

For the nonmagnetic alloys and the magnetic systems described by the Ising models, $c_P$ goes to zero when $T$ goes to 0 K. However, for magnetic alloys described by Heisenberg or generalized-Heisenberg models, $c_P$ goes to a non-zero constant when $T$ goes to 0 K, leading to a divergence of magnetic entropy at 0 K. This issue is due to the use of classical statistics for spins at low temperatures and can be solved, for example, by switching to quantum statistics as described in Sec. 2.4.1, which further complicates the problem. In addition, the calculations in the nonmagnetic alloys are also inconvenient due to the evaluation of the chemical potentials: First, a series of simulations at different temperatures need to be performed for the subsequent thermodynamic integration to obtain $\mu_{avg}$. Then, $\Delta \mu(B \to A)$ needs to be computed via the Widom technique, and finally the chemical potentials are solved from the relations with $\mu_{avg}$ and $\Delta\mu(B \rightarrow A)$. In the next section, we propose a modified algorithm to compute $G_f$ without calculating $\mu_A$.

### 2.4.3 Vacancy formation in magnetic alloys

The Widom-type techique is in fact based on the more general free energy perturbation formalism [118], in which the free-energy difference $G_1 - G_0$ between the system 1 and 0 is computed as

$$G_1 - G_0 = -k_B T \ln < \exp(-\frac{E_1 - E_0}{k_B T}) >_0 \tag{2.92}$$

where $E_1$ and $E_0$ are the energies of the systems 1 and 0 for the same microstate, respectively, and the angle bracket with the subscript 0 denotes the ensemble average in the system 0. $G_1 - G_0$ is equal to $G_f$, if the pure systems 1 and 0 contain the same number of atoms and there is one vacancy in the system 1 but no vacancy in the system 0.

In the magnetic pure $A$ system, the energy difference $E_1 - E_0$ is calculated as

$$E_1 - E_0 = \Delta E(A \rightarrow V) + e_{A,i} \tag{2.93}$$

where $i$ is the lattice site occupied by $A$, $\Delta E(A \rightarrow V)$ is the energy change of deleting this $A$ atom, and $e_{A,i}$ is the energy associated with this atom. To calculate $e_{A,i}$, we virtually put the removed atom with its spin to replace another $A$ atom at a randomly selected lattice site $j$, and calculate $e_{A,i}$ as the sum of its on-site energy and half of the magnetic and nonmagnetic pair-interaction energies with the atoms around the site $j$. In this way, $E_1 - E_0$ can be viewed as the energy change of moving the atom from the site $i$ to the surface.

We have verified with our Hamiltonian that the algorithm presented above in this section (denoted as $e$-Widom algorithm) yields the same results in fcc Fe and Ni as the ones computed using the algorithm presented in the last section (denoted as $\mu$-Widom algorithm). Indeed, for the magnetic pure system, it is possible, though quite complicated, to introduce some auxiliary energy functions in the $\mu$-Widom algorithm to compute the difference between $\mu$ and the auxiliary terms.

In the binary $A - B$ system, if $e_{A,i}$ is calculated the same way as in the pure system, the obtained $G_f$ is different from the $G_f$ computed using $e_{B,i}$. In fact, this is because in a binary system, the energy of a pair interaction can not simply be split into half and attributed evenly for the two atoms forming the pair. The difference $< e_{A,i} > - < e_{B,i} >$, where the angle brackets denote an average, calculated in this way is indeed different from the one obtained from the Widom-type technique [97]. To remedy this issue, we propose to add a correction term $e_{A,corr}$ on $G_f$:

$$e_{A,corr} = T(s_{id,conf} - s_{avg}) + x_B[\Delta\mu_{ex}(B \to A) - \Delta e_d(B \to A) + T(s_{A,d,mag} - s_{B,d,mag})]$$
(2.94)

where $s_{id,conf} = -k_B T(x_A \ln x_A + x_B \ln x_B)$ is the ideal configurational entropy per atom, and $s_{avg}$ is the configurational entropy per atom of the system. In the ordered structures, $s_{avg}$ can be approximated by the ideal configurational entropy in the sublattice. In the disordered structure, $s_{avg}$ can be approximated as $s_{id,conf}$. The term $\Delta\mu_{ex}(B \to A)$ is calculated via the Widom technique:

$$\Delta\mu_{ex}(B \to A) = -k_B T \ln < \exp[-\Delta E(B \to A)] >$$
(2.95)

Note that the factor $\frac{N_A}{N_B+1}$ does not appear here. The term $\Delta e_d(B \to A)$ is

$$\Delta e_d(B \to A) = < e_{A,i} > - < e_{B,i} >$$
(2.96)

where $< e_{A,i} >$ is the average energy per $A$ calculated by considering half of the pair-interaction energy. $s_{A,d,mag} - s_{B,d,mag}$ is the magnetic entropy difference and is formally related to $\Delta e_d(B \to A)$ as

$$s_{A,d,mag} - s_{B,d,mag} = \int \frac{\partial\Delta e_d(B \to A)}{T'\partial T}dT'$$
(2.97)

But in practice, $s_{A,d,mag} - s_{B,d,mag}$ is simply neglected based on the assumption that $A$ and $B$ atoms in the alloy have similar energy variation (versus temperature) and hence similar magnetic entropies, which is the case for the Fe-Ni system.

If $G_f$ is calculated by substitution of $B$ atoms, the corresponding correction to add to $G_f$ is:

$$e_{B,corr} = T(s_{id,conf} - s_{avg}) - x_A(\Delta\mu_{ex}(B \to A) - \Delta e_d(B \to A) + T(s_{A,d,mag} - s_{B,d,mag}))$$
(2.98)

At present we have not yet been able to justify in a very rigorous manner the correction terms presented above. But we have carefully verified that in the nonmagnetic alloys, the results obtained with the $e$-Widom algorithm with the corrections ($G_f + e_{A,corr}$ or $G_f + e_{B,corr}$) are the same as those obtained with the $\mu$-Widom algorithm. The verifications are done with the present EIM by freezing the magnetic evolution, as well as with other nonmagnetic pair-interaction models. At low temperatures, we also find the same $G_f$ in the ordered structures as those predicted by the statistical canonical formalism [62]. We can not directly verify the case of magnetic alloys, but with the successful verifications in the magnetic pure systems and the nonmagnetic binary systems, it is expected that the $e$-Widom algorithm with the corrections should also work in magnetic alloys.

### 2.4.4 Vacancy-mediated diffusion properties

The vacancy-mediated diffusion properties at a given temperature $T$ are obtained as follows. First, the canonical Monte Carlo simulation is performed to attain the equilibrium spin-atom structure. Then, a vacancy is introduced into the system. The diffusion process is simulated by exchanging the vacancy with one of its 12 nearest-neighbour atoms, determined according to the residence-time algorithm:

- The energy $E_{OL}$ of the initial on-lattice system is calculated

- Calculate the 12 migration barriers $E_{m,i} = E_{SP,i} - E_{OL}$, where $E_{SP,i}$ is the energy of the system with the atom $i$, which is one of the 12 first-nearest-neighbour atoms of the vacancy, at the saddle point.

- Calculate the jump rate $w_i = \nu \cdot \exp(-\frac{E_{m,i}}{k_B T})$.

- Calculate the cumulative jump rate $\Gamma_i^{cumu} = \sum_{j=1}^{i} w_j$.

- Take a uniform random variable $u$ in the interval $[0, \Gamma_{12}^{cumu}]$. If $u$ is in the interval $[\Gamma_{i-1}^{cumu}, \Gamma_i^{cumu})$, the vacancy is exchanged with the first-nearest-neighbour atom $i$.

- Update the time $\tau_{MC} = \tau_{MC} + \delta\tau_{MC}$, with $\delta\tau_{MC} = 1/\Gamma_{12}^{cumu}$.

In the Monte Carlo simulation, the introduced vacancy concentration in the system is usually much larger than the equilibrium vacancy concentration, which is determined by the vacancy formation free energy. Consequently, to obtain the diffusion coefficients corresponding to the equilibrium condition, the physical time needs to be rescaled from the time measured in Monte Carlo as [119]

$$\tau = \tau^{MC} \cdot \frac{C_V^{MC}}{C_V^{eq}} \tag{2.99}$$

During the simulation, the individual trajectory $\boldsymbol{r}_i$ of each atom is recorded. The tracer diffusion coefficient $D_A^{A*}$ is calculated from the Einstein relation [116]

$$D^{A*} = \frac{< \boldsymbol{r}_{A*,i}^2 >}{6\tau} \tag{2.100}$$

where $< \boldsymbol{r}_{A*,i}^2 > = \frac{1}{N_{A*}} \sum_{i=1}^{N_{A*}} \boldsymbol{r}_{A*,i}^2$ is the mean square displacement of the tracer atoms $A*$ during the physical time $\tau$. In practice, we consider all $A$ atoms as tracer to obtain good statistics. In the concentrated alloys, a single simulation with $10^6$ vacancy-atom exchanges is sufficient for determining $D^*$. In the dilute alloys, for example with only 1 solute $A$ in the simulation box, $< \boldsymbol{r}_{A,i}^2 >$ should be averaged over a number of independent simulations, which can be performed separately or just obtained as parts of a single long simulation.

In extremely dilute binary alloys, it is of interest to compare the solute diffusion coefficients computed directly from Monte Carlo simulations to the results of the LeClaire model presented in Sec. 2.2.4. In the residence-time algorithm, we can calculate the weighted averages of the accepted migration barriers. These effective migration barriers can then be used for the extended LeClaire model that incorporates the temperature-dependent magnetic effects.

# 3 Thermodynamics of Fe-Ni alloys

---

*This chapter is focused on the thermodynamics of defect-free Fe-Ni alloys. First, energetic, magnetic and vibrational properties of bcc and fcc structures are obtained from DFT calculations, allowing to construct a complete bcc-fcc phase diagram. The DFT-parametrized effective interaction model, combined with Monte Carlo simulations, provides further insights into the impacts of chemical orders and temperature-dependent magnetic excitations and transitions on the phase stability of fcc Fe-Ni alloys. Finally, the effects of Mn and Cr addition on the stability of Fe-Ni ordered structures are discussed.*

---

## 3.1 State of the art

The phase stability of the Fe-Ni system has been extensively investigated experimentally and theoretically. A comprehensively assessed Fe-Ni equilibrium diagram constructed by Swartzendruber *et al.* [120] is presented in Fig. 3.1. Experimentally, the well-established equilibrium solid phases are body-centered-cubic (bcc) and face-centered-cubic (fcc) random alloys, as well as the ordered structure $L1_2$-FeNi$_3$ with a critical ordering temperature, noted hereafter as $T_c^{L1_2}$, of about 783 K [121].

Less clear is the phase diagram below 600 K, where the diffusion is extremely sluggish and thermodynamic equilibrium cannot be achieved under laboratory conditions. One way to access low-temperature thermodynamic data is to study meteoritic systems, which are basically Fe-Ni alloys with small amounts (<1 wt.%) of Co, P, S, and C, slowly cooled down over millions of years [122]. Another alternative is to use irradiated alloys, in which interdiffusion may be enhanced by irradiation-induced defects. In meteoritic specimens, an ordered structure $L1_0$-FeNi was found but disputed to be stable [123] or metastable [122]. This phase was also observed in irradiated samples, and its critical ordering temperature, noted hereafter as $T_c^{L1_0}$, was found to be 593 K [124–126]. A yet more controversial issue is the existence of the metastable ordered Fe$_3$Ni phase, which was suspected to have a $L1_2$ structure but could not be fully confirmed by experiments [127–129].

In view of the scarcity of experimental data and the difficulty to reach phase equilibria below 600 K, parameter-free quantum-mechanical calculations may provide useful information on the ground-state properties of Fe-Ni alloys. Mishin *et al.* [130] investigated various ordered Fe-Ni structures using the full-potential linearized augmented plane wave (FLAPW) method. More ordered structures were later considered in density functional theory (DFT) studies [4, 131]. These calculations show that L1₀-FeNi is also a stable phase in addition to L1₂-FeNi₃, while L1₂-Fe₃Ni is found to be unstable. It was shown that the Fe₃Ni phase with a Z1 superlattice, though still unstable, has lower energy than the L1₂-Fe₃Ni structure [4, 131].

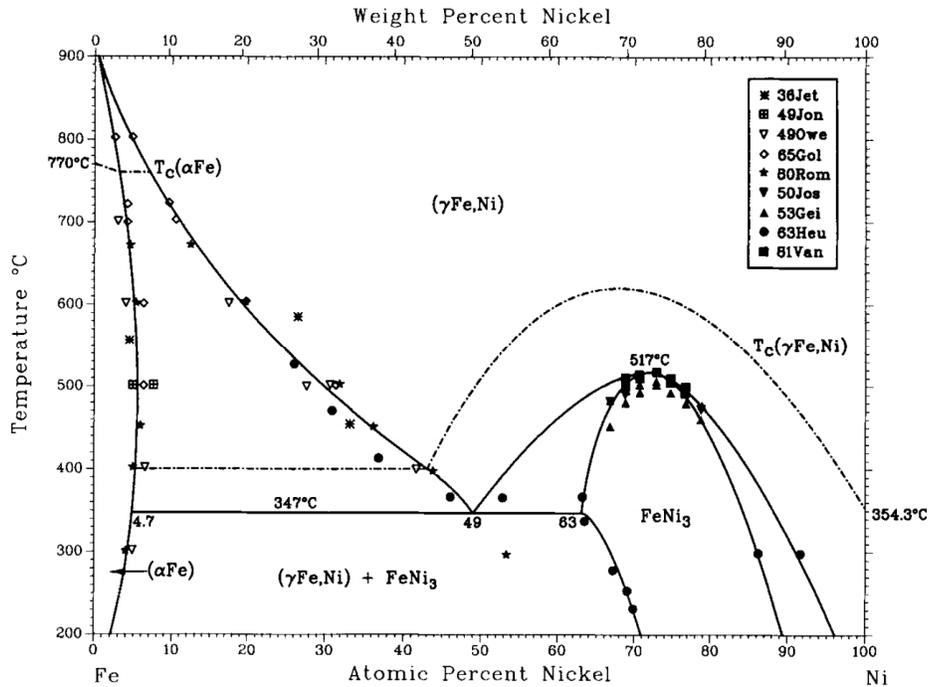

FIGURE 3.1: Critically assessed Fe-Ni phase diagram with experimental data by Swartzendruber *et al.* [120]. The symbols represent the phase boundaries from experiments (see Ref. [120] for references). The *α*-Fe, *γ*-(Fe,Ni) and FeNi₃ in the figure correspond to the bcc and fcc random alloys and L1₂-FeNi₃ ordered structure, respectively.

First principles calculations combined with other methodologies also cast light on the low-temperature phase stability of this system. *Ab initio* derived data were used in combination with available experimental data in the recent CALPHAD modellings to construct the Fe-Ni phase diagram [132, 133]. The L1₀-disorder phase equilibria were studied by Mohri *et al.* [134–137] by coupling FLAPW calculations with cluster variation method (CVM). Magnetic and thermodynamic properties of fcc Fe-Ni alloys were investigated by Lavrentiev *et al.* [56] using *ab initio* parametrized magnetic cluster expansion (MCE) models. By including formation energies of ordered phases from DFT, an interatomic potential based on the embedded atom model (EAM) was developed by Bonny *et al.* [60] for the Fe-Ni system with a globally improved description of bcc and fcc phases.

However, *ab initio* data used in these studies were mostly limited to ordered structures. By extrapolating from properties of ordered states, the effects of chemical disordering on magnetic and thermodynamic properties may not be well captured. An accurate description of disordered structures is also indispensable to correctly predict the order-disorder transitions.

At ambient conditions, Fe and Ni are stable in the bcc and fcc phases, respectively, both being ferromagnetic (FM). The underlying crystal structure of Fe-Ni alloys is dependent on the chemical composition, transforming from bcc to fcc with increasing Ni content. On the other hand, fcc Fe, which exists as a stable paramagnetic (PM) phase above 1183 K, can be stabilized inside fcc Cu matrix at low temperatures, with a non-collinear spin spiral magnetic structure [138, 139]. Such distinct magnetism of pure Fe and Ni on the fcc lattice can lead to complex competing magnetic structures [37], and may eventually influence thermodynamic properties of fcc Fe-Ni alloys. Previous theoretical studies focus on only some specific concentrations [37, 140–142], but a systematic understanding of the concentration evolution of magnetic structures of fcc phases and its effect on thermodynamic properties is still lacking.

At finite temperatures phase stability is determined by the free energy of the system. It is known that configurational entropy of mixing is important in stabilizing disordered alloys over ordered ones. The role of vibrational entropy was highlighted in the EAM potential study [60]. The CVM study by Mohri *et al.* [136] found that the thermal vibration decreases $T_c^{L1_0}$ by about 40 K. In a recent DFT study on the L1$_0$-disorder transformation [143], the predicted $T_c^{L1_0}$ is reduced by about 480 K by including vibrational effects, while electronic and magnetic contributions to the free energy are small. These results suggest that vibrational contribution are relevant to the order-disorder transitions, although the quantitative effect is still unclear.

The effects of magnetism on the chemical order-disorder transitions in fcc Fe-Ni alloys have been studied using magnetic Hamiltonians combined with Monte Carlo simulations [24–26, 56]. It is shown that magnetic interaction has a significant impact on phase stability of the ordered structures [24–26]. However, the Ising and Heisenberg magnetic models adopted in the earlier studies [24–26] are limited to the specific compositions and do not provide a systematic understanding over the whole composition range. Also, the composition dependence of magnetic moments as well as the thermal longitudinal spin fluctuations are not taken into account by these simple models. Recently, Lavrentiev *et al.* developed a generalized-Heisenberg model Hamiltonian for the fcc Fe-Ni system, but the predicted Curie points of random alloys are much lower than the experimental data. Furthermore, these predicted Curie points are also lower than the predicted order-disorder transition temperatures at 50% and 75% Ni, which contradicts with the established phase diagram.

Understanding the effects of the addition of a third element on phase stability of the Fe-Ni based system is important for designing multicomponent austenitic steels. Experimentally, it was observed that the addition of a small amount of Mn (or Cr) in the fcc Fe-Ni alloy with 75% Ni leads to an increase (or a decrease) of its chemical order-disorder transition temperature [144]. In addtion, experiments also show that the Cr atoms in the L1$_2$-FeNi$_3$ structure tend to occupy the Fe sublattice rather than the Ni sublattice [145, 146]. As the Mn-Ni and Cr-Ni systems also exhibit chemical ordering tendency [147–150], there may also be a competition between various ordered structures. Meanwhile, a theoretical investigation on the Cr and Mn effects on magnetism phase stability of the Fe-Ni systems is currently missing.

In summary, phase stability of Fe-Ni alloys at low and intermediate temperatures is not fully clear due to insufficient experimental information, and from the theoretical side, the lack of accurate data of disordered phases at an *ab initio* level. Also, a systematic prediction of vibrational effects on phase stability remains an open issue. On the other hand, a reliable model Hamiltonian enabling a systematic description of magnetic and thermodynamic properties over the whole composition range for the fcc Fe-Ni system is still missing. Such a model is indispensable for predicting thermal chemical and magnetic effects as functions of compositions and temperatures. Finally, the effects of the Mn and Cr addition in the Fe-Ni systems have not been theoretically explored.

The first and second sections of this chapter present a first principles study of ground-state properties and lattice vibration effects of disordered Fe-Ni phases, with a focus on the phase stability of the Fe-Ni system at low and intermediate temperatures. In the first section, we systematically investigate the ground-state magnetic and energetic properties of bcc and fcc Fe-Ni structures. In the second section, we calculate within DFT the vibrational entropies and the free energies of mixing, to construct the complete bcc-fcc phase diagrams below the Curie temperatures. These results are compared with a recent CALPHAD study by Caccimani *et al.* [132] in the light of available experimental data. In the third section, we use *ab initio* parametrized effective interaction model (EIM), combined with Monte Carlo simulations, to study the phase stability of the fcc Fe-Ni alloys for all temperatures. The effects of magnetic excitations and transitions are fully taken into account, and the interplay between the magnetic and chemical degrees of freedom are discussed. Finally, in the fourth section, we study the Mn and Cr additions in the Fe-Ni structures via DFT.

## 3.2 Electronic ground-state properties from DFT

In this section, we discuss the ground-state properties obtained from DFT calculations in the bcc and fcc Fe-Ni systems. Previous first principles investigations in Fe-Ni alloys are mostly limited to the ordered structures, while little theoretical effort is devoted to the disordered ones. We first compare the results in the ordered structures with previous calculated values to validate our calculations, before discussing the energetic and magnetic properties in the disordered structures.

### 3.2.1 Properties of pure and ordered structures

Various collinear magnetic structures of Fe and Ni are considered and the obtained properties are given in Table 3.1. We use AFS and AFD to designate respectively the antiferromagnetic single-layer and double-layer structures. The present results are in excellent agreement with those in Ref. [4]. Our calculations agree with Ref. [151]

that the fcc AFS and AFD states of Fe transform spontaneously into the tetragonal lattices after full relaxation, while the fcc ferromagnetic high-spin (FM-HS) state is a metastable equilibrium position and relaxes towards the fct lattice by setting an initial $c/a > 1$. The tetragonal distortion on the fcc lattice lowers the energy by 0.02 eV/atom for AFS and AFD Fe, and by 0.05 eV/atom for FM-HS Fe.

TABLE 3.1: Equilibrium lattice parameters (in Å), energies (in eV/atom) and average moment magnitudes (in $\mu_B$/atom) of pure Fe and Ni in various magnetic states, compared to the values from the recent first principles calculations [4, 130, 151].

| Phase | $a_0$ | $a_0^{ref}$ | $\Delta E$ | $\Delta E^{ref}$ | $|\mu_{tot}|$ | $|\mu_{tot}^{ref}|$ |
|---|---|---|---|---|---|---|
| Fe bcc NM | 2.755 | 2.755 [4] | 0.475 | 0.475 [4] | 0.000 | 0.000 [4] |
| Fe bcc FM (GS) | 2.832 | 2.831 [4] | 0.000 | 0.000 [4] | 2.205 | 2.199 [4] |
| Fe bcc AFS | 2.793 | 2.791 [4] | 0.443 | 0.444 [4] | 1.318 | 1.290 [4] |
| Fe fcc NM | 3.446 | 3.445 [4] | 0.163 | 0.167 [4] | 0.000 | 0.000 [4] |
| Fe fcc FM-LS | 3.478 | 3.478 [4] | 0.155 | 0.162 [4] | 1.012 | 1.033 [4] |
| | | 3.483 [130] | | 0.129 [130] | | 1.342 [130] |
| Fe fcc FM-HS | 3.637 | 3.631 [4] | 0.153 | 0.153 [4] | 2.589 | 2.572 [4] |
| Fe fct FM-HS | 3.422[a] | 3.418[b] [151] | 0.103 | 0.108 [151] | 2.398 | 2.40 [151] |
| Fe fcc AFS | 3.486 | 3.480 [151] | 0.124 | 0.110 [151] | 1.315 | / |
| Fe fct AFS | 3.419[c] | 3.504 [4] | 0.100 | 0.100 [4] | 1.564 | 1.574 [4] |
| | | 3.423[d] [151] | | 0.076 [130] | | 1.50 [151] |
| | | 3.430[e] [130] | | 0.091 [151] | | |
| Fe fcc AFD | 3.540 | 3.527 [151] | 0.104 | 0.096 [151] | 1.936 | 1.80 [151] |
| Fe fct AFD | 3.447[f] | 3.447[g] [151] | 0.081 | 0.082 [4] | 2.022 | 2.062 [4] |
| | | 3.454[h] [130] | | 0.067 [130] | | 1.99 [151] |
| | | | | 0.077 [151] | | |
| | | | | | | |
| Ni bcc NM | 2.792 | 2.794 [4] | 0.110 | 0.107 [4] | 0.000 | 0.000 [4] |
| Ni bcc FM | 2.800 | 2.802 [4] | 0.094 | 0.092 [4] | 0.558 | 0.569 [4] |
| | | 2.801 [130] | | 0.099 [130] | | 0.540 [130] |
| Ni fcc NM | 3.510 | 3.513 [4] | 0.057 | 0.056 [4] | 0.000 | 0.000 [4] |
| Ni fcc FM (GS) | 3.514 | 3.521 [4] | 0.000 | 0.000 [4] | 0.641 | 0.641 [4] |
| | | 3.519 [130] | | 0.000 [130] | | 0.631 [130] |

[a] $c/a = 1.173$
[b] $c/a = 1.175$
[c] $c/a = 1.073$
[d] $c/a = 1.069$
[e] $c/a = 1.070$
[f] $c/a = 1.088$
[g] $c/a = 1.088$
[h] $c/a = 1.086$

Experimentally, the ground state of Fe is the bcc ferromagnetic (FM) phase. At low temperatures, fcc Fe can be stabilized in fcc Cu matrix in the form of small clusters, exhibiting a noncollinear spin spiral magnetic structure [138, 152]. Regarding Ni, only its fcc phase exists in the nature, being FM below the Curie temperature of around 630 K [153]. The bcc structure of Ni can be stabilized as a thin film on GaAs at 170 K, also showing ferromagnetism below 456 K [154]. Our results confirm the experimental magnetic ground states for bcc Fe, bcc and fcc Ni. Within the collinear approximation, the magnetic ground state of fcc Fe is found to be the AFD phase. This state is predicted to be just slightly higher in energy than the lowest-energy spin spiral state, according to the non-collinear magnetic calculations [3].

We perform DFT calculations in the ordered structures presented in Ref. [4, 130, 131]. The mixing enthalpy per atom of a structure $Fe_{1-x}Ni_x$ is calculated as follows:

$$\Delta H_{mix} = H(Fe_{1-x}Ni_x) - (1-x)H(Fe) - xH(Ni) \qquad (3.1)$$

where $x$ is the Ni concentration of the structure, $H(Fe_{1-x}Ni_x)$ is the enthalpy per atom of the alloy, and $H(Fe)$ and $H(Ni)$ are the enthalpy per atom of Fe and Ni in the reference states, respectively. As shown in Table 3.2, the mixing enthalpies and other properties are in good agreement with those from Ref. [130]. In agreement with Ref. [4, 131], the most stable phase of $Fe_3Ni$ is the Z1 structure with a positive mixing enthalpy. Among these structures, five of them, all being FM, are found to have negative mixing enthalpies. The two most stable structures are the experimentally observed phases $L1_2$-$FeNi_3$ and $L1_0$-$FeNi$. Their configurations are illustrated in Fig. 3.2. The other three structures with negative mixing enthalpies are cI32-$FeNi_7$, $C11_f$-$Fe_2Ni$ and $C11_f$-$FeNi_2$, but they have not yet been found in any experiment. Indeed, these three structures are not thermodynamically stable as we shall see in the next section.

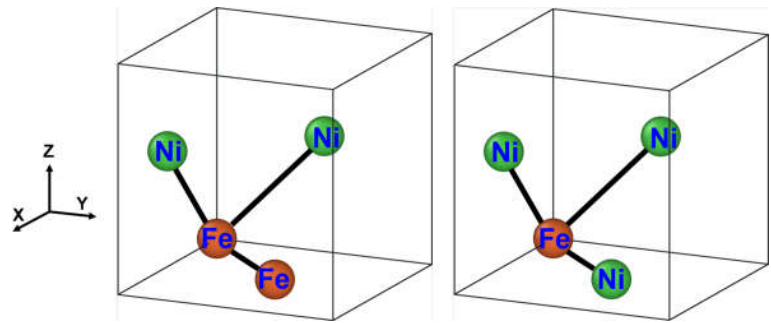

FIGURE 3.2: Cubic unit cells of $L1_0$-FeNi and $L1_2$-$FeNi_3$. In $L1_0$-FeNi, an Fe atom has four 1NN (first nearest neighbour) Fe atoms which are in the same X-Y layer, and eight 1NN Ni atoms which are in the different X-Y layers; symmetrically, an Ni atom has four 1NN Ni atoms and eight 1NN Fe atoms. In $L1_2$-$FeNi_3$, an Fe atom has twelve 1NN Ni atoms; an Ni atom has four 1NN Fe atoms and eight 1NN Ni atoms.

### 3.2.2 Mixing enthalpies of disordered structures

Previous first principles calculations in Fe-Ni alloys were mostly limited to the ordered structures, whereas an accurate description of the phase stability requires an accurate description for both the ordered and disordered phases. In this subsection, we present the ground-state energetic properties of the disordered structures and compare the DFT results with the predictions of previous empirical models.

TABLE 3.2: Equilibrium lattice parameters (in Å), mixing enthalpies (in eV/atom) and average magnetic moments (in $\mu_B$/atom) of Fe-Ni ordered structures compared with Ref. [130]. The mixing enthalpies are calculated with bcc FM Fe and fcc FM Ni as the reference states. All the structures are FM except for the ferrimagnetic structure of $Fe_7Ni$ marked below.

| Formula | Structure | $a_0$ | $a_0^{ref}$ | $H_{mix}$ | $H_{mix}^{ref}$ | $\mu_{tot}$ | $\mu_{tot}^{ref}$ | $\mu_{Fe}$ | $\mu_{Ni}$ |
|---------|-----------|-------|-------------|-----------|-----------------|-------------|-------------------|------------|------------|
| $Fe_7Ni$ | fcc cF32 | 7.029 | 7.029 | 0.126 | 0.109 | 1.435 | 1.361 | 1.573 | 0.466 |
| $Fe_7Ni$ | fcc cF32[a] | 7.045 | 7.034 | 0.111 | 0.092 | 1.015 | 0.942 | 1.095 | 0.458 |
| $Fe_7Ni$ | bcc cI16 | 5.709 | 5.704 | 0.007 | 0.009 | 2.303 | 2.244 | 2.507 | 0.869 |
| $Fe_3Ni$ | fcc $L1_2$ | 3.594 | 3.578 | 0.045 | 0.048 | 2.081 | 2.066 | 2.567 | 0.621 |
| $Fe_3Ni$ | bcc $D0_3$ | 5.712 | 5.718 | 0.030 | 0.034 | 2.163 | 2.162 | 2.609 | 0.827 |
| $Fe_3Ni$ | fcc Z1 | 3.499 | | 0.017 | | 2.035 | | 2.510 | 0.610 |
| $Fe_2Ni$ | fct $C11_f$ | 3.516[b] | 3.525[c] | -0.015 | -0.015 | 1.887 | 1.887 | 2.536 | 0.589 |
| FeNi | sc B1 | 4.737 | 4.745 | 0.568 | 0.560 | 1.862 | 1.870 | 2.954 | 0.769 |
| FeNi | bcc B2 | 2.854 | 2.854 | 0.077 | 0.084 | 1.755 | 1.768 | 2.807 | 0.702 |
| FeNi | fcc $L1_1$ | 3.569 | 3.564 | 0.000 | 0.002 | 1.617 | 1.588 | 2.555 | 0.678 |
| FeNi | fct $L1_0$ | 3.552[d] | 3.556[e] | -0.067 | -0.067 | 1.637 | 1.630 | 2.649 | 0.625 |
| $FeNi_2$ | fct $C11_f$ | 3.554[f] | 3.560[g] | -0.053 | -0.053 | 1.289 | 1.285 | 2.653 | 0.607 |
| $FeNi_3$ | bcc $D0_3$ | 5.637 | 5.643 | 0.019 | 0.024 | 1.111 | 1.102 | 2.852 | 0.531 |
| $FeNi_3$ | fcc $L1_2$ | 3.542 | 3.545 | -0.088 | -0.089 | 1.204 | 1.202 | 2.905 | 0.637 |
| $FeNi_7$ | bcc cI16 | 5.620 | 5.628 | 0.059 | 0.062 | 0.850 | 0.851 | 2.910 | 0.556 |
| $FeNi_7$ | fcc cF32 | 7.058 | 7.065 | -0.035 | -0.038 | 0.913 | 0.906 | 2.862 | 0.635 |

[a] Ferrimagnetic structure, with the spins of the Ni atom and its six nearest-neighbouring Fe atoms aligned in the up direction, and the spin of the remaining Fe atom, which is a second neighbour to the Ni atoms, aligned in the down direction.

[b] $c/a = 3.152$
[c] $c/a = 3.142$
[d] $c/a = 1.007$
[e] $c/a = 1.007$
[f] $c/a = 2.975$
[g] $c/a = 2.985$

Three kinds of disordered phases have been considered: the bcc FM and fcc AFD SQSs below 50% Ni, and the fcc FM SQSs over the complete composition range. We find that the bcc FM SQSs with more than 40% Ni transform into an orthorhombic lattice after relaxation, which is associated with the instability of bcc Fe-Ni alloys with high Ni content. Therefore, the bcc phases above 40% Ni are not be considered in the following. Similar to the case of AFD Fe, fcc AFD SQSs transform spontaneously into the fct lattice, with the $c/a$ ratio decreasing with increasing Ni concentrations.

Fig. 3.3 shows the mixing enthalpies of the SQSs and the most stable ordered structures. Overall, the ordered structures have lower energies than the SQSs. $L1_2$-FeNi$_3$ and $L1_0$-FeNi form the convex hull with bcc Fe and fcc Ni, and thus are the only chemical ground states of Fe-Ni alloys. We note that the energy of $L1_0$-FeNi is just 0.008 eV/atom below the line connecting the energies of bcc Fe and $L1_2$-FeNi$_3$. The most stable state of SQSs is FM, with the bcc-fcc crossover at about 37% Ni. The fct AFD state is more stable than the fcc FM one below 18% Ni, consequently the curve of the fcc disordered ground states consists of the AFD and FM branches. The predicted AFD-FM crossover is higher than the value from another DFT-GGA study [37], because the latter did not consider the tetragonal deformation of the AFD phase and may consequently underestimate its stability.

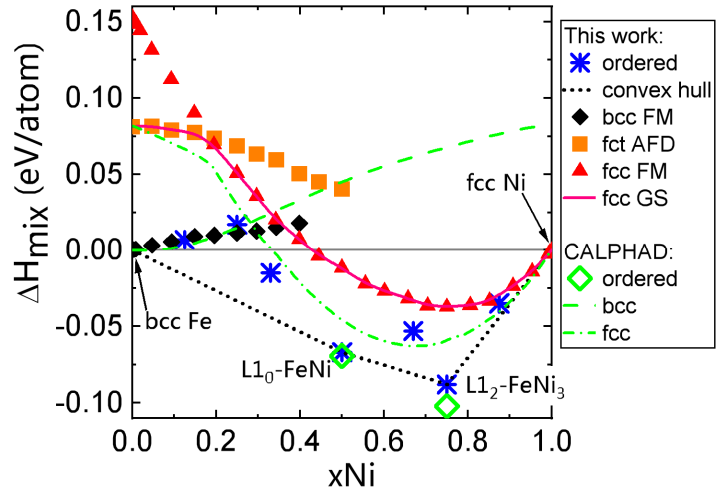

FIGURE 3.3: Mixing enthalpies of ordered and disordered structures, calculated with bcc FM Fe and fcc FM Ni as the reference states. Results from the CALPHAD assessment by Cacciamani *et al.* are shown for comparison [132].

To the best of our knowledge, there is no existing results of the mixing energies of bcc and fcc Fe-Ni disordered phases from first principles studies. Recently, Cacciamani *et al.* [132] performed a comprehensive CALPHAD assessment of the Fe-Ni system. The ground-state mixing enthalpies $\Delta H_{mix}$ of the random solutions and the ordered phases predicted by this CALPHAD assessment are shown in Fig. 3.3 for comparison. The mixing enthalpies for the bcc random solutions below 40% Ni and the ordered structures predicted from CALPHAD and the present study are similar. But the CALPHAD results of fcc random solid solutions are systematically lower than those from DFT, especially in the concentrated concentration range.

Comparison of $\Delta H_{\text{mix}}$ of fcc random solid solutions can also be made with the recent EAM [60] and MCE [56] studies, in which the magnetic ground state of fcc Fe is taken as the reference. Thus a change of the reference of Fe for our results is necessary before direct comparison. The relation between the mixing enthalpies using fct AFD and bcc FM Fe as the reference state is:

$$\Delta H_{\text{mix}}^{\text{fct-Fe}} = H_{\text{mix}}^{\text{bcc-Fe}} - x_{\text{Fe}}(H(\text{fct-Fe}) - H(\text{bcc-Fe})) \qquad (3.2)$$

where $\Delta H_{\text{mix}}^{\text{fct-Fe}}$ and $\Delta H_{\text{mix}}^{\text{bcc-Fe}}$ are respectively the mixing enthalpies taking the fct AFD and bcc FM phases as the reference state of Fe, and $H(\text{fct-Fe})$ and $H(\text{bcc-Fe})$ are the enthalpies of the corresponding phases of Fe.

The calculated mixing enthalpies with fct AFD Fe as the reference are compared with Ref. [56, 60, 132] in Fig. 3.4. Our DFT curve lies between those from the EAM potential and the CALPHAD studies. They show a similar shape, namely concave below 20% Ni and convex above, indicating a change of magnetic order. The curve from the MCE study [56] is rather symmetric with respect to the concentration and does not seem to reflect the change of magnetic order near pure Fe. We note that no *ab initio* data of random alloys were used for the parametrization procedure in these studies [56, 60, 132]. The discrepancy between the present DFT results and the previous predictions thus suggests the need to take into account reliable first principles results of random solutions for an improved prediction .

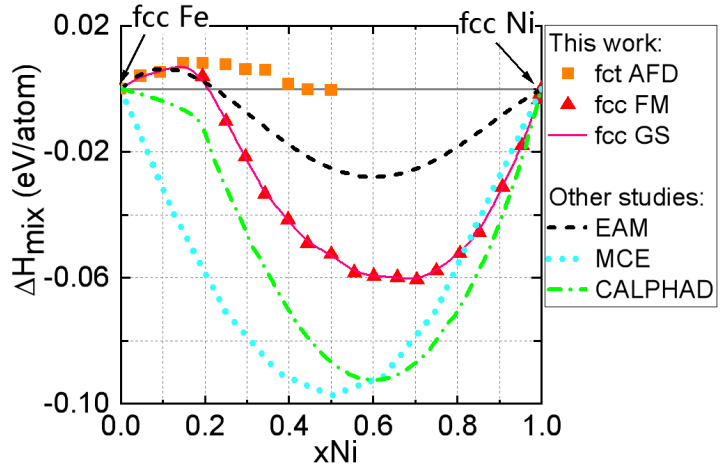

FIGURE 3.4: Comparison of mixing enthalpies of fcc disordered structures between this work and previous calculations (EAM [60], CALPHAD [132], MCE [56]). The reference state of Fe is the fct AFD phase in this study, and the fcc antiferromagnetic state in other approaches [56, 60, 132]. The solid line represents the ground-state mixing enthalpies of fcc disordered structures by DFT.

### 3.2.3    Volume and magnetic moment

Thermodynamic modelling often requires the knowledge of atomic volume [155] and magnetic moment of a given system [133]. Such data were computed from first principles calculations only for bcc FM Fe-Ni random alloys [133], but they are unavailable for fcc Fe-Ni alloys in literature. In this subsection, we present and discuss the atomic volumes and magnetic moments as well as the magneto-volume correlation in the bcc and fcc random alloys.

The average atomic volumes in the SQSs are shown in Fig. 3.5(a). Deviation from Vegard's law is observed in both bcc FM and fct AFD structures. Although pure bcc FM Ni has a smaller Voronoi volume than bcc FM Fe, the volume of bcc SQSs increases upon the addition of Ni, and reaches a maximum at about 15% Ni before decreasing. This is consistent with experiments, in which $a_0$ increases with Ni concentration up to 6% Ni, the maximum concentration of available data [156]. The volume of fct AFD SQSs remains effectively unchanged in the studied range of concentration.

In contrast, the atomic volumes of fcc FM SQSs decrease with increasing Ni content across the whole composition range. We find that the variation of lattice parameter $a_0$ versus Ni concentration is almost linear in fcc FM SQSs, with a slope of -0.011 Å per 10% Ni. The measurements in the fcc alloys at 298 K [157] confirm that $a_0$ above 39% Ni also has such a linear dependence on concentration with a similar slope (-0.012 Å per 10% Ni), though the magnitudes of $a_0$ are systematically larger than our results due to the thermal expansion. The experimental $a_0$ then decreases linearly from 39% to 30% Ni because the experimental samples are no longer homogenously FM [157].

Fig. 3.5(b) reveals more details as to the concentration dependence of the average Voronoi volumes of Fe and Ni. Overall, the variation in the average volumes of Fe and Ni follows the same trend as in the average atomic volume. We note that in the bcc FM SQSs, Ni has a larger Voronoi volume than Fe, while pure bcc FM Ni has a smaller atomic volume than pure bcc FM Fe.

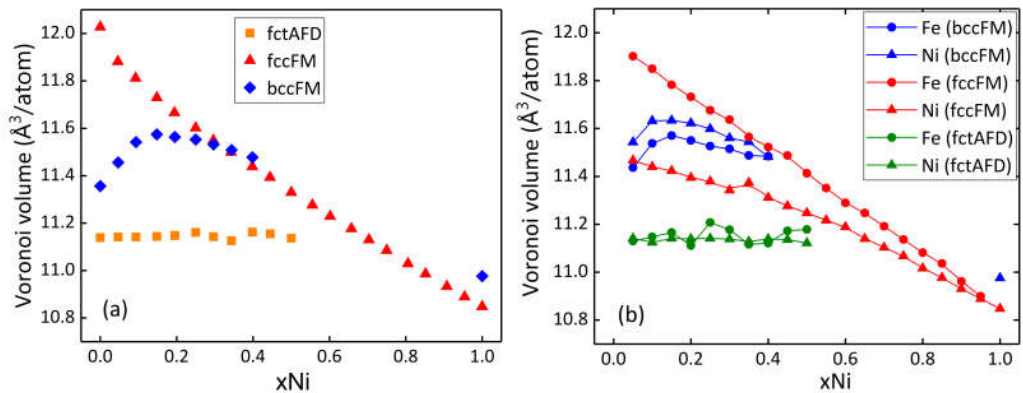

FIGURE 3.5: Average (a) atomic volume and (b) Voronoi volumes of Fe and Ni in the SQSs.

The magnetization results from the present calculations and the experimental measurements [153, 158] are compared in Fig. 3.6. Experimentally, the magnetization at 0 K is extrapolated from a series of measurements of spontaneous magnetization at several temperatures. Our DFT results agree well with the experimental data in bcc alloys, with the maximum magnetization at 10% Ni well reproduced.

The situation is more complicated in fcc alloys due to the change of magnetic order. According to our DFT results, the magnetic ground state of fcc alloys is FM and AFD above and below 18% Ni, respectively. Therefore, the fcc magnetization curve predicted by the present study (red line in Fig. 3.6) consists of the FM and AFD branches. The discrepancy between the DFT and experimental data mainly comes from the frontiers between the FM and AFD ordering tendencies (between 20% and 40% Ni). Indeed, the magnetic structures in the Invar Fe-Ni structures cannot be well described by the FM or AFD structures, and they will be further discussed in the next section.

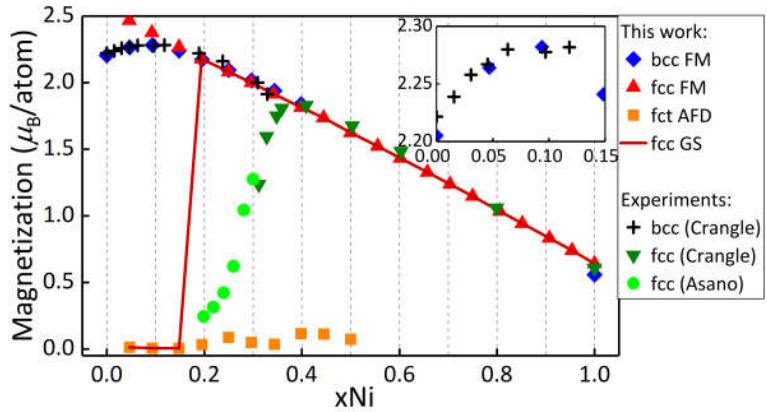

FIGURE 3.6: Magnetization of the present DFT calculations and the measurements by Crangle and Hallam [153] and Asano [158]. The inset displays the enlarged region below 15% Ni for the bcc alloys. The solid line represents the ground-state magnetization of fcc disordered structures.

The variation of magnetization versus concentration in the FM SQSs can be due to the magnetic dilution effect, since the magnetic moment of Fe is larger than that of Ni. It can also be related to the concentration dependence of the Fe and Ni moments. To see the relative effects of the two factors, we show in Fig. 3.7 the magnetic moments per atom and per species in the FM SQSs. In fcc FM SQSs, the concentration dependence of Fe and Ni moments is not significant, compared with the difference between Fe and Ni moments. Therefore, the magnetic dilution effect is dominant, resulting in an almost linear decrease in the magnetization with Ni concentration. In bcc FM SQSs, however, the strong increase in the Fe moment before 10% Ni dominates, leading to a maximum of the magnetization. Above 10% Ni where the increase in the Fe moments becomes weaker, the magnetic dilute effect becomes dominant and the magnetization in bcc FM SQSs decreases with Ni concentration again.

In the following, we discuss about the correlation between magnetism and volume. Nonlinear variations of volume with concentration has been reported in bcc ordered Fe-Ni structures and associated with the nonlinear variation in magnetization [4]. Our results in Fig. 3.5(a) and Fig. 3.7(a) confirm such a correlation in the bcc

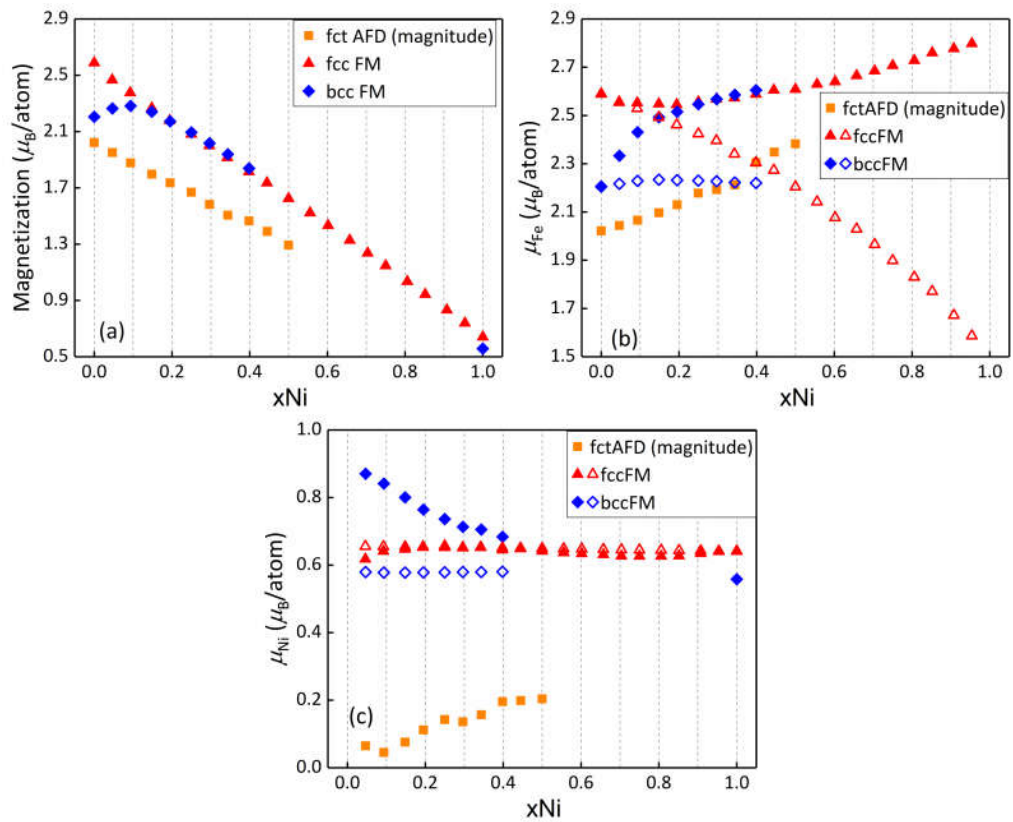

FIGURE 3.7: Average magnetic moments of (a) all atoms, (b) Fe atoms and (c) Ni atoms in the FM SQSs. The average magnetic-moment magnitudes in the AFD SQSs are also shown for comparison. In figures (b) and (c), the filled symbols represent the magnetic moments in the SQSs, while the unfilled ones are the magnetic moments of pure Fe and Ni obtained with the Voronoi volumes in the SQSs.

SQSs. However, we note that the local magnetic moments of Fe and Ni do not necessarily follow the same trend as the Voronoi volumes: in the bcc SQSs, the Voronoi volumes of Fe and Ni reach a maximum at 10% Ni (Fig. 3.5(b)), while the respective average magnetic moments vary monotonously (Fig. 3.7(b) and (c)).

Indeed, the magnetic moment of an atom depends not only on its Voronoi volume but also on the local composition. To separate these two types of effects, we calculate the magnetic moments of pure Fe and Ni using their respective Voronoi volume in the FM SQSs. These results are shown as unfilled symbols in Fig. 3.7. The magnetic moments of Fe in the bcc and fcc FM SQSs are much larger than the ones of pure Fe under the same Voronoi volumes. Consequently, the concentration dependence of the Fe moments in the SQSs is dictated by the local composition rather than the magneto-volume effect. The situation is similar for Ni in bcc FM alloys, whereas the magnetic moment of Ni in fcc FM alloys shows a very weak concentration dependence.

### 3.2.4 Magnetization in Invar alloys

In the previous subsection, we have shown that the computed magnetization agree well with the experimental data except near the Invar region. This discrepancy is related to the fact we have considered only AFD and FM structures, which may not be sufficiently representative for the magnetic structures of Fe-Ni Invar alloys. Indeed, the decrease in the experimental magnetization of fcc Fe-Ni alloys in the Invar region is associated with the antiferromagnetic clusters [153, 158]. Previous calculations in fcc alloy with 35% Ni [142, 159] show that the spin of Fe with few Ni neighbors is antiparallel to the global magnetization. Other non-collinear configurations have also been proposed [140]. Here, we consider some locally antiferromagnetic (LAF) structures with 25% to 35% Ni in order to approach the experimental magnetization values. The structures are constructed based on the atomic configurations of the fcc FM SQSs, by flipping the spins of the Fe atoms with few Ni nearest neighbours. Then the magnetic structures and the atomic positions are allowed to relaxed at fixed volumes.

The magnetization values of the LAF structures are presented in Fig. 3.8. For a given composition, we have investigated the LAF structures with different number of flipped Fe spins. Magnetization is found to decrease with increasing number of flipped Fe spins as expected. The experimental magnetization is covered in the range of calculated values. We find that the energy of the LAF structure increases with the number of flipped Fe spins. Therefore, our results show that the LAF state is less stable than the FM state for fcc Fe-Ni alloys with more than 25% Ni, consistent with previous GGA calculations [37, 142]. On the other hand, previous LDA calculations showed that the magnetic ground state in these structures consists of a small amount of atoms with noncollinear or antiparallel moments [140, 142, 159]. Indeed, Ruban *et al.* [142] showed that the fcc magnetic ground state at 35% Ni at the experimental $a_0$ is predicted to be LAF by LDA but FM by GGA.

In order to verify the effect of exchange-correlation functionals, we perform the same DFT calculations in the FM and LAF structures using the LDA functional. In the LDA calculations, only the atomic positions are optimized, while the equilibrium lattice parameters $a_0$ obtained from the GGA calculations are kept, due to the fact that the LDA functional strongly underestimates the equilibrium $a_0$ of Fe alloys. Our LDA calculations show that the LAF state is energetically more stable than the FM state from 25% to 34% Ni. The magnetization obtained in these LDA calculations is slightly smaller than the value in the same structures in the GGA calculations (Fig. 3.8). However, the LDA results should be treated with caution because LDA fails to predict not only the aforementioned equilibrium $a_0$ but also the strutural and magnetic ground state of Fe [160, 161].

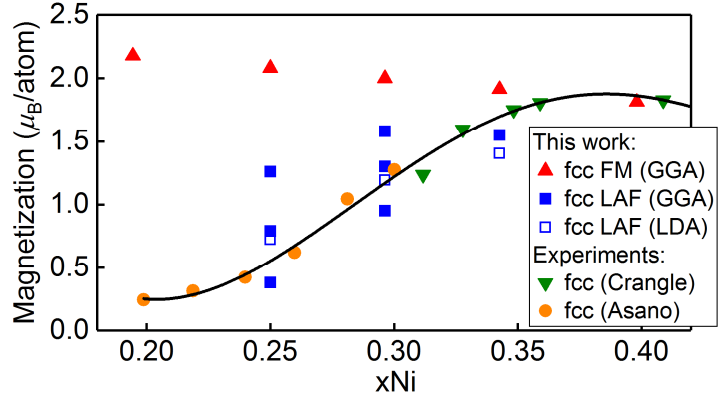

FIGURE 3.8: Magnetization of the fcc FM and LAF structures in the Invar region from the present DFT study, compared to experimental data by Crangle and Hallam [153] and Asano [158]. The black line guides for the experimental trend. The magnetization of the LAF structures calculated from LDA is also shown for comparison, being slightly below the value calculated from GGA. Only one LAF structure for each concentration is recalculated using LDA.

Therefore, our results in the fcc Fe-Ni Invar alloys show that the experimental magnetization values at 0 K can be reproduced using the LAF structures, but the conclusion of the stability of these structures is very dependent on the adopted exchange-correlation functionals. Furthermore, there may be other magnetic structures with similar energy and magnetization but different from the proposed LAF configurations. In particular, the choice of the set of Fe sites with a flipped spin is not unique. Indeed, it was shown in Ref. [37] that there are a large number of magnetic configurations close in energy in Fe-rich Fe-Ni structures in small volumes, resulting in large magnetic entropy and leading to the Invar effects.

## 3.3  Phase stability prediction from DFT

The phase stability under constant pressure is governed by the Gibbs free energy of mixing $\Delta G_{mix}$, which is related to the mixing enthalpy $\Delta H_{mix}$ and the entropy of mixing $\Delta S_{mix}$ as follows:

$$\Delta G_{mix} = \Delta H_{mix} - T\Delta S_{mix} \qquad (3.3)$$

where $\Delta H_{\mathrm{mix}}$ and $\Delta S_{\mathrm{mix}}$ are calculated with respect to bcc Fe and fcc Ni at the same temperature. Approximations are needed to evaluate $\Delta G_{\mathrm{mix}}$. In the first place, we assume $\Delta H_{\mathrm{mix}}$ to be temperature independent, i.e. equal to the ground-state $\Delta H_{\mathrm{mix}}$. The entropy term can have various sources such as the electronic contribution, magnetism, chemical configuration and lattice vibration. The electronic entropy is usually small [143]. The magnetic entropy is expected to be less important in this case, since we focus on the phase stability below the magnetic transition temperatures. Both the electronic and magnetic contributions are consequently ignored in the rest of this section.

The configurational entropy is zero for the perfectly ordered structures, while it is significant for the random alloys. The ideal configurational entropy $\Delta S_{\mathrm{conf}}^{\mathrm{ideal}}$ is often used to approximate the true configurational entropy of disordered alloys, and thus will be used in our estimation of $\Delta G_{\mathrm{mix}}$. As highlighted by an EAM potential study [60], lattice vibration can be another important contribution to entropy and is considered here for the FM alloys.

In the following subsections, we first present the results of the vibrational entropies. From them, we estimate $\Delta G_{\mathrm{mix}}$ and construct the phase diagrams. Finally, we discuss the vibrational effects on the order-disorder transition temperatures.

### 3.3.1 Vibrational entropy

The vibrational properties of the Fe-Ni systems are less studied than ground-state energetic properties. This could be partly due to the expensive computational cost. In the present approach, for example, there are 324 static calculations to extract force constants for a 54-atom SQS. The high computational cost limits the number of studied phases. Thus it is worth discussing which phases are most of our interest. Clearly, the two reference states, bcc FM Fe and fcc FM Ni, and the two ordered structures found in experiments, $L1_2$-FeNi$_3$ and $L1_0$-FeNi, should be considered in the first place. The rest of the ordered structures, though some have negative mixing enthalpies, are shown to lie above the convex hull and thus are not thermodynamically stable. For the disordered structures, we show that the most stable structure is bcc FM and fcc FM below and above 37% Ni respectively, consequently they are the most relevant disordered structures at finite temperatures.

Fig. 3.9 shows the vibrational entropies of mixing $\Delta S_{\mathrm{vib,mix}}$ of ordered and disordered structures as functions of temperatures. Over the whole temperature range, $\Delta S_{\mathrm{vib,mix}}$ is found to be positive for all the studied structures and saturate above 300 K.

Vibrational entropies of mixing of ordered and disordered structures at 300 K are plotted as functions of Ni concentration in Fig. 3.10. The ideal configurational entropy $\Delta S_{\mathrm{conf}}^{\mathrm{ideal}}$ is also presented for comparison, which has the following form:

$$\Delta S_{\mathrm{conf}}^{\mathrm{ideal}} = -k_{\mathrm{B}}(x_{\mathrm{Fe}}\ln x_{\mathrm{Fe}} + x_{\mathrm{Ni}}\ln x_{\mathrm{Ni}}) \tag{3.4}$$

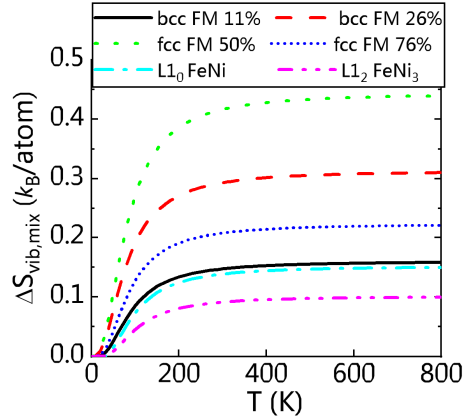

FIGURE 3.9: Vibrational entropies of mixing of ordered and disordered structures as functions of temperatures. Only two selected curves of fcc FM SQSs are shown for the sake of simplicity.

$\Delta S_{\mathrm{conf}}^{\mathrm{ideal}}$ is symmetric with respective to the Ni content, and reaches the maximum of 0.69 $k_B$/atom at 50% Ni. For the disordered structures, $\Delta S_{\mathrm{vib,mix}}$ is of same order of magnitude of $\Delta S_{\mathrm{conf}}^{\mathrm{ideal}}$ and thus is nonnegligible. The fcc disordered structures have larger $\Delta S_{\mathrm{vib,mix}}$ than the ordered ones, suggesting $\Delta S_{\mathrm{vib,mix}}$ should stabilize the disordered phases over ordered ones.

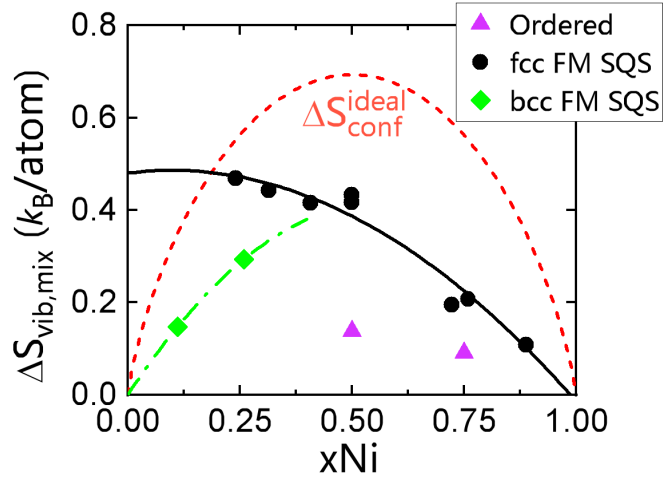

FIGURE 3.10: Vibrational entropies of mixing of ordered and disordered structures as functions of Ni concentrations at 300 K. The ideal configurational entropy is plotted for comparison. The solid and dash-dotted lines are the second order polynomial fits of the fcc and bcc DFT data, respectively.

Recently, Lucas *et al.* [162] obtained the partial phonon densities of states (DOS) of Fe and Ni, defined as the contribution from the given species to the total phonon DOS, at 300 K in fcc Fe-Ni samples through inelastic neutron scattering and nuclear resonant inelastic x-ray scattering. The experimentally derived partial and total vibrational entropies are compared to the DFT values in Fig. 3.11. The overall agreement is reasonably good: the differences between the calculated and measured values for the partial $S_{vib}$ of Fe and Ni and total $S_{vib}$ are less than 0.07, 0.11 and 0.05 $k_B$/atom, respectively. Both our work and the experiment [162] confirm that the chemical ordering reduces $S_{vib}$, but the change in $S_{vib}$ due to the L1$_2$-disorder transition is smaller in the experiment than what our DFT results suggest. Indeed, the ordered samples obtained in the experiment present only a small atomic long-range order parameter (ALRO=0.37) and does not correspond to a fully equilibrated ordered state (ALRO close to 1) [162]. The actual $S_{vib}$ of the perfectly L1$_2$ phase can be further reduced and the actual difference between the perfectly L1$_2$ and the disordered phases would be larger than the experimental difference.

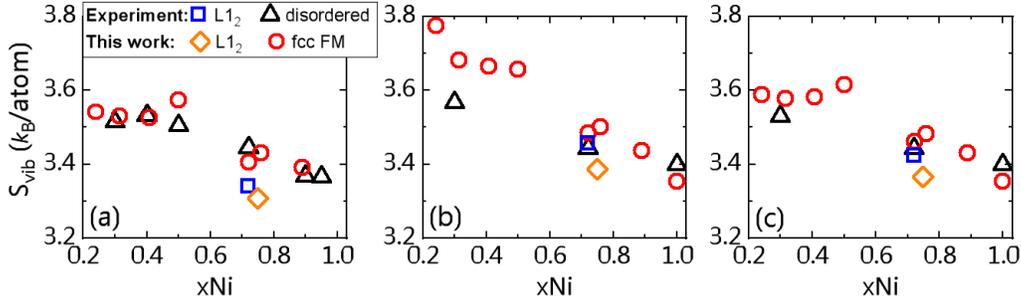

### 3.3.2 Free energy of mixing

We evaluate $\Delta G_{mix}$ using the following form:

$$\Delta G_{mix} = \Delta H_{mix}^{GS} - T(\Delta S_{conf}^{ideal} + \Delta S_{vib,mix})  \qquad (3.5)$$

Here $\Delta H_{mix}^{GS}$ is the ground-state mixing enthalpy, $\Delta S_{conf}^{ideal}$ is equal to the ideal configurational entropy of mixing for the bcc and fcc FM SQSs and zero for the ordered structures, and $\Delta S_{vib,mix}$ is the vibrational entropy of mixing obtained in the last section.

Thermodynamic data from experiments in Fe-Ni alloys were obtained at temperatures generally higher than 1000 K [132], while such data at lower temperatures are often assessed through empirical approaches such as CALPHAD. In a recent CALPHAD modeling of the Fe-Ni system by Cacciamani *et al.* [132], the mixing enthalpies and the activity coefficients of the fcc PM phase fit well with high-temperature experimental results. However, the CALPHAD prediction is not ensured to be accurate for the low-temperature bcc and fcc FM phases, and should be further validated with other methodologies.

Fig. 3.12 shows a comparison between our DFT results and the CALPHAD prediction for the free energies of mixing of bcc and fcc random solutions. For the bcc phase, both methods predict similar $\Delta G_{mix}^{bcc}$ below 6% Ni, the maximum solubility of

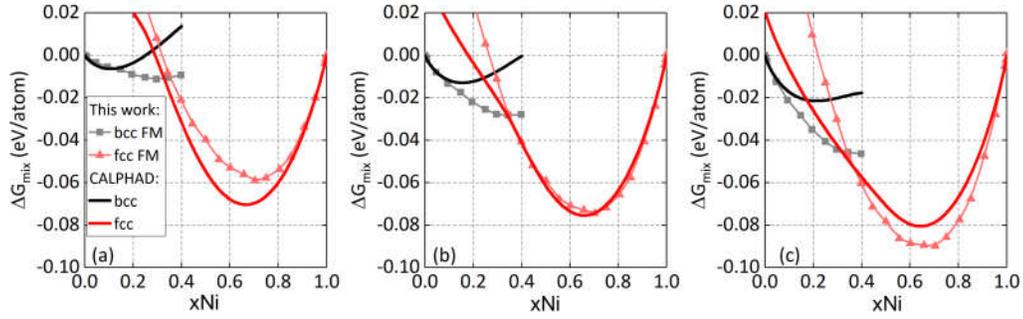

FIGURE 3.12: Comparison of $\Delta G_{\text{mix}}$ of random solutions at (a) 300 K (b) 500 K and (c) 700 K between this study and the CALPHAD assessment by Cacciamani *et al.* [132]. The CALPHAD curves are derived from Ref. [132] using bcc Fe and fcc Ni as references.

Ni in bcc solid solutions [132], but the degree of agreement above 6% Ni decreases with increasing temperatures. As the CALPHAD model was fitted mostly on the thermodynamic data of the fcc phase, its description of the bcc phase is expected to be less accurate and needs to be revisited. For the fcc phase, the results are compared down to the bcc-fcc crossover predicted by DFT near 35% Ni. The best agreement between the CALPHAD and DFT results is observed at 500 K, while the CALPHAD and DFT results differ by at most 0.015 eV/atom at 300 and 700 K. The reason why the best agreement in $\Delta G_{\text{mix}}^{fcc}$ is achieved at 500 K may be understood as follows. Since the CALPHAD assessment mainly utilized experimental data above 1000 K, its prediction is more reliable at high temperatures, while our DFT results are more reliable at low temperatures. Therefore, the CALPHAD results may be more reliable at 700 K, where the magnetic effect neglected in this work becomes non-negligible. However, our approximations are well grounded at lower temperatures, while the extrapolation of CALPHAD from high temperatures above 1000 K is more questionable. Therefore, we believe our DFT calculations give more accurate results at 300 K. In the intermediate temperature range, the two approaches are both valid and agree with each other, which may be seen rather as a cross-validation between the two methods.

### 3.3.3   Complete bcc-fcc Fe-Ni phase diagram

In this subsection, we compare the DFT-predicted phase diagrams with the CALPHAD ones by Cacciamani *et al.* [132] and experimental data. Experimentally, it is not completely clear whether L1$_0$-FeNi is stable [123] or metastable [122], although the present and the previous first principles calculations [4, 130, 131, 134] support the former. Consequently, we first discuss the phase diagrams including only the experimentally well-established phases, namely without L1$_0$-FeNi. Then, we propose a phase diagram including all the theoretically predicted stable phases (including the L1$_0$-FeNi phase), and discuss it in the light of the available experimental data. By comparing the two phase diagrams, we can also gain insights into how L1$_0$-FeNi affects the phase diagram.

The following notations are used for the convenience of discussion. The bcc random solid solution with low Ni content, FM below 1000 K, is termed $\alpha$; $\gamma$ represents the fcc random solid solution, with $\gamma_{PM}$ and $\gamma_{FM}$ specifying the magnetic order; $L1_2$ (resp. $L1_0$) denotes the stoichiometric and off-stoichiometric $L1_2$-FeNi$_3$ (resp. $L1_0$-FeNi) ordered phase. The phase boundary separating a two-phase region $A + B$ and a single-phase region $B$ is noted as $A + B / B$.

We apply the common tangent method [56, 163] to construct the phase diagrams, based on the free energies of mixing of the considered phases. At each temperature, the equilibrium concentrations of the ordered and disordered phases are determined as the points of contact between the free-energy-versus-concentration curves of the two phases and their common tangent.

For the stoichiometric $L1_2$ and $L1_0$ phases, we use the perfectly ordered structures as in many other theoretical studies [56, 164, 165]. For the off-stoichiometric $L1_0$ and $L1_2$ phases (in the range of 40-60% and 60-90% Ni, respectively), we consider the presence of antisites to account for the deviation from the stoichiometric ordered structures, as in the CALPHAD and CVM approaches [132, 133, 136, 137]. For example, for the ordered structures below the stoichiometric Ni concentrations, we assume that the Fe sublattice is fully occupied by the Fe atoms, while the excess Fe atoms are distributed as antisites on the Ni sublattice. With this assumption, however, there are still a large number of possible distributions of the antisites. For simplicity, we considered two relevant types of the antisite distributions: (1) the antisites are randomly distributed; (2) the pairs consisting of two 1NN antisites are randomly distributed. Such arrangements of atoms allow to have a strong long-range order (LRO) and a short-range order (SRO) close to that of the stoichiometric ordered structures. For example, for the generated off-stoichiometric $L1_0$ structure with 60% Ni, its SRO parameters for the first two shells and the LRO parameter are respectively [-0.25, 0.72] and 0.8, compared with [-0.33, 1.00] and 1 for the stoichiometric L10-FeNi, and [0.00, 0.00] and 0 for the SQS at the same concentration. Also, the first type of the distributions is chosen since we are particularly interested in a good description around the order-disorder transitions, and it exhibits the highest configurational entropy within the current assumption. And the second type is considered as they have a lower energy for certain concentrations than the first-type structures, while still having a large configurational entropy. The free energies of mixing of these two types of off-stoichiometric ordered structures are calculated from DFT. Then, the configurations with a lower free energy were used to represent the off-stoichiometric phases at a given concentration: the configurations of the first type have a lower free energy for the $L1_0$ phase below 50% Ni, and for the $L1_2$ phase below and above 75% Ni, while the configurations of the second type are energetically more favourable for the $L1_0$ phase above 50% Ni.

**Phase diagram excluding** L1$_0$

Fig. 3.13 shows the DFT-predicted phase diagram without the L1$_0$ phase compared to the CALPHAD prediction by Cacciamani *et al.* [132] and the representative experimental data. Comprehensive experimental references can be found in Ref. [120, 132].

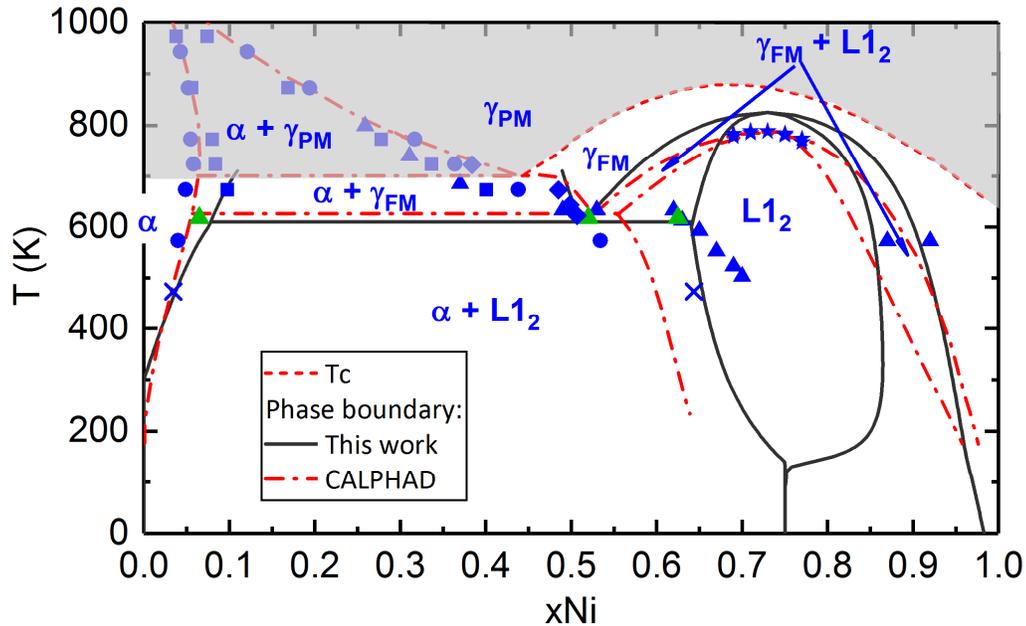

FIGURE 3.13: Comparison of the phase diagrams (without L1$_0$) from the present DFT study, the CALPHAD prediction by Cacciamani *et al.* [132] and the experiments. Experiments: circle [166], triangle [167], star [121], cross [122], square [168], diamond [169]. The shaded zone is the phase regions involving the equilibrium with $\gamma_{PM}$. The triangles in green mark the eutectoid equilibrium. The $T_{Curie}$ curve is from the CALPHAD prediction [132], which agrees well with experimental data summarized in Ref. [170].

Experimentally, the phase equilibria were investigated using samples cooled under laboratory conditions [121, 166–169] and meteorites [122]. It was concluded in Ref. [167] that $\gamma_{FM}$ decomposes eutectoidally into $\alpha$ and L1$_2$ at 618 K (triangles in green in Fig. 3.13), which was widely accepted in the construction of the phase diagrams [120, 122, 132, 171, 172]. The study on meteorites [122] concluded that $\alpha$ equilibrates with L1$_2$ at 473±50 K, but the equilibrium Ni concentration of L1$_2$ is lower than those from Ref. [167]. In another study [166], no ordered structure was detected even at 573 K. The authors therein [166] suggested that the ordered phases may not be easily detected by electron diffraction techniques or may be kinetically restricted in their formation. The above results indicate that there is experimental uncertainty related to the phase boundaries with the ordered structures below 600 K. .

It can be seen from Fig. 3.13 that the major difference between CALPHAD and the experimental results comes from the phase boundaries below 700 K. In particular, the Ni concentration of L1$_2$ of the eutectoid decomposition from CALPHAD is 55.5%, 7% Ni lower than the experimental value.

The DFT-predicted phase diagram is in overall good agreement with the experimental data. The major discrepancy is that the predicted maximum of $T_c^{L1_2}$ is 40 K higher than the measured value of 790 K [121]. We predict the eutectoid reaction occurs at 610 K between $\alpha$ with 7.8% Ni, $\gamma_{FM}$ with 51.6% Ni and L1$_2$ with 64% Ni, in good agreement with Ref. [167] (triangles in green). Compared with the CALPHAD prediction below 700 K, our predicted phase boundaries of the $\alpha$+L1$_2$ and $\gamma_{FM}$ + L1$_2$ two-phase regions agree better with the measurements in Ref. [122, 167, 169]. We note the boundary L1$_2$/L1$_2$+$\gamma_{FM}$ below 500 K is quite different between DFT and CAL-PHAD: the former predicts a retrograde solubility of Ni in L1$_2$ with a maximum of 86% Ni, whereas in the latter the solubility of Ni in L1$_2$ tend to increase with decreasing temperatures.

We note that the $\alpha$+$\gamma_{PM}$ equilibrium (shaded area in Fig. 3.13) are absent in the DFT phase diagram, because the magnetic disordering effects are not considered here.

Though magnetic excitations can occur below $T_{Curie}$, their effects on the phase boundaries between $\gamma_{FM}$ and L1$_2$ are expected to be minor because experimentally $T_{Curie}$ are about 100 K higher than the order-disorder transition temperatures. Indeed, the magnetic disordering/ordering effects on phase stability can be elucidated by comparing the DFT-predicted and experimental phase diagrams. According to the DFT-predicted phase diagram, without the magnetic disorder the $\gamma_{FM}$ single-phase region would be smaller and $T_c^{L1_2}$ would be higher. The latter is consistent with the finding that an extremely high external magnetic field leads to an increase in $T_c^{L1_2}$ [25]. Our results also indicate that without magnetic disorder the solubility of Ni in $\alpha$ increases with increasing temperatures, while the experimental data exhibit a retrograde solubility with a maximum at $\sim$ 700 K, close to $T_{Curie}$ of $\gamma_{FM}$ at the equilibrium concentration. This suggests that the experimental retrograde solubility in $\alpha$ is related to the emergence of magnetic disorder in $\gamma_{FM}$.

The above effects can be understood as follows: since magnetic disorder is more significant for the phase with lower $T_{Curie}$ than the one with higher $T_{Curie}$ at the same temperature, the magnetic entropy stabilizes $\gamma_{FM}$ with respect to both $\alpha$ and L1$_2$ as the experimental $T_{Curie}$ of $\gamma_{FM}$ is lower than the other two phases [121, 173, 174].

**Phase diagram including L1$_0$**

The L1$_0$ phase is known to exist in meteorites [122, 123, 175, 176], and can be obtained by annealing with various special techniques to promote atomic diffusion [177–179]. It was concluded as a stable phase by Reuter *et al.* [123] who found only the equilibrium between $\alpha$ and L1$_0$ in meteorites but didn't observe L1$_2$. However, another meteorite study found that $\alpha$ equilibrates with L1$_2$ instead of L1$_0$ at 473±50 K [122] and thus concluded the L1$_0$ phase to be metastable. But the equilibrium concentration of L1$_2$ obtained in this study [122] is inconsistent with those found in Ref. [167] (see triangles and crosses in Fig. 3.14). The lack of a clear conclusion from meteorite studies is due to the lack of true equilibria in meteorites even after more than $10^8$ years of slow cooling ($< 1$ atomic jump per $10^4$ years at 573 K) [122, 123]. It is shown that under irradiation the L1$_0$ ordering occurs in the initially disordered alloys [124–126]. It is argued that the obtained L1$_0$ phase is not a metastable phase induced by irradiation effects, but rather a stable phase whose formation rate is enhanced by irradiation effects [125, 126].

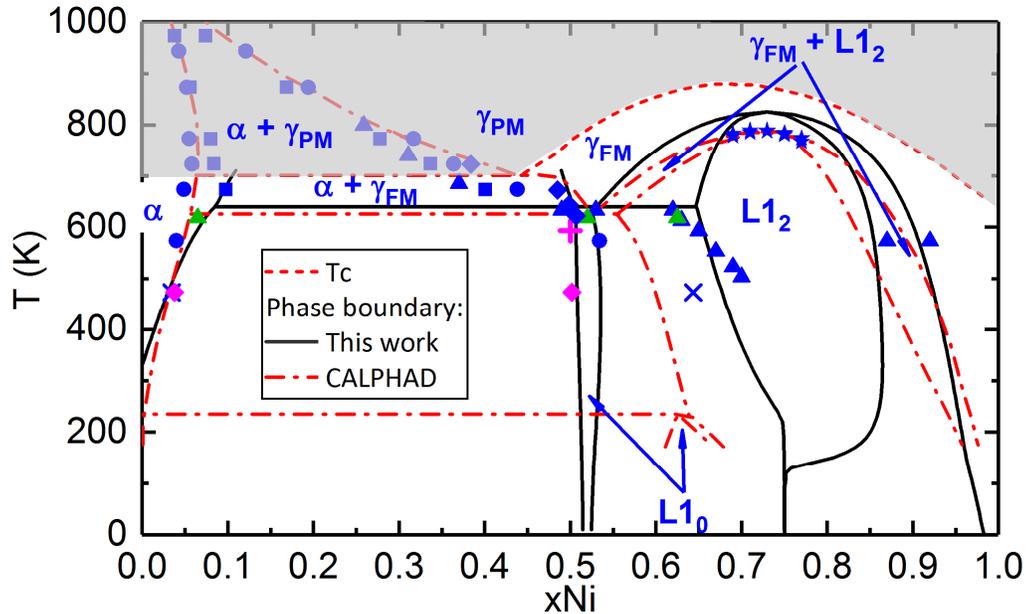

FIGURE 3.14: (Color online) Comparison of the phase diagrams (with L1$_0$) from the present DFT study, the CALPHAD assessment by Cacciamani *et al.* [132] and the experiments [121–124, 126, 166–169]. Diamonds are the lower and upper boundaries of the $\alpha$+L1$_0$ two-phase region from the measurements in meteorites [123]. The plus symbol denotes $T_c^{L1_0}$ obtained from the irradiation experiments [124, 126]. Other notations are the same as in Fig. 3.13.

On the theoretical side, the L1$_0$ phase is predicted to be stable in the recent first principles and phenomenological calculations [4, 130, 131, 134, 180] and is presented in the recent theoretical phase diagrams [60, 132, 168, 181]. Both this study and the CALPHAD assessment by Cacciamani *et al.* [132] suggest the L1$_0$ phase as a stable phase to be included in the phase diagram, as shown in Fig. 3.14.

According to our results, L1$_0$ is stable below $T_c^{L1_0}$ and equilibrates with $\alpha$ and L1$_2$ on the Fe-rich and Ni-rich sides, respectively. However, as the energy of L1$_0$-FeNi lies very close to the line connecting the energies of bcc Fe and L1$_2$-FeNi$_3$ (Fig. 3.3), its stability can be very sensitive to external perturbations (impurity in meteorites, irradiation, etc.) and may not be exactly determined. Therefore, the L1$_0$ phase, either slightly stable or slightly metastable, is highly relevant to be considered in the low-temperature Fe-Ni phase diagram.

For the DFT predicted phase diagram, the major change due to the inclusion of the L1$_0$ phase is the absence of the $\alpha$+L1$_2$ two-phase region. This is because the predicted $T_c^{L1_0}$ of 640 K is higher than the predicted eutectoid reaction temperature of 610 K. The disappearance of the $\alpha$+L1$_2$ equilibrium is expected because L1$_0$-FeNi is increasingly more stable than L1$_2$-FeNi$_3$ with increasing temperatures due to the larger vibrational entropy of mixing of the former (Fig. 3.10). Compared with Fig. 3.13, the predicted phase boundary $\alpha$+L1$_2$/L1$_2$ shifts closer to the experimental data of Ref. [167], while the phase boundary $\alpha/\alpha$+L1$_2$ is nearly unchanged.

The CALPHAD assessment predicts L1$_0$ decomposes into $\alpha$ and L1$_2$, but the decomposition occurs at a temperature less than half of the experimental $T_c^{L1_0}$. Furthermore, $T_c^{L1_0}$ determined from the CALPHAD phase diagram with only fcc phases is 311 K [132], still much lower than the measured $T_c^{L1_0}$ of 593 K. The low $T_c^{L1_0}$ from the CALPHAD prediction is due to the estimation of a lower 0 K mixing enthalpy for fcc disordered phases (Fig. 3.3).

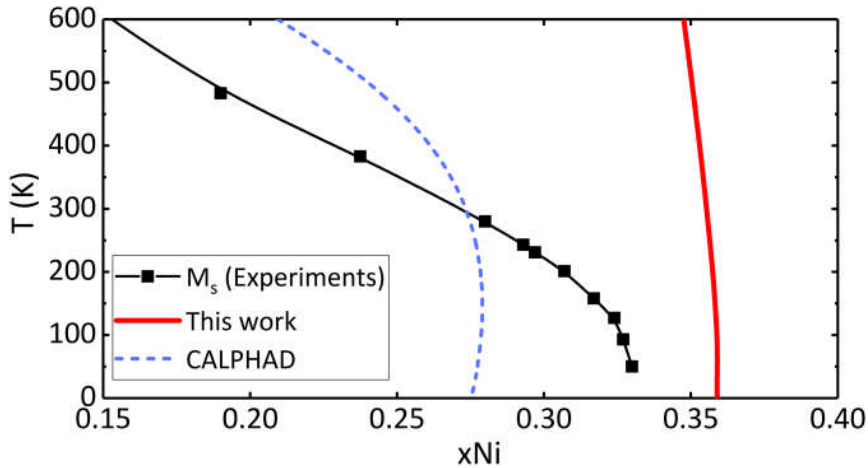

FIGURE 3.15: Comparison of the experimental martensitic start temperature curve $M_S$ [133], and the $T_0$ curves from our DFT calculations, the CALPHAD assessment by Cacciamani *et al.* [132].

As noted in Ref. [133], there is also another inconsistency in the CALPHAD assessment by Cacciamani *et al.* [132] at low temperatures. In Fig. 3.15, we show the experimental martensite start temperature $M_S$ at which martensite starts to form, and theoretical $T_0$, the temperature at which bcc and fcc phases with the same composition have the same $\Delta G_{mix}$. For a given composition, the martensitic transformation starts below $T_0$ so that $\Delta G_{mix}^{fcc}$ is lower than $\Delta G_{mix}^{bcc}$ to provide a driving force to initiate nucleation. Therefore, the $M_S$ curve should be always below the predicted $T_0$ curve. However, this is clearly not the case for the CALPHAD assessment, while the $T_0$ curve by DFT, though seems to be too "stiff", is compatible with the $M_S$ curve.

These results show that low-temperature thermodynamic properties predicted by the present DFT study are in better agreement with existing experimental data than the CALPHAD model by Cacciamani *et al.* [132]. Meanwhile, very recently Ohnuma *et al.* [168] revised the thermodynamic parameters for the Fe-Ni system by taking into account the interaction effect between the chemical and magnetic orderings, based on their experiments in Fe-Ni alloys between 673 K and 973 K. As shown in Fig. 3.16, the predicted L1$_0$-disorder transition temperature is around 556 K, much closer to the experimental value of 593 K than the value predicted by Cacciamani *et al.* [132]. The difference between the phase diagrams by Ohnuma *et al.* [168] and Cacciamani *et al.* [132] concerns mainly temperatures below 600 K. We will discuss the CALPHAD assessment by Ohnuma *et al.* [168] and compare it with our EIM prediction in the next section.

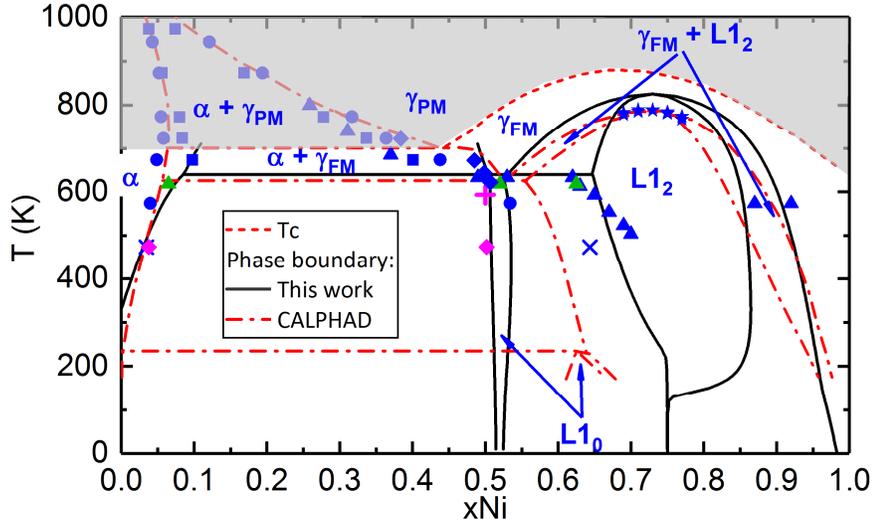

FIGURE 3.16: Fe-Ni phase diagram predicted from the most recent CALPHAD assessment by Ohnuma *et al.* [168]. The figure is taken from Ref [168].

### 3.3.4 Vibrational effects on the order-disorder transitions

The vibrational effects on the order-disorder transitions have been investigated in some theoretical studies. These results are compared with our results and experimental values in Table 3.3.

TABLE 3.3: Order-disorder transition temperatures. The results calculated with and without vibrational effects are shown.

| Method | Exp. | CVM [136] | | EAM [60] | | DFT [143] | | DFT (This work) | |
|---|---|---|---|---|---|---|---|---|---|
| | | with $S_{vib}$ | no $S_{vib}$ | with $S_{vib}$ | no $S_{vib}$ | with $S_{vib}$ | no $S_{vib}$ | with $S_{vib}$ | no $S_{vib}$ |
| $T_c^{L1_2}$ | 790 [121] | - | - | 990 | 1660 | - | - | 830 | 1030 |
| $T_c^{L1_0}$ | 593 [124] | 483 | 523 | 960 | 1800 | 560 | 1040 | 640 | 920 |

All these studies agree that the inclusion of $S_{vib}$ decreases $T_c^{L1_0}$ and $T_c^{L1_2}$, but disagree on the extent of the decrease. In the CVM study [136], $T_c^{L1_0}$ is already lower than the experimental value, and is further reduced by 40 K after introducing vibrational effects. It is noted that phase equilibrium calculated by CVM may not be in the equilibrium state as no local deformation was allowed in their calculations [136]. The EAM potential study [60] stresses the importance of vibrational entropy, but the estimated $T_c^{L1_0}$ with and without vibrational contributions are too high compared with the experimental data, so these results may only be taken qualitatively. It is noted that the EAM study agrees with our results that the reduction in $T_c^{L1_0}$ is larger than in $T_c^{L1_2}$ with the inclusion of vibrational effects. In a recent DFT study on the $L1_0$-disorder transition [143], the free energy, including the configurational, vibrational, electronic and magnetic contributions, is calculated as a function of long-range order. According to this study, the configurational and vibrational effects are the most important and vibrational effects lower $T_c^{L1_0}$ by 480 K compared to the value obtained including merely the configurational contribution. Our results also suggest that lattice vibrations reduce significantly the transition temperatures, with the amount of the decrease in $T_c^{L1_0}$ lying between the results of Ref. [136] and [143].

All the above theoretical results point out a significant role of $S_{vib}$ on the prediction of the chemical order-disorder transition temperatures, though there is no quantitative agreement. Indeed, the predicted chemical order-disorder transition temperatures can be very sensitive to the approximations involved in different approaches. Consequently, it is worth recalling the approximations and possible sources of errors in the present study. Indeed, $\Delta H_{mix}$ is assumed to be constant, while its variation may be non-negligible near the phase transition temperatures according to the CALPHAD prediction by Cacciamani *et al.* [132]. The configurational entropy is calculated only at the low (ordered structures) and high (disordered structures) temperature limits, while the true configurational entropy varies continuously between the two limits as the temperature changes. In addition, vibrational entropy is calculated within the harmonic approximation under constant volume and with perfectly FM order. Furthermore, electronic and magnetic entropies are neglected in this study. Based on these simple assumptions, our predicted phase diagrams below $T_{Curie}$ and our order-disorder transition temperatures are in reasonable agreement with available experimental data, suggesting that the principal sources of entropy are captured, although some possible compensation of errors cannot be totally excluded. A reliable description of various contributions may require very accurate energetic models. On the experimental side, the measurements by Lucas *et al.* [162] do not allow to draw definite conclusions on the vibrational effects because the experimental ordered sample was only partially ordered. Therefore, further experimental and theoretical efforts are needed to elucidate the effects of lattice vibrations.

## 3.4 Phase stability prediction from effective interaction model

In the previous section, we have discussed the phase stability of the bcc and fcc Fe-Ni alloys and constructed a bcc-fcc phase diagram, based directly on our DFT results. In this section, we discuss the phase stability prediction in the fcc Fe-Ni alloys using the effective interaction model (EIM) parametrized on the DFT results. Combined with on-lattice Monte Carlo (MC) simulations, the EIM allows to take into account the effects of magnetic excitations and transitions. It is worth pointing out the advantages and disadvantages of the DFT and the EIM+MC approaches. Compared with the EIM+MC approach, the DFT approach used in the previous section allows a direct evaluation of phase stability and the construction of a bcc-fcc phase diagram.  However, only magnetically ordered structures are considered in this approach, therefore neglecting the effects of magnetic excitations on phase stability.  In addition, it is also difficult to study the magnetochemical interplay using only DFT calculations.  Meanwhile, the EIM+MC approach enables to fully address the magnetic effects and the magnetochemical interplay.  However, it is constrained to the fcc lattice and thus can only be used to study the phase equilibria between the fcc phases.  Also, the mapping from DFT to the EIM involves certainly some inaccuracies.

In the first subsection, we demonstrate the accuracy of the present EIM.  In the second subsection, we present the fcc Fe-Ni phase diagram with a comparison to the DFT and CALPHAD predictions.  Finally, we study the impact of magnetochemical interplay on the phase stability.

We recall that three types of MC simulations are performed for obtaining the equilibrium properties.  The spin MC (SMC) simulations, with only the spins evolving with temperature, are used to obtain the equilibrium magnetic structure for a given chemical configuration.  The spin-atom canonical MC (CMC) simulations are used to obtain the equilibrium spin-atom configuration and the relevant properties for a given composition as a function of temperature.  The semi-grand canonical MC (SGCMC) simulations, which determine the equilibrium composition and spin-atom configuration for a given chemical potential difference, are used as a convenient approach to construct the phase diagrams.

### 3.4.1   Parametrization and accuracy of the EIM

In the following, we first compare the EIM predictions of the ground-state magnetic, energetic and vibrational properties with the DFT results, then we compare the predicted magnetic and chemical transition temperatures with the experimental data.

**Ground-state magnetic, energetic and vibrational properties**

As shown in Table 3.4, the energetic hierarchy of several collinear magnetic states of fcc Fe from the EIM is in good agreement with the DFT results.  The magnetic ground state of Fe predicted by the EIM is a spin spiral with $q = \frac{2\pi}{a}(0, 0, 0.3)$, which has a slightly lower energy than the AFD state.  Experimentally, the magnetic ground state of fcc Fe is a spin spiral with $q = \frac{2\pi}{a}(0.3, 0, 1)$ [138, 139, 152].  First principles calculations in non-collinear magnetic structures predict a spin spiral ground state, but with $q = \frac{2\pi}{a}(0, 0, 0.6)$ [3, 182, 183].  Within the collinear approximation, the ground state is the AFD state [3, 104, 151, 184], with $q = \frac{2\pi}{a}(0, 0, 0.5)$.

We have fitted the EIM parameters of fcc Fe on the DFT results of various magnetic states, in particular on the energy of the AFD state as a function of magnetic-moment magnitude (Fig. 3.17).  It can be seen that this energy landscape is not well captured by the recent magnetic cluster expansion (MCE) model [56] (noted as MCE2014 hereafter).  The MCE2014 model also gives an overestimated equilibrium spin magnitude for the AFD Fe. Note that it is crucial to well reproduce such an energetic landscape, because it dictates the longitudinal spin excitations at finite temperatures, which are significant in fcc Fe and Ni [41, 54, 185]. Furthermore, as shown in the next chapter, the spin magnitude also has a strong influence on the vacancy formation energy. It should therefore be correctly described for an accurate prediction of vacancy formation properties.

TABLE 3.4: Energies (in meV/atom) of several magnetic states (with respect to AFD) of fcc Fe by the DFT and EIM predictions in this study and the predictions from a previous model [56].

|                  | Noncol-AF | AFD | AFS | FM | NM  |
|------------------|-----------|-----|-----|----|-----|
| DFT              | -         | 0   | 19  | 22 | 82  |
| EIM              | -0.5      | 0   | 17  | 29 | 105 |
| EIM in Ref. [56] | -2        | 0   | 0   | 32 | 51  |

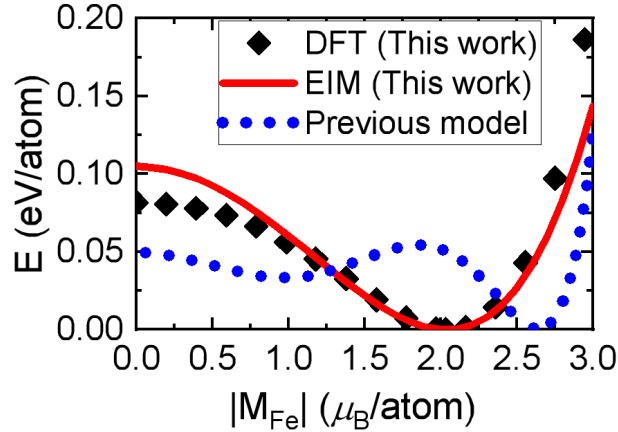

FIGURE 3.17: Energy of AFD fcc Fe as a function of spin magnitude from the DFT and EIM predictions in this study and the prediction from a previous model [56].

The EIM of pure Ni is also fitted on the DFT results of various collinear states and the energy-versus-moment relation in the FM state. In agreement with the DFT results, the magnetic ground state of Ni predicted by the EIM is FM with a magnetic moment of 0.65 $\mu_B$.

It is known that fcc Fe-Ni alloys are chemically disordered at high temperatures, and transform into the ordered phases L1$_0$-FeNi and L1$_2$-FeNi$_3$ at lower temperatures [104, 120, 124, 168]. An accurate description of the phase stability requires therefore an accurate description of both the ordered and disordered phases. In Fig. 3.18(a), we compare the EIM predictions of the mixing enthalpies of fcc ordered and disordered structures with our DFT results. The EIM results are obtained from the SMC simulations at 1 K. In agreement with the DFT results, the mixing enthalpies of other ordered and disordered phases lie above the convex hull consisting of L1$_0$-FeNi, L1$_2$-FeNi$_3$ and pure Fe and Ni. With CMC simulations, we confirm that L1$_0$-FeNi, L1$_2$-FeNi$_3$ and fcc Fe and Ni are the only ground states of the Fe-Ni systems. Specifically,

CMC simulations show that at low temperatures the equilibrium Fe-Ni structure is a phase separation between these phases (namely Fe and L1$_0$-FeNi for nominal Ni content between 0% and 50%, L1$_0$-FeNi and L1$_2$-FeNi$_3$ between 50% and 75% Ni, and L1$_2$-FeNi$_3$ and Ni between 75% and 100% Ni). Such a verification from the CMC simulations is necessary to ensure that L1$_0$-FeNi and L1$_2$-FeNi$_3$ are the most stable structures in the configurational phase space. For example, we find that the chemical ground state at 50% Ni predicted by the EIM2014 model [56] is not L1$_0$-FeNi but another ordered structure, which according to our DFT calculations has a much higher energy than L1$_0$-FeNi.

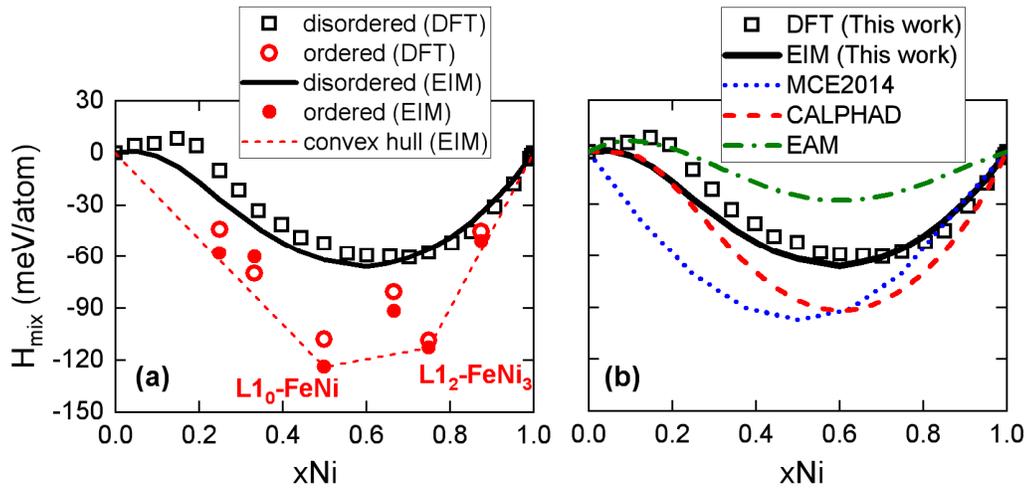

FIGURE 3.18: (a) Comparison of mixing enthalpies of fcc ordered and disordered structures between the DFT and EIM results. (b) Comparison of mixing enthalpies of *disordered* structures with the literature: MCE2014 [56], CALPHAD [132], EAM [186].

As can be seen from Fig. 3.18(b), the agreement between the previous predictions [56, 132, 186] and the present DFT results is limited to the dilute compositions. By contrast, the EIM predictions agree with the DFT data on most of the composition range, except in the Fe-rich region below 25% Ni. This difference between the EIM and the DFT results is related to the different magnetic states of the structures. According to the EIM prediction, the magnetic ground state of the random Fe-Ni structures changes from noncollinear antiferromagnetic to ferromagnetic from 0% to 25% Ni, as indicated in Fig. 3.19. On the other hand, only collinear antiferromagnetic and ferromagnetic structures were considered in the DFT study, while there may exist other magnetic states with lower energies [104]. Therefore, we consider the EIM results are consistent with the DFT ones. In addtion, we do not intend to discuss with the EIM the magnetic ground state of Invar Fe-Ni alloys, which is still an open issue from the DFT point of view [37, 104].

We have shown that the vibrational entropies of mixing in fcc Fe-Ni structures computed by DFT saturate above 300 K [104]. In the current EIM approach, the vibrational entropies of mixing are fitted to the DFT saturated values at 300 K (Fig. 3.20). The fitting is good above 40% Ni. In particular, the relative differences between the ordered and disordered structures at 50% and 75% Ni are well reproduced, which is important for an accurate prediction of the chemical transition temperatures [104].

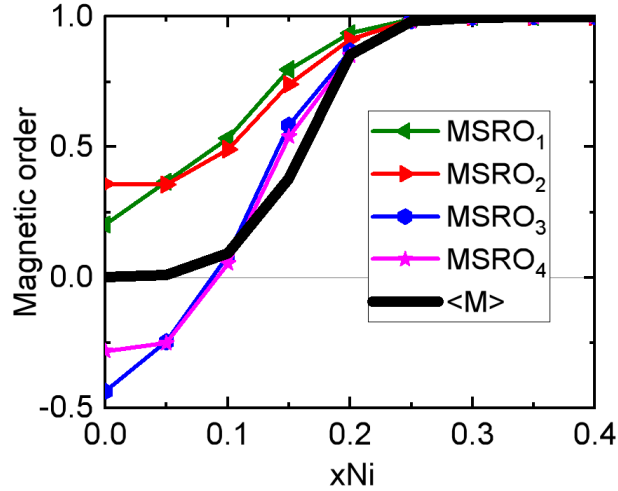

FIGURE 3.19: Reduced magnetization (magnetization over average moment magnitude) and magnetic short-range orders (up to the $4^{th}$ nearest-neighbour shell) of the random structures from the SMC simulations at 1 K.

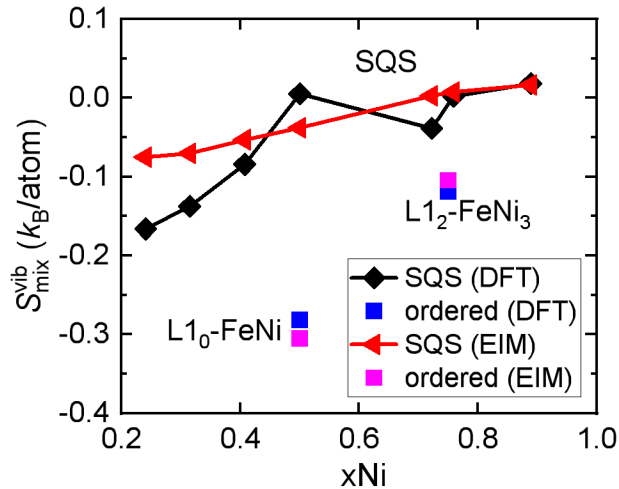

FIGURE 3.20: Comparison of the vibrational entropies of mixing between the DFT results [104] and the EIM predictions.

**Magnetic and chemical transitions**

The calculated Curie point of fcc Ni is 620 K, in good agreement with the measured value of 627 K [187]. The predicted Néel point of fcc Fe is 220 K. Though higher than the experimental value of 67 K [188], this value is low enough to ensure the paramagnetic state of fcc Fe above the room temperature.

As indicated earlier in Fig. 3.19, the fcc random solid solutions with more than 20% Ni have a collinear FM ground state. Fig. 3.21 compares the predicted $T_{\text{Curie}}$ in these structures to the experimental data. The SMC results are obtained with the frozen random chemical configurations, and thus can be interpreted as $T_{Curie}$ measured for the samples quenched from high temperatures. The CMC results are obtained with the equilibrium chemical configurations, and thus correspond to $T_{Curie}$ measured in the slowly cooled samples.

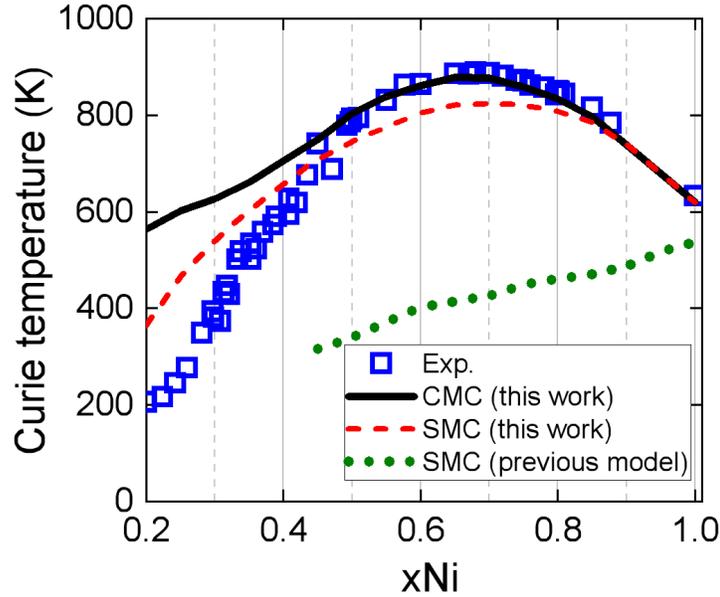

FIGURE 3.21: $T_{Curie}$ of fcc random solid solutions from the experiments [153, 158, 174], the current EIM and the previous model in Ref. [56]. The solid line by the current EIM is obtained with the equilibrium phases from the CMC simulations. The dashed and dotted lines are obtained with the completely random structures from the SMC simulations.

Experimentally, $T_{Curie}$ of the fcc disordered alloys were measured by quenching the samples annealed at temperatures from 923 to 1273 K [153, 158], which should contain some degrees of atomic short-range order (ASRO). Therefore, the equilibrium structures from the CMC simulations are expected be better representative than the fully random structures. Between 45% and 75% Ni, the CMC results are indeed in better agreement with the experimental data than the SMC ones. Below 40% Ni, however, the CMC results show a larger deviation from the experimental results than the SMC ones. As we shall see in the calculated phase diagram presented later, this is because the equilibrium structures in this composition-temperature region are not homogeneous random solutions, but consist of the PM and FM phases with different compositions. Consequently, the SMC results are in better agreement with the experimental data in the structures with less than 40% Ni.

The ordered structures $L1_0$-FeNi and $L1_2$-FeNi$_3$ have a FM ground state, with an experimental $T_{Curie}$ higher than that in the disordered alloys with the same compositions. This is well reproduced by the EIM predictions, which compare favourably with the experimental results as shown in Table 3.5.

TABLE 3.5: Experimental and calculated Curie points (in K) for L1$_0$-FeNi and L1$_2$-FeNi$_3$. The results from the literature were obtained with Monte Carlo simulations [26, 56, 189] or mean field theory [190].

|  | Exp. | This work | Previous calculations |
|---|---|---|---|
| L1$_0$-FeNi | 840 [191] | 845 | 1020 [26], 916 [189], 1000 [56], 1000±200 [190] |
| L1$_2$-FeNi$_3$ | 954 [173], 940 [192] | 968 K | 1180 [26], 1200 [56] |

The chemical order-disorder transition temperatures $T_{chem}$ at 50% and 75% Ni are obtained from the CMC simulations. As shown in Fig. 3.22, the atomic long-range order (ALRO) parameter changes abruptly around 598 K and 766 K at 50% and 75% Ni, respectively. These results agree well with the experimental values of 593 K at 50% Ni [124] and of 770-790 K at 75% Ni [121, 192, 193]. Above $T_{chem}$, the equilibrium structures are found to retain a small degree of ASRO, which decreases slowly with increasing temperatures. We note that the transition temperatures at 50% and 75% Ni are increased by 332 K and 154 K, respectively, if the vibrational contribution is turned off in the EIM. This confirms the conclusion from our DFT study [104] that the inclusion of the vibrational entropy of mixing is necessary to reproduce the experimental transition temperatures.

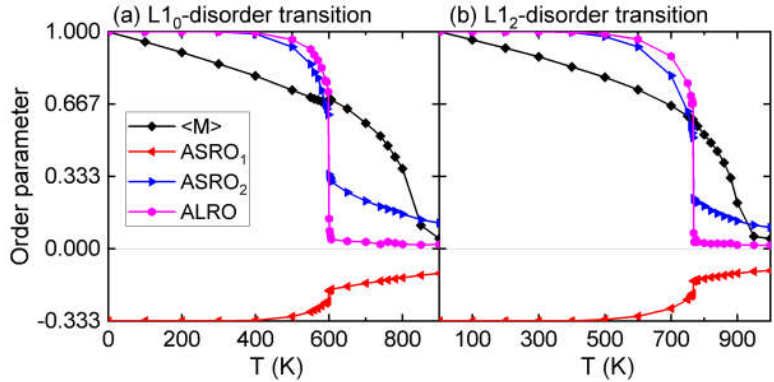

FIGURE 3.22: Temperature dependence of the reduced magnetization, ASRO (of the first two shells) and ALRO parameters in the Fe-Ni alloys with (a) 50% and (b) 75% Ni predicted from CMC simulations with the current EIM.

We note that the present model is considerably improved compared with the recent MCE2014 model for fcc Fe-Ni [56]. The latter does not well capture the magnetic properties and the phase stability, which is one of our motivations to develop the present EIM. According to EIM2014, the predicted $T_{Curie}$ in the random structures are much lower than the experimental data, as shown in Fig. 3.21. Similar predictions are obtained if the equilibrium disordered structures obtained from the CMC simulations are used. Meanwhile, the authors of the MCE2014 model concluded that $T_{chem}$ in the disordered structures with 50% and 75% Ni are 520 K and 730 K respectively [56]. This contradicts with the established Fe-Ni phase diagram where the chemical transitions occur below $T_{Curie}$. Indeed, the contradiction arises from the inconsistent estimation of $T_{chem}$ in Ref. [56]. They were calculated by SMC simulations using the fully ordered and random structures, by comparing the free energies that include the magnetic energies but not the magnetic entropies [56]. Specifically, the free energy in Ref. [56] is evaluated as:

$$F(T) = E^{mag}(T) - TS_{conf} \qquad (3.6)$$

where $E^{mag}$ is the magnetic energy determined from SMC simulations, and $S_{conf}$ is equal to the ideal configuration entropy for the random structures, and zero for the ordered structures. The main drawback of using Eq. 3.6 is that it considers only the increase in the magnetic energy but not the magnetic entropy which decreases the free energy. In fact, due to the low $T_{Curie}$ by MCE2014, the random structures are already PM below the concluded chemical transition temperatures, making the inclusion of magnetic entropy in Eq. 3.6 necessary. To correctly evaluate $T_{chem}$ predicted by EIM2014, we perform the CMC simulations and find that the $T_{chem}$ at 75% Ni is predicted to be 320 K, which is 130 K lower than its predicted $T_c$ this time. However, this value is significantly lower than the experimental results of 770-790 K at 75% Ni [121, 192, 193]. On the other hand, we can not determine a meaningful $T_{chem}$ at 50% Ni with MCE2014, since the model incorrectly predicts another ordered structure other than L1$_0$-FeNi as the chemical ground state.

### 3.4.2   Fcc Fe-Ni phase diagram

The fcc Fe-Ni phase diagram is constructed using the SGCMC simulations with the present EIM. Fig. 3.23 compares the calculated phase diagram with the present DFT results and the most recent CALPHAD assessment by Ohnuma *et al.* [168]. For the convenience of discussion, we denote L1$_0$ and L1$_2$ as the ordered phases around 50% and 75% Ni, respectively, and $\gamma_{FM}$ and $\gamma_{PM}$ as the FM and PM random alloys, respectively.

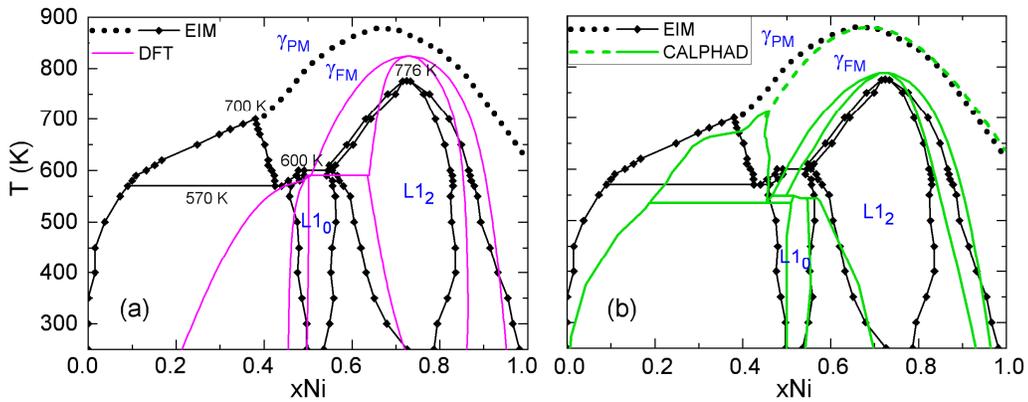

FIGURE 3.23: Fcc Fe-Ni phase diagram predicted by the present EIM, compared to the ones from (a) our DFT study [104] and (b) the CALPHAD assessment by Ohnuma *et al.* [168].

According to the EIM prediction, the phase diagram below 570 K consists of four monophasic regions (Fe-rich $\gamma_{PM}$, L1$_0$, L1$_2$ and Ni-rich $\gamma_{FM}$), which are separated by three corresponding biphasic regions. From 570 to 600 K, the biphasic region $\gamma_{PM}$+L1$_0$ is replaced by the biphasic regions $\gamma_{FM}$+L1$_0$ and $\gamma_{PM}$+$\gamma_{FM}$, which disappear at 600 K and 700 K, respectively. The L1$_0$- and L1$_2$-disorder transitions at 50% and 75% Ni occur at 600 and 775 K, respectively, which are slightly different from the predicted values from CMC simulations. Such differences of less than 10 K are considered to be within the uncertainty.

The major difference between the EIM and DFT phase diagrams is the absence of $\gamma_{PM}$ in the DFT one, which considers only perfectly FM phases. On the other hand, there is no significant difference in the other parts of the phase diagram involving the ordered phases. This is not surprising considering the high Curie temperatures of the ordered phases, which remain rather FM up to the order-disorder transition temperatures.

Ohnuma *et al.* [168] determined experimentally the phase equilibria in Fe-Ni alloys between 673 K and 973 K and revised the thermodynamic descriptions of the relevant phases for the CALPHAD modelling. In particular, the L1$_0$-disorder transition temperature is predicted to be 550 K using the revised CALPHAD parameters, in better agreement with the experimental value of 593 K [124, 194] than the previous CALPHAD prediction of 313 K by Cacciamani *et al.* [132]. The fcc phase diagram calculated using the revised CALPHAD parameters of Ohnuma *et al.* [168] is presented in Fig. 3.23. Despite some differences in the phase boundaries involving the paramagnetic phase, the calculated phase diagrams from EIM and CALPHAD are overall similar. Both predict a small two-phase region between $\gamma_{FM}$ and L1$_0$, and a triangle-shape miscibility gap between the ferromagnetic and paramagnetic random alloys. The miscibility gap is consistent with the observations of chemical and magnetic clusters in the Invar alloys [195–198], in which the Ni-rich and Fe-rich local regions are suggested to be ferromagnetic and paramagnetic respectively [196, 197]. This region will be discussed in more details in the next subsection.

### 3.4.3 Interplay between chemical and magnetic degrees of freedom

It has been shown that the magnetization of the system has a significant impact on the chemical order-disorder transition at 75% Ni [24]. To study how different magnetic states influence the chemical transitions, we need to maintain the magnetic state while allowing the chemical configuration to evolve with temperature. For this purpose, we adopt the adiabatic approximation for the spin relaxation, by assuming that the magnon excitations are faster than the chemical evolution. Note that *without* this approximation, the probability for a given chemical configuration $\sigma$ with a particular magnetic configuration $\boldsymbol{M}$ is proportional to

$$P(\sigma, \boldsymbol{M}) = \exp\left[-\frac{E^{NM}(\sigma) + E^{mag}(\sigma, \boldsymbol{M})}{k_B T}\right] \tag{3.7}$$

where $E^{NM}(\sigma)$ is the nonmagnetic part of our Hamiltonian, and $E^{mag}(\sigma, \boldsymbol{M})$ is the magnetic energy that depends on both the spin and chemical configurations. Within the adiabatic approximation, the probability for a given chemical configuration $\sigma$ to occur becomes proportional to

$$
\begin{aligned}
P_{adia}(\sigma) &= \int \exp\left[-\frac{E^{NM}(\sigma) + E^{mag}(\sigma, \boldsymbol{M})}{k_B T}\right] d\boldsymbol{M} \\
&= \exp\left[-\frac{E^{NM}(\sigma)}{k_B T}\right] \cdot \int \exp\left[-\frac{E^{mag}(\sigma, \boldsymbol{M})}{k_B T}\right] d\boldsymbol{M} \\
&= \exp\left[-\frac{E^{NM}(\sigma)}{k_B T}\right] \cdot Z_\sigma^{mag} \\
&= \exp\left[-\frac{E^{NM}(\sigma) + G_\sigma^{mag}}{k_B T}\right]
\end{aligned}
\tag{3.8}
$$

where $Z_\sigma^{mag}$ by definition is the partition function for the canonical ensemble of the spin configurations for a given chemical structure $\sigma$, and $G_\sigma^{mag} = -k_B T \ln Z_\sigma^{mag}$ is the magnetic free energy for the chemical structure $\sigma$. Therefore, the adiabatic approximation used here is equivalent to integrating the fast magnetic degrees of freedom to obtain a magnetic free energy term that depends only on the chemical configuration (similar to the use of vibrational entropy in our Hamiltonian).

To investigate the effects of the magnetic state on the chemical evolution, we can introduce a spin temperature $T_{spin}$ for calculating $G_\sigma^{mag}$. $T_{spin}$ can be different from the atomic temperature used for the chemical evolution, and it is related to the global magnetic state. In general, it is very difficult to evaluate the magnetic free energy from Monte Carlo simulations at any $T_{spin}$. But it is relatively easy to obtain $G_\sigma^{mag}$ in the two extreme cases, namely the magnetic ground state ($T_{spin} \to 0$) and the ideal paramagnetic state ($T_{spin} \to \infty$). For the first case, $G_\sigma^{mag}$ is simply equal to the minimized magnetic energy, which can be readily calculated during the Monte Carlo simulations.

For the ideal paramagnetic state, the spin-spin correlations are negligible on average, namely $< \boldsymbol{M}_i \cdot \boldsymbol{M}_j >= 0$. Therefore, the magnetic Hamiltonian can be effectively approximated by retaining only the on-site magnetic terms:

$$
E_\sigma^{mag} = \sum_i A_i M_i^2 + B_i M_i^4
\tag{3.9}
$$

The magnetic partition function can be expressed as

$$
\begin{aligned}
Z_\sigma^{mag} &= \int \exp\left[-\frac{\sum_i A_i M_i^2 + B_i M_i^4}{k_B T}\right] d\boldsymbol{M}_1 ... d\boldsymbol{M}_N \\
&= \prod_i z_i^{mag}
\end{aligned}
\tag{3.10}
$$

with $z_i^{mag} = \int \exp\left[-\frac{A_i M_i^2 + B_i M_i^4}{k_B T}\right] d\boldsymbol{M}_i$. The magnetic free energy of the system can then be calculated as

$$
\begin{aligned}
G_\sigma^{mag} &= -k_B T \cdot \ln Z_\sigma^{mag} \\
&= \sum_i -k_B T \ln z_i^{mag} \\
&= \sum_i g_i^{mag}
\end{aligned}
\tag{3.11}
$$

where $g_i^{mag}$ is the magnetic free energy of the $i$-th atom. In the present EIM, the parameters $A_i$ and $B_i$ for an atom depends on its local Ni concentration (and the distance from the vacancy if there is one). There are only a limited sets of $(A_i, B_i)$, and for each of them, the value of $g_i^{mag}$ can be easily evaluated by numerical integration. In practice, $g_i^{mag}$ of all the possible sets of $(A_i, B_i)$ are first evaluated, and it is treated just like an additional on-site nonmagnetic term during the Monte Carlo simulations.

Note that we can not use $T_{spin} \to \infty$ to calculate $g_i^{mag}$, as this leads to the divergence of $z_i^{mag}$. A particular value of $T_{spin}$ needs to be explicitly chosen here. We consider the following criteria for this choice. First, it is preferable to have a universal value of $T_{spin}$ for the evaluation of $g_i^{mag}$. Second, the chosen $T_{spin}$ should be high enough to ensure that the magnetic structure has negligible magnetic long-range and short-range orders (MLRO and MSRO respectively). Third, $T_{spin}$ should not be unreasonably high, so it should be below the melting points of the fcc Fe-Ni alloys, which are around 1700 K [120]. All the random Fe-Ni alloys become paramagnetic above 900 K, and the maximum MSRO among these alloys is 0.20, 0.12, 0.09, 0.08 at 900 K, 1200 K, 1500 K and 1800 K, respectively. In addition, we also check that using 1200 K, 1500 K or 1800 K as $T_{spin}$ leads to very similar chemical transition temperatures at 50% and 75% Ni (the difference being no larger than 10 K). Therefore, we consider 1500 K as a reasonable choice of $T_{spin}$ to calculate $g_i^{mag}$ for the ideal paramagnetic state.

TABLE 3.6: Chemical order-disorder transition temperatures (in K) in the alloys with 75% and 50% Ni, obtained with different magnetic states. The results for the equilibrium magnetic state are obtained with the usual CMC simulations, whereas those for the magnetic ground state and the PM state are obtained within the adiabatic approximation.

| Composition | Equilibrium magnetic state | Magnetic ground state | PM state |
|---|---|---|---|
| 75% Ni | 766 | 885 | 715 |
| 50% Ni | 598 | 555 | 610 |

Table 3.6 shows the chemical transition temperatures in the alloys with 50% and 75% Ni, obtained with different magnetic states. In the alloy with 75% Ni, the predicted transition temperature ranges from 715 K to 885 K depending on the magnetic state of the system. A strong ferromagnetic order as in the magnetic ground state tends to further stabilize the ordered alloy over the disordered one, while the paramagnetic order reduces the phase stability of L1$_2$-FeNi3. On the other hand, the trend is reversed in the alloy with 50% Ni. In addition, the influence of magnetic order on the transition temperature is less important in the alloy with 50% Ni than the one with 75% Ni.

We have shown that there is a coexistence region of the paramagnetic and ferro-magnetic disordered phases in the fcc Fe-Ni phase diagram around 10-40% Ni and 570-700 K. The alloys within this region spontaneously separate into an Fe-rich para-magnetic and a Ni-rich ferromagnetic structures. In order to observe directly the phase separation in a canonical system, the equilibrium structure for a given composition is obtained from CMC simulations. Since the obtained structure may consist of two different phases, the nominal global concentration does not provide information about the compositions of the coexisting phases. The latter are estimated from the distribution of local Ni concentration, which is computed for each fcc site as the atomic fraction of Ni atoms within the fifth coordination shell.

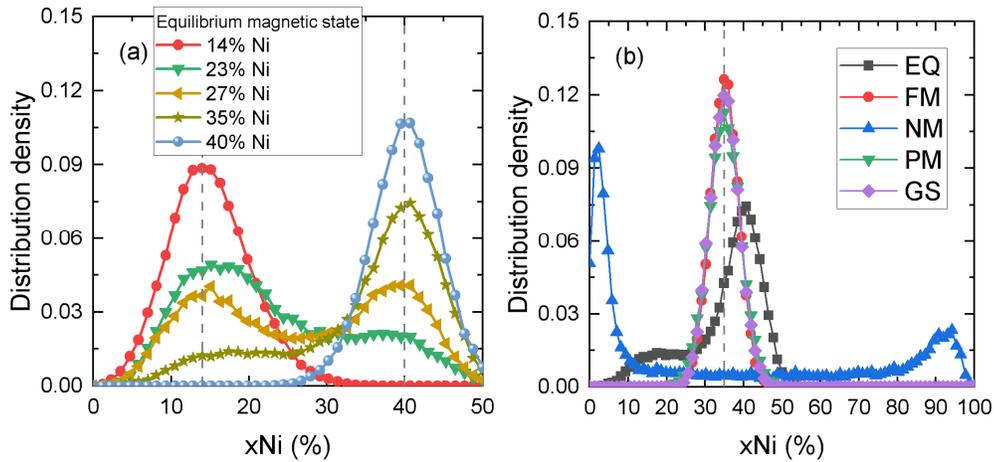

FIGURE 3.24: Distribution of local Ni concentration at 600 K. (a) Equilibrium structures with various Ni content, without constraining the magnetic state. (b) Equilibrium structures with 35% Ni. EQ: equilibrium magnetic state. FM: ferromagnetic. NM: nonmagnetic. PM: paramagnetic within adiabatic approximation ($T_{spin}$=1500 K). GS: magnetic ground state within adiabatic approximation ($T_{spin}$=1 K).

Fig. 3.24(a) shows the distributions in the equilibrium spin-atom structures at 600 K. According to our computed phase diagram, the two-phase composition range at 600 K is between 14% and 40% Ni, as indicated by the vertical lines in Fig. 3.24(a). For the equilibrium structures with 14% and 40% Ni, the distribution exhibits one peak at the nominal Ni concentration, indicating that this is a homogeneous single phase. For the structures with intermediate concentration, the distribution exhibits two peaks indicating the compositions of the two separated phases, namely 14% and 40% Ni.

The phase separation within the $\gamma_{PM}+\gamma_{FM}$ region may be chemically driven, with the magnetic state simply following the composition of the separated phase, or it may be magnetically driven. To elucidate this point, we study the phase equilibrium in the coexistence region by constraining the magnetic state of the system. Four types of con-straints are considered, namely the ferromagnetic (FM), the nonmagnetic (NM), the magnetic ground state (GS), and the fully paramagnetic (PM) states. In the FM state, the spins of all Fe and Ni atoms are aligned to the same direction, and the magnetic-moment magnitudes are fixed to the respective values in the pure phases. In the NM state, the magnetic contribution is turned off by setting all the magnetic-moment magnitudes to zero. The simulations in the magnetic ground state and the PM state are done within the above-mentioned adiabatic manner. Fig. 3.24(b) presents the resulting distributions of local Ni concentration in the equilibrium structures with nominal 35% Ni at 600 K. The distributions obtained in the FM, GS and PM states exhibit one single peak at the nominal concentration, while those obtained in the equilibrium magnetic state and the NM state exhibit two peaks but at different locations. These results indicate that there are three kinds of equilibrium atomic structures. As presented in Fig. 3.25, the equilibrium structure in the PM state is a single-phase random alloy. In the equilibrium spin state, the two-phase atomic structure contains a small region very rich in Fe. The equilibrium atomic structure in the NM state is practically a phase separation between fcc Fe and Ni, which is consistent with our EIM prediction that the nonmagnetic mixing energies of fcc Fe-Ni alloys are positive. These results suggest that the observed phase separation in the $\gamma_{PM}+\gamma_{FM}$ region is driven by magnetic interactions.

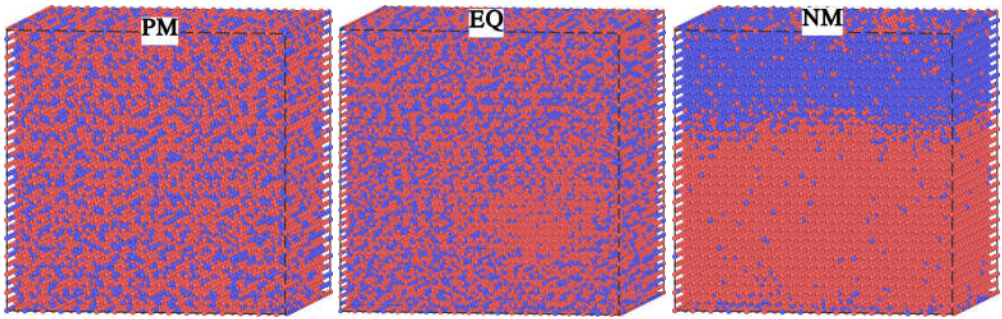

FIGURE 3.25: Equilibrium structures with a nominal concentration of 35% Ni obtained in different magnetic states. The Fe and Ni atoms are marked in red and blue respectively.

On the other hand, the magnetic properties for a given composition depend on both the atomic long-range and short-range orders (ALRO and ASRO respectively). To study quantitatively such dependences, we select 10 different chemical configurations with 75% Ni from the snapshots of the CMC simulations at different temperatures. These structures include both the perfect $L1_2$ ordered structure and the completely disordered one with vanishing atomic short-range order, as well as other configurations with intermediate ALRO and ASRO. Then, SMC simulations are performed to calculate the Curie temperatures for these chemical configurations.

Fig. 3.26 shows the Curie temperatures of these structures as functions of ALRO and the atomic short-range order of the first shell ($ASRO_1$). The ordering of the configurations are the same in the two figures (e.g. the fifth data from the left in (a) and (b), or in (c) and (d), correspond to the same structure). The Curie temperature decreases from 968 K in the perfect L1$_2$ structure to 821 K in the completely disordered one. For structures with vanishing ALRO, their Curie temperatures can still differ by as much as about 80 K, due to the remaining ASRO. Indeed, it is found that the Curie temperatures have a rather linear dependence on ASRO$_1$ (Fig. 3.26(b)).

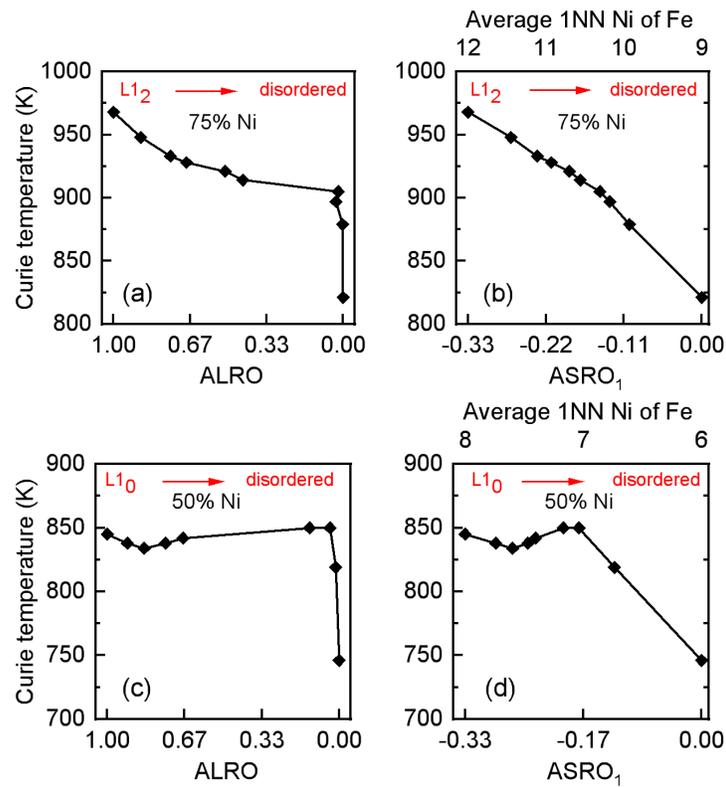

FIGURE 3.26: Curie temperatures as functions of ALRO and $ASRO_1$ in the structures with 75% Ni and 50% Ni. The top X axis in (b) and (d) indicates the corresponding average number of first nearest Ni neighbours for an Fe atom.

A similar investigation is also performed at 50% Ni (Fig. 3.26(c) and (d)). The Curie temperatures are found to be similar among the ordered structures (where ALRO>0), whereas they are more sensitive to the variation in the ASRO$_1$ in the disordered structures (where ALRO=0).

## 3.5 Effects of Cr or Mn addition

In previous sections of this Chapter, we perform a detailed investigation of the phase stability of fcc Fe-Ni alloys. A next step towards the understanding of the thermodynamic properties of Fe-Ni based multicomponent alloys is to elucidate the effects of the additions of other metal elements on the phase stability of the Fe-Ni based systems. In this section, we study the additions of Cr and Mn in the Fe-Ni alloys with 50% and 75% Ni.

### 3.5.1 Properties of Cr-Ni and Mn-Ni ordered structures

The additions of Cr and Mn can impact on the stability of the Fe-Ni ordered structures compared to the disordered ones. The effects may be, at least partially, related to the tendency for the Mn and Cr solutes to form some ordered structures with Fe or Ni. Experimentally, stable fcc ordered structures have been observed in Cr-Ni [199] and Mn-Ni [200] systems, whereas no ordered structure is observed in the Fe-Cr [201] and Fe-Mn [202] systems. In the following we briefly discuss fcc ordered structures in the Cr-Ni and Mn-Ni systems.

Table 3.7 presents the experimental chemical and magnetic transition temperatures for the fcc stable ordered structures in the Fe-Ni, Cr-Ni and Mn-Ni systems. According to the experimental Cr-Ni phase diagram [199], an ordered structure C11$_f$ − CrNi$_2$ is stable below 863 K, with its magnetic transition temperature undetermined. Three stable fcc ordered structures, namely L1$_0$-MnNi, C11$_f$ − MnNi$_2$ and L1$_2$ − MnNi$_3$, are presented in the experimental Mn-Ni phase diagram [200]. The L1$_0$-MnNi and C11$_f$-MnNi$_2$ structures undergo the AF-PM before reaching the respective chemical transition temperatures, while the L1$_2$-MnNi$_3$ structure undergoes the FM-PM and the chemical transitions at the same temperature.

We perform systematic DFT calculations in the ordered structures L1$_0$-$X$Ni, C11$_f$-$X$Ni$_2$ and L1$_2$-$X$Ni$_3$ ($X$=Fe, Cr or Mn). As shown in Table 3.8, the Fe-Ni ordered structures all have a FM ground state. By contrast, the L1$_0$ and C11$_f$ structures are found to be AF in the Cr-Ni and Mn-Ni systems. As illustrated in Fig. 3.27, the magnetic moments of Cr (or Mn) atoms align antiparallelly between the first nearest Cr (or Mn) neighbours, whereas Ni atoms have nearly zero local moments. The magnetic moments of Mn and Ni atoms align parallelly in L1$_2$-MnNi$_3$, whereas in L1$_2$-CrNi$_3$, the magnetic moments of Ni atoms are close to zero and those of Cr atoms align parallelly.

TABLE 3.7: Experimental chemical and magnetic transition temperatures for the fcc stable ordered structures in the Fe-Ni, Cr-Ni and Mn-Ni systems.

| Composition | Chemical transition (K) | Magnetic transition (K) |
|-------------|-------------------------|-------------------------|
| FeNi | 593 [124] | $T_{Curie}$: 793 [173] |
| FeNi$_3$ | 770-790 [121, 192, 193] | $T_{Curie}$: 871 [173] |
| CrNi$_2$ | 863 [199] | / |
| MnNi | 1048 [200] | $T_{Neel}$: 440 [200] |
| MnNi$_2$ | 713 [200] | $T_{Neel}$: 513 [200] |
| MnNi$_3$ | 793 [200] | $T_{Curie}$: 793 [200] |

Our results confirm the experimental observation that L1$_2$-MnNi$_3$ is ferromagnetic, while L1$_0$-MnNi, C11$_f$-MnNi$_2$ are antiferromagnetic [200]. Regarding the C11$_f$-CrNi$_2$ structure, no experimental information is available, but the predicted AF ground state is consistent with the measurements in Cr-Ni alloys showing that the alloys with more than 13% Cr are no longer FM.

TABLE 3.8: Magnetic properties of the ordered structures L1$_0$-$X$Ni, C11$_f$-$X$Ni$_2$ and L1$_2$-$X$Ni$_3$ ($X$=Fe, Cr or Mn). The experimentally confirmed ordered structures are marked in bold. The column MGS denotes the respective magnetic ground state.

| Structure | $X$-Ni | $M_X$ ($\mu_B$) | $M_{Ni}$ ($\mu_B$) | MGS |
|-----------|--------|-----------------|---------------------|-----|
| L1$_0$ | **FeNi** | 2.65 | 0.62 | FM |
| | CrNi | 1.98 | 0.00 | AF between 1NN-Cr |
| | **MnNi** | 3.12 | 0.00 | AF between 1NN-Mn |
| C11$_f$ | FeNi$_2$ | 2.65 | 0.61 | FM |
| | **CrNi$_2$** | 1.69 | 0.00 | AF between 1NN-Cr |
| | **MnNi$_2$** | 3.01 | 0.04 | AF between 1NN-Mn |
| L1$_2$ | **FeNi$_3$** | 2.90 | 0.62 | FM |
| | CrNi$_3$ | 2.25 | -0.05 | FM between Cr |
| | **MnNi$_3$** | 3.23 | 0.53 | FM |

### 3.5.2 Properties of Cr and Mn solutes in Fe-Ni

In this subsection, we focus on the magnetic and energetic properties of one Mn or Cr solute in the Fe-Ni alloys with 50% and 75% Ni in the FM ground state. All the results presented are computed from DFT calculations using fully relaxed 108-atom supercells.

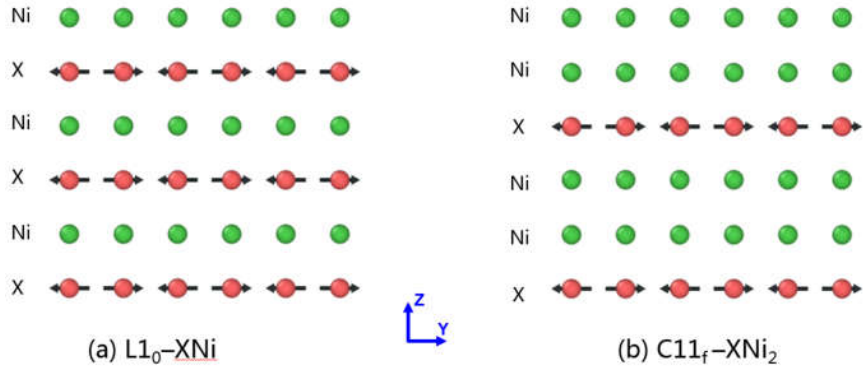

FIGURE 3.27: Illustration of L1$_0$-XNi and C11$_f$-XNi$_2$ (X=Cr or Mn). Mn or Cr atoms (in red) have non-zero local moments, whereas Ni atoms (in green) have nearly zero local moments.

**Spin orientations of Cr and Mn**

First we study the most stable spin orientation for the Cr solute in the FM ordered and disordered alloys with 50% and 75% Ni, as well as in FM Ni as a reference. In each chemical configuration, two different spin orientations are initialized for the Cr solute. We denote FM-Cr for the parallel spin alignment to other atoms, or AF-Cr if otherwise. If both spin alignments are stable (or at least metastable), their energy difference is evaluated.

As shown in Table 3.9, in fcc Ni the AF alignment of the Cr solute is slightly more stable than the FM one , whereas in the Fe sublattice in the L1$_2$-FeNi$_3$ structure the FM alignment of the Cr solute is much more energetically favourable. For the rest of the cases, the FM alignment of the Cr solute is either unstable or very energetically unfavourable. These results suggest that the Cr tends to be AF in the Fe-Ni structures unless its local chemical environment is very rich in Ni. The magnetic moments of Cr in the studied Fe-Ni structures in Table 3.9 are enhanced, compared to the ones in the Cr-Ni ordered structures presented previously in Table 3.8, and to that of fcc Cr which is nonmagnetic [4, 104].

Similar DFT calculations are performed for the Mn solute and the results are shown in Table 3.10. Contrary to the case of the Cr solute, the spin alignment of the Mn solute can be FM or AF in the studied structures. The FM alignment is the most stable in the local chemical environment rich in Ni, while the AF alignment becomes increasing stable with decreasing number of Ni neighbours. The magnetic moments of Mn in the studied Fe-Ni structures in Table 3.10 are similar to ones in the Mn-Ni ordered structures in Table 3.8, but larger than that of fcc Mn, which is antiferromagnetic with a magnetic moment of 1.84 $\mu_B$ [27].

TABLE 3.9: Magnetic properties of a Cr solute in the FM Ni, and Fe-Ni alloys with 50% and 75% Ni. (*m,n*) denotes the numbers of 1NN and 2NN Ni around the solute. MGS denotes the most stable spin orientation of the solute, with FM denoting a parallel alignment of its spin with those of other atoms, and AF denoting the opposite. If both spin orientations of the solute are stable, the energy difference between the two magnetic states is computed. In the ordered structures, we consider placing the solute in the Fe or Ni sublattice. For each SQS, we randomly select 20 different sites (ten occupied by Fe atoms and ten occupied by Ni atoms) to place the solute.

| Location of Cr | (*m,n*) | MGS | $E_{\text{AF-Cr}} - E_{\text{FM-Cr}}$ (eV) | $M_{\text{FM-Cr}}$ ($\mu_B$) | $M_{\text{AF-Cr}}$ ($\mu_B$) |
|---|---|---|---|---|---|
| fcc Ni | (12,6) | AF | -0.046 | 1.969 | 2.044 |
| L1$_2$-FeNi$_3$ (Fe-lat) | (12,0) | FM | 0.481 | 2.292 | 2.481 |
| L1$_2$-FeNi$_3$ (Ni-lat) | (8,6) | AF | (unstable) | / | 2.518 |
| SQS (75%Ni) | (9,4.5) | AF | -0.45±0.13 | 0.7-2.0 | 2.3-2.5 |
| L1$_0$-FeNi (Fe-lat) | (8,0) | AF | (unstable) | / | 2.358 |
| L1$_0$-FeNi (Ni-lat) | (4,6) | AF | (unstable) | / | 2.643 |
| SQS (50%Ni) | (6,3) | AF | (unstable) | / | 2.4-2.6 |

TABLE 3.10: Magnetic properties of a Mn solute in the FM Ni, and Fe-Ni alloys with 50% and 75% Ni. Notations are the same as in Table 3.9.

| Location of Mn | (*m,n*) | MGS | $E_{\text{AF-Mn}} - E_{\text{FM-Mn}}$ (eV) | $M_{\text{FM-Mn}}$ ($\mu_B$) | $M_{\text{AF-Mn}}$ ($\mu_B$) |
|---|---|---|---|---|---|
| fcc Ni | (12,6) | FM | 0.509 | 3.152 | 3.261 |
| L1$_2$-FeNi$_3$ (Fe-lat) | (12,0) | FM | 0.659 | 3.329 | 3.591 |
| L1$_2$-FeNi$_3$ (Ni-lat) | (8,6) | FM | 0.112 | 2.902 | 3.104 |
| SQS (75%Ni) | (9,4.5) | FM | 0.13±0.08 | 3.0-3.2 | 3.1-3.3 |
| L1$_0$-FeNi (Fe-lat) | (8,0) | FM | 0.140 | 3.083 | 3.105 |
| L1$_0$-FeNi (Ni-lat) | (4,6) | AF | -0.470 | 2.844 | 2.995 |
| SQS (50%Ni) | (6,3) | AF | -0.13±0.08 | 3.0-3.2 | 2.9-3.2 |

**Site preference in the Fe-Ni ordered structures**

For a given ordered structure, we define $E_f(X_{\text{Fe}})$ as the formation energy of placing an $X$ solute in the Fe sublattice:

$$E_f(X_{\text{Fe}}) = E(X_{\text{Fe}}) + \mu_{\text{Fe}} - E_0 - \mu_X \tag{3.12}$$

where $E_0$ is the energy of the 108-atom ordered structure, $E(X_{\text{Fe}})$ is the energy of the 108-atom ordered structure with one $X$ atom on the Fe sublattice, $\mu_{\text{Fe}}$ is the chemical potential of Fe in the ordered structure, and $\mu_X$ is the chemical potential of $X$ in a chosen reference state. Similarly, we define $E_f(X_{\text{Ni}})$ as follows

$$E_f(X_{\text{Ni}}) = E(X_{\text{Ni}}) + \mu_{\text{Ni}} - E_0 - \mu_X \tag{3.13}$$

The reference state chosen for the solute $X$ influences the values of $E_f(X_{\text{Fe}})$ and $E_f(X_{\text{Ni}})$, but not their difference, namely

$$\Delta E_f(X) = E_f(X_{\text{Fe}}) - E_f(X_{\text{Ni}}) \tag{3.14}$$

$$= E(X_{\text{Fe}}) - E(X_{\text{Ni}}) + \mu_{\text{Fe}} - \mu_{\text{Ni}} \tag{3.15}$$

A positive value of $\Delta E_f(X)$ indicates that it is more energetically favourable to put the solute $X$ in the Ni sublattice than in the Fe sublattice.

Note that $\mu_{\text{Fe}}$ and $\mu_{\text{Ni}}$ in Eq. 3.15 are the chemical potentials in the ordered structure and can be computed via the statistical method introduced in the Method Chapter (Sec. 2.2.3). Explicitly, Eq. 3.15 can be evaluated as

$$\Delta E_f(X) = E(X_{\text{Fe}}) - E(X_{\text{Ni}}) + \frac{1}{2}\left[E(\text{Fe}_{\text{Ni}}) - E(\text{Ni}_{\text{Fe}})\right] \tag{3.16}$$

where $E(\text{Fe}_{\text{Ni}})$ [or $E(\text{Ni}_{\text{Fe}})$] is the energy of the 108-atom ordered structure with one Fe (or Ni) antisite on the Ni (or Fe) sublattice.

As shown in Table 3.11, $\Delta E_f(\text{Cr})$ and $\Delta E_f(\text{Mn})$ are positive in L1$_0$-FeNi, but they are negative in L1$_2$-FeNi$_3$. Therefore, it is energetically more favourable for the Mn and Cr solutes in L1$_0$-FeNi to be placed in the Ni sublattice than in the Fe sublattice, which is opposite to the cases in L1$_2$-FeNi$_3$.

TABLE 3.11: Formation energy difference of the Cr and Mn solute in the L1$_0$-FeNi and L1$_2$-FeNi$_3$ structures. A positive value of $\Delta E_f(X)$ indicates that it is more energetically favourable to put the solute $X$ in the Ni sublattice than in the Fe sublattice.

|                      | L1$_0$-FeNi | L1$_2$-FeNi$_3$ |
|----------------------|-------------|-----------------|
| $\Delta E_f(\text{Cr})$ | 0.172       | -0.180          |
| $\Delta E_f(\text{Mn})$ | 0.091       | -0.696          |

The quantity $\Delta E_f(X)$ evaluates the energy difference between forming the systems $X_{\text{Fe}}$ (namely one $X$ atom on the Fe sublattice) and $X_{\text{Ni}}$ (with one $X$ atom on the Ni sublattice).  Therefore we are comparing the relative stability of two systems with different numbers of Fe and Ni atoms.  Meanwhile, this is not directly related to the chemical ground state of the system with given numbers of Fe, Ni and $X$ atoms.  For instance, consider the ordered structure with one $X$ solute in the Ni-sublattice.  This structure may not be the most stable configuration, because the energy may be further lowered by exchanging the $X$ solute with a Fe atom in the Fe sublattice, at the expense of creating a Fe antisite in the Ni sublattice.  For simplicity, we only consider the case where the antisite created after the exchange is isolated, namely far away from the $X$ solute.  We represent this exchange process as follows

$$X_{\text{Ni}} \rightarrow X_{\text{Fe}} + \text{Fe}_{\text{Ni}} \tag{3.17}$$

A negative energy change of the process indicates that this exchange is energetically favourable.

The energy changes with different exchange processes are shown in Table 3.12. The results show that the exchange process is only energetically favourable for the case where the Mn solute in L1$_2$-FeNi$_3$ is initially placed in the Ni sublattice.  This suggests the strong preference of the Mn solute for the Fe sublattice, namely the tendency to form the local L1$_2$-MnNi$_3$ structure, even at the expense of creating antisites.

The rest of the exchanges in Table 3.12 are not energetically favourable, but they can be activated at finite temperatures.  The results of Cr in L1$_2$-FeNi$_3$ suggest that moving Cr from the Ni to Fe sublattice costs less energy than the reserved move.  Consequently, our results predict that in an alloy with a dilute amount of Cr and the ratio Fe:Ni equal to 1:3, the Cr concentration is higher in the Fe sublattice than in the Ni sublattice at finite temperatures, in agreement with the experimental observations in $(\text{FeNi}_3)_{1-x}\text{Cr}_x$ alloys [145, 146].  Meanwhile, the concentrations of the Cr and Mn solutes in L1$_0$-FeNi are predicted to be high in the Ni sublattice than in the Fe sublattice.

We note that the results in Table 3.12 are correlated to the ones in Table 3.11, through the following relation:

$$\Delta E_f(X) = \frac{1}{2}[\Delta E(X_{\text{Ni}} \rightarrow X_{\text{Fe}} + \text{Fe}_{\text{Ni}}) - \Delta E(X_{\text{Fe}} \rightarrow X_{\text{Ni}} + \text{Ni}_{\text{Fe}})] \tag{3.18}$$

TABLE 3.12: Energy changes (in eV) associated with difference exchange processes. A negative energy change of the process indicates that this exchange is energetically favourable.

|  | L1$_0$-FeNi | L1$_2$-FeNi$_3$ |
|---|---|---|
| $\Delta E(\text{Cr}_{\text{Fe}} \rightarrow \text{Cr}_{\text{Ni}} + \text{Ni}_{\text{Fe}})$ | 0.088 | 0.465 |
| $\Delta E(\text{Cr}_{\text{Ni}} \rightarrow \text{Cr}_{\text{Fe}} + \text{Fe}_{\text{Ni}})$ | 0.431 | 0.104 |
| $\Delta E(\text{Mn}_{\text{Fe}} \rightarrow \text{Mn}_{\text{Ni}} + \text{Ni}_{\text{Fe}})$ | 0.169 | 0.980 |
| $\Delta E(\text{Mn}_{\text{Ni}} \rightarrow \text{Mn}_{\text{Fe}} + \text{Fe}_{\text{Ni}})$ | 0.351 | -0.411 |

**Effects of Cr and Mn additions on phase stability**

Consider an equilibrium between the ordered and disordered structures for a nominal composition $(FeNi_3)_{1-c}X_c$ (here we consider the case $c = \frac{1}{108} = 0.009$). In this equilibrium, there may be a preference of $X$ for the ordered or the disordered structure, which can be studied by considering the following exchange for the $X$ atom:

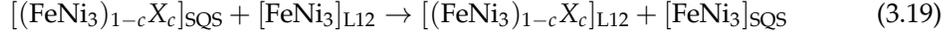

$$[(FeNi_3)_{1-c}X_c]_{SQS} + [FeNi_3]_{L12} \rightarrow [(FeNi_3)_{1-c}X_c]_{L12} + [FeNi_3]_{SQS} \quad (3.19)$$

where the subscripts SQS and L12 denote the state of chemical order in the systems. The energy change associated with this exchange, which is noted hereafter as $\Delta E(X_{SQS} \rightarrow X_{L12})$, indicates a preference of $X$ for the ordered structure, if the value of the energy change is negative. In Table 3.13, we present the energy change per $X$ atom for different exchanges. It can be seen that Mn atoms show a preference for the $L1_2$ structure over the corresponding disordered structure, whereas the trend is the opposite for Cr atoms for the same nominal composition. Meanwhile, both Mn and Cr atoms show a preference for the disordered structure over the $L1_0$ structure.

This energy change is also indicative of the effects of Mn and Cr additions on the chemical order-disorder phase transition. Indeed, $\Delta E(X_{SQS} \rightarrow X_{L12})$ compares the energy difference between $[FeNi_3]_{SQS}$ and $[FeNi_3]_{L12}$, to the energy difference between $[(FeNi_3)_{1-c}X_c]_{SQS}$ and $[(FeNi_3)_{1-c}X_c]_{L12}$. A negative value of $\Delta E(X_{SQS} \rightarrow X_{L12})$ therefore suggests an enhanced energetic stability of the ordered structure due to the $X$ addition. Our results in Table 3.13 show that the energetic stability of the $L1_2$ structure (relative to the disordered phase) is enhanced by the Mn addition, but it is reduced by the Cr addition. On the other hand, the energetic stability of the $L1_0$ structure is reduced by the Mn or Cr addition.

We note that the predicted effects on the energetic stability of the $L1_2$ structure are consistent with the experimental findings, showing that adding 1.5% Cr (or 1.7% Mn) results in a decrease (or an increase) of around 20 K of the $L1_2$-disorder transition temperature [144].

Meanwhile, the present study considers only the energetic stability, whereas the entropic effects also play a role in the finite-temperature phase stability and should be taken into account for a quantitative prediction. In the future work, we plan to include the vibrational and configurational entropies for a more quantitative prediction of the chemical order-disorder transition temperatures. We will also go beyond the dilute limit (up to around 11% Cr and Mn), consider the interactions between solutes and study the potential effects on magnetic transition temperatures.

TABLE 3.13: Energy change (in eV) by exchanging an $X$ atom from the SQS to the ordered structure. A negative value indicates a preference of $X$ for the ordered structure.

|  | $X$=Mn | $X$=Cr |
|---|---|---|
| $\Delta E(X_{\mathrm{SQS}} \rightarrow X_{\mathrm{L12}})$ | -0.181 | 0.160 |
| $\Delta E(X_{\mathrm{SQS}} \rightarrow X_{\mathrm{L10}})$ | 0.135 | 0.144 |

## 3.6 Conclusion

This chapter is focused on the thermodynamics of defect-free Fe-Ni alloys. Energetic, magnetic and vibrational properties of bcc and fcc structures are obtained from DFT calculations. Monte Carlo simulations combined with DFT-parametrized EIM provides further insights into the impacts of magnetic and chemical orders on the phase stability of fcc Fe-Ni alloys. Finally, the effects of Mn and Cr additions in fcc Fe-Ni ordered alloys are discussed in terms of DFT results.

In the first section, we systematically compute the ground-state properties in bcc and fcc Fe-Ni SQSs in the magnetically ordered (AFD and FM) states. We show that the ground-state mixing energies of the disordered alloys predicted from previous empirical models show considerable deviations from our DFT results, pointing out a potential improvement of these empirical models at low temperatures. We find a very good agreement with experimental magnetization in the FM bcc and fcc disordered alloys, confirming the validity of the assumed SQSs for solid solutions. Some locally antiferromagnetic fcc structures are proposed to approach experimental magnetization in the Invar region, while their energetic stability is sensitive to the exchange-correlation functional adopted within DFT.

We confirm that the $L1_0$-FeNi and $L1_2$-FeNi$_3$ structures are the only chemical ground states of Fe-Ni alloys. The vibrational entropies of the two ordered phases and several bcc and fcc FM disordered structures are evaluated as functions of temperature within the harmonic approximation. The comparison with the experimental entropies at 300 K shows an overall good agreement. The vibrational entropies of mixing of all the structures saturates above 300 K, being larger in the SQSs than in the ordered structures. Due to a relatively large difference of vibrational entropies between the ordered and disordered structures, the vibrational entropy, often neglected in previous studies, is found to reduce the predicted order-disorder transition temperatures $T_c^{L1_2}$ and $T_c^{L1_0}$ by 200 and 280 K, respectively. The present study thus indicates a strong effect of vibrational entropy on the chemical order-disorder transitions.

Based on the DFT results, we compute the free energies of mixing for the FM phases, and construct the bcc-fcc phase diagrams below the Curie temperatures. Our predictions agree well with the experimental phase boundaries below 700 K and the measured order-disorder transition temperatures. This suggests that the main contributions to the free energy have been captured. The effects of magnetic disorder on phase stability are also discussed in the light of the theoretical-experimental differences. In particular, it is shown that magnetic disorder can have an impact on the solubility of Ni in bcc Fe-Ni alloys.

The DFT results are used to parametrize an effective interaction model (EIM) for the fcc Fe-Ni system, which naturally takes into account the thermal longitudinal and transversal spin fluctuations. The EIM reproduces accurately the key properties found in DFT, including the ground-state energetic and magnetic properties as well as the vibrational entropies of mixing. Monte Carlo simulations combined with the EIM successfully reproduce the experimental Curie points of the disordered and ordered alloys, as well as the chemical order-disorder transition temperatures. The present EIM can thus be considered as an improved and more reliable model for the fcc Fe-Ni system over the whole range of composition, compared with the previous ones that are either limited to some specific concentrations or give inconsistent phase stability predictions.

The fcc phase diagram predicted by the EIM is in overall agreement with the available experimental data and the most recent CALPHAD assessment. In particular, our model predicts a phase separation between the paramagnetic and ferromagnetic random alloys around 10-40% Ni and 570-700 K. This prediction is supported by the experimental observation of magnetic and chemical clusters in the Invar alloy. We further show that this phase separation is indeed magnetically driven. In addition, the magnetic state of the system is found to have a strong influence on the chemical order-disorder transition temperature at 75% Ni, which increases from 715 K in the ferromagnetic ground state to 885 K in the paramagnetic state. Conversely, the chemical ordering is also found to impact the magnetic transition temperature. We demonstrate in the structures with 50% and 75% Ni that the Curie point is sensitive to both atomic long-range and short-range orders, and tends to increase with increasing chemical ordering.

Finally, we study the effects of a dilute amount of Cr and Mn on properties in the Fe-Ni alloys with 50% and 75% Ni. We show that their spin orientations are sensitive to the local chemical composition. The magnetic moment of the Cr atom tends to align antiparallelly to Fe and Ni atoms, except in the local environment very rich in Ni, where the Cr moment shows a parallel alignment. By contrast, the Mn moment tends to align parallelly in the Ni-rich environment, but it shows an antiparallel alignment in the local environment with around same amount of Fe and Ni neighbours.

We then investigate the site preference of Cr and Mn atoms in the ordered structures. We demonstrate that the Mn and Cr atoms in the $L1_2$ structure show a strong preference for the Fe sublattice over the Ni sublattice, whereas they show the opposite preference in the $L1_0$ structure. The predicted preference of Cr atoms in the $L1_2$ structure is in agreement with the experimental observation.

We find that the $L1_2$ structure has a higher Mn solubility and a lower Cr solubility than the disordered phase with the same composition. Meanwhile, the $L1_0$ structure shows a lower solubility for both Cr and Mn than the disordered phase with the same composition. Our results suggest that the Cr (Mn) addition enhances (reduces) the energetic stability of the $L1_2$ structure, consistent with the experimental trend in the chemical order-disorder transition temperatures. In our future work, we plan to further explore the Cr and Mn effects on the phase stability in Fe-Ni systems, e.g., by going towards higher Cr and Mn concentrations and including vibrational effects.

# 4  Point defect properties in fcc Fe-Ni alloys

*In this chapter, we study the magnetochemical effects on point-defect properties in fcc Fe-Ni alloys. We first present a detailed comparison of the magnetic effects on vacancy formation between fcc Ni, and bcc and fcc Fe, to reveal the role of longitudinal spin fluctuations. Then we perform a systematic investigation of vacancy formation properties as functions of temperature and composition in fcc Fe-Ni alloys. Finally, we discuss via DFT the magnetic effects on the formation and migration of self-interstitial atoms in fcc Fe and Ni.*

## 4.1  State of the art

Vacancy is one of the simplest and most frequently encountered structural defects in transition-metal systems. Vacancy concentration $[V]$ is a key parameter controlling atomic transport via the vacancy mechanism[116]. The equilibrium monovacancy concentration $[V]_{eq}$ can be linked with the vacancy formation free energy $G_f^{\text{non-conf}}$ via the expression [66, 72, 73]

$$[V]_{eq} = \exp(-\frac{G_f^{\text{non-conf}}}{k_B T}) \tag{4.1}$$

where $k_B$ and $T$ are respectively the Boltzmann constant and the absolute temperature, and $G_f^{\text{non-conf}}$ includes all the non-configurational entropy contributions (electronic, vibrational and magnetic) [117]. However, neither $[V]$ nor $G_f$ can always be directly measured in experiments. $[V]$ can be determined from differential dilatometry, but in a thermal-vacancy regime, the technique is only applicable at very high temperatures (sometimes near the melting point) due to the limited experimental resolution [203]. Other methods such as electrical resistivity measurement and positron annihilation spectroscopy usually provide the temperature dependence of $[V]$ rather than its absolute value [204]. For pure systems, $G_f^{\text{non-conf}}$ can also be indirectly estimated as $G_a - G_m$, where $G_a$ is the diffusion activation free energy determined from tracer diffusion experiments [19, 205], and $G_m$ is the vacancy migration free energy coming from resistivity recovery, internal friction or magnetic after-effect experiments [206–209].

In magnetic metal systems, several thermodynamic and kinetic properties are affected by thermal magnetic excitations and the magnetic transition. The additional magnetic degree of freedom makes an accurate prediction of those properties more difficult, particularly from an atomistic-simulation point of view. Experimental measurements of self- and solute diffusion coefficients in bcc Fe revealed strong effects of magnetism [19, 21, 210–212]. These experimental evidences have motivated various recent theoretical studies [13–17, 22, 61]. However, very little is known about such effects on the vacancy-related properties in other magnetic systems (for example fcc Fe, fcc Ni and bcc Cr) that exhibit important longitudinal spin fluctuations at finite temperatures [40, 41, 54, 185].

The thermal magnetic effects can be very system-dependent. First, we focus on fcc Fe and fcc Ni, which are the major constituents for the technologically important austenitic steels. At variation with the bcc-Fe case, experimental data on the vacancy concentration and the vacancy formation free energy are only available for the paramagnetic (PM) states of fcc Fe [8–10, 213, 214] and fcc Ni [203, 215–217], due to the low magnetic transition temperatures (67 K for fcc Fe [188] and 627 K for fcc Ni [187]). On the theoretical side, previous first principles investigations on the vacancy formation and diffusion properties in the PM regime were often informed by results on the ferromagnetic (FM) state of fcc Ni [218–222], and the nonmagnetic (NM) state in the case of fcc Fe [223–225]. It is still an open question whether these results, calculated with ordered or nonmagnetic states, are representative for those in the PM state. For example, the vacancy formation energy obtained for NM fcc Fe is 2.37 eV, much higher than the experimental values (1.40-1.83 eV [8–10, 213, 214]). More advanced descriptions of PM state properties can be achieved via the disordered local moment (DLM) approach [33, 226], the spin-wave method [44], spin dynamics [227–229], spin-lattice dynamics [51, 114] and Monte Carlo simulations based on magnetic model Hamiltonians [56, 58, 59, 103] (see Ref. [108] for a recent review), etc. Most of these studies [33, 44, 51, 56–59, 103, 114, 226–230] only considered defect-free systems. Regarding the vacancy properties in the PM state, most of the investigations addressed bcc Fe [15–17, 22, 61], while there are very few such studies of fcc Fe [36] or fcc Ni. Furthermore, the effects of longitudinal spin fluctuations, which can be important in the PM regime, are often neglected. Taking into account such effects requires, for example in the DLM approach, a more sophisticated statistical treatment combined with constrained local-moment calculations [40, 41, 43], leading to very CPU-demanding calculations. Finally, the previous theoretical studies in fcc Fe and Ni addressed the vacancy properties in the magnetic ground state and the high-temperature ideal PM state, whereas a systematic understanding of the temperature-dependent magnetic effects, which involve simultaneous longitudinal and transversal spin excitations, is missing.

The vacancy formation properties in alloys have been studied for various systems. The investigations in the ordered alloys such as Fe-Al, Ni-Al, Al-Sc and Ti-Al have been performed with the grandcanonical or canonical approaches [62–70]. These approaches are based on the assumption that the vacancy and antisite concentrations are low enough to ignore their interactions. They are thus applicable only at sufficiently low temperatures, namely far below the chemical order-disorder transition temperatures. In addition, to the best of our knowledge, there is no existing theoretical (nor experimental) study addressing the vacancy formation in the ordered Fe-Ni alloys.

On the other hand, vacancy formation properties have been studied in disordered alloys such as Cu-Ni [71, 72], Co-Ni, Fe-Ni, Cr-Ni [94], and Fe-Ni-Cr [35, 93, 95, 231, 232]. Usually, the vacancy formation energy is computed via DFT as a function of local or nominal chemical composition to derive a qualitative trend [35, 93–95, 231, 232]. Quantitative determinations of vacancy formation properties as functions of temperature have been carried out recently for disordered Cu-Ni alloys, using the single-site mean-field approximation [72] or the cluster expansion approach [71].

The studies mentioned above are focused exclusively on either the low-temperature ordered structures with minimal amount of antisites, or the high-temperature random solutions with vanishing short-range order. A consistent, continuous and quantitative description of vacancy formation properties as functions of temperature and thus of chemical long-range and short-range orders, is still missing in the literature. Furthermore, almost all of the aforementioned studies in magnetic systems assume either nonmagnetic or magnetic ground states, systematically ignoring the effects of magnetic excitations and transitions. A simultaneous consideration of chemical and magnetic order/disorder and their coupling effects on vacancy formation, which is even more challenging, has not been studied for any system in the literature.

Self-interstitial atom (SIA) is another important type of point defects. Compared with vacancy, SIA has a rather high formation energy (typically > 3 eV [233]), therefore its concentration under equilibrium conditions is usually negligible. Under nonequilibrium conditions, typically under irradiation, an excess amount of SIAs are created, which are highly mobile and have crucial impact on the microstructure evolution. An accurate prediction of SIA properties is therefore essential for the modelling of irradiated alloys.

Experimentally, the SIA formation energy can not be directly measured, but it can be computed as the difference between the formation energy of a Frenkel pair and that of a vacancy, provided that these quantities are known [234]. The SIA migration energy can be measured using, e.g., resistivity recovery [203, 235, 236]. In this method, the samples are first irradiated at cryogenic temperatures to create frozen Frenkel pairs. They are subsequently annealed at higher temperatures (typically below 200 K [203, 234]), during which interstitials become mobile and their diffusion can be correlated with the measured resistivity variation rate. The measured SIA migration energy in fcc Ni ranges from 0.14 eV to 0.21 eV [203, 237, 238]. Such measurments are not available for fcc Fe, because the bulk is unstable at low temperatures. Dimitrov *et al.* measured the migration energies in fcc Fe-Ni-Cr alloys with different compositions (the Fe content ranging from 9 to 67 wt%) [235, 236]. They concluded a linear relationship for the concentration dependence of the migration energy, suggesting a rather high migration energy of 0.96 eV for fcc Fe [236].

Similar to vacancy properties, the theoretical studies of SIA properties often adopt 0 K magnetic states. For example, the SIA formation and migration properties in Ni were computed with a FM state [239]. The same properties in fcc Fe were investigated using the magnetically ordered states (AFD and AFS) [151, 240].  In particular, the
predicted migration energy in fcc Fe is between 0.2 and 0.25 eV [240], much lower than
the value of 0.96 eV suggested by experiments [236], requiring further investigation
beyond simple magnetic configurations.  In addition, it is also of interest to compare
the magnetic effects on the vacancy and SIA properties.

Based on the results presented above, we can make the following points:

- Previous theoretical studies of magnetic effects on vacancy formation mainly fo-
  cused on bcc Fe.  An understanding of such effects is missing for other magnetic
  systems exhibiting significant longitudinal spin fluctuations.

- Previous investigations of vacancy formation properties in alloys centred only
  on one of the two extreme cases, namely either on the low-temperature ordered
  structures with few antisites, or on the high-temperature random solutions.  There-
  fore, a continuous description of these properties as functions of temperature
  and chemical order is needed.

- Vacancy formation in alloys were previously studied with either the magnetic
  ground state or the ideal PM state.  A more realistic description of magnetic
  effects, including the longitudinal spin fluctuations and the magnetochemical
  coupling effects, is important for alloys with both chemical and magnetic order-
  disorder transitions.

- DFT calculations of SIA properties in fcc Fe and Ni use only magnetically or-
  dered structures, and the effects of magnetic excitations and transitions remain
  unknown.  A comparison of the magnetic effects on the vacancy and SIA prop-
  erties is also of interest.

These points are addressed in this chapter as follows.  The first section is a de-
tailed analysis of the effects of longitudinal and transversal spin excitations and the
magnetic transitions on vacancy formation properties in fcc Ni, and bcc and fcc Fe.
The results between the three systems are systematically compared.  In the second
section, vacancy formation properties are computed in fcc Fe-Ni alloys as functions
of temperature and composition.  The effects of magnetic and chemical orders on va-
cancy formation are discussed.  In the third section, the self-interstitial formation and
migration properties in fcc Fe and Ni are computed from DFT.  They are compared
with the calculated vacancy formation properties, and the migration properties from
experiment.

## 4.2   Vacancy formation properties in pure Fe and Ni

This section is focused on a detailed comparison of the magnetic effects on vacancy
formation between the three pure phases, namely fcc Ni, and bcc and fcc Fe [18].
The results at 0 K are obtained from DFT, while those at finite temperatures are from
EIMs combined with Monte Carlo simulations.  For fcc Fe and Ni, we use the EIMs
parametrized in this thesis; for bcc Fe, we use the EIM parametrized recently by
Schneider *et al.* [22].

In the following, we first give a brief discussion of the magnetic properties in the defect-free systems, to reveal the role of longitudinal spin fluctuations. Then, in the second part, we present the vacancy formation energies at 0 K obtained from DFT. In the final part of this section, we investigate the temperature-dependent vacancy formation properties using the EIMs.

### 4.2.1 Longitudinal spin fluctuations in the defect-free systems

As shown in Table 4.1, the generalized Heisenberg (GH) EIMs reproduce well the experimental Curie temperatures of bcc Fe and fcc Ni, and give a closer prediction of the measured Néel temperature of fcc Fe than that of 450 K from a previous model [56]. Though the predicted Néel temperature is still somehow overestimated, it is ensured that fcc Fe is already paramagnetic at the room temperature.

TABLE 4.1: Experimental and calculated Curie temperature of bcc Fe and fcc Ni and Néel temperature of fcc Fe. GH and CH denote the generalized Heisenberg (i.e., freely evolving spin magnitudes) and classical Heisenberg (i.e., fixed spin magnitudes) simulations, respectively

| System | Exp. | GH | CH |
|--------|------|------|------|
| bcc Fe | 1044 K [187] | 1050 K | 1080 K |
| fcc Fe | 67 K [188] | 220 K | 240 K |
| fcc Ni | 627 K [187] | 620 K | 880 K |

It is known that longitudinal spin fluctuations are more significant in fcc Fe and Ni than in bcc Fe [41, 54, 185]. This point is also confirmed by the EIM results shown in Fig. 4.1. Compared to the bcc-Fe case, fcc Ni and fcc Fe show a stronger variation of the average spin magnitude versus temperature, and a larger dispersion of the spin magnitudes.

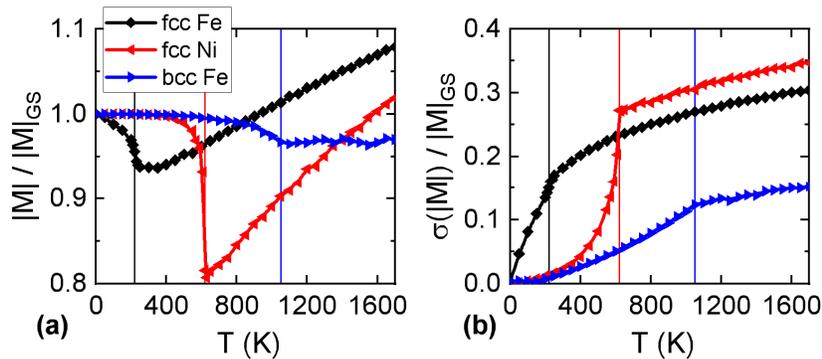

FIGURE 4.1: Variation with temperature of (a) average spin magnitudes and (b) standard deviation of the spin magnitudes, scaled by the respective ground-state moments $|M|_{GS}$. The vertical lines denotes the corresponding magnetic transition temperatures.

Our generalized Heisenberg EIMs naturally account for longitudinal spin fluctuations. To demonstrate the effects of such fluctuations on the magnetic transition temperatures $T_C$, we also perform classical Heisenberg (CH) simulations based on the same EIMs, by constraining the spin magnitudes to the ground-state values. The difference between $T_C^{GH}$ and $T_C^{CH}$ is then due to the fact that the temperature-dependent variation of the average spin magnitude, and the dispersion of spin magnitudes are neglected in CH simulations. The effects of neglecting the variation in the average spin magnitude can be roughly estimated as follows. Since the magnetic transition temperature is proportional to $J_{ij}M^2$, $T_C^{CH}$ can be estimated as

$$T_C^{CH} = T_C^{GH} \times \frac{M^2(0K)}{M^2(T_C^{GH})} \tag{4.2}$$

where $M^2(0K)/M^2(T_C^{GH})$ according to Fig. 4.1(a) is equal to 1.07, 1.14 and 1.53 for bcc Fe, fcc Fe and fcc Ni, respectively. These estimations are consistent with the results obtained from CH simulations (Table 4.1). The actual difference in the transition temperatures between the GH and CH simulations is small for bcc and fcc Fe, but it becomes quite large in fcc Ni, showing the necessity of taking into account the temperature variation of spin magnitudes. On the other hand, if we impose the Ni spin magnitudes (in CH simulations) to be the average value from the GH simulations at the $T_C^{GH}$, the obtained $T_C^{CH}$ is 60 K lower than the $T_C^{GH}$. This suggests a minor but non-negligible effect of the dispersion of atomic spin magnitudes at a given temperature.

### 4.2.2   Vacancy formation energies at 0 K from DFT

The vacancy formation energies $E_f$ in various magnetic states are computed from DFT, as shown in Table 4.2. The dispersion in $E_f$ among different magnetic states is much larger in bcc Fe than in fcc Fe and Ni, suggesting a stronger magnetic effects on $E_f$ in bcc Fe. Please note that such dispersions are related to not only the ordering of local magnetic-moment orientations but also the atomic spin magnitudes. The latter can differ by as much as 0.9 $\mu_B$ between different magnetic states (excluding the NM one) in bcc and fcc Fe.

TABLE 4.2: $E_f$ (in eV) in different magnetic states computed from DFT. The collinear magnetic ground state values are marked in bold.

| System | NM | FM | AFS | AFD | mSQS |
|--------|------|--------|------|--------|------|
| fcc Fe | 2.37 | 1.85 | 2.00 | **1.83** | 2.04 |
| fcc Ni | 1.38 | **1.43** | 1.47 | 1.53 | 1.52 |
| bcc Fe | 0.57 | **2.20** | 0.94 | 1.86 | 1.59 |

To check the dependence of $E_f$ on the atomic spin magnitudes $|M|$, we perform the DFT calculations by constraining all the local $|M|$ to the same values. The results for the three systems are shown in Fig. 4.2.

$E_f$ is less dependent on $|M|$ in fcc Ni than in fcc and bcc Fe. For fcc Ni, $E_f$ in the FM state is lower than in most of other magnetic states with the same $|M|$. Consequently, $E_f$ in fcc Ni at finite temperatures is expected to be larger than that of the FM ground state.

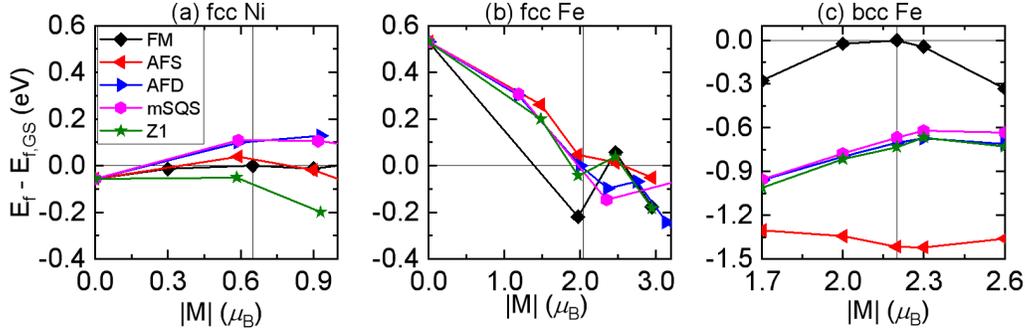

FIGURE 4.2: DFT results of $E_f$ in the three systems as a function of constrained local magnetic moment, with respect to $E_f$ in the collinear magnetic ground states (1.43 eV for FM fcc Ni, 1.83 eV for AFD fcc Fe and 2.20 eV for FM bcc Fe). The vertical lines denote the spin magnitude of the magnetic ground states.

In fcc Fe, $E_f$ for all the spin-orientation orderings tends to decrease with increasing $|M|$ for $|M|$ below $2\mu_B$. Fig. 4.2 also shows that the difference of $E_f$ is relatively small between various spin orderings with the same $|M|$, whereas the variation in $E_f$ with $|M|$ is relatively large for a given spin ordering. This suggests $E_f$ in fcc Fe has a stronger dependence on the atomic spin magnitudes than on the spin orientations. In particular, consider the AFD state and the mSQS. According to the results obtained with optimized magnetic moments (Table 4.2), $E_f$ in the AFD state is 0.2 eV lower than that in the mSQS. From Fig. 4.2, it is clear that $E_f$ of the mSQS and the AFD state is very similar for the same $|M|$. Therefore, the difference of $E_f$ between the two states in Table 4.2 is mainly due to the fact that the optimized $|M|$ is smaller in the mSQS (on average 1.5 $\mu_B$) than in the AFD state (2.0 $\mu_B$). This indicates that $E_f$ in the mSQS with optimized magnetic moments cannot be taken as the value in the PM state, since longitudinal spin fluctuations in the high temperature PM state lead to an increase in $|M|$.

Contrary to the fcc-Fe case, $E_f$ in bcc Fe is much more dependent on the spin-orientation orderings than on the atomic spin magnitudes, as demonstrated in Fig. 4.2(c). Furthermore, the variation in $|M|$ is also expected to be small in bcc Fe: the optimized magnetic moments in the mSQS are on average 2.05 $\mu_B$, rather close to 2.20 $\mu_B$ in the FM ground state; it is also shown in the previous section that the thermal longitudinal spin fluctuations in bcc Fe are small. Therefore, the temperature-dependent vacancy formation properties in bcc Fe are expected to be mainly determined by the arrangement of spin orientations.

We note that our EIMs reproduce satisfactorily the above DFT predictions in Table 4.2 and Fig. 4.2.

Finally, we would like to comment on the effects of exchange-correlation functionals on $E_f$. In this work, we parameterize the EIMs of fcc Fe and Ni based on the GGA-PBE results. It is well known that GGA describes the bulk (in particular magnetic) properties of Fe better than LDA [3]. For Ni, LDA and GGA equilibrium lattice parameters ($a_0$) are 3.416 Å and 3.514 Å, respectively, the latter being closer to the experimental value of 3.52Å [241]. By performing LDA calculations using the GGA $a_0$, we verified that adopting LDA results only introduced a constant shift in $E_f$ for all the magnetic states compared with GGA values, namely the relative difference in $E_f$ between different magnetic states is the same using LDA and GGA. Therefore, the choice of exchange-correlation functionals influences only the parametrization of the nonmagnetic part of the EIMs. This point will be further discussed for Ni in Sec. 4.2.3.

### 4.2.3 Vacancy formation properties at finite temperatures from EIM

We show in Fig. 4.3 the temperature-dependent vacancy formation properties calculated from GH simulations for the three systems. Here, $E_f^{mag}$ is the vacancy formation energy, $S_f^{mag}$ is the magnetic contribution to the vacancy formation entropy, and $G_f^{mag}$ ($= E_f^{mag} - TS_f^{mag}$) is the vacancy formation magnetic free energy (without the vibrational or other entropic contributions). In the following, we first describe the overall behaviour versus temperature and compare our results with experimental data. Then, the effects of magnetism in different temperature ranges are analysed separately.

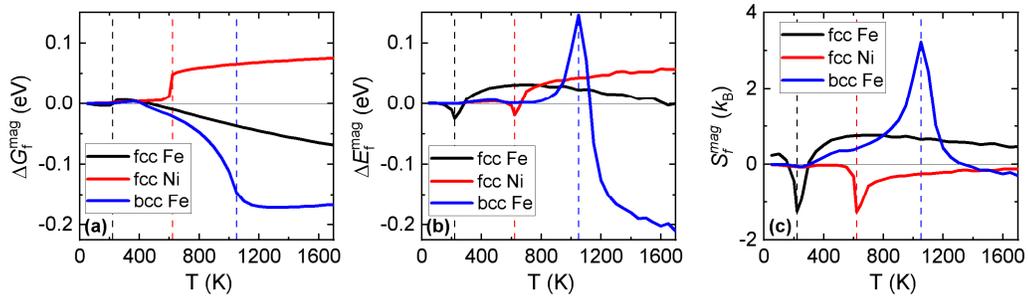

FIGURE 4.3: Magnetic contribution to vacancy formation properties in the three systems calculated from GH simulations (i.e., with both transversal and longitudinal spin fluctuations). $G_f^{mag}$ and $E_f^{mag}$ are given with respect to the ground-state values (1.43 eV for FM fcc Ni, 1.83 eV for AFD fcc Fe and 2.20 eV for FM bcc Fe). The vertical lines denotes the EIM-predicted magnetic transition temperatures.

### Overall behaviour and comparison with experiments

We first consider bcc Fe in which the vacancy activation energy strongly depends on the magnetic state as evidenced by experiments [11, 19]. Theoretical studies reveal that both vacancy formation and migration energies are lower in the fully PM than in the FM state [13–17, 22, 61]. As can be seen in Fig. 4.3, compared with the FM values, the asymptotic $G_f^{mag}$ and $E_f^{mag}$ are reduced by 0.17 and 0.20 eV, respectively.

At variance with the well-known case of bcc Fe, magnetic disorders in fcc Fe and Ni do not lead to a strong decrease of $G_f^{mag}$. In the PM region close to the magnetic transition, $\Delta G_f^{mag} = G_f^{mag}(T) - G_f^{mag}(0K)$ is negligible in fcc Fe but it is positive in fcc Ni. Furthermore, there is a steady variation in $G_f^{mag}$ of fcc Fe and Ni in the PM region, which is different from the saturation behaviour observed in bcc Fe. These distinct behaviours will be further analysed in Sec. 4.2.3.

TABLE 4.3: Vacancy formation energies (in eV) in the collinear magnetic ground state and the PM state from calculations and experiments. For the PM state, we show both $E_f^{mag}$ and $G_f^{mag}$ (given inside the parentheses) calculated at 1500 K. The small difference between $E_f^{mag}$ and $G_f^{mag}$ comes from the longitudinal spin entropy as explained in Sec. 4.2.3.

| | bcc Fe | | fcc Fe | | fcc Ni | |
|---|---|---|---|---|---|---|
| | FM | PM | AFD | PM | FM | PM |
| This work | 2.20 | 2.00 (2.03) | 1.84 | 1.85 (1.78) | 1.43 | 1.48 (1.50) |
| Other calculations | 1.95-2.24 [13–17, 61, 184] | 1.54-1.98 [14–17, 61] | 1.82 [151] | 1.86 [36] | 1.43 [218] | |
| Experiments | 2.0±0.2 [11] | 1.79±0.1 [11] | | 1.40±0.15 [214] | | 1.73±0.07 [217] |
| | 1.81±0.1 [12] | 1.74±0.1 [12] | | 1.7±0.2 [10] | | 1.7±0.1 [216] |
| | 1.60±0.15 [8] | 1.53±0.15 [8] | | 1.83±0.14 [9] | | 1.6±0.1 [216] |
| | | | | | | 1.56±0.04 [242] |

The vacancy formation energies in the magnetic ground state and the PM state are compared with other calculations and experiments in Table 4.3. In bcc Fe, our FM value agrees with the previous calculated results ranging between 2.15-2.23 eV based on the same exchange-correlation functional GGA-PBE [13, 16, 17], whereas the GGA-PW91 prediction gives a lower value of 1.95 eV [184], or 2.13 eV if using the experimental $a_0$ [14]. Our fully PM value is consistent with the previous results, and the dispersion in the calculated values may be due to the different atomic-position relaxation schemes [14–17, 61]. Before comparing with experimental data, it should be noted that the measurement of $E_f$ in bcc Fe is extremely sensitive to the presence of interstitial impurities such as carbon, which could lead to an underestimation of $E_f$ in earlier studies [11]. Compared with the experimental data in bcc Fe, both our ground-state and PM values are on the higher limit. A very relevant quantity to compare is the difference between the GS and PM energies. It summarizes the overall effect of the magnetic transition, and a cancellation of systematic errors can occur in both calculations and experiments. This difference from our prediction is in good agreement with the result of Schepper *et al.* [11].

For fcc Fe, our results agree with the previous *ab initio* predictions [36, 151]. On the experimental side, the measurements on $E_f$ are performed in the PM state. Unlike the bcc-Fe case, the measurement in fcc Fe is less affected by the presence of impurities [8]. The experimental uncertainty is rather due to its limited temperature window of stability [214]. The predicted $E_f$ of PM fcc Fe is consistent with the measured values within the experimental uncertainty. We note that this value is much lower than the one in NM fcc Fe (2.37 eV). The NM state is therefore a poor representative phase to study diffusion properties in fcc Fe, although it has been used in some recent *ab initio* studies on diffusions in PM fcc Fe [223–225].

To the best of our knowledge, there is no theoretical result for the PM $E_f$ and no experimental data for the FM $E_f$ in fcc Ni. The experimental $E_f$ ranges between 1.5 eV and 1.8 eV (see Ref. [117] and references therein). Smedskjaer *et al.* [215] suggested that the experimental discrepancies arise from the different analysis methods and the associated assumptions between positron annihilation spectroscopy experiments, and uncontrolled metallurgical variables for other techniques such as positron lifetime spectroscopy. In the light of this suggestion, the recommended value of $E_f$ in Ref [203] is 1.79 eV. The results from the recent experiments [216, 217, 242] are shown in Table 4.3. Though being higher than the recommended value in Ref [203], our prediction of $E_f$ for PM Ni is within the uncertainty of the experiments [214, 216, 242, 243], including the most recent one using differential dilatometry [242] to measure directly the equilibrium vacancy concentration.

Indeed, the experimental $E_f$ is indirectly deduced from the temperature dependence of the equilibrium vacancy concentrations. Therefore, the latter may allow a more direct comparison. To the best of our knowledge, such measurements on vacancy concentrations have been performed in fcc Ni, but not in bcc and fcc Fe. To determine these concentrations, we have calculated via DFT the vacancy vibrational formation entropy for FM fcc Ni (2.15 $k_B$), and included it to the magnetic free energy of vacancy formation obtained using the EIM. As shown in Fig. 4.4, the calculated equilibrium vacancy concentrations of fcc Ni agree well with the experimental concentrations, which exhibit a larger dispersion at temperatures below 1400 K. We note that $E_f$ deduced as the slope of the experimental curve is very sensitive to this dispersion of the experimental data: the estimated $E_f$ ranges between 1.5 to 1.6 eV if only data above 1400 K are considered, and between 1.45 to 1.80 eV if all data are considered.

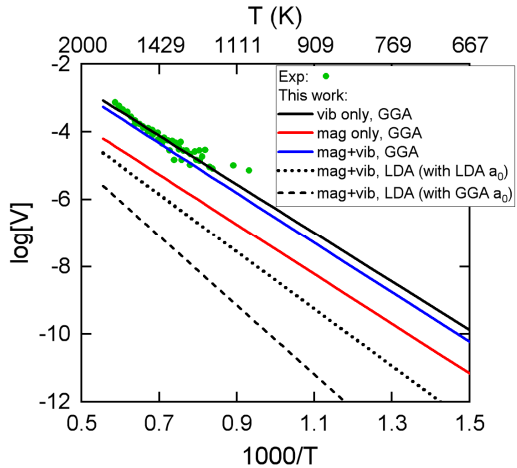

FIGURE 4.4: Calculated $[V]_{eq}$ as a function of temperature, compared to available experimental data in fcc Ni [242, 244–246].

On the theoretical side, $E_f$ of FM Ni is sensitive to the choice of exchange-correlation functional, for instance between the GGA-PBE and the LDA functionals used in DFT calculations [78]. As discussed in Sec. 4.2.2, the choice affects only the EIM prediction of $E_f$ of FM Ni, but not the temperatures evolution $\Delta G_f^{mag}(T)$. To evaluate the LDA-based prediction of equilibrium vacancy concentrations, we calculated the vacancy formation energy (1.65 eV) and the vacancy vibrational formation entropy (0.43 $k_B$) of FM Ni via DFT using LDA, and combine them with $\Delta G_f^{mag}(T)$ predicted by the EIM. Since the experimental $a_0$ of Ni is well reproduced by GGA-PBE but underestimated by LDA, we also applied LDA with the GGA-PBE $a_0$ to calculate the vacancy formation energy (2.02 eV) and the vacancy vibrational formation entropy (0.44 $k_B$) for FM Ni. We note the latter is much lower than the GGA-based value (2.15 $k_B$) and the experimental result (3.3 ±0.5 $k_B$ [242]). As shown in Fig. 4.4, the two LDA-based predictions substantially underestimate the experimental equilibrium vacancy concentrations. This may be due to the large vacancy formation energy and the small vacancy vibrational formation entropy obtained with the LDA functional. Therefore, we conclude that the GGA-based prediction is more consistent with the experimental vacancy concentrations.

Finally, we also compare the relative importance of the vibrational and magnetic effects on the equilibrium vacancy concentration in fcc Ni. As can be seen from Fig. 4.4, neglecting the magnetic effects (namely $G_f = G_f^{mag}(0K) - TS_f^{vib}$) changes very little the predicted vacancy concentrations, whereas neglecting the vibrational contribution (namely $G_f = G_f^{mag}(T)$) would underestimate the vacancy concentration by one order of magnitude. In fact, for the three systems, the vibrational contribution to the vacancy concentration determination is stronger than the magnetic contribution. The latter is small in fcc Fe and Ni, but it is more significant in bcc Fe as neglecting this can lead to an underestimation of vacancy concentration by up to a factor of 5.

**Vacancy formation below magnetic transitions**

The EIMs for bcc Fe and fcc Ni predict the same FM ground state as with DFT, therefore the ground-state $E_f$ for bcc Fe and fcc Ni are correctly reproduced. For fcc Fe, the collinear magnetic ground state by DFT is AFD, whereas the EIM predicts a spin spiral ground state with a slightly lower energy (by 0.5 meV/atom) than the AFD state. The $E_f$ of the spin spiral is found to be slightly higher (by 0.01 eV) than the AFD value.

The vacancy formation properties below the magnetic transitions can be influenced by the transversal and the longitudinal spin fluctuations, and the effects of two types of fluctuations cannot be easily decoupled. In Fig. 4.5, we compare the temperature evolution of $\Delta G_f^{mag}$ predicted by GH and CH simulations.

Below the magnetic transitions, the CH results (without longitudinal spin fluctuations) follow the same trend as the GH values for bcc Fe and fcc Ni, indicating the effects of the longitudinal spin fluctuations are small. This is expected for bcc Fe in which longitudinal spin fluctuations are not significant. For fcc Ni, longitudinal spin fluctuations are rather strong, and the Curie temperature (880 K) from the CH simulations is much higher than the GH value (620 K). However, the CH and GH results of $\Delta G_f^{mag}$ would be similar if they are rescaled to the same Curie temperatures. This is in agreement with the DFT results in Sec. 4.2.2 where $E_f$ in fcc Ni is found to be rather insensitive to the spin magnitudes, especially for the FM state.

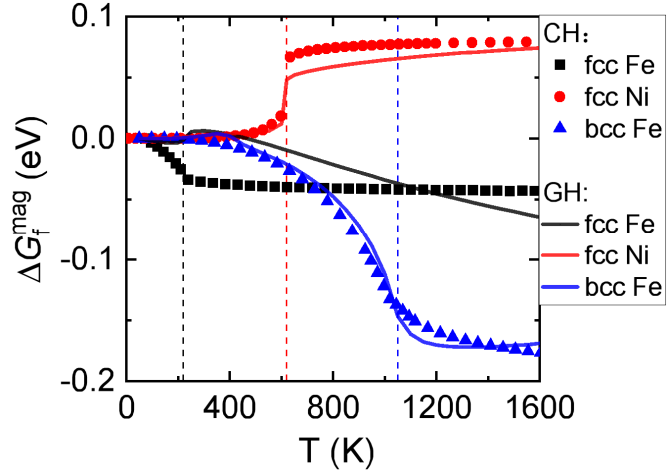

FIGURE 4.5: Comparison of $\Delta G_f^{mag} (= G_f^{mag}(T) - G_f^{mag}(0K))$ as a function of temperature between the GH and CH values. For each pure system, the CH curve is rescaled to have the same magnetic transition temperature as the GH curve. The vertical lines denote the magnetic transition temperatures of the GH curves. $T_{\text{Néel}}$ of fcc Fe is 220 K (240 K) from the GH (CH) calculations; $T_{\text{Curie}}$ of fcc Ni is 620 K (880 K) from the GH (CH) calculations; $T_{\text{Curie}}$ of bcc Fe is 1050 K (1080 K) from the GH (CH) calculations.

For fcc Fe below the Néel temperature, the CH values decrease with temperature while the GH values remain nearly unchanged. This suggests that below the magnetic transition in fcc Fe, the transversal spin fluctuations lead to a decrease in $G_f^{mag}$, while the longitudinal spin fluctuations lead to an increase in $G_f^{mag}$. Indeed, the spin magnitudes tend to decrease with temperature before the Néel temperature (Fig. 4.1), which leads to an increase in $G_f^{mag}$ according to our DFT results in Fig. 4.2(b).

On the other hand, the spin magnitudes tend to increase with temperature above the Néel temperature, so $G_f^{mag}$ is expected to decrease with temperature according to the DFT results in Fig. 4.2(b). As shown later, there is a steady decrease in $G_f^{mag}$ with temperature in PM fcc Fe. This demonstrates that the longitudinal spin fluctuations play a dominant role over the transversal ones in the temperature evolution of $G_f^{mag}$ of fcc Fe.

**Vacancy formation properties in the PM regime**

As shown in Fig. 4.5, $G_f^{mag}$ always saturates in the PM region in the CH simulations. From thermodynamic relations, it can thus be concluded that $E_f^{mag}$ also saturates while $S_f^{mag}$ vanishes at high $T$. Indeed, for CH models, the magnetic contribution to the vacancy formation energy is zero in the ideal PM state, since the thermodynamic average of the spin-spin correlation $< M_i M_j >$ is zero in the ideal PM state.

The steady variation in $G_f^{mag}$ in the PM region in Fig. 4.3(a) is thus related to the longitudinal spin fluctuations and is observed only with the GH models. This behaviour can be quantitatively understood as follows. In the ideal PM region where the spin-spin correlations are negligible, the magnetic Hamiltonian can be approximated by a simplified form, retaining only the on-site terms, namely

$$E_{tot}^{mag} = \sum_i e_i^{mag} = \sum_i A_i M_i^2 + B_i M_i^4 \tag{4.3}$$

where $e_i^{mag} = A_i M_i^2 + B_i M_i^4$ is the on-site magnetic energy for the $i$-th atom. The partition function can be expressed as

$$Z_{tot} = \int e^{-\beta E_{tot}^{mag}} d\boldsymbol{M_1}...d\boldsymbol{M_N} = \prod_i z_i \tag{4.4}$$

$$z_i = \int e^{-\beta e_i^{mag}} d\boldsymbol{M_i} \tag{4.5}$$

The total magnetic energy of the system can then be calculated as

$$G_{tot}^{mag} = -k_B T \cdot ln Z_{tot} = \sum_i g_i^{mag} \tag{4.6}$$

$$g_i^{mag} = -k_B T ln z_i \tag{4.7}$$

where $g_i^{mag}$ is the magnetic free energy of the $i$-th atom. In our EIMs, there are 3 sets of the on-site parameters for each pure system depending on the distance from the vacancy. Consequently, there are three possible values of $g_i^{mag}$ for a pure system containing a vacancy. We use the subscript $i$ equal to 0, 1, and 2 to represent the bulk atoms and those in the first and second neighbouring shells of the vacancy, respectively. The vacancy formation free energy can then expressed as

$$\begin{aligned} G_f^{mag} &= G_{tot,V}^{mag} - \frac{N-1}{N} G_{tot,0}^{mag} \\ &= \sum_{i=1,2} n_i \cdot (g_i^{mag} - g_0^{mag}) \end{aligned} \tag{4.8}$$

where $n_i$ is the coordination number of the $i$-th shell.

The simplified EIMs allow a direct numerical evaluation of $z_i$ and thus all the subsequent quantities. It should be noted that even though the contribution of transversal spin fluctuations to the averaged total energy is negligible in the PM state, its contribution to the total entropy is not zero. This is fully taken into account in Eq. 4.5, where it is equally possible for $\boldsymbol{M_i}$ to take any direction, as expected for the perfectly PM state. In this way, the transversal part of the entropy per atom is maximized and the same for all the atoms independent of their distances from the vacancy. This entropic contribution is thus cancelled out in Eq. 4.8.

The results of $G_f^{mag}$ and $S_f^{mag}$ are compared between the complete and the simplified EIMs for fcc Fe and Ni in Fig. 4.6. In the PM region, $G_f^{mag}$ and $S_f^{mag}$ using the simplified EIMs converge to those from the complete EIMs, confirming that the simplified EIMs are very good representations of the complete ones at high temperatures. The different variation trends in $G_f^{mag}$ between fcc Fe and Ni are related to the signs of

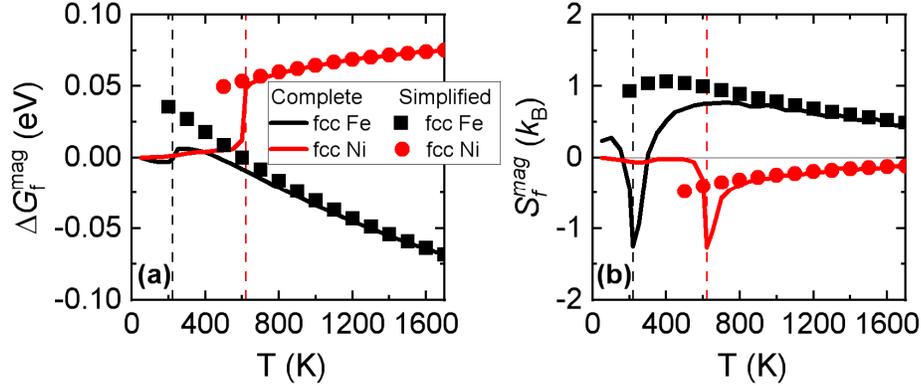

FIGURE 4.6: (a) $\Delta G_f^{mag}$ and (b) $S_f^{mag}$ predicted from the complete and simplified EIMs. The vertical lines denote the EIM-predicted magnetic transition temperatures.

$S_f^{mag}$. This can also be understood with the simplified Hamiltonians that contain only the on-site magnetic energy $A_i M_i^2 + B_i M_i^4$. As shown in Fig. 4.7, for a given temperature (e.g. 1000 K as indicated in the figure), the atoms around the vacancy can explore a larger interval of $|M|$, and thus has a larger entropy than the bulk atoms in fcc Fe, while the case is reversed for fcc Ni.

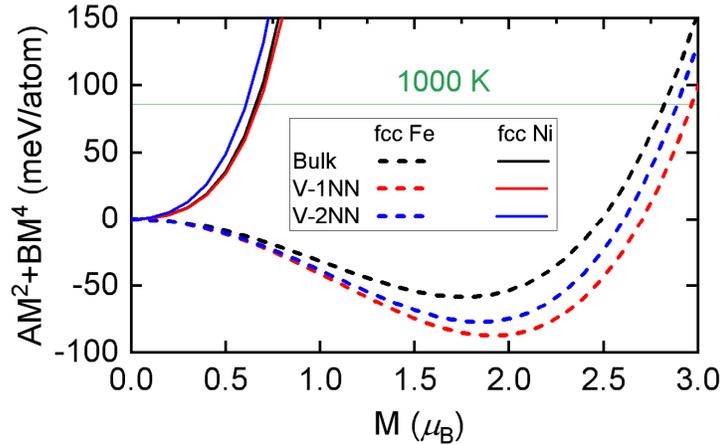

FIGURE 4.7: On-site magnetic energies $A_i M_i^2 + B_i M_i^4$ for the bulk atoms and first and second nearest neighbours of a vacancy.

The current analysis can provide insights into the impact of longitudinal spin variations on vacancy formation at high temperatures. That is, due to the difference of spin magnitudes in the PM state between the atoms near the vacancy and in the bulk, a steady variation of $G_f^{mag}$ with temperature is expected, which is at variance with the predictions from classical Heisenberg models. In this case, it is also non-trivial to apply the well-known Ruch model [247], which proposes

$$\frac{G_f^{mag}(T) - G_f^{mag}(0K)}{G_f^{mag}(PM) - G_f^{mag}(0K)} = 1 - S^2 \tag{4.9}$$

where $S$ is the magnetic long-range order parameter, and $G_f^{mag}(PM)$ is the vacancy formation magnetic free energy in the ideal PM state. $G_f^{mag}(PM)$ can be taken as the saturated value for bcc Fe, but the definition of $G_f^{mag}(PM)$ is ambiguous for fcc Fe and Ni in which $G_f^{mag}(T)$ does not saturate at high temperatures.

**Impact of magnetic transitions on vacancy formation**

It has been shown in Fig. 4.3 that the effect of the magnetic transition is much stronger in bcc Fe than in fcc Fe and Ni. This is mainly related to the change in the exchange interaction energy across the magnetic transitions. To illustrate this, we begin with the CH models where $J_{ij}$ are the same for the atoms in the bulk and near the vacancy, and we focus on the vacancy formation energy $E_f^{mag}$ for simplicity. In this case, the exchange interaction energy $J_{ij}\boldsymbol{M}_i\boldsymbol{M}_j$ does not contribute to $E_f^{mag}$ in the ideal fully PM state. In the magnetic ground state, this contribution to $E_f^{mag}$ is equal to half the ground-state exchange interaction energy per atom. Therefore, the variation in $E_f^{mag}$ across the magnetic transition is equal to half the ground-state exchange interaction energy per atom. The latter value is expected to be larger in bcc Fe than in fcc Fe and Ni in view of the DFT results [4, 104], which show that the energy difference between various magnetic states is more significant in bcc Fe than in fcc Fe and Ni. For a quantitative comparison, we show in Table 4.4 the different contributions to the vacancy formation energies of the three systems. Although there is a compensating variation between the on-site (longitudinal) and exchange (transversal) contributions in the three systems, the latter contribution is much stronger in bcc Fe than in fcc Fe and Ni.

TABLE 4.4: Contributions of different terms in the Hamiltonians to the vacancy formation energy in the magnetic ground state and the PM state, expressed in eV. The PM value is obtained at 1500 K where the transversal contribution is negligible in all the three systems.

| Contribution | bcc Fe | | fcc Fe | | fcc Ni | |
|---|---|---|---|---|---|---|
| | GS | PM | GS | PM | GS | PM |
| NM $\epsilon_i$ | 2.00 | 2.00 | 2.37 | 2.37 | 1.43 | 1.43 |
| On-site $A_i M_i^2 + B_i M_i^4$ | -0.07 | 0.00 | -0.47 | -0.53 | 0.14 | 0.06 |
| Exchange $J_{ij}\boldsymbol{M}_i\boldsymbol{M}_j$ | 0.27 | $\approx 0$ | -0.07 | $\approx 0$ | -0.13 | $\approx 0$ |
| Total | 2.20 | 2.00 | 1.84 | 1.85 | 1.43 | 1.48 |

## 4.3 Vacancy formation properties in fcc Fe-Ni alloys

In the previous section, we have discussed the magnetic effects on vacancy formation in the pure phases of Fe and Ni. In this section, we address such effects in fcc ordered and disordered Fe-Ni alloys. In the first subsection, we validate the EIM by comparing its predictions with the DFT results at 0 K. Then, we discuss the concentration and temperature dependences of vacancy formation properties in the alloys, in the second and third subsections, respectively.

### 4.3.1  Vacancy formation energies at 0 K

As mentioned in Sec. 2.3.2 in the Methods chapter, the EIM applied to fcc Fe-Ni alloys is extended from the EIMs of fcc Fe and Ni. In the previous section, we have demonstrated that the predicted vacancy formation energies $E_f$ and equilibrium vacancy concentrations agree with the available experimental data in fcc Fe and Ni. In this subsection, we validate the EIM for Fe-Ni alloys by comparing its predictions of vacancy formation energy with the DFT results at 0 K.

For the pure phases, it is straightforward to calculate $E_f$ as

$$E_f = E_{tot,V} - \frac{N-1}{N} E_{tot,0} \qquad (4.10)$$

where $E_{tot,V}$ and $E_{tot,0}$ are the total energies of the defective system (with $N-1$ atoms) and the defect-free one (with $N$ atoms), respectively. For alloys, the calculation of $E_f$ requires the knowledge of not only the total energies $E_{tot,V}$ and $E_{tot,0}$, but also the chemical potentials of the constituents, which are not calculated directly from DFT.

As mentioned in Sec. 2.2.3 in the Methods chapter, such problems in the ordered phases can be addressed by the statistical approach that have been applied to various systems [62–66]. Based on the noninteracting-defect approximation, it requires only the input of the total energies of the perfect ordered structures and the ones with an isolated vacancy or antisite on a sublattice. Then, it consists of solving a set of nonlinear equations to obtain point defect formation energies and the chemical potentials. In particular, simple analytical expressions can be derived for the ordered phases with antisite formation energies much lower than vacancy formation energies [63], such as the cases of L1$_2$-FeNi$_3$ and L1$_0$-FeNi.

Using this approach, we have calculated the vacancy and antisite formation energies, with the input energies calculated from DFT and the present EIM, respectively. As shown in Table. 4.5, there is a good agreement of the antisite formation energies between the DFT and EIM predictions. Note that a single value is shown for the antisite formation energy because it is the same in the two sublattices at the 0 K limit [63]. There is also a reasonable agreement between the DFT and EIM predictions of the vacancy formation energies in the Fe and Ni sublattices of L1$_2$-FeNi$_3$ and in the Ni sublattice of L1$_0$-FeNi, while the EIM result in the Fe sublattice of L1$_0$-FeNi is underestimated by 0.34 eV compared with the DFT one.

In concentrated disordered alloys, it is customary to calculate the local vacancy formation energy at the site $i$ as [71, 93–95], $E_f^i$:

$$E_f^i = E_{tot,V_i} - E_{tot,0} + \mu \qquad (4.11)$$

TABLE 4.5: Antisite and vacancy formation energies (in eV) in the Fe and Ni sublattices calculated from DFT and EIM for $L1_2$-FeNi$_3$ and $L1_0$-FeNi at the 0 K limit.

|  | $L1_2$-FeNi$_3$ | | $L1_0$-FeNi | |
| --- | --- | --- | --- | --- |
|  | DFT | EIM | DFT | EIM |
| Antisite | 0.256 | 0.325 | 0.275 | 0.253 |
| Vacancy on the Fe sublattice | 1.392 | 1.273 | 1.897 | 1.561 |
| Vacancy on the Ni sublattice | 1.593 | 1.687 | 1.847 | 1.794 |

where $E_{tot,V_i}$ is the energy of the system with a vacancy at site $i$, and $\mu$ is the chemical potential of the removed atom in the system. Note that $E_f^i$ should not be associated with the chemical species of the removed atom, but rather the local atomic arrangement around the site $i$, which dictates $E_{tot,V_i} - E_{tot,0}$, as well as the composition of the alloy, which dictates $\mu$. In the SQS approach, $\mu$ is often calculated via the Widom substitution [93, 94, 96], which requires a large number of atom substitutions at different sites.

However, if one is interested in the arithmetic average value of vacancy formation energy $< E_f^i >$ in a given phase, the relation $E_{tot,0} = N(x_A\mu_A + x_B\mu_B)$ allows to eliminate $\mu_A$ and $\mu_B$ in $< E_f^i >$ [71]:

$$< E_f^i > = x_A(< E_{tot,V_A} > -E_{tot,0} + \mu_A) + x_B(< E_{tot,V_B} > -E_{tot,0} + \mu_B)$$
$$= x_A < E_{tot,V_A} > +x_B < E_{tot,V_B} > -\frac{N-1}{N}E_{tot,0} \qquad (4.12)$$

where $< E_{tot,V_A} >$ is the average energy of the system with an $A$ atom removed.

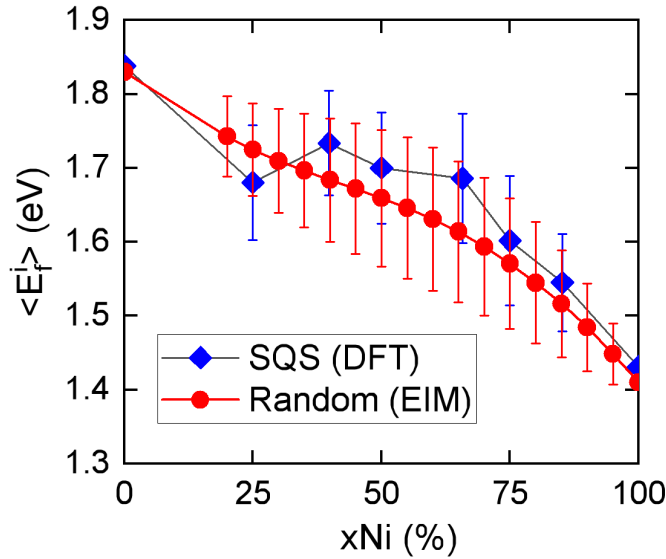

FIGURE 4.8: Average vacancy formation energy as a function of Ni concentration for the random Fe-Ni structures in the FM state. The error bars denote the standard deviations of local vacancy formation energies. The vacancy formation energy in AFD Fe is also presented.

A comparison of $< E_f^i >$ in the random Fe-Ni structures between the DFT and EIM predictions is given in Fig. 4.8. The DFT results are obtained with the 108-site SQSs. For each SQS, we consider in DFT 9 different Fe and Ni sites to obtain $< E_{tot,V_{Fe}} >$ and $< E_{tot,V_{Ni}} >$, respectively. The EIM results are calculated in the 16834-site random structures in the magnetic ground state, and $< E_{tot,V_{Fe}} >$ and $< E_{tot,V_{Ni}} >$ are averaged over all the Fe and Ni sites, respectively. Our DFT results, in agreement with previous DFT studies in fcc Fe-Ni and Fe-Ni-Cr alloys [93, 94], suggest that the average formation energy decreases with increasing Ni concentration, which is also well reproduced by the EIM.

### 4.3.2   Concentration dependence of vacancy formation properties

We investigate the concentration dependence of $G_f^{mag}$ in the equilibrium Fe-Ni alloys at 700 K, 800 K, 1000 K and 1500 K. According to the phase diagram predicted from our EIM, the alloys with 40% to 95% Ni remain ferromagnetic while the rest are paramagnetic at 700 K. In addition, the chemical structures in the alloys with 60% to 80% Ni are ordered whereas the rest are disordered at 700 K. At 800 K, the alloys are chemically disordered over the whole composition range, but those with 55% to 80% Ni are FM while the rest are paramagnetic. The alloys are chemically disordered and paramagnetic at 1000 K and 1500 K.

As shown in Fig. 4.9, $G_f^{mag}$ in the disordered structures tends to decrease with increasing Ni concentration at the studied temperatures. In the composition range of 60% to 80% Ni, the equilibrium phases at 700 K have an L1$_2$ ordered structure. Consequently, the corresponding $G_f^{mag}$ are expected to be larger than those in the disordered structures with similar compositions according to the results of the previous section.

The results in Fig. 4.9 indicate that $G_f^{mag}$ increase with increasing temperature in the disordered structures with more than 30% Ni, while the trend is reversed in the disordered structures below 30% Ni. This can be correlated with our previous results which suggest that $G_f^{mag}$ in Fe decreases but that in Ni increases with increasing temperature [18].

We also show in Fig. 4.9 $G_f^{mag}$ calculated in the disordered structures in the respective magnetic ground states. They are much lower than the alloys in equilibrium magnetic state, and present a minimum near 75% to 80% Ni. Furthermore, as presented in Fig. 4.10, the difference between the two curves of $G_f^{mag}$ at 1500 K reaches a maximum of 0.32 eV around 65% Ni, where the Curie temperature is also the highest. The latter is a sign of the strength of the magnetic exchange interaction, which is also the strongest around 65% Ni according to our model. Indeed, it is shown that the difference between $G_f^{mag}$ in the paramagnetic state and the ground state is closely related to the strength of the magnetic exchange interaction [18].

The calculated $G_f^{mag}$ can be compared to the available experimental vacancy formation energies $E_f$ in Fig. 4.9. To the best of our knowledge, the measurements of $E_f$ in fcc Fe-Ni alloys have been reported only by Caplain and Chambron using magnetic anisotropy measurements [248, 249]. In their first study, measurements were performed in the disordered Fe-Ni samples with 70% Ni quenched from between 873 and 973 K, and $E_f$ was found to be 1.57 eV [248]. In their subsequent study in the disordered samples with 50% to 94% Ni quenched from above the chemical transition temperature, $E_f$ were found to be 1.80 eV regardless of the composition [249]. It is difficult to draw a definitive conclusion regarding the concentration dependence or the values of $E_f$ based solely on these two experiments, which could have large experimental uncertainty as in the cases of pure fcc Fe and Ni. On the other hand, our results of $G_f^{mag}$ fall between $E_f$ from these two sets of measurements, and they are within the uncertainty of the available experimental data over the whole concentration range.

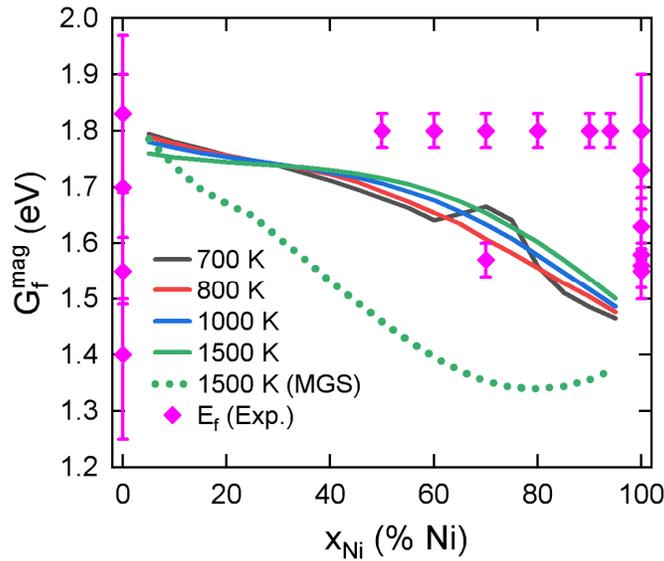

FIGURE 4.9: The predicted $G_f^{mag}$ as a function of Ni concentration at several temperatures, compared to the available experimental vacancy formation energies (pure Fe [8–10, 214], pure Ni [215–217, 242, 245], Fe-Ni alloys [248, 249]). The solid lines denote the results obtained in the equilibrium phases, whereas the triangles denote the results obtained in the disordered structures in the respective magnetic ground states, which are collinear FM above 25% Ni and non-collinear below 25% Ni.

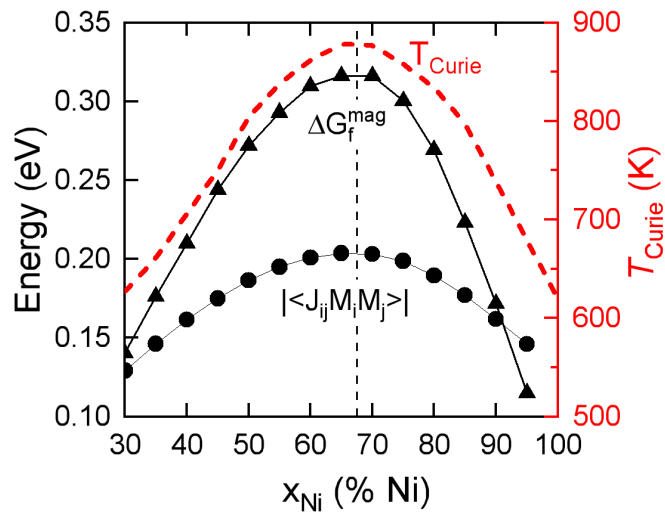

FIGURE 4.10: Difference between $G_f^{mag}$ in the disordered structures at 1500 K with the corresponding equilibrium magnetic states at 1500 K and with the magnetic ground states, the average magnetic exchange interaction energies $< -J_{ij}M_iM_j >$ and the Curie temperatures in the disordered structures.

**Effects of local chemical environment**

The above discussion shows how the vacancy formation magnetic free energy $G_f^{mag}$ in an alloy varies with the nominal concentration. For a given nominal composition and temperature, the preferential site of creating a vacancy is also dependent on the local composition. For a given nominal composition, one can define the local vacancy concentration for any given local chemical environment as

$$C_V^{eq,i} = \frac{N_V^{eq,i}}{N_{site}^i} \tag{4.13}$$

where $N_V^{eq,i}$ and $N_{site}^i$ are the numbers of vacancies and lattice sites with the given local chemical environment $i$, respectively. It can be shown that

$$C_V^{eq,i} = \frac{N_V^{eq} \mathcal{P}_V^i}{N_{site} \mathcal{P}_{site}^i} \tag{4.14}$$

$$= \frac{C_V^{eq} \mathcal{P}_V^i}{\mathcal{P}_{site}^i} \tag{4.15}$$

where $N_V^{eq}$ and $N_{site}$ are respectively the total numbers of vacancies and lattice sites, $C_V^{eq} = N_V^{eq}/N_{site}$ is the global equilibrium vacancy concentration, $\mathcal{P}_V^i$ is the probability of finding a vacancy with the local chemical environment $i$ among all vacancies, and $\mathcal{P}_{site}^i$ is the probability of finding a lattice site with the local chemical environment $i$ among all sites. From Eq. 4.15, one can relate the vacancy formation magnetic free energy of a given local chemical environment, $G_f^{mag,i}$, to the (global) vacancy formation magnetic free energy $G_f^{mag}$ as follows

$$G_f^{mag,i} - G_f^{mag} = -k_B T \ln \frac{C_V^{eq,i}}{C_V^{eq}} \tag{4.16}$$

$$= -k_B T \ln \frac{\mathcal{P}_V^i}{\mathcal{P}_{site}^i} \tag{4.17}$$

In a canonical Monte Carlo simulation, a vacancy is introduced in the system (with the magnetic and chemical configurations already fully relaxed), and it is exchanged with a randomly selected atom according to the Metropolis algorithm. The probabilities $\mathcal{P}_V^i$ and $\mathcal{P}_{site}^i$ can then be measured, respectively, as the percentages of times finding the vacancy and a randomly selected atom with a specific local environment. For a given nominal composition, we perform at least $4 \times 10^7$ vacancy-atom exchanges at each studied temperature.

A first application of the above relation is to compute the solute-vacancy binding free energy in the extremely dilute alloy (namely with only one solute). We start from the usual definition of the binding energy $E_b$ for a 1NN solute-vacancy pair:

$$E_b = E_S + E_V - E_{S+V} - E_0 \tag{4.18}$$

where the four terms on the right-hand side are the energies of the systems (all having the same number of sites) with an isolated solute, an isolated vacancy, a 1NN solute-vacancy pair, and no solute or vacancy, respectively. A positive binding energy as defined in Eq. 4.18 indicates an attractive solute-vacancy interaction. $E_b$ can be rewritten as the difference of the vacancy formation energies, namely:

$$E_b = (E_V + e - E_0) - (E_S + e - E_{S+V}) \tag{4.19}$$

$$= E_f - E_f^{1NN} \tag{4.20}$$

where $e$ is the energy per atom of the solvent, $E_f$ is the vacancy formation energy in the solvent, and $E_f^{1NN}$ is the formation energy of a vacancy as the first nearest neighbour of the solute. Similarly, the binding magnetic free energy of a vacancy-solute pair can be written as the difference of two vacancy formation magnetic free energies:

$$G_b^{mag,i} = G_f^{mag} - G_f^{mag,i} \tag{4.21}$$

where $i$ can be the local environment with a solute in the first or second shell. The predicted magnetic free energies of the Ni-vacancy binding in Fe and the Fe-vacancy binding in Ni are shown in Table 4.6. The solute-vacancy interactions in Fe and Ni in the magnetic ground state are quite weak, being marginally attractive and repulsive, respectively. The magnetic transition in fcc Fe and Ni changes the binding magnetic free energy only slightly, by less than 0.04 eV. According to these results, the Ni-V interaction in fcc Fe and the Fe-V interaction in Ni are not significant.

TABLE 4.6: Solute-vacancy binding free energy (in eV) in fcc Fe and Ni in the magnetic ground state (MGS) and the PM state. The binding energies in the intermediate temperature range lie between the values of the MGS and PM states. In our convention, a positive value indicates an attraction between the vacancy and the solute.

|  | Ni+V in fcc Fe | | Fe+V in fcc Ni | |
| --- | --- | --- | --- | --- |
|  | 1NN | 2NN | 1NN | 2NN |
| MGS | 0.02 | 0.03 | -0.05 | -0.03 |
| PM (1500 K) | 0.01 | -0.03 | -0.04 | -0.07 |

Next we consider the local vacancy formation magnetic free energy $G_f^{mag,i}$ in the alloys with three representative concentrations, namely 25%, 50% and 75% Ni. The local environment is specified by the local Ni concentration in the first two shells of the site

$$c = \frac{n_{1NN\text{-}Ni} + n_{2NN\text{-}Ni}}{n_{1NN} + n_{2NN}} \qquad (4.22)$$

where $n_{1NN\text{-}Ni}$ and $n_{2NN\text{-}Ni}$ are respectively the numbers of 1NN and 2NN Ni atoms, and $n_{1NN}$ and $n_{2NN}$ are the numbers of 1NN and 2NN atoms, namely 12 and 6, respectively.

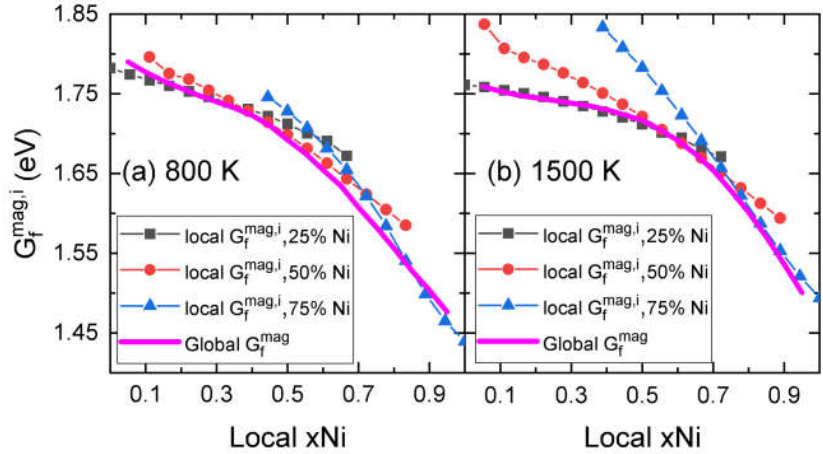

FIGURE 4.11: Local vacancy formation magnetic free energy $G_f^{mag,i}$ as a function of local Ni concentration at 800 K and 1500 K. The local $G_f^{mag,i}$ (the solid lines with symbols) are computed in the alloys with three different nominal compositions. Some of them do not cover the whole $X$ range because of the low probability ($<10^{-5}$) of visiting the corresponding sites for the given nominal composition. For comparison, we also plot the global vacancy formation magnetic free energies (the solid lines without symbols) as functions of nominal Ni concentration at the corresponding temperatures.

As shown in Fig. 4.11, the local vacancy formation magnetic free energy $G_f^{mag,i}$ decreases with increasing local Ni concentration. This behaviour is similar to the one observed previously in the dependence of the global vacancy formation magnetic free energy on the nominal Ni concentration. Indeed, the latter (the dashed line in Fig. 4.11) is found to overlap with different parts of the curves of $G_f^{mag,i}$ depending on the nominal concentration. This is expected, because a majority of the sites have local compositions similar to the nominal one, consequently the corresponding local $G_f^{mag,i}$ are representative of the global $G_f^{mag}$. Meanwhile, it is noted that the local $G_f^{mag,i}$ in the alloy with higher nominal Ni concentration decreases more rapidly with increasing local Ni concentration.

Specifying the local environment by the local Ni concentration in the first two shells does not distinguish between the respective environments in the first and second shells. In Fig. 4.12, we plot the local $G_f^{mag,i}$ as functions of the numbers of 1NN and 2NN Ni atoms. Interestingly, the local $G_f^{mag,i}$ presented in Fig. 4.12 do not always decrease with increasing number of Ni neighbours, despite the decreasing trend of global $G_f^{mag}$. For instance, in the alloy with 25% Ni, $G_f^{mag,i}$ increase with increasing number of 2NN Ni for a given number of 1NN Ni less than five. The same situation occurs in the alloy with 50% Ni for a given number of 1NN Ni less than three. Meanwhile, $G_f^{mag,i}$ in all of the three alloys in Fig. 4.12 increases with the number of 1NN Ni when the latter exceeds a certain value and when there are less than two 2NN Ni.

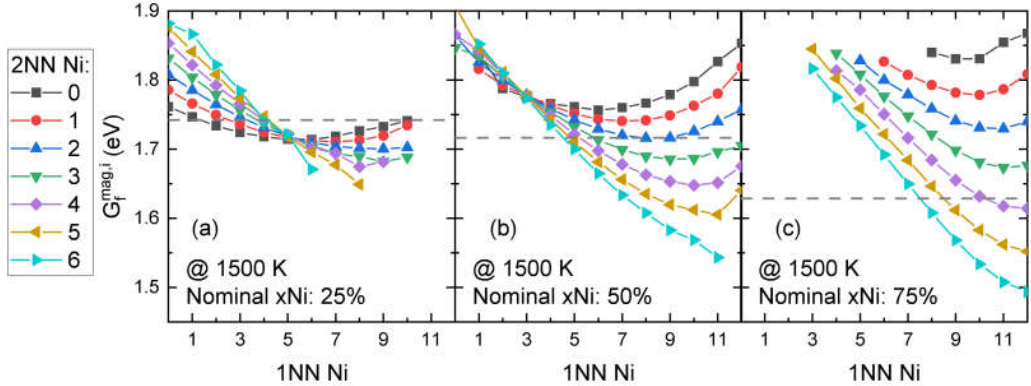

FIGURE 4.12: Local vacancy formation magnetic free energy $G_f^{mag,i}$ as a function of local chemical environment in the alloy with (a) 25% Ni, (b) 50% Ni and (c) 75% Ni at 1500 K. The $X$ axis denotes the number of 1NN Ni, and each curve represents the results with a given number of 2NN Ni. The vertical dashed line denotes the (global) vacancy formation free energy. Some curves do not cover the whole $X$ range because of the low probability ($<10^{-5}$) of visiting the corresponding sites for the given nominal composition.

### 4.3.3 Temperature dependence of vacancy formation properties

We study the temperature dependence of $G_f^{mag}$ in the Fe-Ni alloys with 50% and 75% Ni, where the system undergoes first an order-disorder chemical transition and then a ferromagnetic-paramagnetic transition. The temperature gap separating the chemical and magnetic transitions is about 200 K and 80 K in the alloys with 50% and 75% Ni, respectively. In Fig. 4.13, we present the Monte Carlo results using classical statistics for spins. In the equilibrium phases, $G_f^{mag}$ first increase with increasing temperature, and then decrease abruptly across the chemical transition temperatures and finally increase slowly.

This variation of $G_f^{mag}$ is clearly related to the changes of magnetic and chemical orders in the equilibrium phases. To separate the magnetic and chemical contributions (as much as possible since they are coupled), we calculate $G_f^{mag}$ in the structures where the chemical configurations are frozen while the magnetic configurations are equilibrated with temperature. It can be seen that $G_f^{mag}$ in the equilibrium phases with 50% and 75% Ni follow closely those in the corresponding ordered structures up to 500 K and 600 K, respectively. This can be correlated with our previous results showing that these alloys remain fairly ordered at 500 K and 600 K, with an ALRO>0.96. Near the chemical transition temperatures, $G_f^{mag}$ in the equilibrium phases deviate the trends in the ordered structures, but approach those in the disordered phases. As $G_f^{mag}$ are higher in the ordered structures than in the respective disordered ones, the chemical transitions thus lead to a decrease in $G_f^{mag}$.

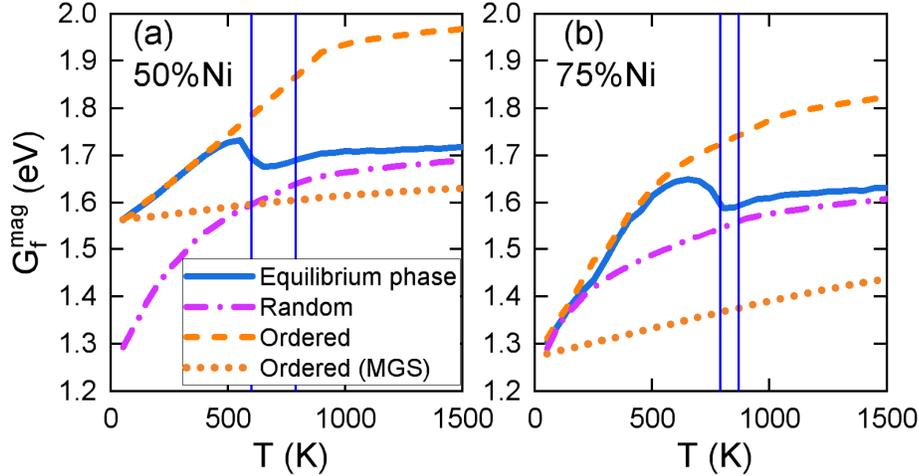

FIGURE 4.13: $G_f^{mag}$ as a function of temperature in the alloys with (a) 50 and (b) 75% Ni. The solid lines are obtained in the equilibrium structures, and the vertical lines denote the corresponding chemical and magnetic transition temperatures. The dash-dotted (or dashed) lines are obtained in the random (or ordered) structures where the chemical order is frozen and only the magnetic order evolves with temperature. The dotted lines are obtained in the ordered structures in the magnetic ground state (both chemical and magnetic configurations are frozen).

Fig. 4.13 indicates that $G_f^{mag}$ in the ordered and disordered structures increase with increasing temperature. Such variations are related not only to magnetic excitations, but also to the changes of weights in the local vacancy formation energies. Indeed, even if the structures are chemically and magnetically frozen, $G_f^{mag}$ still tend to increase with temperature. For example, $G_f^{mag}$ in L1$_2$-FeNi$_3$ with the magnetic ground state can be calculated as

$$G_f^{mag} = -k_B T \ln[0.25 \cdot \exp(-\frac{E_f^{\text{Fe-lat}}}{k_B T}) + 0.75 \cdot \exp(-\frac{E_f^{\text{Ni-lat}}}{k_B T})] \qquad (4.23)$$

where $E_f^{\text{Fe-lat}}$ and $E_f^{\text{Ni-lat}}$ are the vacancy formation energies in the Fe and Ni sublattices given in Table 4.5, respectively. As suggested by the expression, the lower $E_f^{\text{Fe-lat}}$ has a dominant weight in the evaluation of $G_f^{mag}$ at low temperatures, but $E_f^{\text{Fe-lat}}$ and $E_f^{\text{Ni-lat}}$ eventually have similar weights at high temperatures. As a result, $G_f^{mag}$ in L1$_2$-FeNi$_3$ increases from $E_f^{\text{Fe-lat}}$ at low temperature, to the arithmetic average of all local vacancy formation energies at the high-temperature limit, namely:

$$G_f^{mag} = 0.25E_f^{\text{Fe-lat}} + 0.75E_f^{\text{Ni-lat}}, \text{ if } T \rightarrow +\infty \qquad (4.24)$$

The dotted lines in Fig. 4.13 denote $G_f^{mag}$ in L1$_0$-FeNi and L1$_2$-FeNi$_3$ in the respective magnetic ground states. Comparing these results to $G_f^{mag}$ in the same ordered structures but with the equilibrium magnetic configurations, it can be concluded that the variation of $G_f^{mag}$ in the latter cases are mainly due to the magnetic effects. More specifically, this is primarily related to the transversal spin fluctuations, because there is no strong variation with temperature in the average magnetic-moment magnitudes as shown in Fig. 4.14. We note that the results in Fig. 4.13 are obtained using classical statistics for spins, which is known to exaggerate the magnetic excitations at low temperatures.

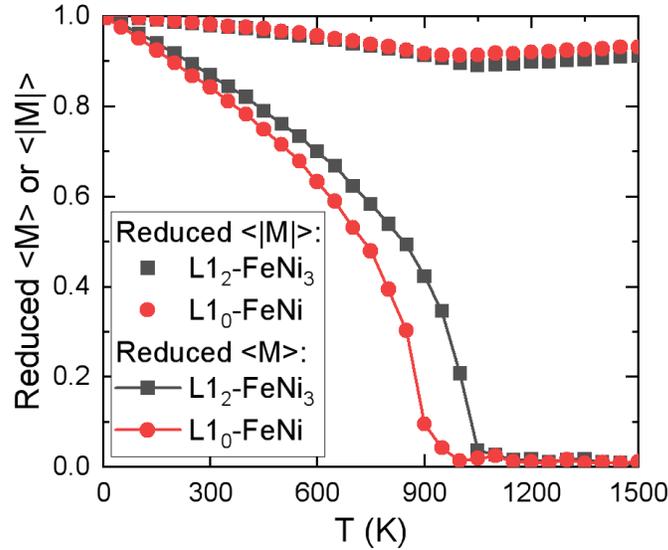

FIGURE 4.14: Reduced magnetization $< M >$ and reduced average magnitude of magnetic moments $< |M| >$ as functions of temperature in the ordered structures L1$_2$-FeNi$_3$ and L1$_0$-FeNi.

## 4.4 Self-interstitial atom (SIA) formation and migration in fcc Fe and Ni

Self-interstitial atom (SIA) is another important type of point defects. Compared with vacancy, SIA usually has a much higher formation energy but a lower migration energy. The knowledge of SIA properties is particularly important for predicting the atomic diffusion and the microstructural evolution in irradiated materials which contain an excess amount of SIAs. As austenitic steels in the operational temperature range are typically in the paramagnetic state, it is therefore crucial to go beyond the use of magnetic ground state and investigate the effects of magnetic states on SIA properties. In this section, we employ DFT calculations to elucidate the magnetic effects on the SIA properties in fcc Fe and Ni. We first identify the most stable SIA configuration by comparing the formation energies of various SIA configurations. Then we discuss the formation and migration properties of the most stable SIA configuration.

### 4.4.1   Formation properties

**Relative stability of SIA configurations**

The SIA configurations considered are presented in Fig. 4.15. They include the <100>, <110>, <111> types of dumbbells, and the octahedral and tetrahedral SIAs. Note that the directions that are crystallographically equivalent may not be equivalent for some magnetic orderings. For instance, the [100] and [001] dumbbells are equivalent in the FM and NM structures. But they become non-equivalent in the AFS and AFD structures, where the spin orientations are the same in the same XY plane but differ in the different XY plane.

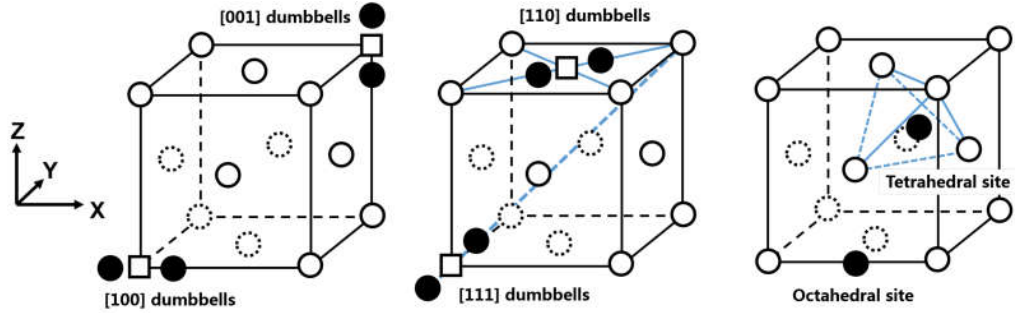

FIGURE 4.15: High symmetry SIA configurations in the fcc lattice. The atoms in the fcc lattice sites are represented by the unfilled circles (the dashed ones are atoms behind the front plane). The SIAs and vacancies are represented by the filled circles and the unfilled squares, respectively. The dumbbells are identified by their axis direction.

The SIA formation energy $E_f$ in the pure system is calculated as follows:

$$E_f = E_{tot,SIA} - \frac{N+1}{N} E_{tot,0} \qquad (4.25)$$

where $E_{tot,0}$ is the energy of the supercell without defect, and $E_{tot,SIA}$ is the energy of the supercell inserted with an additional atom (SIA). The DFT calculations with the perfect and defective supercells are performed by fully optimizing the atomic positions, the supercell shape and volume. The calculated SIA formation energies in fcc Fe and Ni with different magnetic states are shown in Table 4.7. The results in AFD Fe are obtained with supercells containing 128 sites, and the rest are obtained with supercells containing 108 sites. Note that more than one value is shown for some types of SIAs in AFS and AFD Fe. As mentioned above, this is because there are non-equivalent SIA configurations due to the underlying magnetic structures.

Before commenting on the obtained formation energies, we first compare them with previous theoretical results. The formation energies of various SIA configurations in fcc Fe (in the AFD, AFS and FM states) were calculated in a previous DFT study by Klaver *et al.* [151], using the GGA functional parametrized by Perdew and Wang [250] and spin interpolation of the correlation potential provided by the improved Vosko-Wilk-Nusair scheme [251]. The agreement between the present study and Ref. [151] in AFS Fe is very good (difference<0.03 eV). Our results for a given SIA configuration in AFD Fe are higher by 0.2 to 0.3 eV than those in Ref. [151], but the predicted relative stability between different configurations is the same. Regarding fcc Ni, the SIA formation energies in its FM state were studied by Tucker *et al.* [239]. The results in FM Ni presented here differ by less than 0.02 eV from those in Ref. [239] for any given configuration, except for the <110> dumbbell of which our predicted value is 0.1 eV lower than that in Ref. [239].

TABLE 4.7: SIA formation energies (in eV) in fcc Fe and Ni. For AFS and AFD Fe, there are SIA configurations that are of the same type but non-equivalent. Therefore, more than one value of $E_f$ are shown for some types of SIA configurations. The results in magnetic SQSs are also shown (see text for details).

| | Fe | | | | Ni | | |
|---|---|---|---|---|---|---|---|
| | AFD | AFS | NM | mSQS | FM | NM | mSQS |
| <100> | 3.348, 3.605 | 3.542, 3.637 | 3.511 | 3.615±0.151 | 4.091 | 3.841 | 3.726±0.040 |
| tetra | 4.010, 4.146 | 4.312 | 4.596 | / | 4.716 | 4.428 | / |
| octa | 4.494 | 4.341 | 4.683 | 4.331±0.161 | 4.272 | 4.001 | 3.911±0.071 |
| <111> | 4.010 | 4.550 | 4.761 | / | 4.712 | 4.362 | / |
| <110> | 5.101 | 4.737, 4.774 | 4.947 | / | 4.878 | 4.420 | / |

It can be seen from Table 4.7 that the <100> dumbbell formation energy in fcc Fe ranges from 3.35 to 3.64 eV depending on the magnetic state. These values are much lower than those of other types of configurations, ranging from 4.01 to 5.10 eV. In fcc Ni, the formation energy of the <100> dumbbell is between 3.84 and 4.09 eV, lower than the values of the other configurations (4.36 to 4.88 eV) except the octahedral one. In fact, the octahedral configuration is not stable: further DFT calculations in fcc NM Fe, and fcc FM and NM Ni show that there are negative phonon frequencies associated with the octahedral configuration, which is in fact a saddle-point configuration for the migration of the <100> dumbbell (via a pure translation), as will be further discussed in the next subsection.

We intend to estimate the effect of paramagnetism on the formation energies of the <100> dumbbell and the octahedral interstitial. The formation energies are computed from DFT using magnetic SQSs of fcc Fe and Ni, in which the atomic positions are fixed to the ones obtained in the magnetic ground state (AFD Fe or FM Ni), and the cell shape and volume are optimized. For Fe, the initialized spin orientations are able to be kept in the converged magnetic SQSs, with the spin magnitudes fully optimized. For Ni which exhibits a more itinerant character of magnetism, the initialized spin orientations cannot be kept in the converged magnetic SQSs. Therefore, we perform constrained-moment DFT calculations, fixing the spin orientations and magnitudes (fixed to 0.6 $\mu_B$) on atoms other than the interstitials.

In Table 4.7, we present the average value and the standard deviation of the formation energies for a given SIA configuration with nine and five different magnetic configurations for Fe and Ni, respectively. The average formation energies of the <100>

dumbbell and the octahedral interstitial in the mSQSs of Fe are within the ranges of the values in the magnetically ordered states. Meanwhile, the average formation energies of the <100> dumbbell and the octahedral interstitial in the mSQSs of Ni are lower than the ones in the FM and NM states. According to these results, the magnetic transition has a nonnegligible effect on the SIA formation energies, but it does not change significantly the relative energy difference between different SIA configurations. The <100> dumbbell is therefore expected to be the most stable and relevant SIA configuration at finite temperatures.

**Formation properties of the <100> dumbbell**

Our results above confirm that the most stable SIA configuration is the <100> dumbbell in both fcc Fe and Ni, irrespective of the magnetic state. This is in agreement with previous DFT studies in fcc Fe and Ni [151, 218, 252], and experimental evidence in fcc Ni [253]. Therefore, in the following discussion we focus on the properties associated with the <100> dumbbell, with a comparison to vacancy formation properties.

As shown in Table 4.8, the spin magnitudes of the atoms forming the <100> dumbbell are much reduced compared with the values of bulk atoms in fcc Fe, whereas those of the atoms nearby are increased or decreased, depending on if they are tensile or compressive sites, by up to 0.4 $\mu_B$. For comparison, a vacancy can also induce a perturbation of 0.3-0.4 $\mu_B$ on the magnetic-moment magnitudes of its neighbours. Regarding Ni, the magnetic-moment magnitudes of all atoms including the SIAs, are in the range of 0.62 to 0.69 $\mu_B$. The same conclusion is also found in other SIA configurations, and the vacancy-containing structure of Ni. Therefore, the magnetic-moment magnitude of Ni is insensitive to the presence of isolated point defects.

TABLE 4.8: Magnetic-moment magnitudes (in $\mu_B$) of atoms in the <100> dumbbell configuration of fcc Fe and Ni.

|                           | Fe        |           | Ni        |
|---------------------------|-----------|-----------|-----------|
|                           | AFD       | AFS       | FM        |
| <100> dumbbell            | 0.06-0.53 | 0.51-0.78 | 0.69      |
| 1NN (tensile sites)       | 2.25-2.39 | 1.99-2.03 | 0.62      |
| 1NN (compressive sites)   | 1.60-1.67 | 1.10-1.23 | 0.64-0.66 |
| Bulk atoms                | 2.03      | 1.56      | 0.65      |

In Table 4.9 are presented the SIA and vacancy formation energies in the low-temperature and high-temperature limits, estimated in the magnetic ground state and the magnetic SQS. In Fe, the magnetic disorder is suggested to lead to an increase in both the SIA and vacancy formation energies, with the effect slightly stronger on the SIA formation. By contrast, the magnetic disorder in Ni results in a decrease of 0.365 eV in the SIA formation energy, but a small increase of 0.09 eV in the vacancy formation energy. These results indicate that the magnetic effects are much stronger on the SIA formation than the vacancy formation in fcc Ni. A more quantitative conclusion requires to account for the effects of thermal longitudinal spin variations which are not treated at present.

TABLE 4.9: SIA and vacancy formation energies (in eV) in the magnetic ground state and the magnetic SQS.

| defect | Fe | | Ni | |
|---|---|---|---|---|
| | AFD | mSQS | FM | mSQS |
| <100> dumbbell | 3.348 | 3.615 | 4.091 | 3.726 |
| vacancy | 1.83 | 2.04 | 1.43 | 1.52 |

The vibrational entropies of the <100> dumbbell formation are computed within the harmonic approximation for the magnetic ground state and the nonmagnetic state of fcc Fe and Ni, as shown in Table 4.10. The vibrational entropy of SIA formation differs by 1.5 $k_B$ between the AFD and NM states of Fe, but it is very similar in FM and NM Ni. The vibrational entropy of SIA formation in fcc Ni is found to be particularly large, compared with the values of SIA and vacancy formation in fcc Fe and that of vacancy formation in fcc Ni. Our results for the SIA formation in fcc Ni are in agreement with the values from the DFT study by Tucker [254] (harmonic: 12.7 $k_B$) and an EAM study [255] (harmonic: 12.7 $k_B$; anharmonic: 13.1 $k_B$). The difference between the SIA and vacancy formation vibrational entropies is around 11 $k_B$, namely the vibrational contribution to the SIA equilibrium concentration is $\exp(\frac{\Delta S_f}{k_B}) = 6 \times 10^4$ times larger than to the vacancy equilibrium concentration. However, due to the high SIA formation energy, the SIA equilibrium concentration is still several orders of magnitudes lower than the vacancy equilibrium concentration in fcc Ni. For instance, if we use the SIA formation energy in the magnetic SQS of Ni and the SIA formation entropy in FM Ni, the SIA equilibrium concentration at 1500 K is $10^{-7}$, compared with the computed vacancy equilibrium concentration of $10^{-4}$ (similar to the measured value) at the same temperature.

TABLE 4.10: SIA and vacancy formation entropies (in $k_B$) in fcc Fe and Ni.

| defect | Fe | | Ni | |
|---|---|---|---|---|
| | AFD | NM | FM | NM |
| <100> dumbbell | 4.75 | 3.20 | 13.00 | 12.70 |
| vacancy | 3.05 | 2.77 | 2.15 | 1.52 |

### 4.4.2 Migration properties

The rate of self-diffusion via vacancy or SIA is determined by the point-defect concentration and the related migration barrier. The point-defect concentration is controlled by the point-defect formation free energy under equilibrium conditions, or the external point-defect source under nonequilibrium conditions (e.g. under irradiation). Due to the high SIA formation energy, the SIA equilibrium concentration is usually extremely low, but interstitial-mediated diffusion can be important for radiation-induced diffusion, segregation or precipitation [116].

In this subsection, we study the migration of the most stable SIA configuration, namely the <100> dumbbell. In Fig. 4.16, we illustrate three commonly considered migration mechanisms [239, 255, 256]. In a rotation-translation (RT) jump, the SIA moves to a 1NN site with a rotation of the <100> dumbbell axis. In a pure-translation (PT) jump, the SIA moves to a 2NN site by passing directly the octahedral site, without changing the axis orientation. In a pure-rotation (PR) jump, the SIAs rotate around the original site. Unlike the RT and PT jumps, the PR jumps alone do not lead to long-distance diffusion, therefore in the following we only investigate systematically the magnetic effects on the migration energies of the RT and PT jumps. We consider the PR jumps only in the end to discuss the dimensionality of the diffusion.

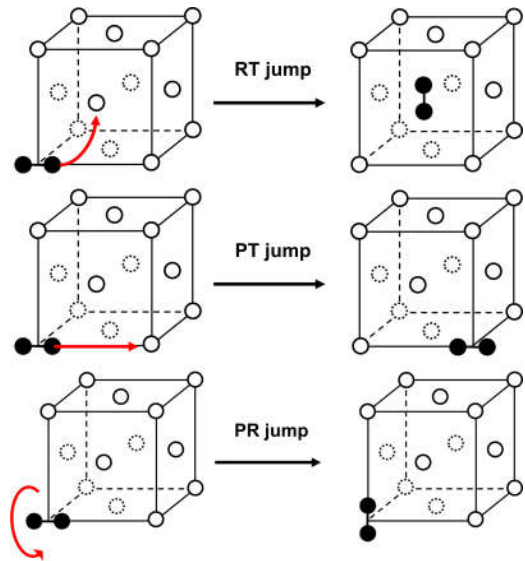

FIGURE 4.16: Schematics of rotation-translation (RT), pure-translation (PT) and pure-rotation (PR) jumps involved in the SIA migration.

### Effects of spin-orientation variation

The migration barriers obtained from DFT are presented in Table 4.11. The saddle-point configurations in the magnetically ordered states (FM, AFS, AFD, NM) are calculated using the NEB method. The saddle-point configurations in the magnetic SQSs are obtained with the same atomic positions as in AFD Fe or FM Ni, but with volume fully relaxed. The calculations in the AFS state and magnetic SQSs of Ni are performed by fixing the magnetic-moment magnitudes to 0.6 $\mu_B$, because these magnetic structures are unstable without magnetic constraint.

The migration barrier $E_m$ in Ni with different magnetic states differs by less than 0.05 eV for a given mechanism. $E_m$ of the RT jump is about 0.04 eV lower than that of the PT jump for the same magnetic state of Ni. As this difference is small, both mechanism are relevant for the interstitial-mediated diffusion in Ni. Our results in Ni are in agreement with the previous DFT results in FM Ni [239], and are consistent with the experimental values of 0.14-0.21 eV measured from magnetic relaxation and electrical resistivity methods [203, 237, 238].

TABLE 4.11: Migration barriers (in eV) in fcc Fe and Ni.

| system | magnetic state | RT jump | PT jump |
|--------|----------------|---------|---------|
| Ni | FM | 0.14 | 0.18 |
|  | NM | 0.10 | 0.16 |
|  | AFS | 0.10 | 0.14 |
|  | mSQS | 0.14±0.06 | 0.18±0.08 |
| Fe | AFD | 0.13, 0.16, 0.30, 0.42, 0.56 | 0.89, 1.15 |
|  | AFS | 0.22, 0.25, 0.32 | 0.70, 0.78 |
|  | NM | 0.33 | 1.17 |
|  | mSQS | 0.28±0.07 | 0.72±0.13 |

Contrasted to the cases in Ni, the magnetic state is found to have a significant effect on $E_m$ in fcc Fe. For the RT jump, $E_m$ varies from 0.13 eV to 0.56 eV, with the most of the values ranging from 0.1 to 0.4 eV. $E_m$ of the RT jump is therefore much lower than the PT value ranging between 0.7 to 1.1 eV. Therefore, the present DFT results suggest that the RT jump is the most favourable mechanism for the SIA migration in fcc Fe.

**Effects of spin-magnitude variation**

The above results reveal that effects of magnetic orderings (i.e. spin orientations) on SIA migration energy. In the following, we also discuss the role of thermal longitudinal variations on SIA migration energy.

Using constrained-moment calculations, we fix the spin magnitudes to the same value for all the atoms, except for the SIAs on which the spin magnitudes are optimized. The migration barriers are not obtained from the NEB calculations. Instead, we use the atomic positions obtained in the previous calculations (the ones used to obtain the results in Table 4.11), and optimize only the shape and volume of the supercells. The migration barriers are computed as the energy difference of the initial and saddle-point configurations. The obtained $E_m$ are presented in Fig. 4.17.

For a given magnetic ordering in Ni, the variation in $E_m$ due to the change of spin magnitudes is less than 0.05 eV, for both RT and PT mechanisms. Therefore, $E_m$ of the RT and PT jumps in Ni are expected to be insensitive to the thermal transversal and longitudinal fluctuations.

The spin-magnitude effects on $E_m$ are stronger in fcc Fe. In particular, $E_m$ in AFD Fe varies from 0.1 to 0.35 eV depending on the spin magnitudes. In addition, it is noted that the results of $E_m$ in AFD do not approach the NM value with decreasing spin magnitudes. This indicates that the lattice relaxation is not negligible when changing the spin magnitudes in AFD Fe. To verify this, we also optimize the atomic positions of the initial and final configurations, and compute the saddle-point configuration via the NEB method. The results in AFD Fe are represented by the diamond symbols in Fig. 4.17, which suggest that the lattice relaxation can indeed induce a difference of up to 0.2 eV in $E_m$.

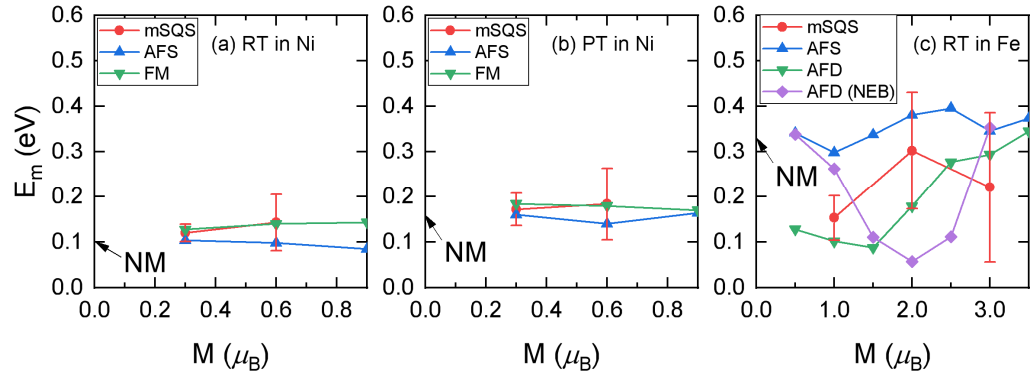

FIGURE 4.17: $E_m$ as functions of spin magnitudes for (a) the rotation-translation jump in Ni, (b) the pure-translation jump in Ni, and (c) the rotation-translation jump in Fe. The arrows mark $E_m$ in the NM systems.

**Dimensionality of the diffusion**

The above results in different magnetic states show that the RT jump is the most favourable diffusion mechanism in fcc Fe, whereas the RT and PT are both likely to occur in fcc Ni considering their similar migration energies. We note that the diffusion via the RT jumps alone is three-dimensional, whereas the diffusion via the PT jumps alone is one dimensional. The PT jumps can contribute to the three-dimensional diffusion when combined with the RT jumps, or the PR jumps, if the latter can also be easily activated. To investigate this possibility in fcc Ni, we compute the migration energies for the PR jumps in the FM and NM states. They are found to be 0.91 eV and 0.71 eV in the FM and NM states, respectively, much higher than the ones for the PT and RT jumps. The PR jumps are thus much harder to activate compared to the PT and RT jumps.

To conclude, the three-dimensional SIA diffusion in fcc Fe occurs primarily via the RT jumps, while the three-dimensional SIA diffusion in fcc Ni involves both RT and PT jumps. In the latter case, it can be expected that correlation factors, even for the self-diffusion, depend on temperature because of the activations of both PT and RT jumps.

**Comparison with experiments**

Our results above suggest that with all magnetic effects taken into account, the migration barrier of Ni should be within the range of 0.1 to 0.25 eV, whereas that of fcc Fe should be within the range of 0.1 to 0.5 eV. The results in Ni are consistent with the experimental values of 0.14-0.21 eV (high voltage electron microscope [237], magnetic aftereffect and electrical resistivity recovery [203, 238]). On the other hand, there is no measurement of $E_m$ in fcc Fe to the best of our knowledge. Dimitrov *et al.* [235, 236] measured $E_m$ in fcc Fe-Ni-Cr alloys using the electrical resistivity measurements. Their samples were first irradiated at cryogenic temperatures to create excess frozen defects. Then the irradiated samples were annealed at higher temperatures (180-220 K [235, 236]), during which interstitials became mobiles and their diffusion can be correlated with the measured resistivity variation rate. Note that the same procedure can not be performed in pure fcc Fe due to its instability at low temperatures.

TABLE 4.12: Experimental activation enthalpies of SIA migration in fcc Fe-Ni-Cr alloys [235, 236] and fcc Ni [203, 237, 238].

| Fe (wt%) | Ni (wt%) | Cr (wt%) | $E_m$ (eV) |
|---|---|---|---|
| 67 | 25 | 8 | 0.83±0.06 |
| 65 | 25 | 10 | 0.83±0.06 |
| 62 | 25 | 13 | 0.87±0.06 |
| 59 | 25 | 16 | 0.89±0.06 |
| 64 | 20 | 16 | 0.92±0.03 |
| 59 | 25 | 16 | 0.89±0.06 |
| 39 | 45 | 16 | 0.70±0.02 |
| 9 | 75 | 16 | 0.50±0.04 |
| 0 | 100 | 0 | 0.14-0.21 |

The experimental migrational enthalpies in fcc Fe-Ni-Cr alloys [235, 236] and fcc Ni [203, 237, 238] are presented in Table 4.12. Based on these data, a linear relation has been proposed [236]:

$$E_m = 0.96 + 0.65x_{Cr} - 0.80x_{Ni} \tag{4.26}$$

where $E_m$ is the interstitial migration barrier in eV, and $x_{Cr}$ and $x_{Ni}$ are the atomic fractions of Cr and Ni, respectively. However, this relation of $E_m$ implies that the interstitial migration barrier in fcc Fe is 0.96 eV, whereas it seems very unlikely to go beyond 0.5 eV according to our DFT results. This discrepancy requires further investigation of migration properties in fcc Fe-Ni-Cr alloys. For instance, in dilute bcc Fe-P, the fast diffusion of the P interstitial ($E_m \approx$ 0.2 eV) can be significantly reduced due to the trapping of a Fe-P dumbbell near a substitutional P atom (with a dissociation energy > 1 eV) [257]. Consequently, similar scenario of forming highly stable defect-solute complexes may also be a possible cause that the measured effective migration energy is higher than the calculated value.

## 4.5 Conclusion

In this chapter, the magnetic effects of vacancy formation properties are investigated for the Fe-Ni systems, using DFT calculations and the DFT-parametrized EIMs. The same effects on the SIA formation and migration properties are studied via DFT calculations.

In the first section, the magnetic effects on vacancy formation properties in fcc Ni, and in bcc and fcc Fe are examined in details. DFT calculations reveal a larger dispersion of the vacancy formation energies $E_f$ in bcc and fcc Fe than in fcc Ni. The dispersion is related to the spin orientations and magnitudes. The spin orientations have a dominant influence on $E_f$ in bcc Fe, while $E_f$ in fcc Fe is more dependent on the spin magnitudes; in fcc Ni both types of effects are small.

Based on an efficient Monte Carlo scheme, the vacancy formation properties are determined as functions of temperature for the three systems. Overall, the predicted vacancy formation energies and equilibrium vacancy concentrations are in good agreement with the experimental data. The vibrational contribution to the vacancy formation is stronger than the magnetic one for all the three systems, but the magnetic contribution is nonnegligible in bcc Fe.

The impact of the magnetic transition is more significant in bcc Fe than in fcc Fe and Ni. The substantial decrease of $E_f$ across the magnetic transition in bcc Fe is mainly due to the transversal spin variations, coherent with the strong dependence of $E_f$ on the spin orientations in bcc Fe as predicted by DFT. Consistently, a weaker dependence of $E_f$ on the spin orientation in fcc Fe and Ni leads to a smaller variation of the vacancy properties below and above the magnetic transition.

In fcc Fe, we note a significant effect of longitudinal spin excitations on the magnetic free energy of vacancy formation $G_f^{mag}$, resulting in its steady decrease above the Néel point. These are consistent with the DFT results demonstrating a stronger dependence of $E_f$ on the spin magnitude rather than the spin ordering in fcc Fe. Interestingly, the predicted $E_f$ in the paramagnetic state is close to the AFD-state value, but it is much lower than $E_f$ in nonmagnetic fcc Fe. Our results therefore suggest that the paramagnetic fcc Fe is better represented by the AFD state than the nonmagnetic one, which has been used in some recent studies on diffusion properties in fcc Fe. In fcc Ni, the magnetic transition leads to an increase in $G_f^{mag}$, while the longitudinal spin fluctuations above the Curie temperature lead to a steady increase in $G_f^{mag}$.

Our results reveal that the features of vacancy formation in fcc Fe and Ni are well distinct from those in bcc Fe, pointing out a relevant effect of longitudinal spin excitations on the vacancy formation properties, which can not be well captured by a classical Heisenberg model.

In the second section, the EIM of fcc Fe-Ni alloys is extended from those of fcc Fe and Ni. Its predictions of $E_f$ in the ordered and disordered structures are validated by the DFT results. Monte Carlo simulations are performed to study the temperature and concentration dependences of vacancy formation properties.

$G_f^{mag}$ in the alloys with 50% and 75% Ni tend to increase with increasing temperature, except in the temperature range close to the chemical transitions where an abrupt decrease in $G_f^{mag}$ is observed. Further analysis reveals that $G_f^{mag}$ increases with magnetic disorder but decreases with chemical disorder. Although the variation in $G_f^{mag}$ is more visible across the chemical transition than the one across the magnetic transition, in the whole temperature range the overall increase due to the magnetic excitations is larger than the decrease resulting from the loss of long-range order.

We investigate $G_f^{mag}$ over the whole composition range for several temperatures. For a given temperature, $G_f^{mag}$ in the disordered alloys decrease with increasing Ni concentration. We find a large difference between $G_f^{mag}$ in the concentrated random alloys with the paramagnetic state and the magnetic ground state. The maximum of the difference is found in the alloys with around 65% Ni, where the magnetic exchange interaction is the strongest. Our results are within the uncertainty of the vacancy formation energies from two measurements in alloys, requiring further experimental investigation.

In the third section, the SIA formation and migration properties in pure fcc Fe and Ni are computed from DFT, using magnetically ordered and disordered states.

Among the various SIA configurations studied, the <100> dumbbell is found to be the most stable for both fcc Fe and Ni, in agreement with previous theoretical and available experimental results. The subsequent calculations and discussions are therefore focused on only the properties of this SIA configuration.

Magnetic disorder is suggested to increase the SIA formation energy in Fe, while this is opposite in Ni. The magnetic state in Ni is shown to have stronger impact on the SIA than the vacancy formation energy, whereas this impact in Fe is similar and important for both SIA and vacancy. The results of the vibrational entropies of SIA formation in Fe and Ni do not suggest a strong dependence on the magnetic state. It is noted that Ni has a SIA formation entropy of 13 $k_B$, much larger than its vacancy formation entropy. For comparison, the SIA formation entropy in Fe is estimated to 3-5 $k_B$, higher than its vacancy formation entropy of $\approx$3 $k_B$.

We consider mainly the pure-translation (PT) and the rotation-translation (RT) mechanisms for the migration of the <100> dumbbell. In Ni, the migration energy $E_m$ is insensitive to the spin orientations and magnitudes. $E_m$ of the RT and PT jumps are estimated to be 0.10-0.14 eV and 0.14-0.18 eV, respectively, in agreement with the measured values. Both mechanisms are thus relevant to the SIA diffusion, with the RT jump more favourable.

Meanwhile, the magnetic effects in Fe are stronger than in Ni, leading to a wider dispersion in $E_m$. Most of the values for the RT jumps are in the range of 0.15 to 0.4 eV. They are considerably lower than $E_m$ of the PT jump ranging from 0.7 to 1.2 eV. Therefore, our results suggest that only the RT mechanism is responsible for the SIA migration in Fe. The calculated $E_m$ in Fe is much lower than the value of 0.96 eV extrapolated from the measured $E_m$ in fcc Fe-Ni-Cr alloys. The experimental-theoretical discrepancy therefore requires further investigation into SIA migration properties in dilute fcc Fe alloys.

# 5 Vacancy-mediated diffusion properties in fcc Fe-Ni alloys

---

*This chapter is dedicated to the prediction of vacancy-mediated diffusion properties in fcc Fe-Ni alloys. We compare the computed tracer diffusion coefficients and activation energies to the experimental data, which are available only in the paramagnetic disordered alloys, to show the accuracy of our prediction. We discuss the temperature and concentration dependences of the diffusion properties, to elucidate the effects of magnetic and chemical orders as well as the effects of Ni concentration.*

---

## 5.1 State of the art

Diffusion processes are relevant for the kinetics of microstructural evolutions such as precipitation, dissolution, segregation and thermal oxidation [103, 258–261]. One of the most common and fundamental atomic mechanisms of diffusion in crystals is the vacancy mechanism, which consists of consecutive nearest-neighbour atom-vacancy exchanges. It is often considered as the dominant diffusion mechanism for matrix and substitutional atoms under equilibrium conditions [116].

The rate of diffusion can be described in terms of tracer diffusion coefficients. They can be measured experimentally using the radiotracer technique, in which a thin layer of radiotracers for given elements is deposited on the surface of bulk material and the movements of radiotracers are followed [116, 262–264]. In atomistic modelling, one can follow the movements of the tagged atoms and derive the tracer diffusion coefficient from the Einstein relation [116]:

$$D^{A*} = \frac{< r^2_{A*,i} >}{6\tau} \tag{5.1}$$

where $< r^2_{A*,i} >$ is the mean square displacement of the tagged atoms $A*$ in the time interval $\tau$.

Tracer diffusion coefficients are closely related to vacancy concentration and other diffusion properties such as kinetic correlation factors, migration and diffusion activation energies. These quantities can be sensitive to the chemical order in the system. One of the earliest experimental evidence was obtained by Kuper *et al.* [265] in bcc Cu-Zn. According to their findings, the self-diffusion coefficients obey an Arrhenius behaviour only in the fully disordered phase, whereas they show a striking dependence on the degree of chemical long-range order below the B2-disorder transition temperature, and a slight dependence of chemical short-range order [265]. To account for the non-Arrhenius behaviour in diffusion coefficient, Girifalco [266] developed a formalism in which the diffusion coefficient takes the following form:

$$D = D_0 \exp[-\frac{Q_a(T)}{k_B T}] \qquad (5.2)$$

where $D_0$ is a constant while $Q_a(T)$ is the temperature-dependent diffusion activation energy. In Girifalco's model, $Q_a$ is expressed as

$$Q_a(S) = Q_a(S=0) \cdot [1 + \alpha \cdot S^2] \qquad (5.3)$$

where $S$ is the chemical long-range order parameter (equal to 1 for the fully ordered state, and 0 for the disordered state) for the given temperature, and $\alpha$ is a system-dependent constant. If $Q_a$ in the fully ordered and disordered states are known, then

$$\alpha = \frac{Q_a(S=1) - Q_a(S=0)}{Q_a(S=0)} \qquad (5.4)$$

If the temperature dependence of the chemical long-range order $S(T)$ is also known, one can then use Eq. 5.3 to obtain $Q_a(T)$ and hence $D(T)$.

Diffusion properties can also be sensitive to the magnetic state of the system, e.g. in the well-documented case of bcc Fe. Experimentally, it is observed that the self- and solute diffusion coefficients and the corresponding diffusion activation energies in ferromagnetic bcc Fe show significant differences from the paramagnetic-state values [19–21]. The self- and solute diffusion in Fe have also been extensively studied via various computational approaches. Following Girifalco's treatment for nonmagnetic alloys, Ruch *et al.* [247] proposed a similar magnetic model for ferromagnetic unary systems:

$$Q_a(S) = Q_a^{\text{PM}} \cdot [1 + \alpha \cdot S^2] \qquad (5.5)$$

where $Q_a^{\text{PM}}$ is the diffusion activation energy in the paramagnetic state, $S$ is the reduced magnetization at the given temperature. Ruch's model has been adopted in some DFT studies [13, 14, 267] using the DFT-calculated activation energies and experimental data of reduced magnetization as a function of temperature in bcc Fe. One limitation of this approach is that the inputs $S(T)$ can not be obtained directly from DFT, which may hinder the application of the Ruch' model to other magnetic systems where such experimental data are not available. Another approach to account for magnetic effects is spin-lattice dynamics, which simulates atomic movements and spin fluctuations within a unified framework [53]. Recently, it has been applied to the investiation of vacancy formation and self-diffusion properties in bcc Fe [15, 52] with magnetic interaction described by a Heisenberg model. Meanwhile, it should be noted that spin-lattice dynamics is computationally heavy to have significant numbers of atom-vacancy exchanges during simulations, which could be a problem when it is applied to, e.g. , to the study of solute diffusion. In addition, spin-lattice potentials are currently available only for pure systems (e.g. Fe [15, 51, 52] and Co [49, 50]) to our best knowledge, the parametrization of a spin-lattice potential for alloys remains highly challenging [53]. An alternative approach is the use of effective interaction models with explicit spin and atomic variables, in combination with on-lattice Monte Carlo simulations. Its previous applications were mostly focused on magnetic effects in defect-free pure and alloy systems [24, 29, 54–58], while the investigation of the self- and solute diffusion in bcc Fe has been perfomed only very recently [22].

Meanwhile, very few theoretical studies has been devoted to the understanding of magnetic effects on diffusion properties in other magnetic systems, in contrast to the extensive efforts to the case of bcc Fe. Effects of magnetic excitations and transitions are often neglected in the simulations of self- and solute diffusion in magnetic metals other than bcc Fe. For instance, self- and solute diffusion properties in paramagnetic fcc Ni are usually investigated using the ferromagnetic ground state [218, 268]. The diffusion in paramagnetic fcc Fe is even studied using a nonmagnetic state, resulting in a self-diffusion activation energy much higher (around 1 eV) than experimental value [223, 225]. The paramagnetic fcc Fe and Ni are known to exhibit significant longitudinal spin fluctuations [54, 108], but these effects on diffusion have not been discussed in literature.

Regarding fcc Fe-Ni alloys, the vacancy-mediated diffusion properties have been recently examined using molecular dynamics [94, 165, 269–272] with EAM-type empirical interatomic potentials [60, 273, 274]. The accuracy of these interatomic potentials can be questionable. For example, their predicted chemical order-disorder transition temperatures are already beyond the experimental melting points [60]. The predicted vacancy properties between these EAM potentials are shown to be rather different [94]. Due to the use of a fixed vacancy concentration in the molecular dynamics studies [165, 269–272], no comparison with experimental data was performed to demonstrate the accuracy of the results. The effects of chemical order-disorder transitions as well as magnetic excitations and transitions are not investigated in these studies.

The objective of this Chapter is to address effects of magnetism and magnetochemical interplay on diffusion in fcc Fe-Ni alloys, using DFT calculations and Monte Carlo simulations combined with the DFT-parametrized EIM. We intend to extend our understanding of such effects, which were previously limited to bcc Fe, to other magnetic metals with significant longitudinal spin fluctuations such as fcc Fe and Ni, and to concentrated magnetic alloys exhibiting strong magnetochemical interplay. This Chapter is organized as follows. In the first section, we discuss the self- and (Ni- or Fe-) solute diffusion properties in fcc Fe and Ni over the whole temperature range. In the second section, we present the diffusion properties in the alloy with 75% Ni over the whole temperature range, and discuss the effects of magnetic and chemical transitions. In the third section, we compare the computed tracer diffusion coefficients with experimental data for the whole composition range, which are available only in the paramagnetic and disordered alloys.

## 5.2   Temperature dependence of diffusion properties in fcc Fe and Ni

### 5.2.1   Diffusion properties from DFT

**Migration energies in pure Fe and Ni**

The migration energies calculated from DFT for fcc Fe and Ni with different magnetic states are shown in Table 5.1. In the AFS and AFD states, the spin orientations are set to be the same within a XY plane, while the spin orientations along the Z axis follow the order $\uparrow\downarrow\uparrow\downarrow$ in the AFS state, and the order $\uparrow\uparrow\downarrow\downarrow$ in the AFD state. In Table 5.1, we distinguish between the vacancy-atom exchanges within a XY plane (the intra-plane jump) and the ones across the XY planes (the inter-plane jump). The results of the magnetic SQSs (mSQSs) of Fe and Ni are averaged over 30 jumps with different local magnetic configurations. The magnetic-moment magnitudes are optimized in the mSQS of Fe, while they are fixed to 0.6 $\mu_B$ in the mSQS of Ni.

TABLE 5.1: Migration energies (in eV) in fcc Fe and Ni with different magnetic states.

| Magnetic state | fcc Fe | fcc Ni |
|---|---|---|
| AFD (in plane) | 0.740 | / |
| AFD (cross plane) | 0.770, 1.010 | / |
| AFS (in plane) | 0.526 | / |
| AFS (cross plane) | 1.217 | / |
| FM | 0.155, 0.811 | 1.096 |
| NM | 1.399 | 0.967 |
| mSQS | 0.922 | 1.011 |

In Fe, we find a particularly low value of the migration energy (0.155 eV) in the FM state, in agreement with the value of 0.133 eV calculated by Klaver *et al.* [151]. Indeed, they showed that this fct FM structure of Fe is mechanically unstable and relaxes towards the bcc FM structure upon a small orthorhombic perturbation, and that this low migration energy is another possible sign of the structural instability [151]. In the AFS and AFD structures of Fe, the migration energies of the intra-plane jumps are lower than those of the inter-plane jumps, but it should be noted that a fully three-dimensional diffusion would require both types of jumps in these structures.

The results in Table 5.1 show that the migration energy is much more sensitive to the magnetic state in fcc Fe than in Ni. This point is further supported by the analysis of the distribution of migration energies in the mSQSs. As shown in Fig. 5.1, the migration energies in Fe are distributed over a wide range of values, while those in Ni are narrowly centred around the average value.

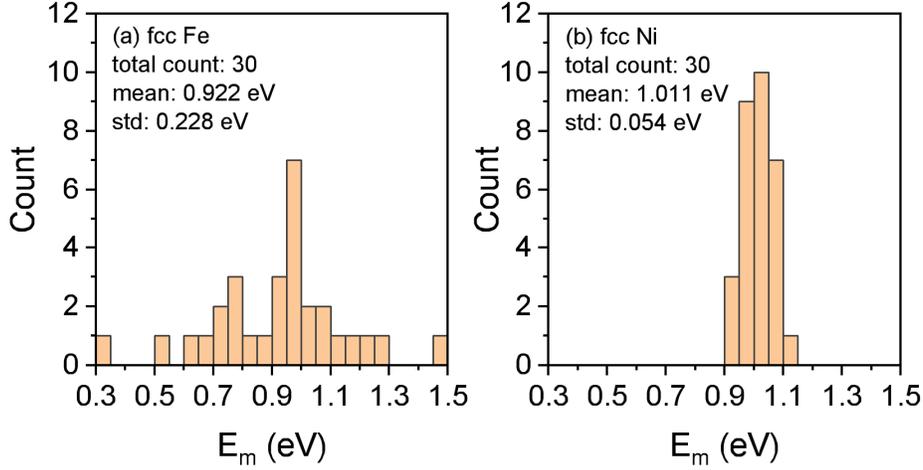

FIGURE 5.1: Distribution of migration energies in the magnetic SQSs of fcc Fe and Ni.

We find that the effects of magnetic-moment magnitudes on migration energies are also more significant in fcc Fe than in Ni. In Fig. 5.2, we present the migration energies in Fe and Ni obtained by fixing the magnetic-moment magnitudes in the stable and saddle-point configurations. In Ni, the migration energy has a weak dependence on the magnetic-moment magnitude, varying less than 0.1 eV in the studied range. By contrast, the migration energy in Fe show a clear decreasing trend with increasing magnetic-moment magnitude. These DFT results suggest that the thermal longitudinal spin variations should have little influence on the migration energies in Ni, whereas they are expected to result in a decrease the migration energies in Fe with increasing temperature.

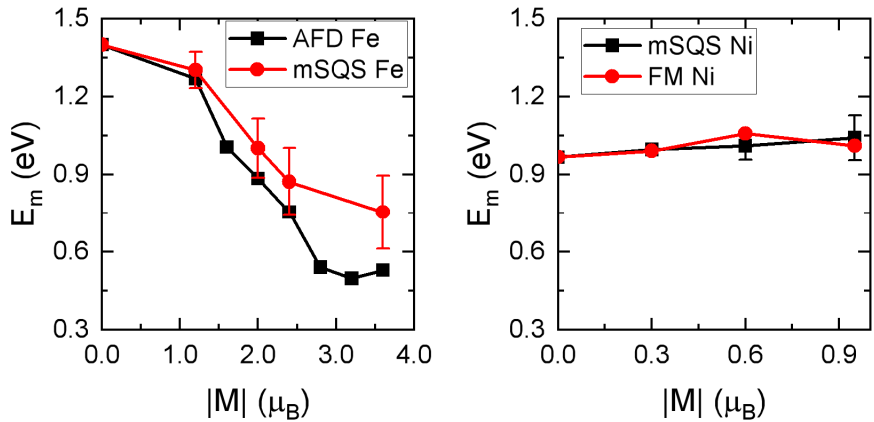

FIGURE 5.2: Migration energies as functions of magnetic-moment magnitude in Fe and Ni. All the magnetic-moment magnitudes in the stable and saddle-point configurations are constrained to the same value for each data point.

**Solute-vacancy binding and solute migration in fcc Fe and Ni**

The rates of the self- and the solute diffusions in a dilute alloy are influenced by the solute-vacancy and solute-solvent interactions. Here we consider the cases of Ni solute in fcc Fe, and Fe solute in fcc Ni, respectively. We present in Table 5.2 the first nearest-neighbour (1NN) solute-vacancy binding energies, and in Table 5.3 the migration energies of various jumps around a solute. The 1NN Ni-vacancy interaction is generally attractive in the magnetically ordered structures of Fe, whereas the interaction in the some magnetically disordered structures of Fe can be either repulsive or attractive. By contrast, the 1NN Fe-vacancy interaction in Ni is weakly repulsive for all the studied magnetic structures. In FM Ni, the migration energies of the Ni-vacancy exchanges around the Fe solute are not very different from those of the Ni-vacancy exchanges in the bulk, and they are larger than the ones of the Fe-vacancy exchanges. In AFD Fe, the migration energies exhibit a relatively large dispersion for the same type of atom-vacancy exchanges because of the different local magnetic environments.

TABLE 5.2: Binding energy (in eV) for a solute and a first nearest-neighbour vacancy in fcc Fe and Ni. The binding energy is positive if the vacancy-solute interaction is attractive. We also include the binding energies in some magnetically disordered structures of fcc Fe and Ni.

| Magnetic state | V-Ni in Fe | V-Fe in Ni |
|---|---|---|
| AFD (same plane) | 0.054 | / |
| AFD (different plane) | 0.042, -0.283 | / |
| AFS (same plane) | 0.114 | / |
| AFS (different plane) | 0.049 | / |
| FM | 0.085 | -0.016 |
| NM | 0.116 | -0.055 |
| mag. disordered | -0.334 to 0.223 | -0.047 to -0.009 |

TABLE 5.3: Migration energies (in eV) of the atom-vacancy exchanges in AFD Fe and FM Ni. $w_0$-$w_4$ denote the different atom-vacancy exchanges considered in the five-frequency model by LeClaire [100]. An illustrate of these exchanges is presented in Fig. 5.3.

|  | in AFD Fe | in FM Ni |
|---|---|---|
| $w_0$ | 0.740-1.010 | 1.096 |
| $w_1$ | 0.991-1.100 | 1.148 |
| $w_2$ | 0.885-1.174 | 0.963 |
| $w_3$ | 0.746-1.129 | 1.091 |
| $w_4$ | 0.695-1.089 | 1.093 |

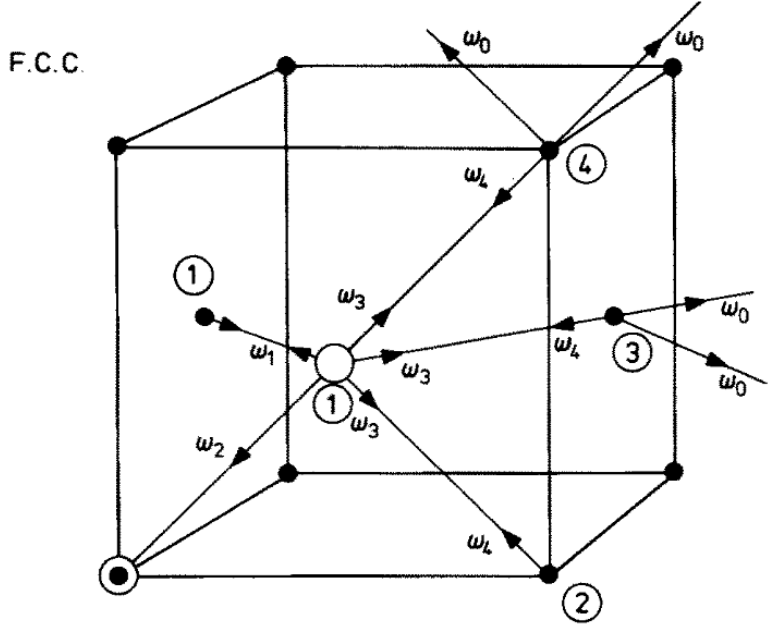

FIGURE 5.3: Vacancy jumps involved in the five-frequency model by LeClaire [100]. The unfilled circle is the vacancy, the partially filled circle is the solute $B$ and filled circles are the $A$ atoms. $w_2$ is the solute-vacancy exchange, $w_1$, $w_3$ and $w_4$ are the solvent-vacancy exchanges around the solute, and $w_0$ is the solvent-vacancy exchanges away from the solute.

**Attempt frequencies and vacancy formation vibrational entropies**

The DFT-computed migration energies of the Fe-vacancy and Ni-vacancy exchanges in various magnetic and chemical configurations are used to parametrize the EIM, which allows to predict the migration properties for a given magnetic and chemical configuration. The tracer diffusion coefficient can be calculated in Monte Carlo simulations via the Einstein relation [116, 119]:

$$D^{A*} = \frac{<r_{A*,i}^2>}{6\tau} \tag{5.6}$$

$$= \frac{<r_{A*,i}^2> C_V^{eq}}{6\tau_{MC} C_V^{MC}} \tag{5.7}$$

with $C_V^{MC} = 1/N_{\text{site}}$,

$$\tau_{MC} \propto \frac{N_{jump}}{\nu_0} \tag{5.8}$$

$$C_V^{eq} = \exp[-\frac{G_f^{mag}}{k_B T}] \exp(\frac{S_f^{vib}}{k_B}) \tag{5.9}$$

The attempt frequency $\nu_0$ in Eq. 5.8 and the vacancy formation vibrational entropy in Eq. 5.9 are not predicted from the EIM. Instead, they are computed from the phonon calculations using DFT as explained in the Methods chapter.

TABLE 5.4: Attempt frequencies $\nu_0$ (in THz) of jumping atoms, and vacancy formation vibrational entropies $S_f^{\text{vib}}$ (in $k_B$) in the bulk systems with different magnetic states.

| | Bulk system | | | | |
|---|---|---|---|---|---|
| | AFD Fe | AFS Fe | NM Fe | FM Ni | NM Ni |
| $\nu_0(\text{Fe})$ | 1.85 | 2.80 | 9.20 | 3.77 | / |
| $\nu_0(\text{Ni})$ | 24.33 | 5.95 | 21.83 | 22.19 | 13.51 |
| $S_f^{\text{vib}}$ | 3.05 | 2.41 | 2.77 | 2.15 | 1.53 |
| $\exp(S_f^{\text{vib}}/k_B)$ | 21.12 | 11.13 | 15.96 | 8.58 | 4.62 |
| $\nu_0(\text{Fe}) \cdot \exp(S_f^{\text{vib}}/k_B)$ | 39 | 31 | 147 | 32 | / |
| $\nu_0(\text{Ni}) \cdot \exp(S_f^{\text{vib}}/k_B)$ | 514 | 66 | 348 | 190 | 62 |

As shown in Table 5.4, the attempt frequency of Fe atoms is lower than that of Ni atoms in all the studied cases. The results of attempt frequencies are sensitive to the magnetic state and can differ by a factor of 2 to 5. Indeed, the attempt frequencies and the vacancy formation vibrational entropies are related to the second derivatives of energy with respect to position displacement, hence they are very sensitive to the conditions and simulation setting applied in DFT calculations. It should be therefore noted that these quantities can be a source of discrepancy when comparing our predictions with experimental data. Indeed, we neglect the magnon-phonon coupling and use the AFD-state (respectively FM-state) values of the attempt frequencies and the vacancy formation vibrational entropy in Fe (respectively in Ni) for the whole temperature range. These values are computed within the harmonic approximation, neglecting the anharmonic vibrational effects that may occur at very high temperatures [117, 275, 276].

### 5.2.2   Diffusion properties from EIM

**Tracer diffusion coefficients**

The tracer diffusion coefficients for the self- and solute diffusion in fcc Fe and Ni are computed from Monte Carlo simulations using the EIM. They are compared to the experimental results in Fig. 5.4.

The experimental measurements in fcc Fe were performed between 1185 K and 1667 K, within which the fcc phase of Fe is stable and paramagnetic [277–279]. Our predictions of $D^{\text{Fe}^*}$ and $D^{\text{Ni}^*}$ agree well with the experimental results, with our results of $D^{\text{Ni}^*}$ being on the higher limits of the dispersed measured values [278, 279].

The experimental data in fcc Ni [262, 264] are available at lower temperatures, which are however still above its Curie point of 627 K [187]. Our results of $D^{\text{Fe}^*}$ and $D^{\text{Ni}^*}$ agree with the experimental data in the temperature range above 1000 K, but the results of $D^{\text{Ni}^*}$ show a deviation at lower temperatures.

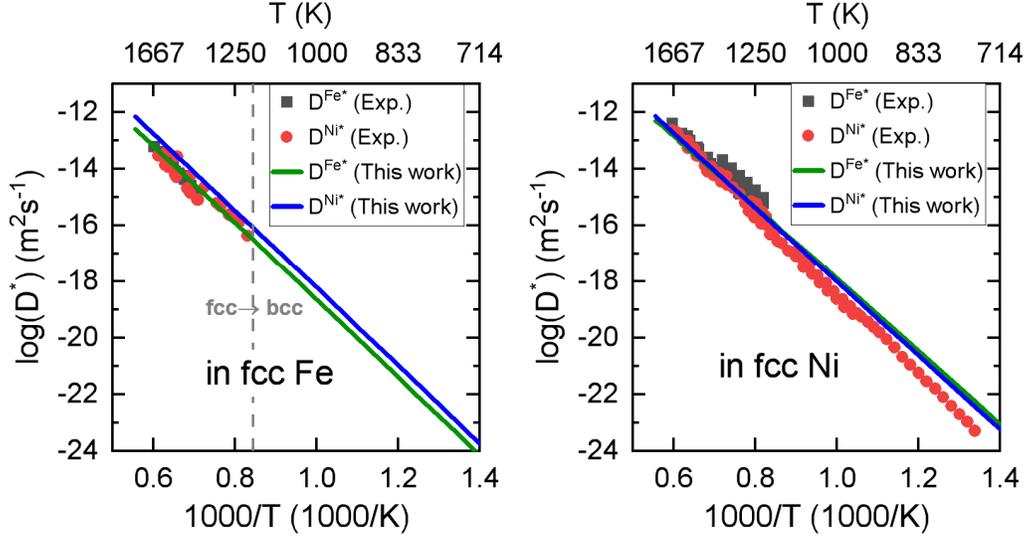

FIGURE 5.4: Tracer diffusion coefficients of Fe and Ni in fcc Fe and Ni. References for the experimental data: $D^{Fe*}$ in fcc Fe [277], $D^{Ni*}$ in fcc Fe [278, 279], $D^{Fe*}$ and $D^{Ni*}$ in fcc Ni [262, 264].

In Table 5.5, we present the calculated pre-exponential factor $D_0$ and activation energy $Q_a$ obtained from an Arrhenius fit $D^* = D_0 * \exp[-\frac{Q_a}{k_B T}]$ in the experimental temperature range.

The computed pre-exponential factor and the activation energy for the self-diffusion in fcc Fe are in good agreement with the experimental results, while the results for the Ni diffusion in Fe are lower than the experimental results. Our agreement in the self-diffusion activation energy can be contrasted with the previously DFT-predicted value using the NM state of fcc Fe [223, 225]. Indeed, if one estimates the self-diffusion activation energy in PM fcc Fe from DFT using one of the magnetic states shown in Table 5.6, the NM-state result would show the largest deviation from the experimental value. Compared with our results using EIM, this high value of the self-diffusion activation energy in NM Fe is due to both the overestimation of both formation and migration energies. This highlights again the fact that defect properties obtained with the NM state of fcc Fe is not representative of those in the PM state [223–225].

On the other hand, our computed activation energies for the self- and Ni diffusion in Ni are 0.2-0.3 eV lower than the experimental data. This theoretical-experimental difference may be due to, e.g. the anharmonic effects [117], and worth further investigation. Meanwhile, we note that a previous DFT study [218] predicted results closer to the experimental data. In that DFT study [218] performed with the FM state of Ni, the GGA functional was used to compute all the quantities (vacancy vibrational formation entropy, migration energy, attempt frequency, etc) except the vacancy formation energy, which was calculated using the LDA functional and is 0.2 eV higher than the GGA value obtained in our study. However, we have shown using the GGA functional consistently for both vacancy formation energy and vibrational formation entropy yields much better agreement with of the experimental equilibrium vacancy concentration than using the LDA functional (see Sec. 4.2.3).

TABLE 5.5: The calculated and experimental pre-exponential factor $D_0$ and activation energy $Q_a$ in fcc Fe and Ni, as obtained from an Arrhenius fit $D^* = D_0 * \exp[-\frac{Q_a}{k_B T}]$.

| | This work | | Experiments | | Other calculations | |
|---|---|---|---|---|---|---|
| | $D_0$ $(cm^2/s)$ | $Q_a$ (eV) | $D_0$ $(cm^2/s)$ | $Q_a$ (eV) | $D_0$ $(cm^2/s)$ | $Q_a$ (eV) |
| $D^{Fe^*}$ in Fe | 0.09 | 2.75 | 0.1-1.3 [264] | 2.75-2.95 [264] | 0.95 [225] | 3.75-3.79 [223, 225] |
| $D^{Ni^*}$ in Fe | 0.25 | 2.76 | 0.77-6.92 [278–280] | 2.91-3.36 [278–280] | / | / |
| $D^{Fe^*}$ in Ni | 0.05 | 2.50 | 2.1-4.1 [262, 281] | 2.8-2.9 [262, 281] | 0.05 [218] | 2.71 [218] |
| $D^{Ni^*}$ in Ni | 0.16 | 2.64 | 0.4-3.4 [264] | 2.81-3.03 [264, 280] | 0.02 [218] | 2.72 [218] |

TABLE 5.6: Self-diffusion activation energies $Q_a$ (in eV) in fcc Fe and Ni with different magnetic states, computed as the sum of the migration and vacancy formation energies. The experimental results available in the PM state of Fe and Ni are shown.

| Magnetic state | fcc Fe | fcc Ni |
|---|---|---|
| AFD | 2.57-2.83 | / |
| AFS | 2.53-3.22 | / |
| FM | 2.05, 2.61 | 2.53 |
| NM | 3.77 | 2.35 |
| mSQS | 2.96 | 2.53 |
| Exp. (PM) | 2.75-2.95 [264] | 2.81-3.03 [264] |

**Average migration energies and correlation factors**

In the following, we discuss the temperature dependences of average migration energies and correlation factors for Fe and Ni atoms in fcc Fe and Ni, computed using quantum statistics for the magnetic degree of freedom. The ($\tau$-weighted) average migration energy for the $A$-vacancy exchanges is computed within the residence time algorithm as follows:

$$E_m(A) = \frac{\sum_i^N \tau_i \cdot E_{m,i}(A)}{\sum_i^N \tau_i} \tag{5.10}$$

where $\tau_i$ and $E_{m,i}(A)$ are respectively the time increment and the migration energy for the $i$-th vacancy-$A$ exchange.

As shown in Fig. 5.5, the average migration energies of Ni atoms $E_m$(Ni) in Fe and Ni change by less than 0.03 eV in the whole temperature range. The magnetic effects on the Ni diffusion are therefore shown to be small in both Fe and Ni. By contrast, the average migration energies of Fe atoms $E_m$(Fe) show stronger variations with temperatures. In Fe, $E_m$(Fe) increases by up to 0.05 eV with temperature up the Néel point, above which it decreases by up to 0.15 eV with temperature up to 1800 K. Meanwhile, $E_m$(Fe) in Ni always decreases with temperature, with the total variation being 0.15 eV and 0.1 eV in the temperature range below and above the Curie temperature, respectively. The decreases of $E_m$(Fe) in Fe and Ni above the magnetic transition temperatures can be attributed to two different sources. The first one is the large dispersion of the individual barriers of Fe-vacancy exchanges (Fig. 5.1). This large dispersion implies that the $\tau$-weighted average $E_m$ converges to the arithmetic average of the individual barriers $E_{m,i}$ only at very high temperature, hence presenting a variation even above the magnetic transition temperature. The second source is related to the strong dependence of the Fe migration barriers on the spin magnitudes, which exhibit significant fluctuations and tend to increase with temperature in paramagnetic Fe and Ni. The decreasing trend of $E_m(\text{Fe})$ with temperature is consistent with our DFT results in Fig. 5.2 showing that $E_m(\text{Fe})$ tend to decrease with increasing magnetic-moment magnitude.

We note that the difference between $E_m(\text{Fe})$ in Fe and $E_m(\text{Fe})$ in Ni is reduced from 0.20 eV to 0.05 eV from 100 K to 1800 K. Meanwhile, the difference between $E_m(\text{Fe})$ and $E_m(\text{Ni})$ in Ni is increased from 0.10 eV to 0.32 eV in the whole temperature range.

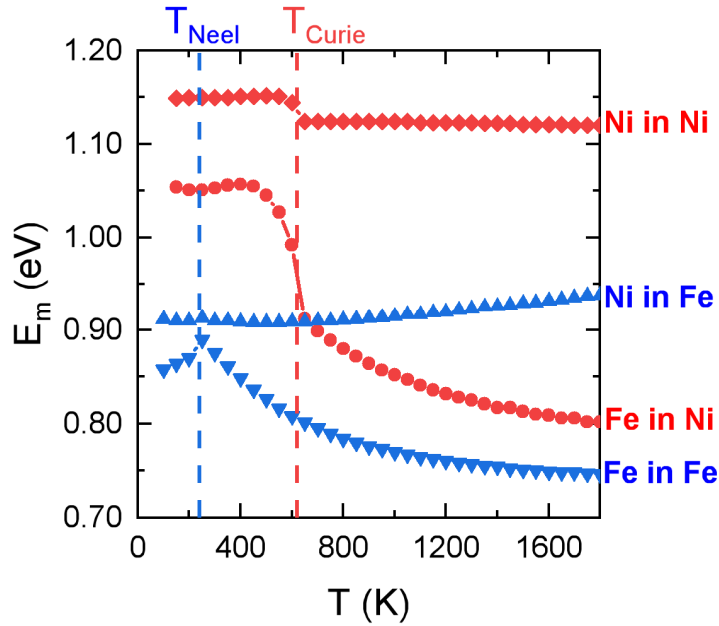

FIGURE 5.5: Average migration energies for Fe and Ni atoms as functions of temperature in fcc Fe and Ni. Quantum statistics is used for the magnetic degree of freedom.

The correlation factor characterizes the deviation of the diffusion behaviour from the random walk and can be computed as [116]:

$$f_A = \frac{D^{A*}}{D_{\text{random}}} \tag{5.11}$$

$$= \frac{< r^2_{A*,i} >}{< n > d^2} \tag{5.12}$$

where $< r^2_{A*,i} >$ and $< n >$ are respectively the mean square displacement of $A$ atoms and the average numbers of $A$-vacancy exchanges for a given time interval, and $d$ is the first-nearest-neighbour distance. The above relation enables to compute the correlation factor of the diffusion for a given species from Monte Carlo simulations.

Before presenting the results computed from Monte Carlo simulations, we would like to first review some theoretical results, which help to understand the computed results later. It is known that the self-diffusion correlation factor $f_0$ in the fcc lattice is equal to 0.7815 [282]. Meanwhile, the solute diffusion correlation factor $f_2$ can be computed with LeClaire's five-frequency model [100] (see Sec. 2.2.4). Please note that LeClaire's model was originally developed without considering the magnetic degrees of freedom. By assuming vacancy-solute interaction negligible beyond the first-nearest neighbour distance, we can express $f_2$ as a function of five frequencies $w_i$ ($i$=0-4):

$$f_2 = \frac{2w_1 + 7w_3 F}{2w_1 + 2w_2 + 7w_3 F} \tag{5.13}$$

where $w_i = \nu_i \exp(-\frac{E_{m,i}}{k_B T})$ and $F$ is a function of $\alpha = \frac{w_4}{w_0}$ [101, 102]:

$$F = 1 - \frac{10\alpha^4 + 180.5\alpha^3 + 927\alpha^2 + 1341\alpha}{7(2\alpha^4 + 40.2\alpha^3 + 254\alpha^2 + 597\alpha + 436)} \tag{5.14}$$

Here $w_2$ is the jump frequency for the solute-vacancy exchange, and the rest of $w_i$ are the different jump frequencies for the solvent-vacancy exchanges. We assume that the attempt frequency is the same for all solvent atoms, namely $\nu_0 = \nu_1 = \nu_3 = \nu_4$, but it is different from $\nu_2$ of the solute atom.

In the low-temperature limit ($T \to 0$), $f_2$ is determined dominantly by $E_{m,i}$. If $E_{m,2}$ is the smallest among $E_{m,i}$, therefore $w_2 \gg w_1$ and $w_2 \gg w_3$, hence $f_2 \to 0$ according to Eq. 5.13. If $E_{m,2}$ is the largest among $E_{m,i}$, then $w_2 \ll w_1$ and $w_2 \ll w_3$, hence $f_2 \to 1$.

In the high-temperature limit ($T \to +\infty$), the difference in $E_{m,i}$ has a negligible contribution to $f_2$, which is determined predominantly by $\nu_0$ and $\nu_2$. In this case, Eq. 5.13 becomes

$$f_2 = \frac{7.15}{7.15 + 2\frac{\nu_2}{\nu_0}} \tag{5.15}$$

Therefore, we have

$$f_2 = f_0 = 0.7815, \text{ if } \nu_2 = \nu_0$$
$$0 < f_2 < f_0, \text{ if } \nu_2 > \nu_0$$
$$1 > f_2 > f_0, \text{ if } \nu_2 < \nu_0$$

Now we discuss the Monte Carlo results of the correlation factors in Fig. 5.6. In Ni, the self-diffusion correlation factor $f_{Ni}$ is practically equal to the theoretical value of $f_0 = 0.7815$ [horizontal line in Fig. 5.6(a)] over the whole temperature range. Using the average migration energies shown in Fig. 5.6(b) as inputs, we find that LeClaire's model reproduces well the Fe-solute diffusion correlation factor $f_2$ computed directly from Monte Carlo simulations, hence verifying the validity of LeClaire's model in Ni.

The strong temperature dependence of $f_2$ in Fig. 5.6(a) can be understood in terms of LeClaire's model. According to our previous discussion, $f_2$ depends on $E_{m,i}$ at low temperatures and $\nu_2/\nu_0$ at high temperatures. As shown in Table 5.7, $f_2$ in Ni is equal to 0 and 0.95 respectively in the low-T and high-T limits. Therefore, $f_2$ in Ni tends to increase with temperature, except across the Curie temperature where the strong decrease in $E_{m,2}$ results in the decrease in $f_2$.

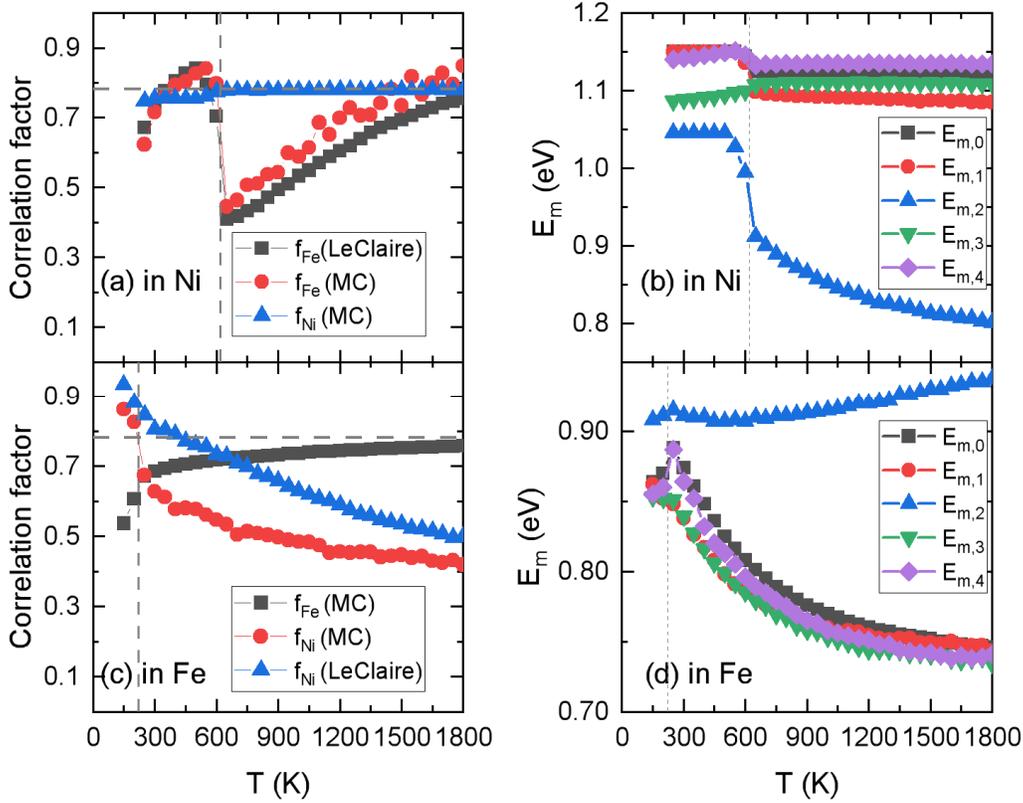

FIGURE 5.6: Diffusion correlation factors and average migration energies for Fe and Ni atoms in fcc Fe and Ni. Correlation factors in (a) fcc Ni and (c) fcc Fe. The average migration energies for different vacancy-atom exchanges involved in LeClaire's model in (b) fcc Ni and (d) fcc Fe. The horizontal lines in (a) and (c) are the theoretical self-diffusion correlation factor (0.7815). The vertical lines denote the magnetic transition temperatures. Quantum statistics is used for the magnetic degree of freedom.

In Fe, the self-diffusion correlation factor $f_{Fe}$ is lower than the theoretical value of 0.7815, but it converges towards this value at high temperatures. Indeed, the theoretical self-diffusion correlation factor is derived based on the assumption that all atoms have the same migration energy, which is not necessarily true in magnetic systems. Here we propose two possible causes: the timescales of magnon excitations and atomic jump duration, and the anisotropy of the magnetic structure. If atomic jumps occur much faster than magnon excitations, which is the approximation we take, the migration energy is then dependent on the local magnetic environment. Otherwise, if magnon excitations are assumed to be faster, the self-diffusion migration energy should be the same for all atom-vacancy exchanges in ferromagnetic or paramagnetic structures. Meanwhile, if the magnetic structure is anisotropic, the migration energy can be dependent on the jump direction, even if magnon excitations are assumed to be faster. For example, in the AFD ground state of fcc Fe, the migration energy is different between the intra-plane and inter-plane jumps.

TABLE 5.7: Solute correlation factors in Ni and Fe in the low-temperature and high-temperature limits according to LeClaire's model. The results in Fig.5.6(b) and (d) are used as inputs.

| | \multicolumn{2}{c}{Fe in Ni} | | \multicolumn{2}{c}{Ni in Fe} | |
| | $f_2$ | comment | $f_2$ | comment |
|---|---|---|---|---|
| $T = 0$ | 0 | $E_{m,2}$ is the smallest $E_{m,i}$ | 1 | $E_{m,2}$ is the largest $E_{m,i}$ |
| $T = +\infty$ | 0.95 | $\frac{\nu_2}{\nu_0} = 0.17$ | 0.21 | $\frac{\nu_2}{\nu_0} = 13.2$ |

For the self-diffusion correlation factor in Fe, its deviation from the theoretical value of 0.7815 below the Néel temperature is due to the anisotropy of the magnetic structure, and also possibly to the approximation of faster atomic jump attempts, whereas its deviation above the Néel temperature is possibly due to the approximation of faster atomic jump attempts. Meanwhile, we note that such a deviation does not appear in fcc Ni in Fig. 5.6(a), or in bcc Fe [283] despite using the same approximation of faster atomic jump attempts. Therefore, we still need to further understand the origin of this deviation in fcc Fe and why it is not presented in fcc Ni and bcc Fe.

In Fe, the Ni-solute correlation factor computed from Monte Carlo simulations are significantly different from the ones predicted from LeClaire's model above the Néel temperature, which is worth further investigation.

TABLE 5.8: Solute-vacancy binding free energy (in eV) in fcc Fe and Ni in the magnetic ground state (MGS) and the PM state. The binding energies in the intermediate temperature range lie between the values of the MGS and PM states. In our convention, a positive value indicates an attraction between the vacancy and the solute.

| | \multicolumn{2}{c}{Ni+V in fcc Fe} | | \multicolumn{2}{c}{Fe+V in fcc Ni} | |
| | 1NN | 2NN | 1NN | 2NN |
|---|---|---|---|---|
| MGS | 0.02 | 0.03 | -0.05 | -0.03 |
| PM (1500 K) | 0.01 | -0.03 | -0.04 | -0.07 |

## 5.3   Temperature dependence of diffusion properties in Fe-Ni alloys

In the previous section, we have discussed the self- and solute diffusion properties as functions of temperature in fcc Fe and Ni. In this section, we focus on the temperature dependence of diffusion properties in the concentrated fcc Fe-Ni alloys. We discuss first the migration energies obtained from DFT calculations at 0 K, then the diffusion properties as functions of temperature computed from Monte Carlo calculations using the EIM. Classical statistics is used to treat the magnetic degree of freedom over the whole temperature range for alloys.

### 5.3.1 Migration energies at 0 K

In this subsection, we present the migration energies in Fe-Ni alloys with the FM ground state, obtained from the DFT-NEB calculations.

The results in the FM SQSs are shown in Fig. 5.7. Both the average migration energies of Fe and Ni increase with increasing Ni concentration, with the former exhibiting a larger variation. The migration energies of Fe atoms are generally lower than those of Ni atoms in the SQSs, though the ranges of the individual values have some overlap in the concentrated SQSs. It should be noted that the effective migration energies in the alloys at finite temperatures cannot be easily concluded based on these DFT results. Indeed, since the atomic jumps with lower migration energies occur more frequently, the effective migration energy is generally lower than the arithmetic-average migration energy, though the difference between them is expected to reduce with increasing temperature. In addition, magnetic excitations and transitions can influence the concentration dependence of the migration energies.

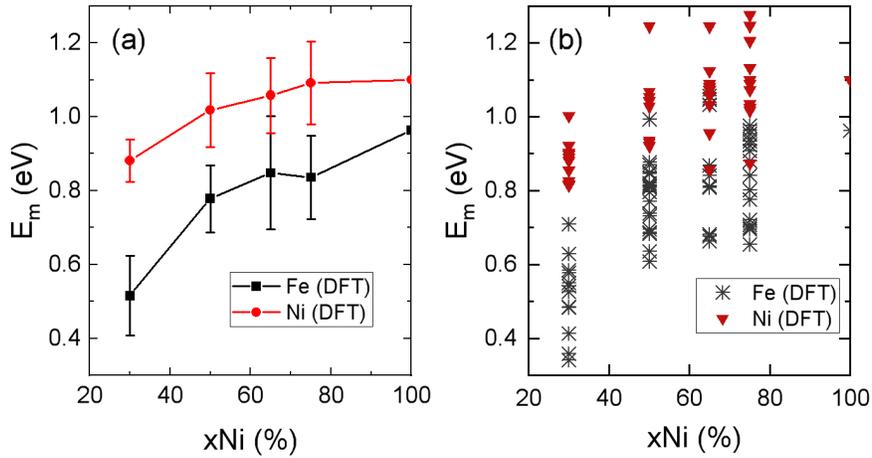

FIGURE 5.7: Migration energies of Fe and Ni in the FM SQSs, calculated from the DFT calculations. The average and standard deviations of the migration energies are shown in (a) and the dispersions of the individual values are shown in (b). We note that these results are reproduced reasonably well by our model (see Appendix A).

In ordered structures, the long-distance diffusion is often associated with jump cycles [284, 285]. During a jump cycle, antisites are created in the first few vacancy-atom exchanges, followed by the vacancy-atom exchanges that recover antisites created previously. Therefore, the vacancy migrates to a new lattice site without creating antisites after a jump cycle. Here we compute the migration energies involved in these jump cycles in the ordered structures as they should be the most relevant ones at lower temperatures.

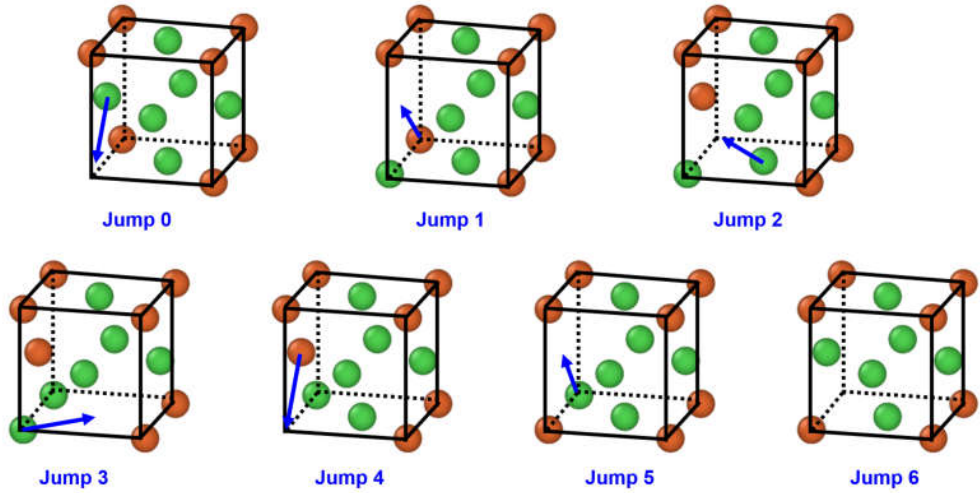

FIGURE 5.8: A six-jump cycle in the ordered structure $L1_2$-FeNi$_3$. This corresponds
to the cycle 1 of $V_{Fe}$ in Table 5.9. In a perfect $L1_2$-FeNi$_3$, the Fe atoms (in red) sit
on the corners of the cubic cell, and the Ni atoms (in green) sit on the face centres.
Initially the vacancy is on the Fe sublattice, with no local antisite; After six jumps, the
vacancy is on a 2NN Fe sublattice site, with no local antisite.

In the perfectly ordered structure $L1_2$-FeNi$_3$, all the 1NN and 3NN atoms of an Fe
atoms are Ni atoms, whereas all its 2NN and 4NN atoms are Fe atoms. For a vacancy
initially on the Fe sublattice, we therefore consider the jump cycles that allow the
vacancy to migrate to a nearest 2NN or 4NN Fe sublattice site. An illustration of a
six-jump cycle is given in Fig. 5.8.

The migration energies involved in the jump cycles in $L1_2$-FeNi$_3$ are shown in
Fig. 5.9, and the results are summarized in Table 5.9. The individual migration barri-
ers for the Fe-vacancy exchanges are between 0.8 to 1.0 eV, whereas the those for the
Ni-vacancy exchanges are either between 0.8 to 1.0 eV for the exchanges recovering
antisites, and around 1.3 eV for the exchanges creating antisites.

Similarly, we compute the migration energies in some jump cycles in $L1_0$-FeNi. As
shown in Table 5.10, the migration energies of the Fe-vacancy exchange are found to
be between 0.6 and 0.9 eV, lower than those of the Ni-vacancy exchanges ranging from
1.0 to 1.3 eV. The migration energies are 0.6-1.0 eV and 0.8-1.3 eV for the Fe-vacancy
and Ni-vacancy exchanges, respectively. These ranges of values are therefore similar
to those found in the SQSs shown in Fig. 5.7(b).

It should be noted that the effective migration barrier in a multi-step path is shown
to be equal to the global barrier, defined as the energy difference between the highest
saddle point and the most stable configuration along the path, while the correlation
factor depends on the number of jumps involved [286]. It can be seen from Table 5.9
and 5.10 that the effective migration barriers of different cycles are higher than that of
a direct 1NN jump in both $L1_0$ and $L1_2$ structures.

TABLE 5.9: Characteristics of some jump cycles in the ordered structure $L1_2$-FeNi$_3$. $V_{Fe}$ and $V_{Ni}$ denote the sublattices occupied initially by the vacancy before a jump cycle. The jump length is the distance between the initial and final positions of the vacancy in a jump cycle. The number of jumps and the maximum number of antisites created during a jump cycle are given. The migration energies are expressed in eV. The global migration energy is defined as the highest saddle-point energy minus the lowest stable-configuration energy in the whole cycle.

| $L1_2$-FeNi$_3$ | $V_{Fe}$ | | $V_{Ni}$ | | |
|---|---|---|---|---|---|
| Jump type | Cycle 1 | Cycle 2 | Cycle 1 | Cycle 2 | 1NN jump |
| Jump length | 2NN | 4NN | 1NN | 1NN | 1NN |
| Number of jumps | 6 | 7 | 6 | 5 | 1 |
| Max antisite number | 3 | 3 | 3 | 2 | 0 |
| $E_m$(Fe) | 0.840-0.923 | 0.930-1.021 | 0.904-0.997 | 0.827-0.926 | / |
| $E_m$(Ni) | 0.820-1.316 | 0.820-1.313 | 0.922-1.339 | 0.922-1.339 | 1.208 |
| $E_m$(global) | 1.726 | 2.071 | 1.436 | 1.267 | 1.208 |

TABLE 5.10: Characteristics of some jump cycles in the ordered structure $L1_0$-FeNi. Same notations as in Table 5.9.

| $L1_0$-FeNi | $V_{Fe}$ | | | $V_{Ni}$ | | |
|---|---|---|---|---|---|---|
| Jump type | Cycle 1 | Cycle 2 | 1NN jump | Cycle 1 | Cycle 2 | 1NN jump |
| Jump length | 1NN | 1NN | 1NN | 1NN | 1NN | 1NN |
| Number of jumps | 5 | 7 | 1 | 5 | 7 | 1 |
| Max antisite number | 2 | 3 | 0 | 2 | 3 | 0 |
| $E_m$(Fe) | 0.687-0.845 | 0.786-0.973 | 0.854 | 0.585-0.899 | 0.585-0.899 | / |
| $E_m$(Ni) | 1.156-1.306 | 1.156-1.306 | / | 1.178-1.302 | 0.981-1.306 | 1.045 |
| $E_m$(global) | 1.569 | 1.569 | 0.854 | 1.531 | 1.534 | 1.045 |

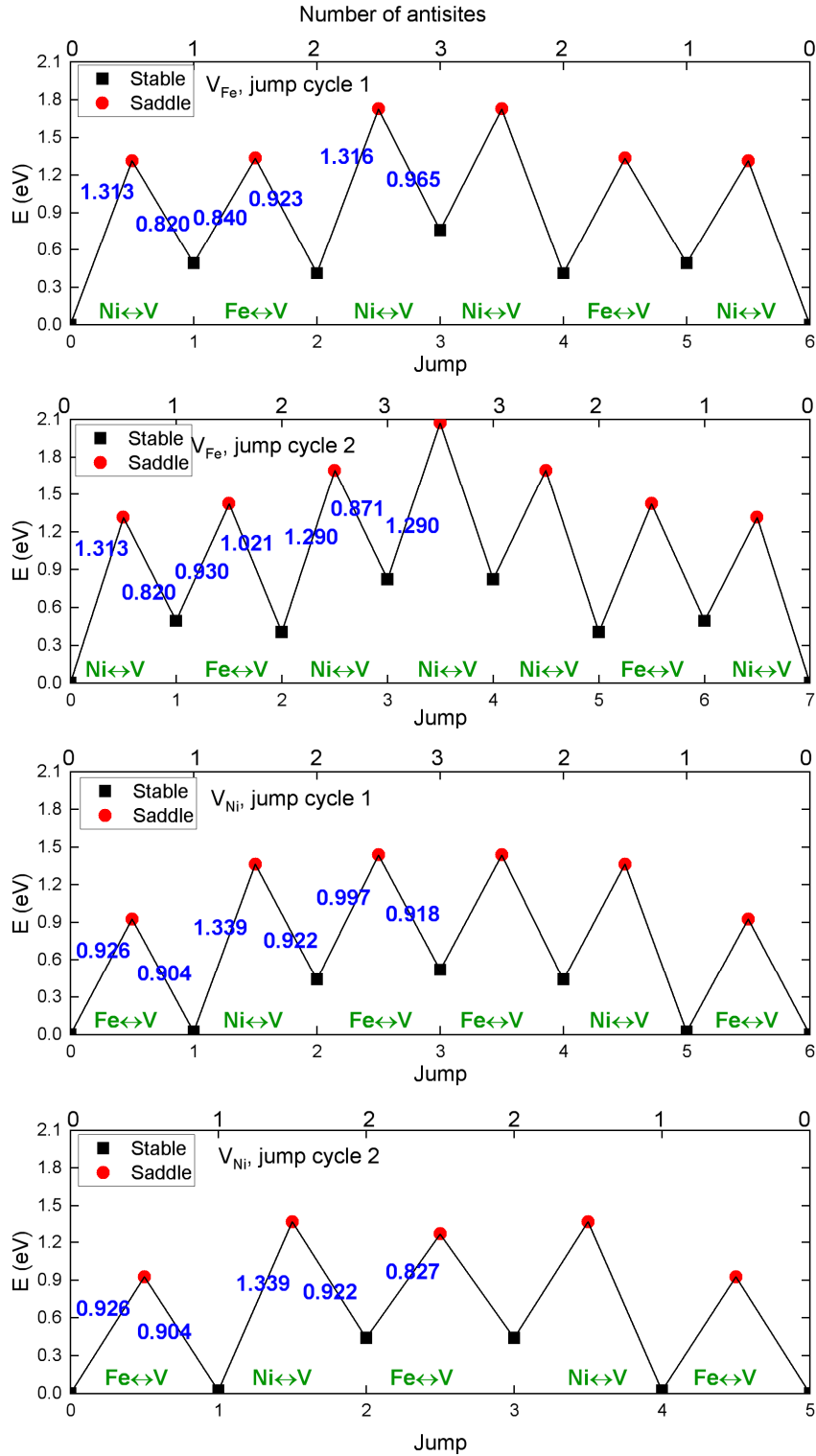

FIGURE 5.9: Energies of the stable and saddle-point configurations in different jump cycles in the ordered structure L1$_2$-FeNi$_3$. The migration energies are given for the jumps in the first half of the jump cycle during which the antisites are created; the migration energies in the second half of the jump cycle (in which antisites are recovered) can be symmetrically obtained. The upper X axis denotes the number of antisites in the stable configurations.

### 5.3.2 Diffusion properties at finite temperatures

In this subsection, we discuss the migration properties in Fe-Ni alloys obtained with Monte Carlo simulations. We have previously calculated via DFT the attempt frequencies of Fe and Ni atoms ($\nu_{Fe}$ and $\nu_{Ni}$) and the vacancy formation vibrational entropy $S_f^{vib}$ in AFD Fe and FM Ni. Based on these DFT results, we perform linear interpolations with respect to Ni concentration to obtain the corresponding values in the alloys. As shown in Fig. 5.10, the absolute variations of $\nu_{Fe}$ and $\nu_{Ni}$ with respect to Ni concentrations are small, whereas the ratio $\nu_{Ni}/\nu_{Fe}$ reduces by about a factor of two from pure Fe to pure Ni, which is mainly due to the increase of $\nu_{Fe}$ with Ni content. For the vacancy formation entropy, we use the linear interpolation directly on $\exp[S_f/k_B]$ (stars in Fig. 5.10). We note that these values are similar to the ones (diamonds in Fig. 5.10) obtained from the linearly interpolated results of $S_f$.

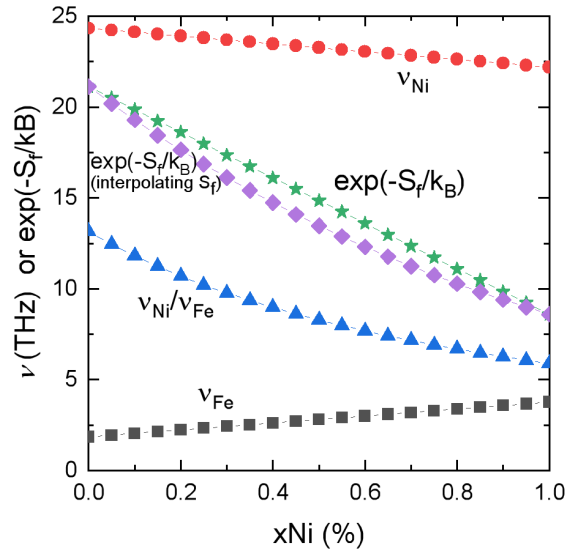

FIGURE 5.10: Linear interpolation for the prefactor terms ($\nu_{Fe}$, $\nu_{Ni}$ and $\exp[S_f/k_B]$) in Fe-Ni alloys. We also show the results of $\exp[S_f/k_B]$ based on the linear interpolation of $S_f$, and the ratio $\nu_{Ni}/\nu_{Fe}$ based on the linearly interpolated values of $\nu_{Fe}$ and $\nu_{Ni}$.

The computed tracer diffusion coefficients $D^{Fe*}$ and $D^{Ni*}$ as functions of temperature are compared to the experimental data in Fig. 5.11. We note that the alloys are paramagnetic and disordered in the temperature range shown in Fig. 5.11. For a given composition, the computed $D^{Fe*}$ and $D^{Ni*}$ are found to be similar, in agreement with experiments. Our results are in good agreement with the experimental ones below 1400 K, but they deviate to lower values at higher temperatures, which might be related to anharmonic effects. We find a better agreement with experiments in the dilute alloys (in particular for $D^{Ni*}$) than in more concentrated alloys. The largest theoretical-experimental difference is found in the alloys with 60.5% Ni, where our model underestimates the experimental $D^{Fe*}$ at 1578 K by just a factor of six.

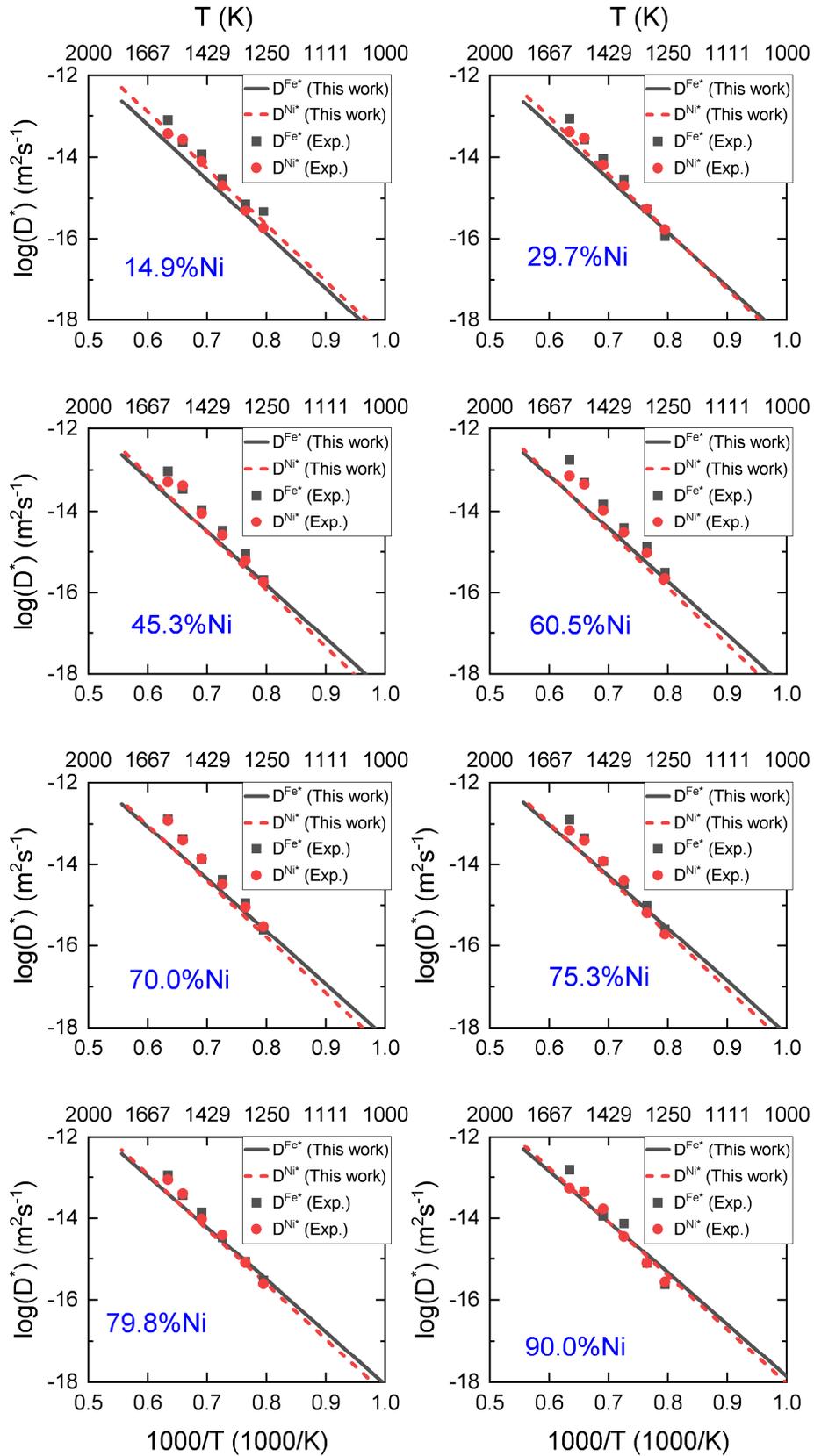

FIGURE 5.11: Tracer diffusion coefficients of Fe and Ni as functions of temperature in the fcc Fe-Ni alloys with different Ni concentrations. Experimental data from Ref. [262].

**Temperature dependence of diffusion properties in alloy with 75% Ni**

In the following, we provide a detailed discussion on the temperature dependence of the diffusion-related properties in the Fe-Ni alloy with 75% Ni. In Fig. 5.12(a), we present the tracer diffusion coefficients over a large temperature range, from 300 K to 1800 K. Overall, the tracer diffusion coefficients of Fe and Ni are similar. This point can be seen more clearly in Fig. 5.12(b), where the ratio of the two coefficients is found to be around 1 to 2. The tracer diffusion coefficient of Fe is generally larger than that of Ni because of the lower migration barriers for the Fe-vacancy exchanges. With increasing temperature, it can be expected that the difference in the migration energies of Fe and Ni plays a less important part in the ratio $D^{Fe^*}/D^{Ni^*}$. The latter should gradually converge to the ratio of the attempt frequencies $\nu_{Fe}/\nu_{Ni}$, which explains the decreasing trend of $D^{Fe^*}/D^{Ni^*}$ above the Curie temperature. As shown in Fig. 5.12(d), the diffusion correlation factor of Fe is lower than that of Ni, which is correlated with the higher migration energies of Fe atoms that result in a higher probability of an Fe atom to jump back to a previous position.

In Fig. 5.12(e), we show the tracer diffusion coefficients near the chemical and magnetic transition temperatures. A change of slope is noted around the chemical order-disorder transition temperature, above which the diffusion in the disordered alloy is enhanced compared to the case in the ordered structure. On the other hand, the magnetic transition has no significant effect on the diffusion properties. Based on the Arrhenius fits on the tracer diffusion coefficients from 600 K to 750 K, and the ones from 900 K to 1100 K, respectively, the diffusion activation energy as marked in Fig. 5.12(e) is found to be reduced by 0.55 eV across the chemical and magnetic transitions.

Meanwhile, we note that the notion of diffusion activation energy $Q_a$ is really meaningful only when the diffusion coefficients follow the Arrhenius law, that is $D^* = D_0 \exp[-\frac{Q_a}{K_B T}]$ with $D_0$ and $Q_a$ being constant within a temperature range. This can be verified by computing a diffusion activation energy as the derivative of $D^*$ with respect to $T$. Concretely,

$$E_a = -\frac{\partial \ln D^*}{\partial \frac{1}{K_B T}} \tag{5.16}$$

where we use the notation $E_a$ to distinguish it from the activation energy $Q_a$ that is commonly defined for a Arrhenius behaviour. In a temperature regime where the Arrhenius law is satisfied, $E_a$ is constant and equivalent to the commonly defined $Q_a$.

As shown in Fig. 5.12(f), $E_a$ of the Fe and Ni diffusion undergo strong variations below 1000 K. Above 1000 K, $E_a$ of the Ni diffusion is constant whereas $E_a$ of the Fe diffusion shows a slight variation with temperature. The results in Fig. 5.12(f) suggest that the diffusion in the ordered alloy exhibits a strong deviation from the Arrhenius law.

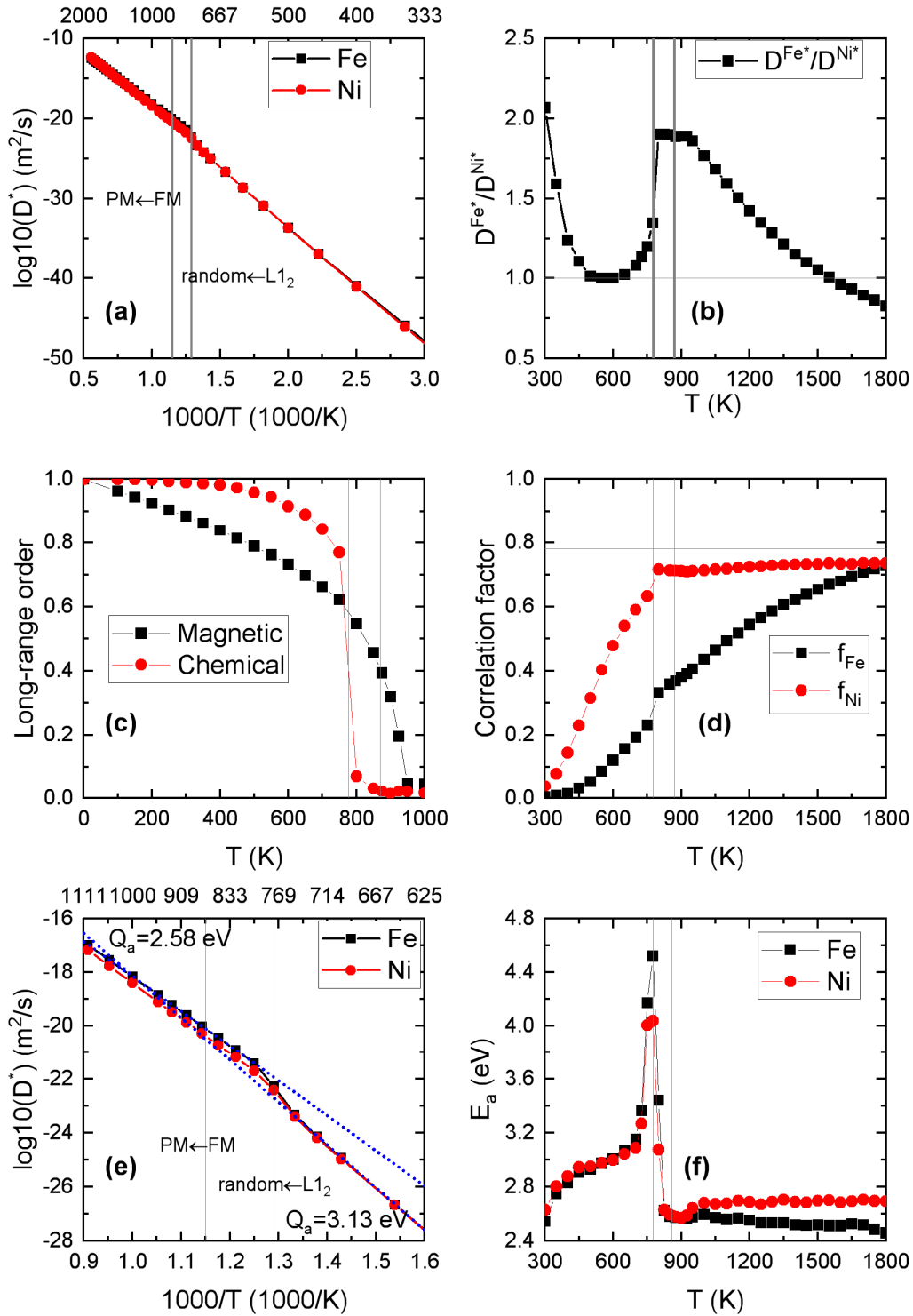

FIGURE 5.12: Temperature dependence of diffusion-related properties in the Fe-Ni alloy with 75% Ni. (a) Tracer diffusion coefficients; (b) Ratio of the tracer diffusion coefficients $D^{Fe*}/D^{Ni*}$; (c) Magnetic and chemical long-range order parameters; (d) correlation factors of Fe and Ni tracer diffusion; (e) Tracer diffusion coefficients near the transition temperatures; (f) Diffusion activation energies of Fe and Ni, calculated as $-\frac{\partial \ln D^*}{\partial \frac{1}{k_B T}}$. The vertical lines in all figures denote the chemical and magnetic transition temperatures.

To better understand this non-Arrhenius behaviour, we write the tracer diffusion coefficient $D^*$ in the following form:

$$D^* = D_0 \exp\left[-\frac{G_a}{k_B T}\right] \tag{5.17}$$

where $D_0$ is a constant, and $G_a$ can be interpreted as the diffusion activation free energy including magnetic entropy. Please note that $G_a$ is different from $E_a$ and their relation is

$$E_a = \frac{\partial \frac{G_a}{T}}{\partial \frac{1}{T}} = G_a - T\frac{\partial G_a}{\partial T} \tag{5.18}$$

which is analogue to the relation between energy and free energy ($E = \frac{\partial \frac{G}{T}}{\partial \frac{1}{T}}$). Note that $E_a$ is a constant (i.e. an Arrhenius behaviour) if $G_a$ is a linear function of $T$ (including being a constant).

Using the form of Eq. 5.17, we can derive $G_a$ based on the calculated diffusion coefficients. The results of $G_a$ for the Ni and Fe diffusion are shown in Fig. 5.13(c) and (d), respectively. It can be seen that below the chemical order-disorder transition temperature of 776 K, $G_a$ for the Fe and Ni diffusion decrease with increasing temperature. The total decrease of $G_a$ between 300 K and 776 K is about 0.3 eV and 0.2 eV for the Fe and Ni diffusion, respectively.

The variation of $G_a$ with temperature and the origin of the non-Arrhenius behaviour can be better understood by decomposing $G_a$ as a sum of other quantities such as the average migration energy $E_m$, vacancy formation magnetic free energy $G_f^{mag}$, and correlation factors $f_{Fe}$ and $f_{Ni}$. Then, their individual variations with temperature can be analysed and their respective contributions to $G_a$ can be concluded. However, $G_a$ may not be easily written in a simple function of other quantities, especially in the ordered structure where a very careful examination is needed.

In the first place, we compare $G_a$ with the sum of $E_m$ and $G_f^{mag}$ [represented as up-triangle in Fig. 5.13(a) and (b)]. It can be seen that the sums $E_m + G_f^{mag}$ reproduce well the global variations of $G_a$ for the Fe and Ni diffusion. The agreement between $E_m + G_f^{mag}$ and $G_a$ is rather good for the Ni diffusion but less so for the Fe diffusion. This can be related to the stronger correlation for the Fe diffusion than for the Ni diffusion, as can be seen from Fig. 5.12(d). Indeed, including additionally the correlation effects [down-triangle in Fig. 5.13(a)] leads to a very good agreement with $G_a$ for the Fe diffusion, but it results in an overestimation $G_a$ for the Ni diffusion around 400 K. The cause of this overestimation requires more careful examination and shall be discussed in the end.

The results in Fig. 5.13(a) and (b) show that $G_a$ can be well approximated as the sum of the three contributions, namely $E_m$, $G_f^{mag}$, and kinetic correlation. For both Fe and Ni diffusion, the overall decrease of $G_a$ from 300 K to 800 K can be attributed primarily to the strong decrease of $E_m$, and also partly to the overall increase of $G_f^{mag}$ and the weaker kinetic correlation. For instance, for the Fe diffusion from 300 K to 800

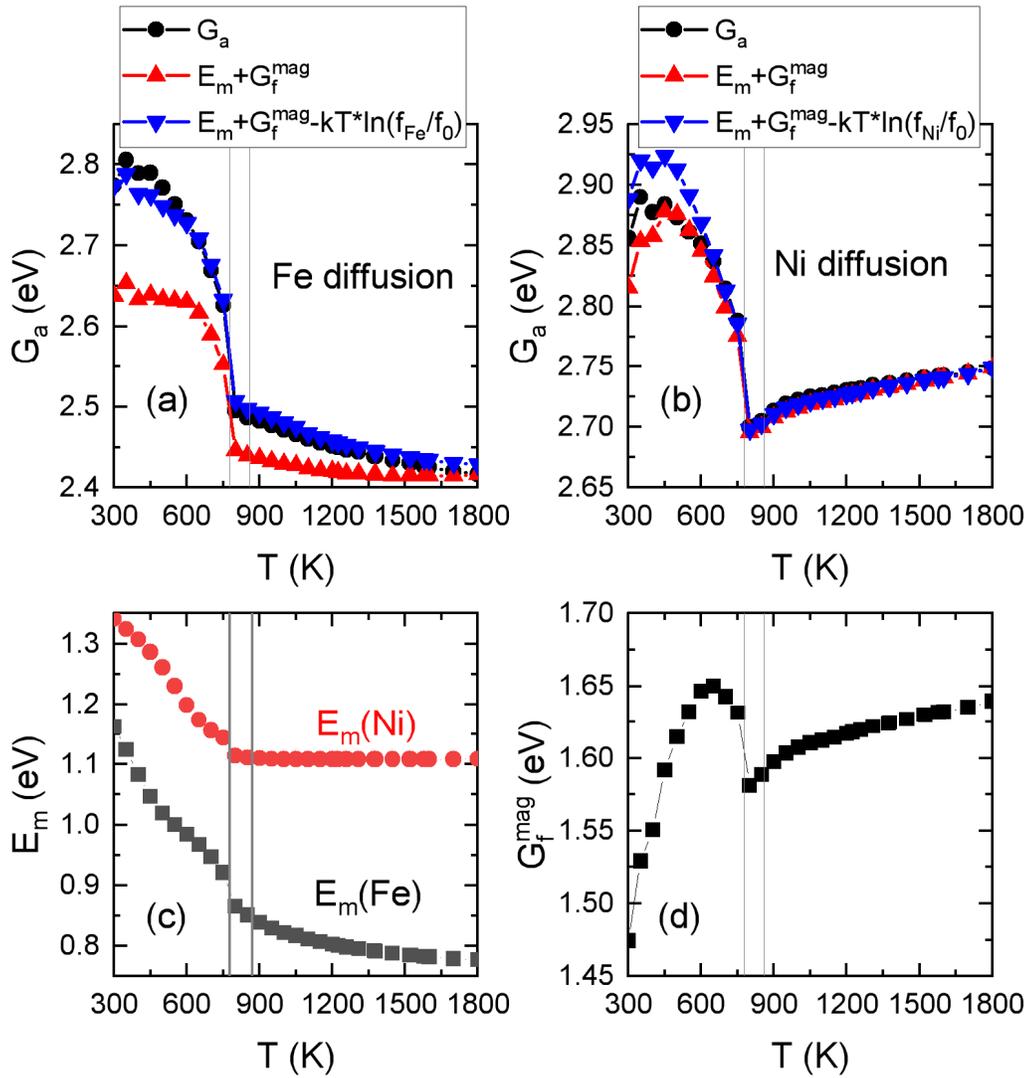

FIGURE 5.13: Temperature dependence of (a) diffusion activation free energy for Fe diffusion, (b) diffusion activation free energy for Ni diffusion, (c) average migration energies of Fe and Ni atoms, and (d) vacancy formation magnetic free energy, in the alloy with 75% Ni.

TABLE 5.11: Variations of $G_a$ and its constituent parts (in eV) for the Fe and Ni diffusion in the alloy with 75% Ni over the corresponding temperature ranges.  The chemical order-disorder transition temperature is 776 K.

| | Fe or Ni | Fe | | | Ni | | |
|---|---|---|---|---|---|---|---|
| | $\Delta G_f^{mag}$ | $\Delta E_m$ | $\Delta(-k_B T \ln \frac{f}{f_0})$ | $\Delta G_a$ | $\Delta E_m$ | $\Delta(-k_B T \ln \frac{f}{f_0})$ | $\Delta G_a$ |
| 300-800 K | 0.11 | -0.30 | -0.08 | -0.29 | -0.23 | -0.05 | -0.17 |
| 300-650 K | 0.18 | -0.20 | -0.05 | -0.07 | -0.17 | -0.05 | -0.03 |

K, the variation of $E_m$ is larger (by a factor of 3 to 4) than those due to $G_f^{\text{mag}}$ and kinetic correlation, as can be seen from Table 5.11.

As shown previously in Fig. 5.12(c), the chemical configuration remains fairly ordered up to around 650 K. The increase of $G_f^{\text{mag}}$ in Fig. 5.12(d) before 650 K is therefore mainly due to the magnetic excitations, whereas its decrease from 650 K to 800 K can be attributed to the loss of the chemical long-range order. By contrast, $E_m$ for the Fe and Ni diffusion in Fig. 5.12(d) always decrease with temperature below 800 K. Therefore, magnetic excitations have opposite effects on $G_f^{\text{mag}}$ and $E_m$, whereas chemical disorder leads to a decrease in both $G_f^{\text{mag}}$ and $E_m$. Meanwhile, it should be noted that the effects of magnetic excitations at low temperatures, and hence the associated variations in the above-mentioned quantities may be exaggerated due to the use of classical statistics instead of quantum statistics for the magnetic degree of freedom. The analysis of the quantum effects is not yet available here but is in our plan.

Finally, we have shown that including kinetic correlation effects leads to a better estimation of $G_a$ of the Fe diffusion for the whole temperature range, but to an overestimation of $G_a$ for the Ni diffusion around 400 K. Indeed, Fig. 5.12(c) shows that there is very few antisites in the structure around 400 K. In such a chemically ordered structure, $G_a$ may not be well approximated simply as the sum of $G_f^{\text{mag}}$, $E_m$ and kinetic correlation effects. In the perfectly ordered L1$_2$ structure, all the first-nearest neighbours of Fe atoms are Ni atoms, whereas Ni atoms have first-nearest Fe and Ni neighbours. Therefore, Fe atoms can only diffuse as antisites in the Ni sublattice, whereas the Ni diffusion can be antisite-assisted or occur only in the Ni sublattice. In this case, the antisite-vacancy interaction also need to be considered and the inclusion of its effect on diffusion is nontrivial and requires a carefuly examination [287]. In addition, as the creation of antisites occurs with a very low probability around 400 K, the accurate simulation of antisite-assisted diffusion may require a significant number of atom-vacancy exchanges (the results presented here are obtained $10^8$ atom-vacancy exchanges). Hence the convergence of the results at very low temperature still needs to be checked.

## 5.4 Concentration dependence of diffusion properties in Fe-Ni alloys

In this section, we first discuss the concentration dependence of the calculated diffusion properties at temperatures from 700 K to 1500 K. Then we compare our results with the experimental data available only in the PM disordered Fe-Ni alloys.

### 5.4.1 Results and discussion

This subsection is focused on the concentration dependence of diffusion properties in Fe-Ni alloys at four different temperatures, namely 700 K, 800 K, 1000 K, and 1500 K.

The atomic and magnetic order parameters of the equilibrium alloys at these temperatures are shown in Fig.5.14.  At 700 K, the alloys with 45% to 90% Ni are FM, while those with 65 to 80% Ni have an L1$_2$ ordered structure.  At 800 K, the alloys are disordered over the whole range of composition, whereas those with 50% to 80% Ni are FM. At 1000 K and 1500 K, the alloys over the whole concentration range are PM and disordered.

As shown in Fig. 5.15(a), the average migration energies of Fe atoms present a small minimum around 25% Ni at the four studied temperatures.  The migration energies at lower temperatures have a stronger dependence on concentration: they change from 0.75 eV to 0.95 eV at 700 K, whereas the difference between the migration energies in the whole composition range is around 0.05 eV at 1500 K. The migration energy always decreases with increasing temperature for a given composition.  The decrease of the migration energy from 700 K to 1500 K is relatively small (about 0.05 eV) in the Fe-rich alloys below 45% Ni, whereas in the Ni-rich alloys the variation is larger, being between 0.1 and 0.15 eV. The migration energy varies smoothly with concentration at 1000 K and 1500 K, where the alloys are PM and disordered.  Meanwhile, the results of 700 K and 800 K exhibit some kinks because of the strong atomic and magnetic long-range orders in the concentrated alloys.

Such kinks are also observed in the results for the Ni-vacancy exchanges at 700 K and 800 K, as shown in Fig. 5.15(b).  But overall, the migration energies of Ni tend to increase with increasing Ni concentration, from 0.9 eV in fcc Fe to 1.1 eV in fcc Ni. The results in the PM disordered alloys at 1500 K suggest that the migration energies of Ni are more sensitive to the Ni concentration than the migration energies of Fe.  In Fig. 5.15(c), we also show the average migration energy for the atom-vacancy exchanges, namely the average vacancy migration energy.  It can be seen that the vacancy migration energy in the PM disordered alloys is even more dependent on the Ni concentration, increasing from 0.8 eV in fcc Fe to 1.1 eV in fcc Ni.

As shown in Fig. 5.16(a), the Fe diffusion at 700 K and 800 K is strongly correlated in the composition range of 50% to 80% Ni, where the alloys are still FM. The correlation of the Fe diffusion is less significant at 1000 K and 1500 K, but a minimum of the correlation factor is noted at around 70% Ni, where the atomic and magnetic short-range orders are strongest as can be seen in Fig. 5.14(c) and (d).  By contrast, Fig. 5.16(b) indicates that Ni diffusion is more correlated in Fe-rich alloys. Its correlation factor is mainly dependent on the Ni concentration, while the effects of the atomic and magnetic orders are less important compared to the case of the Fe diffusion.

The above results of the migration energies and the correlation factors suggest that magnetic and chemical orders have stronger effects on the Fe diffusion than on the Ni diffusion, whereas the concentration dependence is more pronounced for the Ni diffusion than the Fe diffusion in the PM disordered alloys with vanishing magnetic and chemical orders (namely at 1500 K).

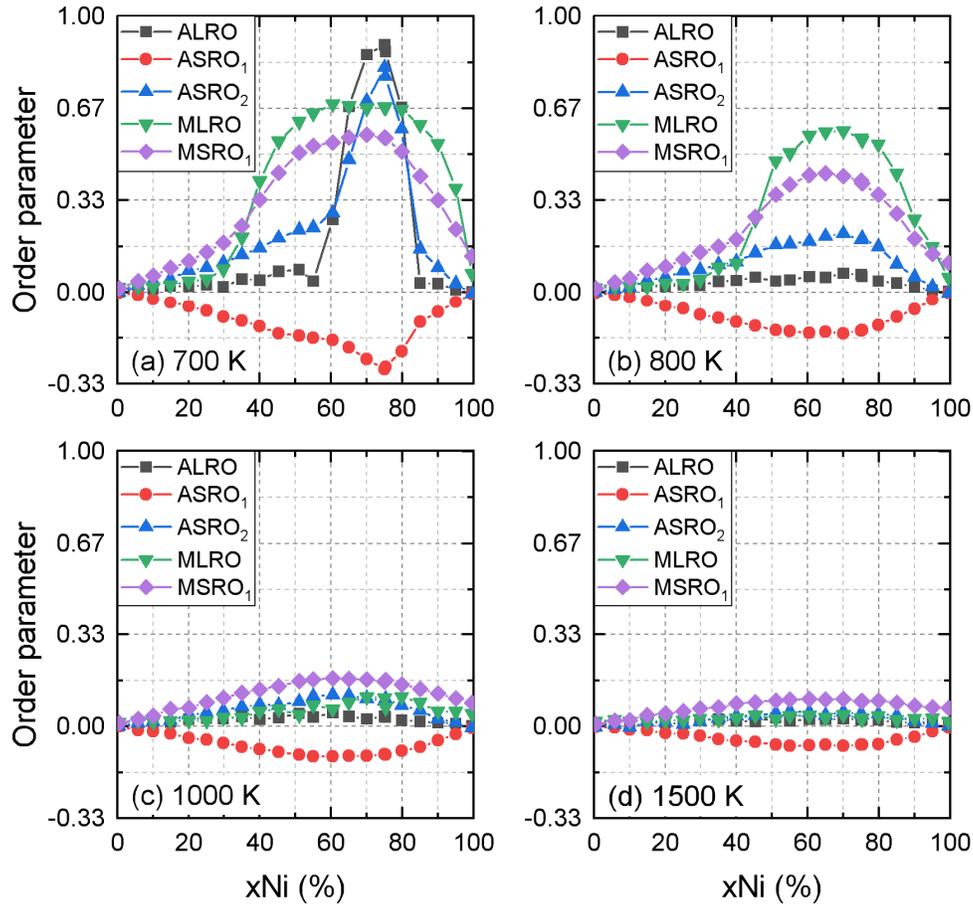

FIGURE 5.14: Atomic and magnetic order parameters as functions of Ni concentration at different temperatures. ALRO or MLRO: atomic or magnetic long-range order. ASRO$_i$ or MSRO$_i$: atomic or magnetic short-range order in the $i$-th shell.

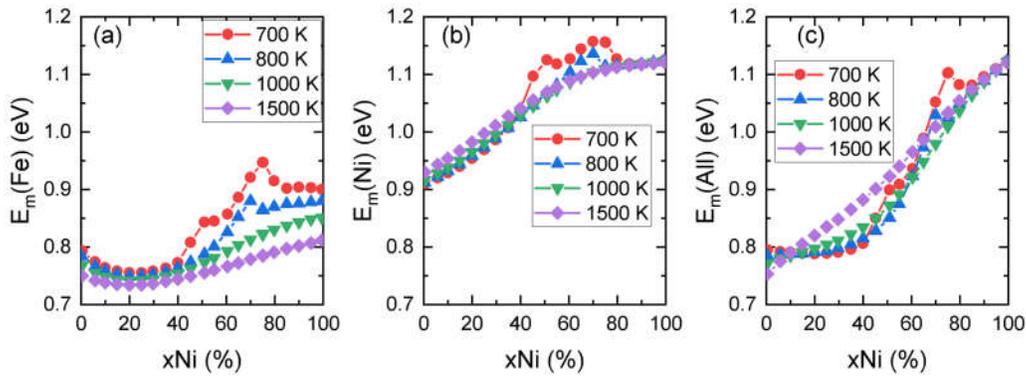

FIGURE 5.15: Concentration dependence of average migration energies for (a) Fe-vacancy (b) Ni-vacancy and (c) all atom-vacancy exchanges at different temperatures.

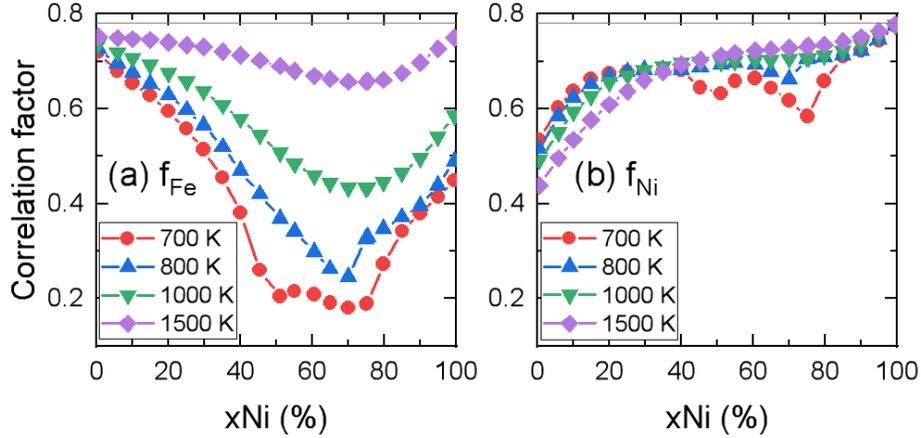

FIGURE 5.16: Concentration dependence of kinetic correlation factors for Fe and Ni diffusion at different temperatures.

### 5.4.2   Comparison with experimental data

In this subsection, we compare the calculated diffusion properties with the available tracer diffusion experiments (see Ref. [262] and the references therein). The calculated and experimental tracer diffusion coefficients in fcc Fe-Ni alloys are shown in Fig. 5.17. For the experimental data at a given temperature around a given composition, those of the Fe diffusion differ by less than one order of magnitude, whereas those of the Ni diffusion exhibit larger dispersions, with an experimental uncertainty of one to two orders of magnitude. Our results of the Fe and Ni diffusions are within the dispersion of the experimental data over the whole composition range for the three temperatures.

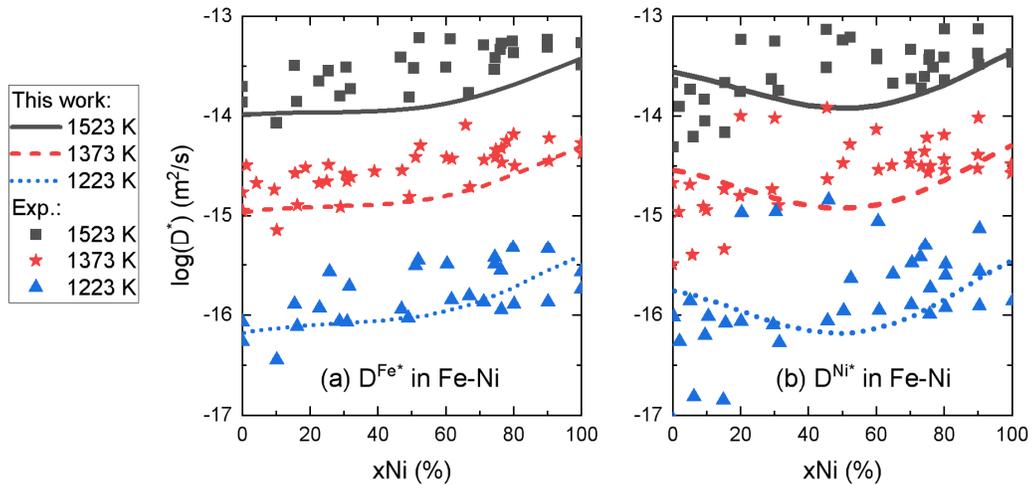

FIGURE 5.17: Calculated tracer diffusion coefficients of Fe and Ni compared to experimental data (see Ref. [262] and the references cited therein).

Experimentally, it is customary to assume an Arrhenius behaviour for the temperature dependence of the tracer diffusion coefficients, namely $D^* = D_0 \exp(-\frac{Q_d}{k_B T})$.

Then, the diffusion prefactor $D_0$ and the activation energy $Q_a$ can be derived respectively as the intercept and the slope of the linear fit on the $\log D^*$ versus $1/T$, based on the following relation

$$\log D^* = \log D_0 - \frac{Q_a \log e}{k_B T} \qquad (5.19)$$

The prefactors and the activation energies derived from our results and the experimental ones by Million *et al.* [262] are presented in Fig. 5.18. Despite the agreement of the computed tracer diffusion coefficients with the experimental data, there are some theoretical-experimental discrepancies of the prefactors and the activation energies, in particular in the Ni-rich alloys. In the following, we propose some possible experimental and theoretical causes to explain these differences.

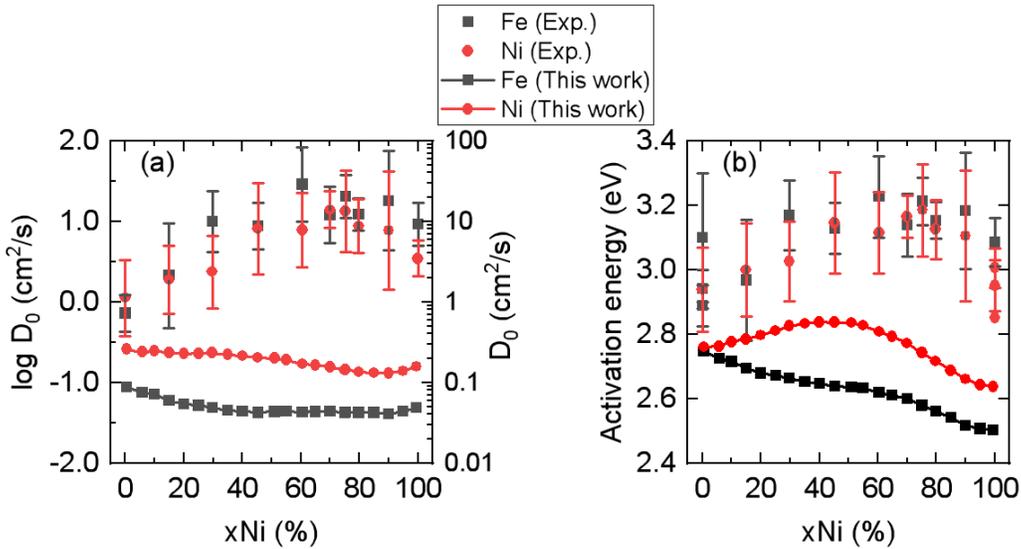

FIGURE 5.18: Calculated diffusion prefactors and activation energies of Fe and Ni compared to the values obtained by Million *et al.* [262]. We use for the Arrhenius fit our results of the tracer diffusion coefficients in 1200 K to 1600 K, which is similar to the experimental temperature range of 1258 K to 1578 K used by Million *et al.* [262]. The left and right Y axis in (a) show the values of $\log D_0$ and $D_0$ respectively.

First, the prefactor and the activation energy determined from the Arrhenius fit are extremely sensitive to the small variations in the tracer diffusion coefficients. To demonstrate this point, we perform the Arrhenius fit on the experimental data in different temperature ranges measured by Million *et al.* [262]. As shown in Fig. 5.19, the derived activation energy for a given composition can be very different depending on the portion of the experimental data used for the fit.

It should be noted that the separation of the prefactor and the activation energy from the tracer diffusion coefficient is reasonable only if the temperature dependence of the tracer diffusion coefficients exhibits a precise Arrhenius behaviour. This can be checked, for example, by performing the Arrhenius fit locally (namely $Q_a = -\partial \ln D^*/\partial \frac{1}{T}$) instead of using the whole temperature range. We find that our calculated tracer diffusion coefficients above 1200 K follow well the Arrhenius behaviour. We perform such fits on the experimental data of Million *et al.* [262], and the results of the activation energies are shown in Fig. 5.20. It can be seen that the derived activation energies change dramatically with temperature, with some clear tendencies: the activation energies of the Fe diffusion in the alloys have minima at 1378 K or 1448 K, whereas those of the Ni diffusion decrease with temperature above 1448 K. In view of the significant variations of the derived activation energies with temperature, it is clear why the activation energies in Fig. 5.19 can be very different when using data in different temperature ranges. Therefore, one may seriously question the validity of the Arrhenius fit to interpret the experimental data. In addition, this Arrhenius or non-Arrhenius behaviour may be difficult to confirmed experimentally, due to the sensitivity of the derived activation energies to the measured diffusion coefficients, for which a high experimental precision is required.

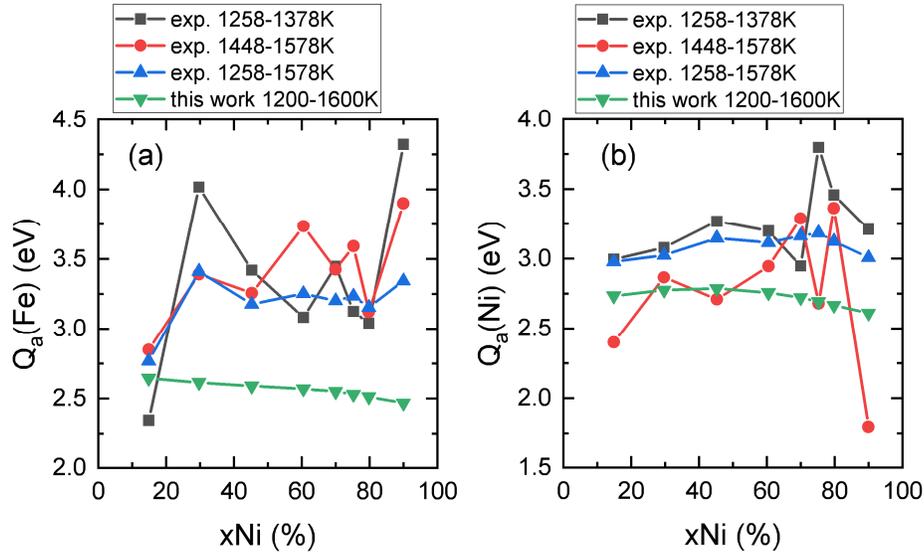

FIGURE 5.19: Diffusion activation energies of (a) Fe and (b) Ni, fitted to different data sets. The squares, circles and up-triangles are the results fitted to the experimental data in different temperature ranges by Million *et al.* [262].

On the theoretical side, it has been shown that the equilibrium vacancy concentrations in fcc metals such as Al, Cu and Ni do not follow the Arrhenius law, due to the strong anharmonic effects at high temperatures [117, 276]. Such effects in Ni are found to be important at temperatures above 1100 K [117]. One may therefore expect a non-Arrhenius behaviour of the diffusion coefficients at least in Ni-rich alloys due to the anharmonic effects. Such effects on the vacancy formation and diffusion properties are neglected in the present work, which can be a possible source of errors. The agreement between the calculated and experimental tracer diffusion coefficients suggests that neglecting the anharmonic effects may still lead to a reasonable estimation of tracer diffusion coefficients. Meanwhile, the derived prefactor and activation energy can be very sensitive to the deviation of the experimental tracer diffusion coefficients from the Arrhenius law.

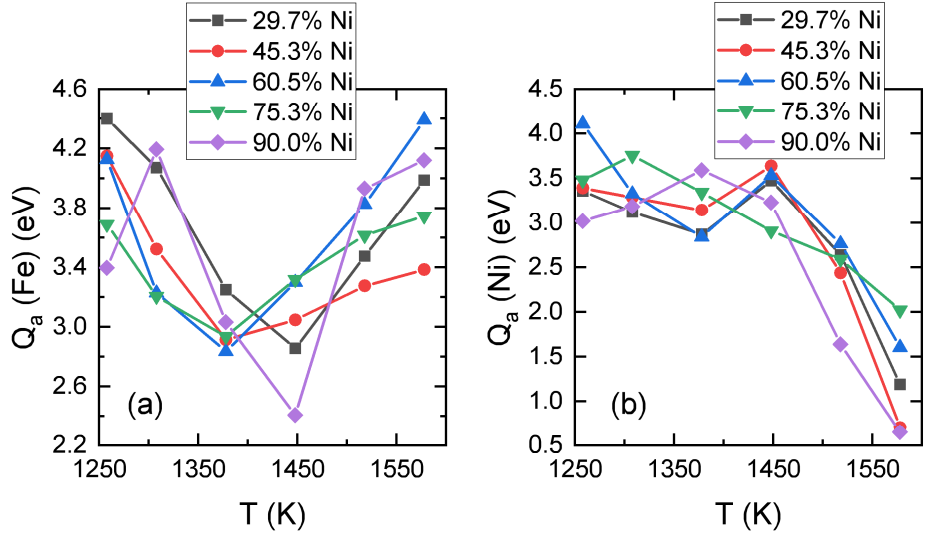

FIGURE 5.20: Diffusion activation energies of (a) Fe and (b) Ni, derived as $Q_a = -\frac{\partial \log D^*}{\partial \frac{1}{T}}$ from the experimental data of Million *et al.* [262].

## 5.5 Conclusion

In this Chapter, we investigate vacancy-mediated diffusion properties in fcc Fe-Ni alloys using DFT calculations and Monte Carlo simulations combined with the DFT-parametrized EIM. Diffusion properties such as tracer diffusion coefficients, kinetic correlation factors, migration energies are systematically computed for a broad range of temperature and composition and compared to available experimental data.

In the first place, we study the self- and (Ni or Fe) solute diffusion in fcc Fe and Ni. The predicted tracer diffusion coefficients in Fe and Ni are in good agreement with experimental data, which are only available above the magnetic transition temperatures. In both fcc Fe and Ni, the average migration energy for the Ni diffusion shows a weak temperature dependence, varying by less than 0.05 eV. By contrast, $E_m$ for the Fe diffusion in fcc Fe and Ni at high temperatures are reduced by 0.10 and 0.25 eV, respectively, compared to the magnetic ground-state values. In addition, the longitudinal spin fluctuations in fcc Fe and Ni also contribute to the variation of $E_m$ for the Fe diffusion in the paramagnetic regime. In addition, we note a strong effect of the magnetic transition on the Fe-diffusion correlation factor in Ni, which is associated with the strong variation of the Fe migration energy across the Curie temperature

The magnetic effects on the migration energies can be compared to our results of vacancy formation properties in Fe and Ni. They consistently reveal that the magnetic transition should have a large effect on a property that depends strongly on spin orientation arrangements. Meanwhile, a steady variation with temperature in the paramagnetic regime can arise from strong longitudinal spin fluctuations, if the property is sensitive to the variation of spin magnitudes.

Then, we compute the tracer diffusion coefficients as functions of temperature in the alloys over a broad range of composition. We find a overall good agreement with the experimental data, which are available only at high temperatures, where the alloys are paramagnetic and chemically disordered. We examine in details the diffusion properties in the alloy with 75% Ni. The tracer diffusion coefficients obey the Arrhenius law only in the paramagnetic random alloy. In the ferromagnetic ordered phase, the non-Arrhenius behaviour exhibits as the strong decrease in the diffusion activation free energy $G_a$ with temperature, which is related to the magnetic excitations and change of chemical long-range order. This decrease in $G_a$ with temperature can be mainly attributed to the decrease in the average migration energy, and also partly to the variations in $G_f^{mag}$ and kinetic correlation effects. The diffusion activation energy is found to be significantly reduced across the chemical transition temperature, but not much influenced across the Curie temperature.

We investigate the concentration dependence of the diffusion properties in the alloys. Overall, the average migration energies of the Fe and Ni atoms tend to increase with increasing Ni content in disordered alloys. For a given composition, the Fe atoms have a smaller average migration energy and a smaller attempt frequency than the Ni atoms. This compensating effect of the migration energy and the attempt frequency leads to similar tracer diffusion coefficients of Fe and Ni in a random alloy. We find that the Fe diffusion shows a stronger correlation in the concentrated disordered alloys, whereas the Ni diffusion shows a stronger correlation in the disordered alloys very rich in Fe.

The calculated tracer diffusion coefficients are in agreement within the uncertainty of the available experimental data. Meanwhile, the calculated prefactors and diffusion activation energies are found to be lower than the experimental values resulting from the Arrhenius fits on the tracer diffusion coefficients. This discrepancy could be related to the non-Arrhenieus behaviour in experimental data due to, e.g., anharmonicity or impurities.

# 6 Conclusions and perspectives

The objective of this thesis is to elucidate the effects of magnetism and the magneto-chemical interplay on thermodynamics, defects formation and atomic diffusion. We address these problems in fcc Fe-Ni alloys for the whole composition range, because (1) they constitute the technologically important austenitic steels and other multicomponent alloys, (2) they undergo successive chemical and magnetic transitions within a small temperature window, therefore potentially presenting a strong magnetochemical coupling, and (3) they exhibit significant longitudinal spin excitations whose effects are often neglected and need to be clarified.

We adopt a DFT-based modelling strategy in this work. We employ the DFT method to compute the energetic, magnetic and vibrational properties of the systems with various magnetic and chemical configurations. The DFT results provide not only a fundamental understanding of various properties, but also reliable inputs to parametrize an effective interaction model (EIM) for fcc Fe-Ni alloys over the whole composition range. On-lattice Monte Carlo simulations combined with the EIM enable to predict the temperature-dependent properties, with the simultaneous magnetic and chemical evolutions fully taken into account.

The first part of the thesis centres on the thermodynamics of defect-free Fe-Ni alloys, which is indispensable for the subsequent modelling of defect formation and diffusion.

We perform systematic DFT calculations in bcc and fcc Fe-Ni structures. Overall, the computed magnetization and vibrational entropies in the ferromagnetic disordered alloys are in agreement with available experimental data, confirming the accuracy of the DFT approach. Based on the DFT results, we compute the free energies of mixing for the ferromagnetic phases, and construct the bcc-fcc phase diagrams below the Curie temperatures, which are in good agreement with available experimental data. We demonstrate an important role of vibrational entropy on the prediction of chemical phase transitions.

To include the effects of magnetic excitations and to go beyond magnetic transitions, we parametrize an EIM with explicit magnetic and chemical variables and couple it with Monte Carlo simulations. We show that the EIM reproduces successfully not only the low-temperature properties as predicted by DFT, but also the experimental chemical order-disorder transition temperatures and the Curie temperatures.

According to our DFT and EIM predictions, the vibrational entropy has a stronger impact on the chemical order-disorder transitions than the magnetic entropy. From a methodological point of view, our results suggest that it is possible and convenient to construct a reliable phase diagram using solely DFT calculations, if the studied region is below magnetic transitions.

The accurate EIM allows to investigate the magnetochemical interplay on the phase stability. A phase separation between the paramagnetic and ferromagnetic disordered alloys is found in the composition range of 10-40% Ni at 600-700 K, which is shown to be magnetically driven. We also show that the chemical transition temperatures decrease with increasing magnetic disorder. Conversely, the magnetic transition temperatures decrease with increasing chemical long-range and short-range disorders.

Additionally, we study the addition of a Cr or Mn solute in the chemically ordered and disordered Fe-Ni alloys with 50% and 75% Ni. Although the spin alignment of the solute is dependent on the local chemical composition, overall the Cr solute tends to have an antiparallel spin alignment to other atoms, whereas the trend for the Mn solute is the opposite. In the $L1_2$ structure, both Mn and Cr solutes show a strong preference for the Fe sublattice over the Ni sublattice, contrary to their sublattice preference in the $L1_0$ structure. We show that the Cr (Mn) addition enhances (reduces) the energetic stability of the $L1_2$ structure, consistent with the experimental trend in the chemical order-disorder transition temperatures.

The second part of the work is focused on point-defect properties in fcc Fe-Ni alloys. In particular, we develop efficient Monte Carlo schemes to achieve a continuous and consistent prediction of vacancy formation properties as functions of temperature, especially for concentrated magnetic alloys.

The comparative study in bcc Fe, fcc Fe and Ni demonstrates that the magnetic effects on vacancy formation properties are very system dependent. The spin orientations have a stronger influence on the vacancy formation energy than the spin magnitudes in bcc Fe, whereas this is the opposite in fcc Fe. Meanwhile, the vacancy formation energy in fcc Ni is not very sensitive to the magnetic state. The magnetic transition induces a decrease in the vacancy formation magnetic free energy $G_f^{mag}$ in bcc Fe, nearly no change in fcc Fe, and an increase in fcc Ni. Unlike in bcc Fe, the significant longitudinal spin fluctuations in fcc Fe and Ni lead to a steady variation of $G_f^{mag}$ above the magnetic transition temperatures.

We compute $G_f^{mag}$ as a function of temperature in the alloys with 50% and 75% Ni, revealing that magnetic disorder tends to increase $G_f^{mag}$, while chemical disorder has an opposite effect. For the whole temperature range, the total variation of $G_f^{mag}$ due to magnetic disorder is larger than the variation caused by chemical disorder. Meanwhile, the chemical transition is found to have a more noticeable effect than the magnetic transition on $G_f^{mag}$. We study the concentration dependence of $G_f^{mag}$, showing that $G_f^{mag}$ in the disordered alloys decrease with increasing Ni content.

Finally, we study the SIA properties in fcc Fe and Ni. The magnetic state in Ni is shown to have a stronger impact on the SIA formation energy than the vacancy formation energy, whereas the impacts on the SIA and vacancy formation energies in Fe are similarly important. Meanwhile, the underlying magnetic states do not change the relative stability between different SIA configurations. The magnetic state has a strong effect on the SIA migration energy in fcc Fe, while the effect in Ni is small. We show that the rotation-translation mechanism is the most favourable mechanism for the SIA diffusion in Fe, whereas both rotation-translation and pure-translation mechanisms should be considered for the SIA diffusion in Ni.

The third and final part of the work is devoted to vacancy-mediated diffusion in fcc Fe-Ni alloys.

We study the self- and (Ni- or Fe-) solute diffusions in fcc Fe and Ni. The average migration energies $E_m$ for the Ni diffusion in fcc Fe and Ni are insensitive to the magnetic state of the system, showing little variation with temperature. By contrast, $E_m$ for the Fe diffusion in paramagnetic fcc Fe and Ni are reduced by up to 0.10 and 0.25 eV, respectively, compared to the magnetic ground-state values. In addition, the longitudinal spin fluctuations in fcc Fe and Ni are also found to contribute to the decrease of $E_m$ for the Fe diffusion in the paramagnetic regime.

The above results of diffusion properties in Fe and Ni can be compared to our results on vacancy formation in Fe and Ni. They consistently reveal that the magnetic transition should have a large effect on a property that depends strongly on spin orientation arrangements. Meanwhile, a steady variation with temperature in the paramagnetic regime can be expected due to longitudinal spin fluctuations, if the given property is sensitive to the variation of spin magnitudes.

We examine the temperature dependence of the diffusion properties in the alloy with 75% Ni. The tracer diffusion coefficients obey the Arrhenius law only in the paramagnetic random alloy. In the ferromagnetic ordered phase, the non-Arrhenius behaviour exhibits as the strong decrease in the diffusion activation free energy $G_a$ with temperature, which is mainly caused by the changes in magnetic and chemical orders. This decrease in $G_a$ with temperature is mainly attributed to the decrease in the average migration energy, and also in a minor extent to the variations in $G_f^{mag}$ and kinetic correlation effects. The diffusion activation energy is found to be significantly reduced across the chemical transition temperature, but not much influenced across the Curie temperature.

We also investigate the concentration dependence of the diffusion properties in the alloys. Overall, the average migration energies of the Fe and Ni atoms tend to increase with increasing Ni content in disordered alloys. For a given composition, the Fe atoms have a smaller average migration energy and a smaller attempt frequency than the Ni atoms. This compensating effect of the migration energy and the attempt frequency leads to similar tracer diffusion coefficients of Fe and Ni in a random alloy. In addition, we find that the Fe diffusion shows a stronger kinetic correlation in the concentrated alloys, whereas the Ni diffusion shows a stronger correlation in the alloys very rich in Fe. The calculated tracer diffusion coefficients are in agreement with the available measured data within the experimental uncertainty.

In this thesis, we investigate the magnetochemical effects on the thermodynamic, defect formation and diffusion properties in fcc Fe-Ni alloys over the whole composition range. Based solely on DFT results, either directly or mapped to the EIMs, and without using any empirical or experimental inputs, our approach enables to achieve an accurate and consistent prediction that is in overall good agreement with the available experimental data. This demonstrates the accuracy and the predictive power of the present modelling approach, which can be applied to the investigation of the magnetic effects on various properties and kinetic processes in other magnetic alloys.

Despite our efforts in this thesis, the understanding of the effects of magnetism and its interplay with other degrees of freedom on various properties is far from complete. Some perspectives of this work are proposed as follows.

First, it would be straightforward to extend the present EIM to study the diffusion of impurities (e.g., Mn, Cr, Co, Cu and C). A good starting point may be to predict the impurity diffusion in fcc Fe and Ni. Then, one may also investigate the magnetochemical interplay on the impurity diffusion, e.g., in the Fe-Ni alloy with 75% Ni across the magnetic and chemical transitions.

This thesis is primarily focused on the vacancy formation and diffusion properties, whereas the magnetic effects on the SIA properties are only limited to pure Fe and Ni and lack a quantitative prediction. In order to gain insights into kinetic processes in nonequilibrium conditions such as irradiation induced segregation, it would be of interest to include also the SIAs in the present EIM and to have a more comprehensive understanding of magnetic effects.

As suggested by our DFT study, the additions of Mn and Cr may have significant effects on the magnetic and chemical orders in Fe-Ni alloys. It is therefore very interesting to obtain a quantitative understanding of the effects on phase stability, by extending the present EIM to other ternary alloys such as Fe-Ni-Cr and Fe-Ni-Mn. Such models could then be further extended for the investigation of other properties (e.g. those studied in this thesis) in the ternary systems and eventually towards multicomponent alloys such as Fe-Ni-Cr-Mn systems.

Finally, the present treatment of vibrational entropy is implicit. It is equivalent to integrate the fast vibrational degree of freedom into the nonmagnetic interaction terms in the EIM. We adopt this treatment mainly for its simplicity, whereas this may not be well justified considering the timescales of magnon and phonon excitations are not very different. Therefore, an exciting but highly challenging task would be to include explicit vibrational variables in the EIM and to explore the potential magnon-phonon coupling effects.

# A Parametrization of effective interaction models

We first present in Sec. A.1 the model parameters for the on-lattice configuration, where all the atoms are in the lattice sites. The on-lattice configuration can contain one or no vacancy. Then we present in Sec. A.2 the model parameters for the saddle-point configuration, where the migrating atom is at the saddle point while the rest of the atoms are in the lattice sites. Finally we present in Sec. A.3 the quality of the fitting by comparing the fitted results with the DFT results.

## A.1 Parameters for the on-lattice configurations

The Hamiltonian for the on-lattice (OL) configuration has the following form:

$$E_{OL} = \sum_i \sigma_i \cdot [A_i M_i^2 + B_i M_i^4 + \sum_j \sigma_j \cdot J_{ij} \boldsymbol{M}_i \boldsymbol{M}_j + \epsilon_i + \sum_j \sigma_j \cdot (V_{ij} + \alpha_{ij} T)] \qquad \text{(A.1)}$$

where $i$ is the $i$-th fcc lattice site, $\sigma_i$ is the occupation variable and is equal to 1 (0) for an occupied (vacant) lattice site, and $\sum_j$ is a sum over all the neighbours up to the fourth-neighbour shell. In the magnetic part, $M_i$ is the magnetic moment, $A_i$ and $B_i$ are the on-site magnetic parameters, $J_{ij}$ is the exchange interaction parameters. In the nonmagnetic part, $\epsilon_i$ is the on-site nonmagnetic parameter, $V_{ij}$ and $\alpha_{ij}$ are the nonmagnetic interaction parameters, and $T$ is the absolute temperature.

We note that each pair interaction is in total counted twice when calculating $E_{OL}$. For example, the term $V_{ij}$ of a pair $ij$ appears twice when the sum $\sum_i$ goes over $i$ and $j$.

All the parameters depend on the distance between the site $i$ and the vacancy. The parameters are dependent on the local chemical environment, i.e. the local concentration of Ni and the presence of a vacancy around the atom $i$. A detailed description of the parameters is given in Table A.1.

TABLE A.1: Description of the model parameters. The first column presents the parameters in the Hamiltonian in Eq. A.1. The second column expresses these parameters as functions of the local environment. The third column describes the parameters and the notations. The units of the numerical values for magnetic parameters $J_{ij}$, $A_i$ and $B_j$ are meV/$\mu_B^2$, meV/$\mu_B^2$ and meV/$\mu_B^4$, respectively. The units of the numerical values for all nonmagnetic parameters are meV.

| Parameter | Expression | Description |
|---|---|---|
| $A_i$ | $A_i = A_{i,0}^{nV} + A_{i,1}^{nV} * c_i$ | $A_i$ is a linear function of $c_i$ (Ni concentration in the nearest-neighbour shell of $i$). $A_{i,0}^{nV}$ and $A_{i,1}^{nV}$ are two constants depending on the distance $nV$ from the vacancy. $nV$ is denoted as 1 for the nearest-neighbour distance, 2 for the second nearest-neighbour distance, and 0 for the farther distances (in this case $i$ is considered as in the bulk). |
| $B_i$ | $B_i = B_i^{nV}$ | $B_i$ depends only on $nV$. |
| $J_{ij}$ | $J_{ij} = J_{(ij)^k,0}^{nV} + J_{(ij)^k,1}^{nV} * c_i$ | $J_{ij}$ is a linear function of $c_i$. $k$ denotes the $k^{th}$ nearest-neighbour shell. The constants $J_{(ij)^k,0}^{nV}$ is symmetric with respect to $ij$ ($J_{(ij)^k,0}^{nV} = J_{(ji)^k,0}^{nV}$), and so is $J_{(ij)^k,1}^{nV}$. Consequently, $J_{ij} \neq J_{ji}$ if $c_i = c_j$ |
| $\epsilon_i$ | $\epsilon_i = \epsilon_X^{nV}$ | $\epsilon_i$ depends only on $nV$. |
| $V_{ij}$ | $V_{ij} = V_{(ij)^k}^{nV}$ | $V_{ij}$ depends on $(ij)^k$ and $nV$. $V_{(ij)^k}^{nV}$ is symmetric with respect to $ij$ [i.e. $V_{(ij)^k}^{nV} = V_{(ji)^k}^{nV}$]. |
| $\alpha_{ij}$ | $\alpha_{ij} = \alpha_{(ij)^k,0} + \alpha_{(ij)^k,1} * c_i$ | $\alpha_{ij}$ is a linear function of $c_i$. $\alpha_{(ij)^k,0}$ and $\alpha_{(ij)^k,1}$ are symmetric with respect to $ij$. Consequently, $\alpha_{ij} \neq \alpha_{ji}$ if $c_i \neq c_j$. |

TABLE A.2: Magnetic parameters $A_{X,0}^0$, $B_X^0$, $J_{(Fe-Fe)^k,0}^0$ and $J_{(Ni-Ni)^k,0}^0$ for defect-free Fe and Ni.

| Parameter | $A_{X,0}^0$ | $B_X^0$ | $J_{(X-X)^1,0}^0$ | $J_{(X-X)^2,0}^0$ | $J_{(X-X)^3,0}^0$ | $J_{(X-X)^4,0}^0$ |
|---|---|---|---|---|---|---|
| $X = $ Fe | -37.9255 | 6.1116 | -0.3058 | -2.9979 | 0.2574 | 0.8652 |
| $X = $ Ni | 73.1841 | 277.7885 | -19.9994 | -8.6825 | 1.6658 | -4.9932 |

TABLE A.3: Additional magnetic parameters for defect-free Fe-Ni alloys. A value of 0 means that this parameter was assumed to be zero and not used in the fitting.

| Parameter | $A^0_{X,1}$ | $J^0_{(X-X)^1,1}$ | $J^0_{(X-X)^2,1}$ | $J^0_{(X-X)^3,1}$ | $J^0_{(X-X)^4,1}$ |
|---|---|---|---|---|---|
| $X$ = Fe | -49.1235 | -3.9148 | 4.8989 | 0 | 0 |
| $X$ = Ni | 0 | 0 | 0 | 0 | 0 |

| Parameter | $J^0_{(Fe-Ni)^1,n}$ | $J^0_{(Fe-Ni)^2,n}$ | $J^0_{(Fe-Ni)^3,n}$ | $J^0_{(Fe-Ni)^4,n}$ |
|---|---|---|---|---|
| $n=0$ | -4.6266 | -4.5693 | -0.1020 | -2.0075 |
| $n=1$ | -8.2601 | 15.7138 | 0 | 0 |

TABLE A.4: Nonmagnetic parameters for defect-free Fe-Ni alloys.

| Parameter | $\sigma^0_X$ | $V_{(X-X)^1}$ | $V_{(X-X)^2}$ | $V_{(X-X)^3}$ | $V_{(X-X)^4}$ | $V^0_{(Fe-Ni)^1}$ | $V^0_{(Fe-Ni)^2}$ | $V^0_{(Fe-Ni)^3}$ | $V^0_{(Fe-Ni)^4}$ |
|---|---|---|---|---|---|---|---|---|---|
| $X$ = Fe | -8044.964 | 4.7174 | -4.7331 | -0.7931 | -1.2585 | 5.4638 | 4.9789 | 0.7396 | 1.5661 |
| $X$ = Ni | -5518.851 | 6.2255 | 6.4987 | 0.3030 | -1.6027 | | | | |

TABLE A.5: Nonmagnetic vibrational parameters for defect-free Fe-Ni alloys.

| Parameter | $\alpha_{(Fe-Ni)^1,0}$ | $\alpha_{(Fe-Ni)^2,0}$ | $\alpha_{(Fe-Ni)^3,0}$ | $\alpha_{(Fe-Ni)^4,0}$ |
|---|---|---|---|---|
| $X$ = Fe or Ni | $2.9949\times10^{-3}$ | $-2.5950\times10^{-3}$ | $1.4889\times10^{-3}$ | $-2.3568\times10^{-3}$ |

| Parameter | $\alpha_{(Fe-Ni)^1,1}$ | $\alpha_{(Fe-Ni)^2,1}$ | $\alpha_{(Fe-Ni)^3,1}$ | $\alpha_{(Fe-Ni)^4,1}$ |
|---|---|---|---|---|
| $X$ = Fe or Ni | $-0.1544\times10^{-3}$ | $4.6870\times10^{-3}$ | $-2.6055\times10^{-3}$ | $-0.1000\times10^{-3}$ |

TABLE A.6: Magnetic parameters for atoms around the vacancy in pure Fe and Ni. $d$ is the nearest-neighbor distance between the atom and the vacancy. *Bulk* means that the parameter remains the same as that of the bulk atoms in Table A.2.

| Parameter | | $A^D_{X,0}$ | $B^D_X$ | $J^D_{(X-X)^1,0}$ | $J^D_{(X-X)^2,0}$ | $J^D_{(X-X)^3,0}$ | $J^D_{(X-X)^4,0}$ |
|---|---|---|---|---|---|---|---|
| $X$=Fe | $d=1$ | -47.9065 | 6.5375 | -0.7776 | -2.5729 | *Bulk* | *Bulk* |
| | $d=2$ | -45.4071 | 6.6602 | 0.3547 | -3.6241 | *Bulk* | *Bulk* |
| $X$=Ni | $d=1$ | 72.8230 | 250.7870 | -19.8272 | -13.4330 | *Bulk* | *Bulk* |
| | $d=2$ | 114.5447 | 316.7424 | -22.9430 | -7.7159 | *Bulk* | *Bulk* |

TABLE A.7: Magnetic parameters for atoms around the vacancy in Fe-Ni alloys. $d$ is the nearest-neighbor distance between the atom and the vacancy. *Bulk* means that the parameter remains the same as that of bulk atoms in Table A.3; *Bulk(0)* means additionally that the corresponding same parameter of bulk atoms in Table A.3 is zero (namely not used in the fitting).

| Parameter | | $A_{X,1}^d$ | $J_{(X-X)^1,1}^d$ | $J_{(X-X)^2,1}^d$ | $J_{(X-X)^3,1}^d$ | $J_{(X-X)^4,1}^d$ |
|---|---|---|---|---|---|---|
| $X$=Fe | $d$=1 | -54.4202 | *Bulk* | *Bulk* | *Bulk(0)* | *Bulk(0)* |
|  | $d$=2 | -35.7534 | *Bulk* | *Bulk* | *Bulk(0)* | *Bulk(0)* |
| $X$=Ni | $d$=1 | *Bulk(0)* | *Bulk(0)* | *Bulk(0)* | *Bulk(0)* | *Bulk(0)* |
|  | $d$=2 | *Bulk(0)* | *Bulk(0)* | *Bulk(0)* | *Bulk(0)* | *Bulk(0)* |

| Parameter | $J_{(Fe-Ni)^1,0}^d$ | $J_{(Fe-Ni)^2,0}^d$ | $J_{(Fe-Ni)^3,0}^d$ | $J_{(Fe-Ni)^4,0}^d$ | $J_{(Fe-Ni)^1,1}^d$ | $J_{(Fe-Ni)^2,1}^d$ | $J_{(Fe-Ni)^3,1}^d$ | $J_{(Fe-Ni)^4,1}^d$ |
|---|---|---|---|---|---|---|---|---|
| $d$=1 | -5.9182 | -3.7765 | *Bulk* | *Bulk* | *Bulk* | *Bulk* | *Bulk(0)* | *Bulk(0)* |
| $d$=2 | -6.7239 | -1.3378 | *Bulk* | *Bulk* | *Bulk* | *Bulk* | *Bulk(0)* | *Bulk(0)* |

TABLE A.8: Nonmagnetic parameters $\sigma_X^1$ and $V_{(Fe-Ni)^1}^1$ of atoms as first-nearest neighbors of the vacancy. The nonmagnetic parameters of atoms beyond the first-nearest-neighbor shell are the same as those of bulk atoms.

| $\sigma_{Fe}^1$ | $\sigma_{Ni}^1$ | $V_{(Fe-Ni)^1}^1$ | $V_{(Fe-Ni)^1}^1$ |
|---|---|---|---|
| -7851.1051 | -5393.5547 | 8.8990 | 0.7200 |

## A.2 Parameters for the saddle-point configurations

An $N$-site saddle-point (SP) configuration consists of $N-2$ on-lattice (OL) atoms, two first-nearest-neighbor vacant sites, and one SP atom. We express the total energy of the SP configuration as the sum of two parts:

$$E_{tot} = E_{OL} + E_{SP} \tag{A.2}$$

where $E_{OL}$ is the energy due to the interactions between the on-lattice atoms, and $E_{SP}$ is the energy due to the interactions of the SP atom with its neighbouring atoms on the lattice. $E_{OL}$ can be determined with the model parameters presented in the previous section. Note that for OL atoms with two vacancies around, their model parameters depend on the distance from the nearest vacancy. The expression of $E_{SP}$ is as follows:

$$E_{SP} = A_{SP}M_{SP}^2 + B_{SP}M_{SP}^4 + \sum_j J_{SP,j}\boldsymbol{M}_{SP}\boldsymbol{M}_j + \epsilon_{SP} + \sum_j V_{SP,j} \tag{A.3}$$

where $M_{SP}$ is the magnetic moment of the saddle-point atom, and $A_{SP}$, $B_{SP}$, $J_{SP,j}$, $\epsilon_{SP}$ and $V_{SP,j}$ are the corresponding magnetic and nonmagnetic parameters. The sum $\sum_j$ goes over the atoms within the fourth-neighbour shell of the SP atom. Note that the number of neighbours and the distance from neighbours for the SP atom are different from those of the OL atoms.

The parameters are presented in Table A.9-A.11.

We note that each pair interaction is in total counted *only once* when calculating $E_{OL}$. For example, the term $V_{SP,j}$ of a pair $SP - j$ appears only once in $E_{SP}$.

TABLE A.9: Magnetic and nonmagnetic on-site parameters for the SP atom.

|          | $A_{SP}$  | $B_{SP}$ | $\epsilon_{SP}$ |
|----------|-----------|----------|-----------------|
| $SP$=Fe  | -205.2016 | 15.6446  | -7707.2197      |
| $SP$=Ni  | 74.0210   | 301.1836 | -5883.8414      |

TABLE A.10: Magnetic exchange interaction parameters $J_{SP,OL}$ between the SP atom and the OL atom separated by a given distance.

| $SP$ | $OL$ | 1NN      | 2NN      | 3NN      | 4NN     |
|------|------|----------|----------|----------|---------|
| Fe   | Fe   | -5.3855  | -8.0161  | 1.5916   | -0.6457 |
| Fe   | Ni   | -24.5305 | 5.8653   | 9.1694   | -9.4957 |
| Ni   | Ni   | -71.6976 | -35.0767 | 13.9183  | -8.3676 |
| Ni   | Fe   | -13.2241 | -3.7589  | -18.7414 | 5.8729  |

TABLE A.11: Nonmagnetic parameters $V_{SP,OL}$ between the SP atom and the OL atom separated by a given distance.

| $SP$ | $OL$ | 1NN      | 2NN     | 3NN     | 4NN      |
|------|------|----------|---------|---------|----------|
| Fe   | Fe   | -28.7755 | 47.2913 | 30.9704 | -29.1062 |
| Fe   | Ni   | -34.6618 | 14.367  | 62.7312 | -19.1265 |
| Ni   | Ni   | 147.0236 | 54.6353 | 57.2937 | -56.6513 |
| Ni   | Fe   | 65.3323  | 12.8306 | 60.7932 | -67.5517 |

## A.3 Quality of the fittings

In this section, we compare the EIM predictions to the DFT results. All the EIM predictions here are computed using the DFT magnetic configurations as inputs.

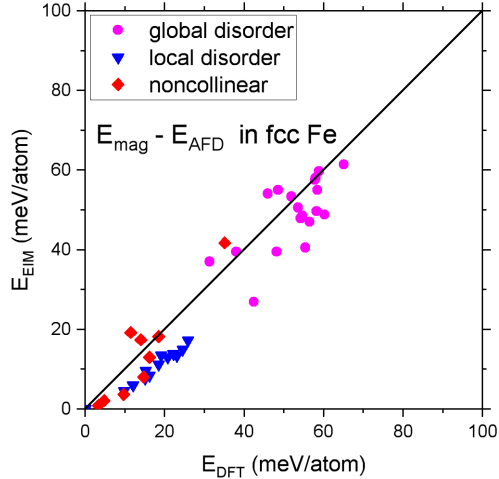

(a) Energy difference between various magnetic states of pure Fe.

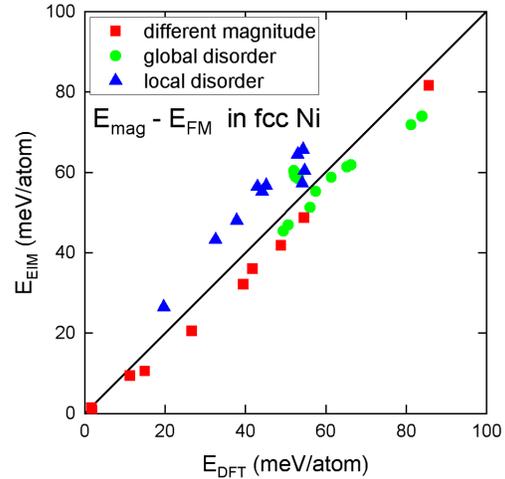

(b) Energy difference between various magnetic states of pure Ni.

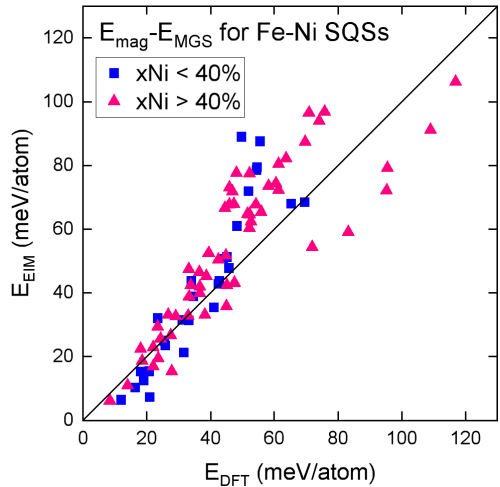

(c) Energy difference between various magnetic states of Fe-Ni alloys. We consider different magnetic states for each chemical configuration.

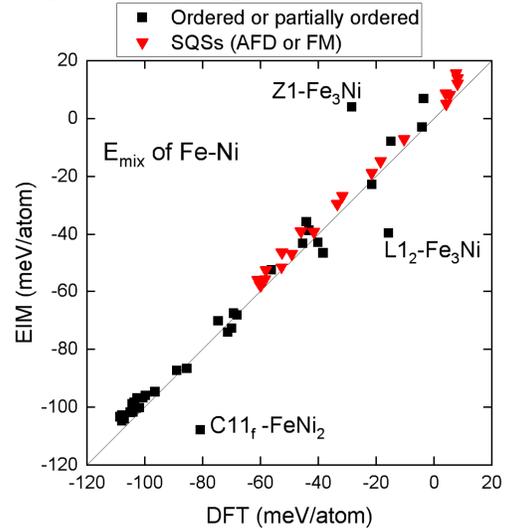

(d) Mixing energies per atom (with respect to AFD Fe and FM Ni) of Fe-Ni alloys in the respective magnetic ground states.

FIGURE A.1: Comparison between the DFT results and the model predictions (using DFT configurations as input) for vacancy-free on-lattice Fe-Ni configurations. The straight lines show x=y.

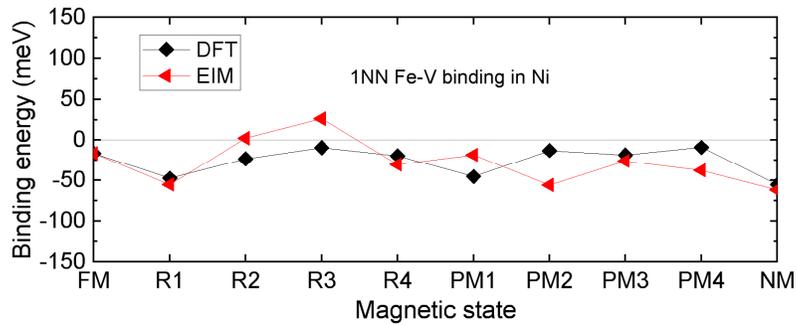

(a) 1NN Fe-vacancy binding energy in different magnetic configurations of Ni. FM: ferromagnetic state. R1-R4: globally FM state with local magnetic disorder. PM1-PM4: globally disordered magnetic state. NM: nonmagnetic state.

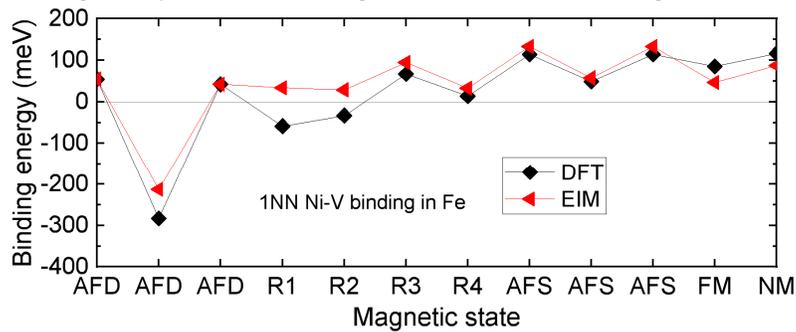

(b) 1NN Ni-vacancy binding energy in different magnetic configurations of Fe. AFD: antiferromagnetic double-layer state. AFS: antiferromagnetic single-layer state. Other notations are the same as in (a). In the AFD (or AFS) state, there are several non-equivalent Ni-vacancy binding configurations.

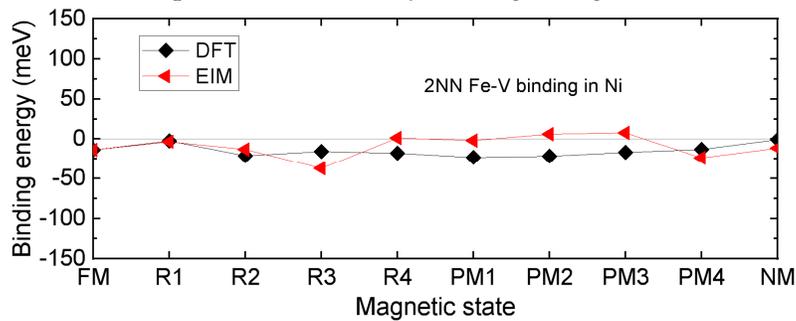

(c) 2NN Fe-vacancy binding energy in different magnetic configurations of Ni.

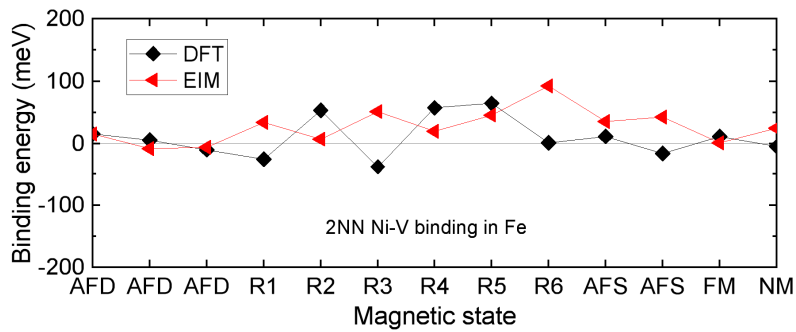

(d) 2NN Ni-vacancy binding energy in different magnetic configurations of Fe.

FIGURE A.2: Comparison between the DFT and the model predictions (using DFT configurations as input) for solute-vacancy binding energies in Fe and Ni.

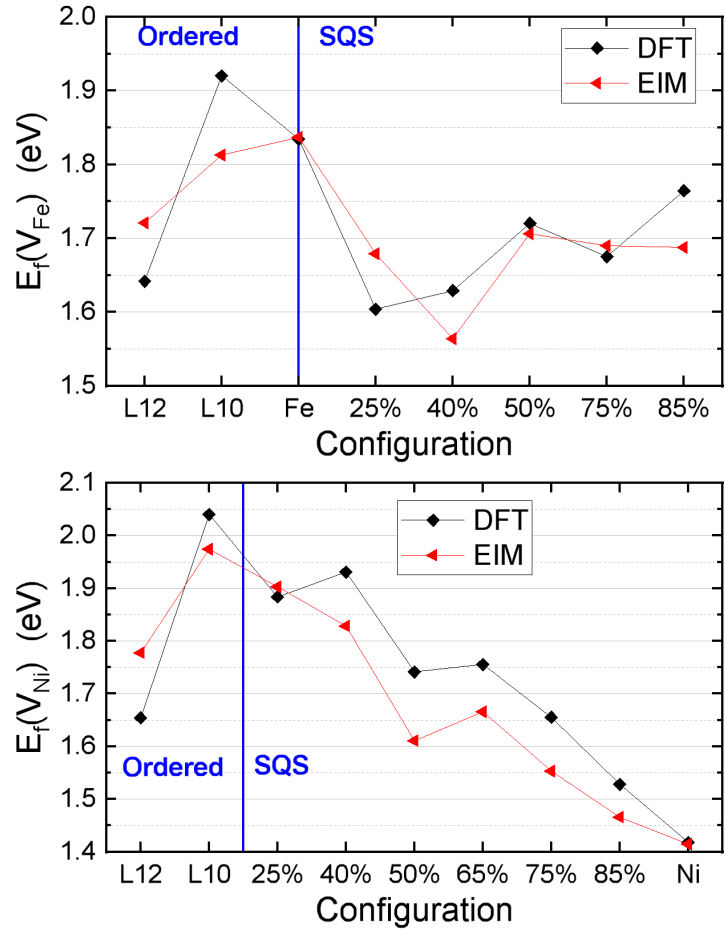

FIGURE A.3: Comparison between the DFT and the model predictions (using DFT configurations as input) for Fe- and Ni-vacancy formation energies in different Fe-Ni configurations in the magnetic ground states. The percentages in the X axis denote the Ni concentration in the SQSs. The result of each SQS is obtained by averaging over at least eight sites having different local chemical configurations. Please note that here $E_f(V_{\text{Fe}})$ and $E_f(V_{\text{Ni}})$ are computed using $e_{\text{AFD-Fe}}$ (energy per atom of AFD Fe) and $e_{\text{FM-Ni}}$ (energy per atom of FM Ni) as the chemical potentials. These results of $E_f(V_{\text{Fe}})$ and $E_f(V_{\text{Ni}})$ should be distinguished from those in the main text where the chemical potentials in the alloys are used.

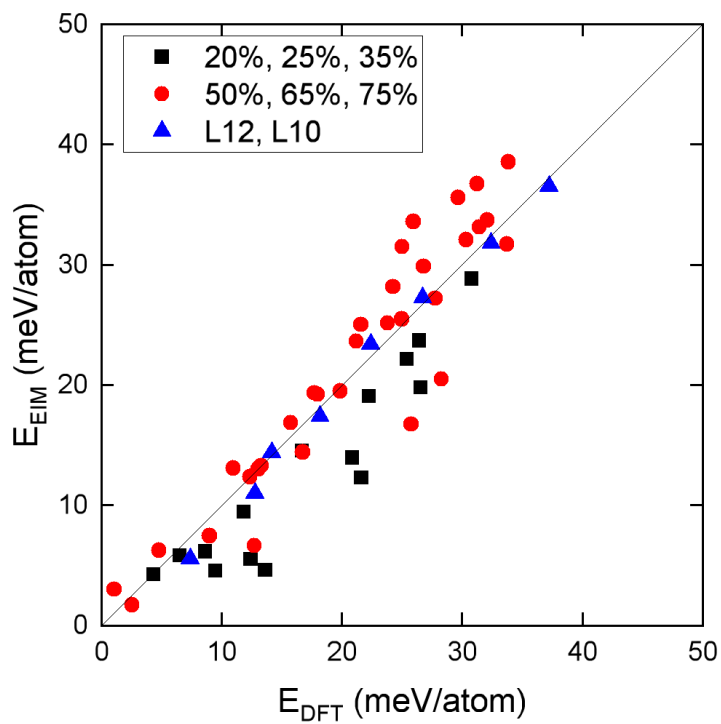

FIGURE A.4: Comparison between the DFT and the model predictions (using DFT configurations as input) of energies per atom in vacancy-containing on-lattice Fe-Ni configurations in various magnetic states.

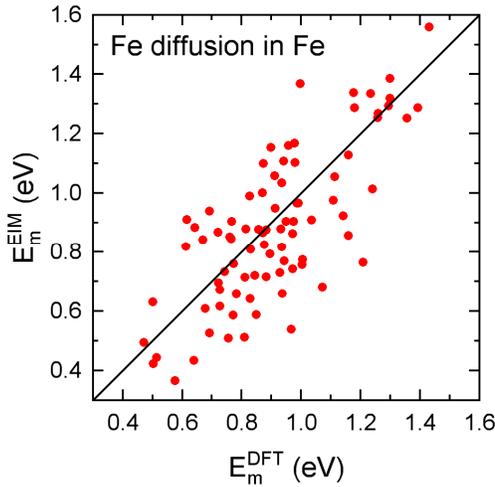

(a) Migration barriers of Fe-vacancy exchanges in various magnetic states of pure Fe.

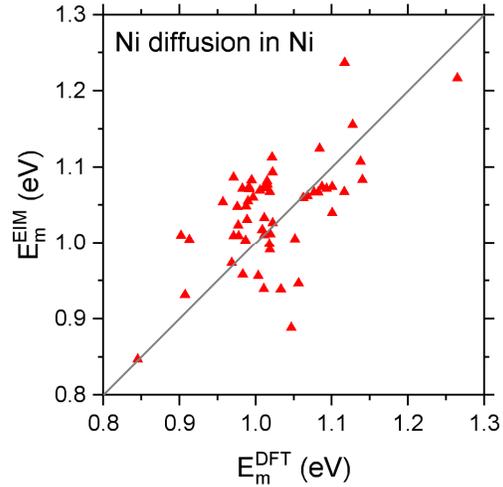

(b) Migration barriers of Ni-vacancy exchanges in various magnetic states of pure Ni.

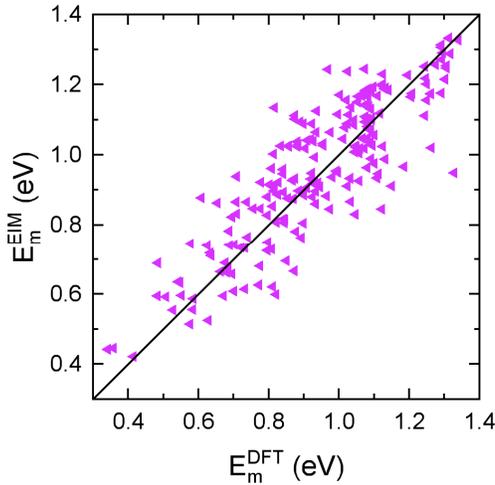

(c) Migration barriers of Fe- and Ni-vacancy exchanges in chemically ordered and disordered Fe-Ni configurations in various magnetic states. The standard error is 0.10 eV.

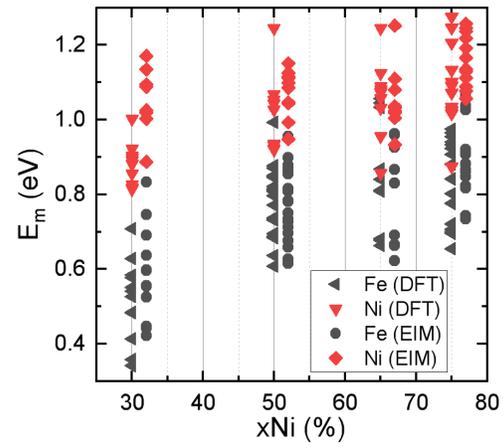

(d) Migration barriers of Fe- and Ni-vacancy exchanges in Fe-Ni disordered structures (SQSs) in the ferromagnetic ground states. Four SQSs (with 30%, 50%, 65% and 75% Ni) are considered and the migration barriers in different sites are computed. To facilitate the DFT-EIM comparison, the EIM results are slightly shifted towards the right-hand side of the DFT results.

FIGURE A.5: Comparison between the DFT results and the model predictions (using DFT configurations as input) for saddle-point Fe-Ni configurations.

# B Publications

### 1. Effects of magnetic excitations and transitions on vacancy formation: Cases of fcc Fe and Ni compared to bcc Fe

**K. Li**, C.-C. Fu, A. Schneider, *Phys. Rev. B, 104, 104406 (2021)*.

**Abstract:** Vacancy is one of the most frequent defects in metals. We study the impacts of magnetism on vacancy formation properties in fcc Ni, and in bcc and fcc Fe, via density functional theory (DFT) and effective interaction models combined with Monte Carlo simulations. Overall, the predicted vacancy formation energies and equilibrium vacancy concentrations are in good agreement with experimental data, available only at the high-temperature paramagnetic regime. Effects of magnetic transitions on vacancy formation energies are found to be more important in bcc Fe than in fcc Fe and Ni. The distinct behaviour is correlated to the relative roles of longitudinal and transversal spin excitations. At variance with the bcc-Fe case, we note a clear effect of longitudinal spin excitations on the magnetic free energy of vacancy formation in fcc Fe and Ni, leading to its steady variation above the respective magnetic transition temperature. Below the Néel point, such effect in fcc Fe is comparable but opposite to the one of the transversal excitations. Regarding fcc Ni, although neglecting the longitudinal spin excitations induces an overestimation of the Curie temperature by 220 K, no additional effect is visible below the Curie point. The distinct effects on the three systems are closely linked to DFT predictions of the dependence of vacancy formation energy on the variation of local magnetic-moment magnitudes and orientations.

### 2. Ground-state properties and lattice-vibration effects of disordered Fe-Ni systems for phase stability predictions

**K. Li**, C.-C. Fu, *Phys. Rev. Mater. 4, 023606 (2020)*.

**Abstract:** By means of density functional theory, we perform a focused study of both body-centered-cubic (bcc) and face-centered-cubic (fcc) Fe-Ni random solid solutions, represented by special quasirandom structures. The whole concentration range and various magnetic configurations are considered. Excellent agreement on the concentration dependence of magnetization is found between our results and experimental data, except in the Invar region. Some locally antiferromagnetic fcc structures are proposed to approach experimental values of magnetization. Vibrational entropies of ordered and disordered systems are calculated for various concentrations, showing an overall good agreement with available experimental data. The vibrational entropy systematically contributes to stabilize disordered rather than ordered structures and is not negligible compared to the configurational entropy. Free energy of mixing is estimated by including the vibrational and ideal configurational entropies. From them, low- and intermediate-temperature Fe-Ni phase diagrams are constructed, showing a better agreement with experimental data than the one from a recent thermodynamic assessment for some phase boundaries below 700 K. The determined order-disorder transition temperatures for the $L1_0$ and $L1_2$ phases are in good agreement with the experimental values, suggesting an important contribution of vibrational entropy.

### 3. Temperature-dependent magnetochemical effects on thermodynamics and vacancy formation in fcc Fe-Ni alloys: an atomistic study

**K. Li**, C.-C. Fu, M. Nastar, F. Soisson, M. Y. Lavrentiev, *to be submitted*.

**Abstract:** We investigate thermodynamic and vacancy formation properties of fcc Fe-Ni alloys over a broad composition-temperature range, via density functional theory (DFT) and DFT-parametrized effective interaction model (EIM) with explicit chemical and spin variables. Monte Carlo simulations combined with the EIM enable to take into account the effects of thermal spin fluctuations and chemical evolutions. The EIM reproduces successfully the experimental chemical order-disorder transition temperatures and the Curie temperatures, and the predicted phase diagram is in agreement with the most recent CALPHAD assessment. The EIM predicts a phase separation between paramagnetic and ferromagnetic disordered phases around 10-40% Ni at 600-700 K, which is shown to be magnetically driven. We also find a strong interplay between the magnetic and chemical orders. We calculate vacancy formation magnetic free energy across the chemical and magnetic transition temperatures in the alloys with 50% and 75% Ni, revealing distinct effects of magnetic and chemical orders on vacancy formation. The computed vacancy formation magnetic free energies are compared to experimental data which are only available in paramagnetic random alloys.

### 4. Combining DFT and CALPHAD for the development of on-lattice interaction models: The case of Fe-Ni system

Y. Wang, **K. Li**, F. Soisson, C. S. Becquart, *Phys. Rev. Mater. 4, 113801 (2020)*

**Abstract:** We present a model of pair interactions on rigid lattice to study the thermodynamic properties of iron-nickel alloys. The pair interactions are fitted at 0 K on ab initio calculations of formation enthalpies of ordered and disordered (special quasirandom) structures. They are also systematically fitted on the Gibbs free energy of the $\gamma$ Fe-Ni solid solution as described in a CALPHAD (CALculation of PHAse Diagrams) study by Cacciamani et al. This allows the effects of finite temperature, especially those of magnetic transitions, to be accurately described. We show that the ab initio and CALPHAD data for the $\gamma$ solid solution and for the FeNi$_3$-L1$_2$ ordered phase can be well reproduced, in a large domain of composition and temperature, using first and second neighbor pair interactions which depend on temperature and local alloy composition. The procedure makes it possible to distinguish and separately compare magnetic, chemical, and configuration enthalpies and entropies. We discuss the remaining differences between the pair interaction model and CALPHAD, which are mainly due to the treatment of the short-range order and configurational entropy of the solid solution. The FCC phase diagram of the Fe-Ni system is determined by Monte Carlo simulations in the semigrand canonical ensemble and is compared with experimental studies and other models. We especially discuss the stability of the FeNi-$L1_0$ phase at low temperature.

### 5. Predicting magnetization of ferromagnetic binary Fe alloys from chemical short-range order

V.-T. Tran, C.-C. Fu, **K. Li**, *Comput. Mater. Sci. 172, 109344 (2020)*

**Abstract:** Among the ferromagnetic binary alloys, body centered cubic (bcc) Fe-Co is the one showing the highest magnetization. It is known experimentally that ordered Fe-Co structures show a larger magnetization than the random solid solutions with the same Co content. In this work, based on density functional theory (DFT) studies, we aim at a quantitative prediction of this feature, and point out the role of the orbital magnetic moments. Then, we introduce a DFT-based analytical model correlating local magnetic moments and chemical compositions for Co concentrations ranging from 0 to 70 at.%. It is also extended to predict the global magnetization of both ordered and disordered structures at given concentration and chemical short range orders. The latter model is particularly useful for interpreting experimental data. Based on these models, we note that the local magnetic moment of a Fe atom is mainly dictated by the Co concentration in its first two neighbor shells. The detailed local arrangement of the Co atoms has a minor effect. These simple models can fully reproduce the difference in magnetization between the ordered and disordered Fe-Co alloys between 30% and 70% Co, in good agreement with experimental data. Finally, we show that a similar model can be established for another bcc binary Fe alloy, the Fe-Ni, also presenting ferromagnetic interactions between atoms.

### 6. Magnetic and atomic short-range order in $Fe_{1-x}Cr_x$ alloys

I. Mirebeau, V. Pierron-Bohnes, C. Decorse, E. Rivière, C.-C. Fu, **K. Li**, G. Parette, N. Martin, *Phys. Rev. B 100, 224406 (2019)*

**Abstract:** We study the magnetic short range order (MSRO) in $Fe_{1-x}Cr_x$ ($0 \leq x \leq 0.15$) where an inversion of atomic short range order (ASRO) occurs at $x_C = 0.11(1)$. Our combination of neutron diffuse scattering and bulk magnetization measurements offers a comprehensive description of these local orders at a microscopic level. In the dilute alloys (x<0.04), the Cr atoms bear a large moment $\mu_{Cr} = -1.0(1)\mu_B$, antiparallel to the Fe ones ($\mu_{Fe}$). They fully repel their Cr first and second neighbors, and perturb the surrounding Fe moments. With increasing $x$, near neighbor Cr-Cr pairs start to appear and the Cr moment magnitude decreases, while $\mu_{\text{Fe}}$ shows a rounded maximum for $x_1 = 0.06(1) < x_{\text{C}}$. Above $x_{\text{C}}$, ASRO turns to local Cr segregation, thereby increasing magnetic frustration. First principles calculations reproduce the observed moment variations but overestimate the magnitude of the Cr moment. In order to reconcile theory with experiment quantitatively, we propose that the magnetic moments start canting locally, already above $x_1$. This picture actually anticipates the spin glasslike behavior of Cr-rich alloys. The whole study points out the subtle interplay of MSRO and ASRO, yielding an increasing frustration as $x$ increases, due to competing Fe-Cr and Cr-Cr interactions and Cr clustering tendency.

# C Résumé substantiel

Le magnétisme est un ingrédient indispensable pour prédire les propriétés du Fe et des alliages à base de Fe. Il joue un rôle crucial dans la stabilité relative des différentes phases de Fe [3, 4] et dans la transition de phase α-γ dans Fe [5–7]. Dans Fe cubique centré (cc), il est démontré que l'état magnétique a des impacts significatifs sur d'autres propriétés telles que l'énergie de formation de lacunes [8–18], coefficients de diffusion et énergies d'activation [13–15, 19–22]. Dans les alliages à base de Fe, le magnétisme peut avoir une forte interaction avec les ordres chimiques [23–29].

Dans ce travail, nous nous concentrons sur la thermodynamique, la formation de défauts et la diffusion atomique dans les alliages Fe-Ni. Le système Fe-Ni présente des propriétés magnétiques et mécaniques distinctes selon la composition chimique. Par exemple, le permalloy, avec environ 80% Ni, est utilisé dans le blindage magnétique grâce à sa haute perméabilité magnétique [30]. Un autre exemple bien connu est celui des alliages Invar Fe-Ni, qui possèdent des coefficients de dilatation thermique extrêmement faibles [31], et sont largement utilisés dans les outils de précision, les sources laser et les dispositifs sismographiques où une stabilité dimensionnelle élevée avec la température est requise [32]. Les alliages Fe-Ni cfc présentent respectivement une tendance antiferromagnétique et ferromagnétique dans le régime riche en Fe et riche en Ni, avec une dépendance de composition non linéaire des températures de Curie. Les alliages avec environ 50% et 75% Ni sont ferromagnétiques et chimiquement ordonnés à basse température. Avec l'augmentation de la température, les transitions chimiques et magnétiques successives se produisent dans une petite fenêtre de température, présentant donc un fort couplage magnétochimique. Expérimentalement, la plupart des mesures sur les propriétés thermodynamiques, de formation de lacunes et de diffusion ont été effectuées à des températures élevées, où les alliages sont déjà paramagnétiques et chimiquement désordonnés. Les effets des ordres magnétiques et chimiques sur la stabilité de phase et les propriétés des défauts restent largement inexplorés expérimentalement et théoriquement.

Plusieurs problèmes restent à résoudre dans la modélisation des alliages Fe-Ni. Le premier est l'absence de modèles décrivant bien la stabilité des phases magnétique et chimique sur toute la plage de composition. Le potentiel interatomique classique existant pour les applications de dynamique moléculaire surestime les températures de transition chimique expérimentales de plus de 1000 K [60]. En ce qui concerne les modèles d'interaction efficaces, ils n'étaient auparavant paramétrés que pour des régions de composition limitées (e.g. Fe et Ni cfc [54], les alliages Fe-Ni avec 70-80% Ni [24]). Récemment, de tels modèles ont été développés pour les alliages Fe-Ni cfc sur l'ensemble de la composition [56, 57], mais leurs prédictions de stabilité de phase sont en contradiction avec le diagramme de phase expérimental établi. Malgré les difficultés de paramétrage, des modèles magnétiques précis sont souhaitables car ils permettent d'étudier, par exemple, l'influence mutuelle entre les ordres magnétique et chimique, et l'importance relative des effets vibrationnels et magnétiques sur la transition de phase. De plus, les modèles d'alliages Fe-Ni sans défaut constituent la base de modèles plus sophistiqués qui incluent des défauts et d'autres éléments.

Le deuxième problème est le manque de compréhension des effets des fluctuations de spin longitudinales sur les propriétés des défauts. La plupart des efforts théoriques ont été consacrés à élucider les effets magnétiques sur les propriétés des défauts Fe cc [13–17, 22, 61], où les effets magnétiques dominants sont attribués aux fluctuations de spin transversales. On sait très peu de choses sur d'autres systèmes magnétiques plus itinérants (par exemple, Cr cc, Fe et Ni cfc) qui présentent d'importantes fluctuations de spin longitudinales à des températures finies [54]. De telles fluctuations au-dessous et au-dessus de la température de transition magnétique pourraient avoir des effets distincts de ceux des fluctuations de spin transversales.

Le troisième est lié au fait que même en l'absence de magnétisme, il n'y a pas de formalisme établi pour calculer les propriétés de formation de défauts dans les alliages avec de différéntes ordres chimiques. En effet, les approches existantes sont dédiées exclusivement soit à des structures ordonnées à basse température avec une quantité minimale d'antisites [62–70], soit à des alliages aléatoires à haute température avec un ordre chimique à courte portée en voie de disparition [71–73]. Un cadre unifié abordant les propriétés de formation de défauts en fonction de la température à travers les températures de transition chimique et magnétique, fait toujours défaut dans la littérature.

Enfin, il existe un manque de compréhension physique quant à la façon dont la formation de défauts et les propriétés de diffusion sont influencées par les ordres magnétiques et chimiques sous-jacents dans les alliages. En raison du manque de données expérimentales, il serait intéressant de démontrer les effets des transitions chimiques et magnétiques sur les propriétés des défauts, et de comparer l'importance relative des effets chimiques et magnétiques. On ne sait pas non plus, par exemple, comment les effets magnétiques sur les propriétés des défauts peuvent différer entre les alliages avec des ordres et des compositions chimiques distincts.

L'objectif de cette thèse est d'élucider les effets du magnétisme et de l'interaction magnétochimique sur la thermodynamique, la formation de défauts et la diffusion atomique. Nous abordons ces problèmes dans les alliages Fe-Ni cfc pour toute la gamme de composition, car (1) ils constituent les aciers austénitiques importants sur le plan technologique et d'autres alliages multi-composants, (2) ils subissent des transitions chimiques et magnétiques successives dans une petite fenêtre de température, donc potentiellement présentant un fort couplage magnétochimique, et (3) ils présentent des excitations de spin longitudinales importantes dont les effets sont souvent négligés et doivent être clarifiés.

Nous adoptons une stratégie de modélisation basée sur la DFT dans ce travail. Nous utilisons la méthode DFT pour calculer les propriétés énergétiques, magnétiques et vibrationnelles des systèmes avec diverses configurations magnétiques et chimiques. Les résultats DFT fournissent non seulement une compréhension fondamentale de diverses propriétés, mais également des entrées fiables pour paramétrer un modèle d'interaction efficace (EIM) pour les alliages Fe-Ni cfc sur toute la plage de composition. Des simulations Monte Carlo sur réseau combinées à l'EIM permettent de prédire les propriétés dépendantes de la température, les évolutions magnétiques et chimiques simultanées étant pleinement prises en compte.

La première partie de la thèse est centrée sur la thermodynamique des alliages Fe-Ni sans défaut, ce qui est indispensable pour la modélisation ultérieure de la formation et de la diffusion des défauts.

Nous effectuons des calculs DFT systématiques dans les structures Fe-Ni cc et cfc. Dans l'ensemble, la magnétisation et les entropies vibrationnelles calculées dans les alliages ferromagnétiques désordonnés sont en accord avec les données expérimentales disponibles, confirmant la précision de l'approche DFT. Sur la base des résultats DFT, nous calculons les énergies libres de mélange pour les phases ferromagnétiques et construisons les diagrammes de phases cc-cfc en dessous des températures de Curie, qui sont en bon accord avec les données expérimentales disponibles. Nous démontrons un rôle important de l'entropie vibrationnelle sur la prédiction des transitions de phases chimiques.

Pour inclure les effets des excitations magnétiques et aller au-delà des transitions magnétiques, nous paramétrons un EIM avec des variables magnétiques et chimiques explicites et le couplerons avec des simulations de Monte Carlo. Nous montrons que l'EIM reproduit avec succès non seulement les propriétés à basse température telles que prédites par DFT, mais aussi les températures expérimentales de transition ordre-désordre chimique et les températures de Curie.

D'après nos prédictions DFT et EIM, l'entropie vibrationnelle a un impact plus important sur les transitions ordre-désordre chimiques que l'entropie magnétique. D'un point de vue méthodologique, nos résultats suggèrent qu'il est possible et pratique de construire un diagramme de phase fiable en utilisant uniquement des calculs DFT, si la région étudiée est en dessous des transitions magnétiques.

L'EIM précis permet d'étudier l'interaction magnétochimique sur la stabilité de phase. Une séparation de phase entre les alliages désordonnés paramagnétiques et ferromagnétiques est trouvée dans la gamme de composition de 10-40% Ni à 600-700 K, qui est montrée comme étant entraînée magnétiquement. Nous montrons également que les températures de transition chimique diminuent avec l'augmentation du désordre magnétique. Inversement, les températures de transition magnétique diminuent avec l'augmentation des troubles chimiques à longue et courte portée.

De plus, nous étudions l'ajout d'un soluté de Cr ou de Mn dans les alliages Fe-Ni chimiquement ordonnés et désordonnés avec 50% et 75% Ni. Bien que l'alignement de spin du soluté dépende de la composition chimique locale, dans l'ensemble, le soluté de Cr a tendance à avoir un alignement de spin antiparallèle avec les autres atomes, alors que la tendance pour le soluté de Mn est à l'opposé. Dans la structure L1$_2$, les solutés de Mn et de Cr montrent une forte préférence pour le sous-réseau Fe par rapport au sous-réseau Ni, contrairement à leur préférence de sous-réseau dans la structure L1$_0$. Nous montrons que l'ajout de Cr (Mn) améliore (réduit) la stabilité énergétique de la structure L1$_2$, en accord avec la tendance expérimentale des températures de transition ordre-désordre chimique.

La deuxième partie du travail est axée sur les propriétés de défauts ponctuels dans les alliages Fe-Ni cfc. En particulier, nous développons des schémas de Monte Carlo efficaces pour obtenir une prédiction continue et cohérente des propriétés de formation de lacunes en fonction de la température, en particulier pour les alliages magnétiques concentrés.

L'étude comparative dans Fe cc, Fe et Ni cfc démontre que les effets magnétiques sur les propriétés de formation de lacunes sont très dépendants du système. Les orientations de spin ont une influence plus forte sur l'énergie de formation de lacunes que les amplitudes de spin dans Fe cc, alors que c'est l'inverse dans Fe cfc. L'énergie de formation de lacunes dans Ni cfc est peu sensible à l'état magnétique. La transition magnétique induit une diminution de l'énergie libre magnétique de formation de lacunes $G_f^{mag}$ en Fe cc, presque aucun changement en Fe cfc, et une augmentation en Ni cfc. Contrairement à Fe cc, les fluctuations de spin longitudinales importantes dans Fe et Ni cfc conduisent à une variation constante de $G_f^{mag}$ au-dessus des températures de transition magnétique.

Nous calculons $G_f^{mag}$ en fonction de la température dans les alliages à 50% et 75% Ni, révélant que le désordre magnétique a tendance à augmenter $G_f^{mag}$, tandis que le désordre chimique a un contraire effet. Pour toute la plage de température, la variation totale de $G_f^{mag}$ due au désordre magnétique est plus grande que la variation causée par le désordre chimique. La transition chimique s'avère avoir un effet plus notable que la transition magnétique sur $G_f^{mag}$. Nous étudions la dépendance à la concentration de $G_f^{mag}$, montrant que $G_f^{mag}$ dans les alliages désordonnés diminue avec l'augmentation de la teneur en Ni.

Enfin, nous étudions les propriétés SIA dans les Fe et Ni cfc. On montre que l'état magnétique dans Ni a un impact plus fort sur l'énergie de formation de SIA que l'énergie de formation de lacunes, tandis que les impacts sur les énergies de formation de SIA et de lacunes dans Fe sont tout aussi importants. Les états magnétiques sous-jacents ne modifient pas la stabilité relative entre les différentes configurations SIA. L'état magnétique a un effet important sur l'énergie de migration SIA dans le Fe cfc, tandis que l'effet dans Ni est faible. Nous montrons que le mécanisme de rotation-translation est le mécanisme le plus favorable pour la diffusion de SIA dans Fe, alors que les mécanismes de rotation-translation et de traduction pure devraient être considérés pour la diffusion de SIA dans Ni.

La troisième et dernière partie du travail est consacrée à la diffusion médiée par les lacunes dans les alliages Fe-Ni cfc.

Nous étudions la diffusion dans les Fe et Ni cfc. Les énergies moyennes de migration $E_m$ pour la diffusion de Ni dans les Fe et Ni cfc sont insensibles à l'état magnétique du système, montrant peu de variation avec la température. En revanche, $E_m$ pour la diffusion de Fe dans les Fe et Ni paramagnétiques sont réduits jusqu'à 0,10 et 0,25 eV, respectivement, par rapport aux valeurs magnétiques de l'état fondamental. De plus, les fluctuations de spin longitudinales de Fe et Ni cfc contribuent également à la diminution de $E_m$ pour la diffusion de Fe dans le régime paramagnétique.

Les résultats ci-dessus des propriétés de diffusion dans Fe et Ni peuvent être comparés à nos résultats sur la formation de lacunes dans Fe et Ni. Ils révèlent systématiquement que la transition magnétique devrait avoir un effet important sur une propriété qui dépend fortement des arrangements d'orientation de spin. Une variation constante avec la température dans le régime paramagnétique peut être attendue en raison des fluctuations de spin longitudinales, si la propriété donnée est sensible à la variation des magnitudes de spin.

Nous examinons la dépendance à la température des propriétés de diffusion dans l'alliage avec 75% Ni. Les coefficients de diffusion du traceur n'obéissent à la loi d'Arrhenius que dans l'alliage aléatoire paramagnétique. Dans la phase ordonnée ferromagnétique, le comportement non-Arrhenius présente une forte diminution de l'énergie libre d'activation de diffusion $G_a$ avec la température, qui est principalement causée par les changements d'ordre magnétique et chimique. Cette diminution de $G_a$ avec la température est principalement attribuée à la diminution de l'énergie moyenne de migration, et aussi dans une moindre mesure aux variations de $G_f^{mag}$ et aux effets de corrélation cinétique. L'énergie d'activation de la diffusion est significativement réduite à travers la température de transition chimique, mais peu influencée à travers la température de Curie.

Nous étudions également la dépendance à la concentration des propriétés de diffusion dans les alliages. Dans l'ensemble, les énergies moyennes de migration des atomes Fe et Ni ont tendance à augmenter avec l'augmentation de la teneur en Ni dans les alliages désordonnés. Pour une composition donnée, les atomes de Fe ont une énergie moyenne de migration et une fréquence de tentative plus faibles que les atomes de Ni. Cet effet compensateur de l'énergie de migration et de la fréquence de tentative conduit à des coefficients de diffusion de traceur similaires de Fe et Ni dans un alliage aléatoire. De plus, nous trouvons que la diffusion Fe montre une corrélation cinétique plus forte dans les alliages concentrés, alors que la diffusion Ni montre une corrélation plus forte dans les alliages très riches en Fe. Les coefficients de diffusion du traceur calculés sont en accord avec les données mesurées disponibles dans l'incertitude expérimentale.

Dans cette thèse, nous étudions les effets magnétochimiques sur les propriétés thermodynamiques, de formation de défauts et de diffusion dans les alliages Fe-Ni sur toute la gamme de composition. Basée uniquement sur les résultats DFT, directement ou mappés sur les EIM, et sans utiliser d'entrées empiriques ou expérimentales, notre approche permet d'obtenir une prédiction précise et cohérente qui est globalement en bon accord avec les données expérimentales disponibles. Cela démontre la précision et le pouvoir prédictif de la présente approche de modélisation, qui peut être appliquée à l'étude des effets magnétiques sur diverses propriétés et processus cinétiques dans d'autres alliages magnétiques.

Malgré nos efforts dans cette thèse, la compréhension des effets du magnétisme et de son interaction avec d'autres degrés de liberté sur diverses propriétés est loin d'être complète. Quelques perspectives de ce travail sont proposées comme suit.

Premièrement, il serait simple d'étendre le présent EIM pour étudier la diffusion des impuretés (par exemple, Mn, Cr, Co, Cu et C). Un bon point de départ peut être de prédire la diffusion des impuretés dans les Fe et Ni cfc. Ensuite, on peut également étudier l'interaction magnétochimique sur la diffusion des impuretés, par exemple, dans l'alliage Fe-Ni avec 75% Ni à travers les transitions magnétiques et chimiques.

Cette thèse est principalement axée sur les propriétés de formation de lacunes et de diffusion, alors que les effets magnétiques sur les propriétés SIA ne sont limités qu'au Fe et Ni purs et manquent de prédiction quantitative. Afin de mieux comprendre les processus cinétiques dans des conditions de non-équilibre telles que la ségrégation induite par l'irradiation, il serait intéressant d'inclure également les SIA dans le présent EIM et d'avoir une compréhension plus complète des effets magnétiques.

Comme suggéré par notre étude DFT, les ajouts de Mn et Cr peuvent avoir des effets significatifs sur les ordres magnétiques et chimiques dans les alliages Fe-Ni. Il est donc très intéressant d'obtenir une compréhension quantitative des effets sur la stabilité de phase, en étendant le présent EIM à d'autres alliages ternaires tels que Fe-Ni-Cr et Fe-Ni-Mn. De tels modèles pourraient ensuite être étendus pour l'étude d'autres propriétés (par exemple celles étudiées dans cette thèse) dans les systèmes ternaires et éventuellement vers des alliages multi-composants tels que les systèmes Fe-Ni-Cr-Mn.

Enfin, le présent traitement de l'entropie vibrationnelle est implicite. Cela équivaut à intégrer le degré de liberté vibrationnel rapide dans les termes d'interaction non magnétique dans l'EIM. Nous adoptons ce traitement principalement pour sa simplicité, alors que cela peut ne pas être bien justifié étant donné que les échelles de temps des excitations des magnon et des phonons ne sont pas très différentes. Par conséquent, une tâche passionnante mais très difficile serait d'inclure des variables vibrationnelles explicites dans l'EIM et d'explorer les effets potentiels de couplage magnon-phonon.